\documentclass[%
 reprint,
superscriptaddress,
 amsmath,amssymb,
 aps,
 prmaterials,
floatfix,
]{revtex4-2}
\usepackage{graphicx}
\usepackage{amsmath}
\graphicspath{ {images/} }
\usepackage{dcolumn}
\usepackage{bm}
\usepackage{xcolor}
\usepackage[normalem]{ulem}
\usepackage{xr}
\usepackage{booktabs}
\usepackage{makecell}

\makeatletter
\newcommand*{\addFileDependency}[1]{
  \typeout{(#1)}
  \@addtofilelist{#1}
  \IfFileExists{#1}{}{\typeout{No file #1.}}
}
\makeatother

\begin{document}


\title{Thermal expansion mechanisms, heat transport, and band-gap renormalization in strongly anharmonic Rb$_2$SO$_4$, Rb$_2$Ca$_2$(SO$_4$)$_3$, and Ba$_3$Lu(BO$_3$)$_3$ from machine-learned interatomic potentials}


\author{Ekaterina Spirande}
\altaffiliation{These authors contributed equally to this work.}
\affiliation{Skolkovo Institute of Science and Technology, Bolshoi bulvar 30, build.1, 121205, Moscow, Russia}
\affiliation{Digital Materials LLC, Bolshoi bulvar 30, build.1, 121205, Moscow, Russia}

\author{Rustam Arabov}
\altaffiliation{These authors contributed equally to this work.}
\affiliation{Skolkovo Institute of Science and Technology, Bolshoi bulvar 30, build.1, 121205, Moscow, Russia}

\author{Sofya Demina}
\affiliation{Institute of Silicate Chemistry, Makarova Emb. 2, Saint Petersburg, 199034, Russia}

\author{Nikita Rybin}
\affiliation{Moscow Engineering Physics Institute, Kashirskoe Highway 31, 115409, Moscow, Russia}
\affiliation{Skolkovo Institute of Science and Technology, Bolshoi bulvar 30, build.1, 121205, Moscow, Russia}
\affiliation{Digital Materials LLC, Bolshoi bulvar 30, build.1, 121205, Moscow, Russia}
\email{n.rybin@skoltech.ru}

\author{Alexander Shapeev}
\affiliation{Skolkovo Institute of Science and Technology, Bolshoi bulvar 30, build.1, 121205, Moscow, Russia}
\affiliation{Digital Materials LLC, Bolshoi bulvar 30, build.1, 121205, Moscow, Russia}
\email{a.shapeev@skoltech.ru}

\date{\today}

\begin{abstract}

Rb$_2$SO$_4$, Rb$_2$Ca$_2$(SO$_4$)$_3$, and Ba$_3$Lu(BO$_3$)$_3$ are structurally complex materials whose thermal expansion has been interpreted, on the basis of high-temperature x-ray diffraction, in terms of specific polyhedral rearrangements. We study their lattice dynamics and its consequences for thermal and electronic properties using molecular dynamics with actively learned moment tensor potentials trained on density-functional theory data. The anharmonicity measure $\sigma^A$ (compares the anharmonic contribution to the interatomic forces with the total forces acting on atoms) reaches $\approx$0.3--0.5 for $\beta$-Rb$_2$SO$_4$, Rb$_2$Ca$_2$(SO$_4$)$_3$, and Ba$_3$Lu(BO$_3$)$_3$ and $\approx$0.75 for orientationally disordered $\alpha$-Rb$_2$SO$_4$, and the imaginary harmonic modes of Rb$_2$Ca$_2$(SO$_4$)$_3$ and $\alpha$-Rb$_2$SO$_4$ are stabilized at finite temperature. The simulations reproduce the measured thermal expansion and support the mechanisms proposed from diffraction: expansion along $c$ in $\beta$-Rb$_2$SO$_4$ arises from both elongation and straightening of corrugated Rb chains, BO$_3$ reorientation accompanies the anisotropic expansion of Ba$_3$Lu(BO$_3$)$_3$, and bending of polyhedral linkages limits expansion along the weakly expanding directions and in cubic Rb$_2$Ca$_2$(SO$_4$)$_3$. Green--Kubo lattice thermal conductivities differ by more than an order of magnitude between compounds with comparable $\sigma^A$, showing that the anharmonicity measure alone does not determine heat transport. Phonon-induced band-gap renormalization, including lattice thermal expansion, reaches 2.7 and 1.7~eV at 1250~K in Rb$_2$Ca$_2$(SO$_4$)$_3$ and Ba$_3$Lu(BO$_3$)$_3$ and 1.4~eV at 650~K in $\beta$-Rb$_2$SO$_4$, with zero-point renormalizations of 0.64, 0.70, and 0.31~eV for $\beta$-Rb$_2$SO$_4$, Rb$_2$Ca$_2$(SO$_4$)$_3$, and Ba$_3$Lu(BO$_3$)$_3$, respectively.

\end{abstract}

\maketitle

\section{Introduction}
\begin{figure}[h!]
    \begin{minipage}[h]{1\linewidth}
        \center{\includegraphics[trim={0cm 0cm 0cm 0cm},clip,width=1\linewidth]{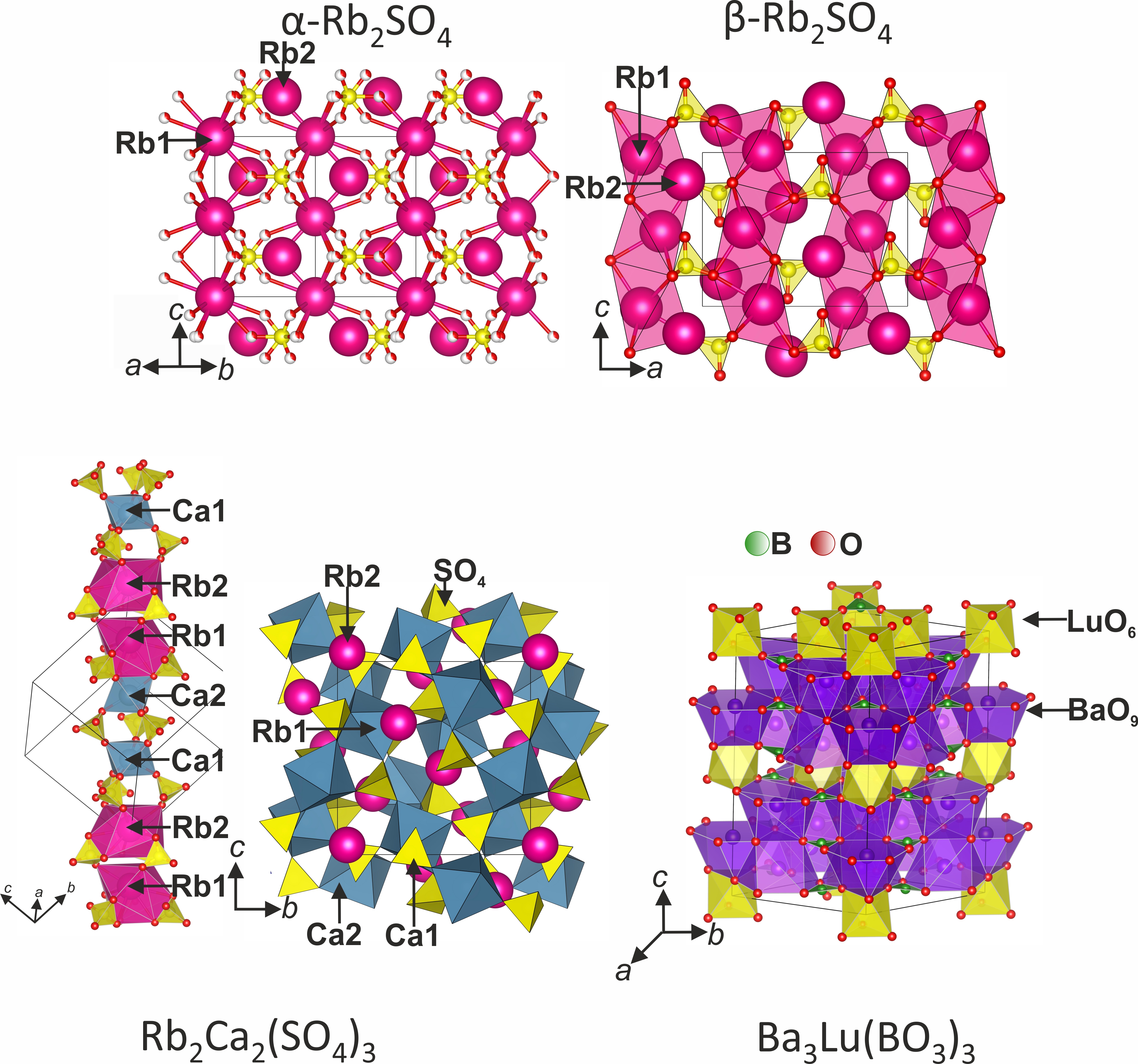}}
    \end{minipage}

    \caption{Crystal structures of $\alpha$-Rb$_2$SO$_4$, $\beta$-Rb$_2$SO$_4$, Rb$_2$Ca$_2$(SO$_4$)$_3$, and Ba$_3$Lu(BO$_3$)$_3$. In Ba$_3$Lu(BO$_3$)$_3$, Ba--O polyhedra are shown in purple, LuO$_6$ octahedra in yellow, and boron atoms of the BO$_3$ groups as small green spheres. In the sulfates, Rb atoms and Rb--O polyhedra are shown in magenta, CaO$_6$ octahedra in blue-gray, and SO$_4$ tetrahedra in yellow. Oxygen atoms are shown in red, and black lines mark the unit cells.}
    \label{fig:structures}
\end{figure}

Borate and sulfate crystals built from rigid anionic groups linked by metal--oxygen polyhedra are an important class of optical materials, in particular for nonlinear optics in the ultraviolet range~\cite{halasyamani2018must}. The polar borate Ba$_3$Lu(BO$_3$)$_3$ and the noncentrosymmetric langbeinite-type sulfate Rb$_2$Ca$_2$(SO$_4$)$_3$ belong to this class and have been studied as potential nonlinear optical materials~\cite{ba3lu_nlo, rb2ca2_nlo}. Rb$_2$SO$_4$ is a related arcanite-type sulfate that transforms at 936~K from orthorhombic $\beta$-Rb$_2$SO$_4$ to hexagonal $\alpha$-Rb$_2$SO$_4$ with orientationally disordered SO$_4$ tetrahedra~\cite{shablinskii2025structural}, which makes it a model system for anharmonic dynamics of molecular anions. The crystal structures of these compounds are shown in Fig.~\ref{fig:structures}; their unit cells contain between 14 ($\alpha$-Rb$_2$SO$_4$) and 96 (Ba$_3$Lu(BO$_3$)$_3$) atoms. For optical crystals operating at high power or elevated temperatures, the anisotropy of thermal expansion, the ability to dissipate heat, and the temperature dependence of the band gap, which sets the fundamental absorption edge, are key characteristics.

The thermal expansion of all three compounds has recently been measured by high-temperature powder x-ray diffraction~\cite{biryukov2023thermal,shablinskii2025structural}. Rb$_2$SO$_4$ and Ba$_3$Lu(BO$_3$)$_3$ expand strongly anisotropically, with the largest expansion along $c$, whereas cubic Rb$_2$Ca$_2$(SO$_4$)$_3$ expands isotropically. Following the crystal-chemical approach to thermal expansion~\cite{bubnova2013high}, these studies attributed the expansion to specific structural rearrangements: straightening of corrugated chains of Rb atoms accompanied by rotation of SO$_4$ tetrahedra in Rb$_2$SO$_4$, bending of polyhedral linkages and rocking of SO$_4$ tetrahedra in Rb$_2$Ca$_2$(SO$_4$)$_3$, and reorientation of BO$_3$ triangles combined with anisotropic deformation of BaO$_9$ polyhedra in Ba$_3$Lu(BO$_3$)$_3$. Because diffraction yields only time- and space-averaged structures, such mechanisms are inferred from changes in average atomic positions and displacement parameters and are difficult to quantify. Molecular dynamics simulations, in contrast, give access to instantaneous atomic configurations and allow the contributions of individual structural mechanisms to the lattice expansion to be separated directly.

Lattice dynamics is most commonly described within the harmonic or quasiharmonic approximation using force constants from density-functional theory (DFT)~\cite{baroni2001phonons, ld_theor}, but this description fails when the potential-energy surface is strongly anharmonic~\cite{hellman2011lattice, anh_ld_theor}, as expected for crystals in which rigid anionic groups reorient and for phases that are dynamically unstable in the harmonic approximation. The degree of anharmonicity can be quantified by the anharmonicity measure $\sigma^A(T)$, which compares the anharmonic contribution to the interatomic forces with the total forces~\cite{knoop2020anharmonicity, knoop2023anharmonicity}. Finite-temperature properties of such materials must instead be obtained by sampling the actual atomic motion, for example, with the temperature-dependent effective potential (TDEP) method and its stochastic variant (sTDEP) for phonons~\cite{hellman2011lattice, hellman2013temperature, stdep_paper}, the Green--Kubo formalism for the lattice thermal conductivity (LTC)~\cite{green1954markoff, kubo1957statistical, carbogno2017ab}, and configurational averaging for the phonon-induced band-gap renormalization (BGR)~\cite{giustino2017electron, zacharias2016one, monserrat2018electron, karsai2018electron}. Such sampling is prohibitively expensive with \textit{ab initio} molecular dynamics (MD) for unit cells of up to 96 atoms, but becomes feasible with machine-learned interatomic potentials (MLIPs), which reproduce DFT energies and forces at a small fraction of the cost and, when trained with active learning, remain reliable for strongly anharmonic materials~\cite{PhysRevMaterials.9.063801}. In particular, moment tensor potentials (MTPs)~\cite{mtp, mlip} trained with D-optimality-based active learning~\cite{podryabinkin2017active, gubaev2019accelerating} have been applied to lattice dynamics~\cite{ladygin2020lattice}, lattice thermal conductivity~\cite{mortazavi2020machine, rybin2024moment, arabov2026thermal, arabov2026tuning}, phonon-induced BGR~\cite{arabov2026thermal, arabov2026tuning}, and quantum nuclear effects~\cite{solovykh2025path}.

In this work, we train MTPs for $\beta$-Rb$_2$SO$_4$, $\alpha$-Rb$_2$SO$_4$, Rb$_2$Ca$_2$(SO$_4$)$_3$, and Ba$_3$Lu(BO$_3$)$_3$ and use them to study the anharmonic lattice dynamics, lattice thermal expansion, LTC, and BGR of these compounds. We first quantify the anharmonicity with $\sigma^A(T)$ and the sTDEP phonon dispersions. We then compare the calculated thermal expansion with the diffraction data and quantify the structural mechanisms proposed in Refs.~\cite{biryukov2023thermal,shablinskii2025structural} by decomposing the expansion of chains of connected atoms into contributions from changing interatomic distances and changing directions of the connections. Finally, we compute the LTC with the Green--Kubo method and obtain the BGR from MD and sTDEP sampling, including the effects of lattice thermal expansion and zero-point motion.

\section{Computational Methods}\label{sec:methods}

\subsection{Machine-learned interatomic potentials}\label{sec:active_learning}

Interatomic interactions were described by moment tensor potentials (MTPs)~\cite{mtp} of level 16 with a cutoff radius of 5.5~\AA{} (5.0~\AA{} for Ba$_3$Lu(BO$_3$)$_3$) as implemented in the MLIP package~\cite{mlip}; separate potentials were trained for $\beta$-Rb$_2$SO$_4$, $\alpha$-Rb$_2$SO$_4$, Rb$_2$Ca$_2$(SO$_4$)$_3$, and Ba$_3$Lu(BO$_3$)$_3$. Reference energies, forces, and stresses were computed with VASP~\cite{VASP19961} using the projector augmented-wave (PAW) method~\cite{PAW, gga}, the PBE functional~\cite{pbe} with the D3 dispersion correction~\cite{grimme2010consistent}, a plane-wave cutoff of 520~eV, and generalized Monkhorst--Pack $k$-point grids with a minimum real-space distance of 21.7~\AA~\cite{wisesa2016efficient, wang2021rapid}. Initial training sets were generated from short \textit{ab initio} MD (AIMD) simulations in the $NPT$ ensemble, starting from the DFT-relaxed experimental structures of $\beta$-Rb$_2$SO$_4$~\cite{weber1989determination}, Rb$_2$Ca$_2$(SO$_4$)$_3$~\cite{boujelben2007langbeinite}, Ba$_3$Lu(BO$_3$)$_3$~\cite{ilyukhin1993crystal}, and $\alpha$-Rb$_2$SO$_4$~\cite{shablinskii2025structural} (Table~S3). Because the experimental structure of $\alpha$-Rb$_2$SO$_4$ describes orientationally disordered SO$_4$ tetrahedra by partially occupied oxygen sites, we constructed several fully ordered supercells by assigning to each SO$_4$ group one of the orientations of the apex- and edge-oriented models proposed experimentally~\cite{shablinskii2025structural} and used them as starting configurations.

The training sets were then extended by active learning based on the D-optimality criterion~\cite{podryabinkin2017active, gubaev2019accelerating}: during $NPT$ MD with the current MTP, configurations with extrapolation grade $\gamma > 2$ were collected (and the simulation was stopped if $\gamma > 10$), the most informative of them were selected with the MaxVol algorithm~\cite{goreinov2010submatrix}, computed with DFT, and added to the training set, and the MTP was refitted until no configurations with $\gamma > 2$ appeared. The final training sets contain 1246--1846 configurations (Table~S2), and the root-mean-square errors are 0.8--1.3~meV/atom in energy (4.2~meV/atom for $\alpha$-Rb$_2$SO$_4$) and 0.05--0.09~eV/\AA\ in forces (Fig.~S1).

\subsection{Anharmonic lattice dynamics}\label{sec:ld_methods}

Harmonic phonons were computed with the finite-displacement method~\cite{supercell-method} as implemented in phonopy~\cite{phonopy}, using displacements of 0.03~\AA. Finite-temperature effective phonons were obtained with the TDEP method~\cite{hellman2011lattice, hellman2013temperature}, which describes the system by the effective harmonic Hamiltonian
\begin{equation}
\hat{H}=U_0+\sum_i\frac{\mathbf p_i^2}{2m_i}
+\frac{1}{2}\sum_{i,j}\sum_{\alpha,\beta}
\Phi_{\alpha\beta}^{ij}u_{i\alpha}u_{j\beta}
\label{eq:tdep_hamilt}
\end{equation}
and determines the force constants $\Phi_{\alpha\beta}^{ij}$ by minimizing
\begin{equation}
\Delta F=\frac{1}{N_t}\sum_{t=1}^{N_t}\left\|\mathbf F_t+\bar{\Phi}\,\mathbf U_t\right\|_2^2
\label{eq:delta_F_tdep}
\end{equation}
over $N_t$ sampled configurations, where $\mathbf F_t$ and $\mathbf U_t$ are the vectors of all force and displacement components, $\bar{\Phi}$ is the force-constant matrix, and $u_{i\alpha}$ are displacements from the relaxed atomic positions in the cell expanded to temperature $T$ according to the MD thermal expansion (Sec.~\ref{sec:lte_methods}).

We used the stochastic self-consistent variant of TDEP (sTDEP)~\cite{stdep_paper}: starting from force constants fitted to an $NVT$ MD trajectory, supercells were generated with atoms displaced along the normal modes $s$ with Gaussian amplitudes of variance
\begin{equation}
A_{is}^2=\frac{\hbar\left(n_s+1/2\right)}{m_i\,\omega_s}\;\xrightarrow{\;k_{\mathrm B}T\gg\hbar\omega_s\;}\;\frac{k_{\mathrm B}T}{m_i\,\omega_s^2},
\label{eq:amplitudes}
\end{equation}
where $\omega_s$ is the frequency of mode $s$ and $k_{\mathrm B}$ is the Boltzmann constant. The MTP forces of these supercells were added to the fitting set, and the force constants were refitted until the phonon density of states converged (see the Supplementary Material). Classical sampling was used to analyze anharmonicity and phonon renormalization, whereas quantum sampling with Bose--Einstein occupations $n_s$ was used for the BGR (Sec.~\ref{sec:bgr_methods}). Only second-order force constants were fitted, using for each compound the largest pair cutoff compatible with the supercell (7.8--10.6~\AA; Table~S4).

The anharmonicity measure~\cite{knoop2020anharmonicity}
\begin{equation}
\sigma^A(T) = \sqrt{\frac{\sum_{i,\alpha}\langle(F^A_{i\alpha})^2\rangle_T}{\sum_{i,\alpha}\langle F_{i\alpha}^2\rangle_T}}
\label{eq:anh-measure}
\end{equation}
was evaluated from $NVT$ MD trajectories in the thermally expanded cells, where $F_{i\alpha}$ are the MTP forces and $F^A_{i\alpha}=F_{i\alpha}+\sum_{j\beta}\Phi_{\alpha\beta}^{ij}u_{j\beta}$ are their anharmonic parts with respect to the converged classical sTDEP force constants. Reference values for Si and SrTiO$_3$ are taken from Ref.~\cite{knoop2020anharmonicity}, where $\sigma^A$ was evaluated with respect to 0~K harmonic force constants. Because effective force constants absorb part of the anharmonicity, our definition typically yields smaller $\sigma^A$, so the comparison with these references is conservative.

\subsection{Thermal expansion and structural analysis}\label{sec:lte_methods}

Lattice thermal expansion was obtained from $NPT$ MD simulations performed with LAMMPS~\cite{LAMMPS}, using Langevin temperature control and an anisotropic barostat for Rb$_2$SO$_4$ and Ba$_3$Lu(BO$_3$)$_3$ and an isotropic barostat for cubic Rb$_2$Ca$_2$(SO$_4$)$_3$; the $\alpha$ and $\beta$ phases of Rb$_2$SO$_4$ were simulated with their respective MTPs. For each temperature, several supercell sizes were simulated with eight independent trajectories per size (see the Supplementary Material for details).

The lattice parameters, the unit-cell volume, and all structural quantities discussed below were fitted as functions of temperature $T$ and inverse number of atoms $N^{-1}$ by Gaussian process regression (GPR)~\cite{gp_rasmussen} with the kernel~\cite{spirande2025automated,miryashkin2023bayesian}
\begin{equation}
\begin{split}
k\left((T_1,N_1), (T_2,N_2)\right)
={}&\theta_0^2+\theta_1^2
\exp\!\left[-\frac{(T_1-T_2)^2}{2\theta_2^2}\right]\\
&\times\exp\!\left[-\frac{1}{2\theta_3^2}
\left(\frac{1}{N_1}-\frac{1}{N_2}\right)^2\right],
\end{split}
\label{eq:gp_kernel}
\end{equation}
where $\theta_0,\ldots,\theta_3$ are hyperparameters obtained by maximizing the log marginal likelihood. The dependence on $N^{-1}$ allows extrapolation to the thermodynamic limit $N\rightarrow\infty$, and all quantities reported below refer to this limit. The linear and volumetric thermal expansion coefficients, $\alpha_i=L_i^{-1}\,dL_i/dT$ and $\alpha_V=V^{-1}\,dV/dT$, were obtained from the analytic temperature derivatives of the GPR models; the experimental lattice parameters and volumes~\cite{biryukov2023thermal,shablinskii2025structural} were fitted in the same way with a temperature-only kernel that includes a temperature-dependent noise term (see the Supplementary Material).

To test the mechanisms proposed from diffraction, we followed interatomic distances, bond angles, and the orientations of the BO$_3$ and SO$_4$ groups along the MD trajectories. Following Ref.~\cite{shablinskii2025structural}, we call a chain of connected atoms spanning a lattice period a rod; writing its end-to-end length as $D=\sum_i \ell_i u_i$, where $\ell_i$ is the length of segment $i$ and $u_i$ is the cosine of its angle to the rod axis, the thermal change
\begin{equation}
\Delta D=\sum_i \Delta\ell_i\,\bar u_i+\sum_i \bar\ell_i\,\Delta u_i
\label{eq:rod_decomposition}
\end{equation}
separates exactly into a contribution from changing segment lengths and one from changing segment directions (rod straightening or bending), where bars denote averages of the initial and final values (details in the Supplementary Material).

\subsection{Lattice thermal conductivity}\label{sec:ltc_methods}

We computed the LTC with the equilibrium Green--Kubo method~\cite{green1954markoff,kubo1957statistical}. Within classical statistics, it includes anharmonicity to all orders and does not assume well-defined phonon quasiparticles, which matters for the orientationally disordered $\alpha$-Rb$_2$SO$_4$. The thermal conductivity tensor is
\begin{equation}
\kappa_{\alpha\beta}=\frac{V}{k_{\mathrm B}T^2}\int_0^{\infty}\left\langle J_{\alpha}(0)J_{\beta}(t)\right\rangle dt,
\label{eq:green_kubo}
\end{equation}
where $V$ is the cell volume, $T$ is the temperature, and $J_{\alpha}$ is a Cartesian component of the heat flux. The heat flux was computed from the per-atom energies and many-body per-atom virials of the MTP following Ref.~\cite{lammps-mlip-gk}; the explicit expression is given in the Supplementary Material. For each compound and temperature, the cell was fixed to the average $NPT$ cell, and 25--45 independent $NVE$ trajectories of 200--800~ps were run in supercells of 840--2052 atoms (Table~S5). The running integrals of the diagonal components of the heat-flux autocorrelation function were averaged over a plateau window (Table~S5), and the uncertainties are standard errors over trajectories. We report the three diagonal components for orthorhombic $\beta$-Rb$_2$SO$_4$, their average for cubic Rb$_2$Ca$_2$(SO$_4$)$_3$, and the in-plane average $(\kappa_{xx}+\kappa_{yy})/2$ together with $\kappa_{zz}$ for hexagonal $\alpha$-Rb$_2$SO$_4$ and Ba$_3$Lu(BO$_3$)$_3$.

\subsection{Band-gap renormalization}\label{sec:bgr_methods}

We computed the BGR of $\beta$-Rb$_2$SO$_4$, Rb$_2$Ca$_2$(SO$_4$)$_3$, and Ba$_3$Lu(BO$_3$)$_3$ by averaging DFT band gaps over configurations sampled at finite temperature~\cite{zacharias2016one,karsai2018electron}. $\alpha$-Rb$_2$SO$_4$ was excluded because its orientational disorder would additionally require averaging over disorder configurations. The BGR is defined as ${\rm BGR}(T)=\langle E_g\rangle_T-E_g^0$, where $E_g$ is the fundamental gap obtained from the electronic density of states, $\langle\cdot\rangle_T$ denotes the average over configurations, and $E_g^0$ is the gap of the static 0~K structure. ${\rm BGR^{MD}}$ is obtained from snapshots of classical $NVT$ MD with the MTPs in supercells with the 0~K lattice parameters, at 250--1250~K (250--650~K for $\beta$-Rb$_2$SO$_4$). ${\rm BGR^{MD+LTE}}={\rm BGR^{MD}}+[E_g^0(a(T))-E_g^0]$ additionally includes the lattice thermal expansion (LTE), i.e., the gap change of the static cell with the MD lattice parameters $a(T)$ (Sec.~\ref{sec:lte_methods}). ${\rm BGR^{q}}$ includes zero-point motion and Bose--Einstein statistics: its configurations are generated by sTDEP sampling with the quantum amplitudes of Eq.~(\ref{eq:amplitudes}) in the thermally expanded cells (Sec.~\ref{sec:ld_methods}) at 0--1050~K (0--650~K for $\beta$-Rb$_2$SO$_4$), and its value at 0~K is the zero-point renormalization (ZPR). Band gaps were computed with FHI-aims~\cite{fhiaims} and the PBE functional~\cite{pbe} using \textit{light} settings, at the $\Gamma$ point for the supercells; the supercells, numbers of configurations, and convergence tests are given in the Supplementary Material.

\section{Results}

\subsection{Anharmonicity and phonon renormalization}\label{sec:ld}

\begin{figure}[htbp]
\centering
\includegraphics[width=\linewidth]{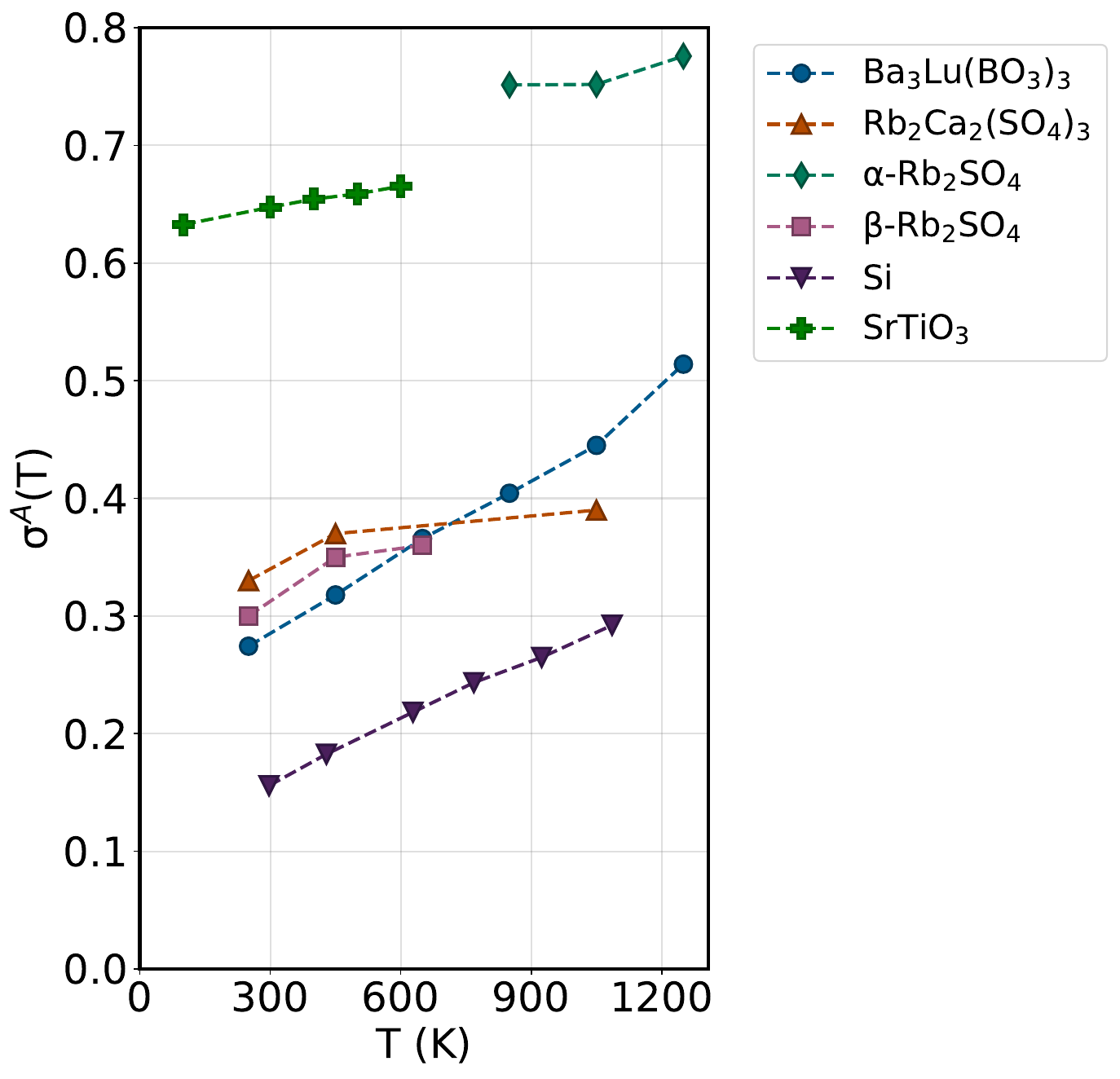}
\caption{Anharmonicity measure $\sigma^{A}(T)$ of $\alpha$- and $\beta$-Rb$_2$SO$_4$, Rb$_2$Ca$_2$(SO$_4$)$_3$, and Ba$_3$Lu(BO$_3$)$_3$ from MD with the MTPs, compared with Si (harmonic case) and SrTiO$_3$ (anharmonic case) from Ref.~\cite{knoop2020anharmonicity}, which were evaluated with respect to 0~K harmonic force constants.}
\label{fig:anharmonicity_measure}
\end{figure}

Figure~\ref{fig:anharmonicity_measure} shows the anharmonicity measure $\sigma^A(T)$ of the four phases together with reference values for Si and SrTiO$_3$~\cite{knoop2020anharmonicity}. All four phases are more anharmonic than Si ($\sigma^A\approx0.16$--$0.29$ at 300--1100~K), a material well described by the harmonic approximation. For $\beta$-Rb$_2$SO$_4$ and Rb$_2$Ca$_2$(SO$_4$)$_3$, $\sigma^A\approx0.30$--$0.39$ and depends only weakly on temperature, whereas for Ba$_3$Lu(BO$_3$)$_3$ it increases steadily from 0.27 at 250~K to 0.51 at 1250~K, in parallel with the growing rocking of the BO$_3$ triangles (Sec.~\ref{sec:lte}). $\alpha$-Rb$_2$SO$_4$ has $\sigma^A\approx0.75$--$0.78$ at 850--1250~K, above the values of 0.63--0.67 reported for the strongly anharmonic perovskite SrTiO$_3$. We attribute this to the reorientation of the SO$_4$ tetrahedra in the orientationally disordered $\alpha$ phase (Sec.~\ref{sec:lte}), a motion that no harmonic model can describe. Since our $\sigma^A$ is defined with respect to effective rather than 0~K harmonic force constants (Sec.~\ref{sec:ld_methods}), these comparisons are conservative. Anharmonic forces are therefore substantial in all four phases, and the harmonic approximation is insufficient to describe their finite-temperature lattice dynamics.

\begin{figure*}[t!]
\centering
\includegraphics[width=0.7\linewidth]{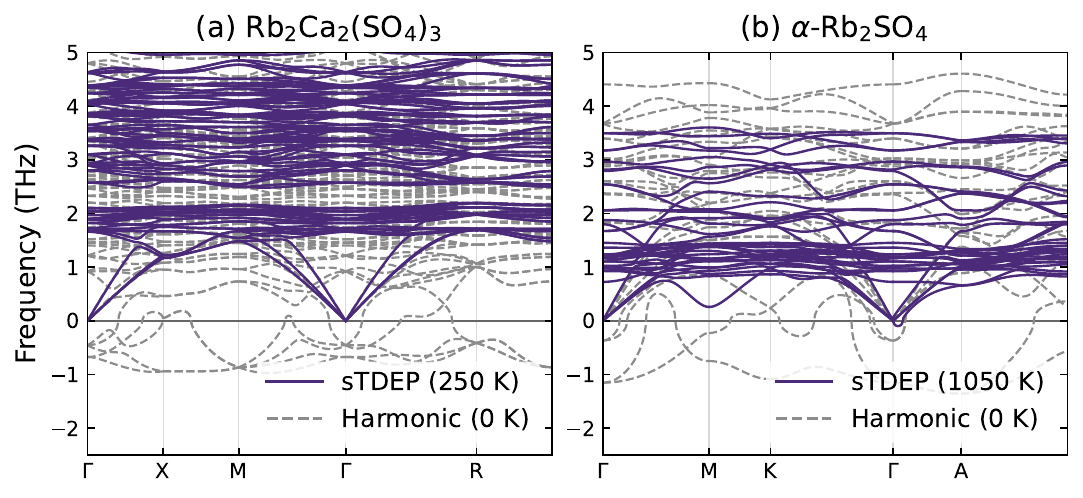}
\caption{Low-frequency phonon dispersions of (a) Rb$_2$Ca$_2$(SO$_4$)$_3$ and (b) $\alpha$-Rb$_2$SO$_4$ in the harmonic approximation (dashed) and from classical sTDEP at 250 and 1050~K, respectively (solid).}
\label{fig:phonon_renormalization_all}
\end{figure*}

Figure~\ref{fig:phonon_renormalization_all} illustrates how anharmonicity changes the phonon spectra. The harmonic dispersions of Rb$_2$Ca$_2$(SO$_4$)$_3$ and $\alpha$-Rb$_2$SO$_4$ contain imaginary modes, i.e., these experimentally observed phases are dynamically unstable in the harmonic approximation. In the sTDEP dispersions at 250 and 1050~K, respectively, the low-frequency modes are strongly renormalized and the imaginary branches are shifted to positive frequencies; only a small imaginary pocket of about 0.1~THz remains near $\Gamma$ along $\Gamma$--A in $\alpha$-Rb$_2$SO$_4$. The harmonic spectra of Ba$_3$Lu(BO$_3$)$_3$ and $\beta$-Rb$_2$SO$_4$ contain no imaginary modes, but their low-frequency modes are also renormalized at 250~K (Fig.~S5). Thus, as indicated by $\sigma^A$, the harmonic approximation fails for all four phases, not only for those with harmonic instabilities.

\subsection{Lattice thermal expansion}\label{sec:lte}

\begin{figure*}[p!]
\centering
\begin{tabular}{@{}c@{}}
\includegraphics[width=0.84\textwidth]{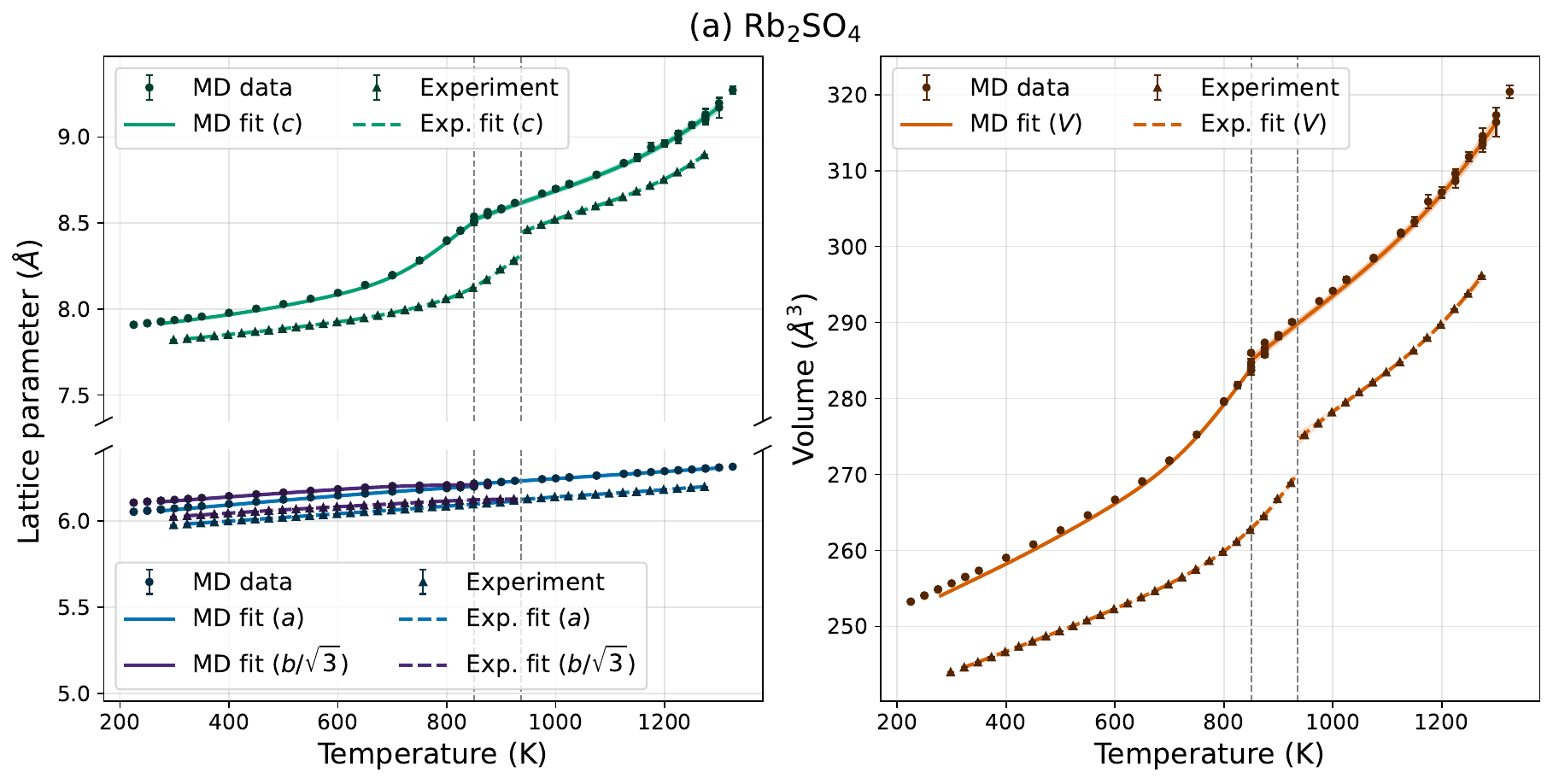} \\[0.5ex]
\includegraphics[width=0.84\textwidth]{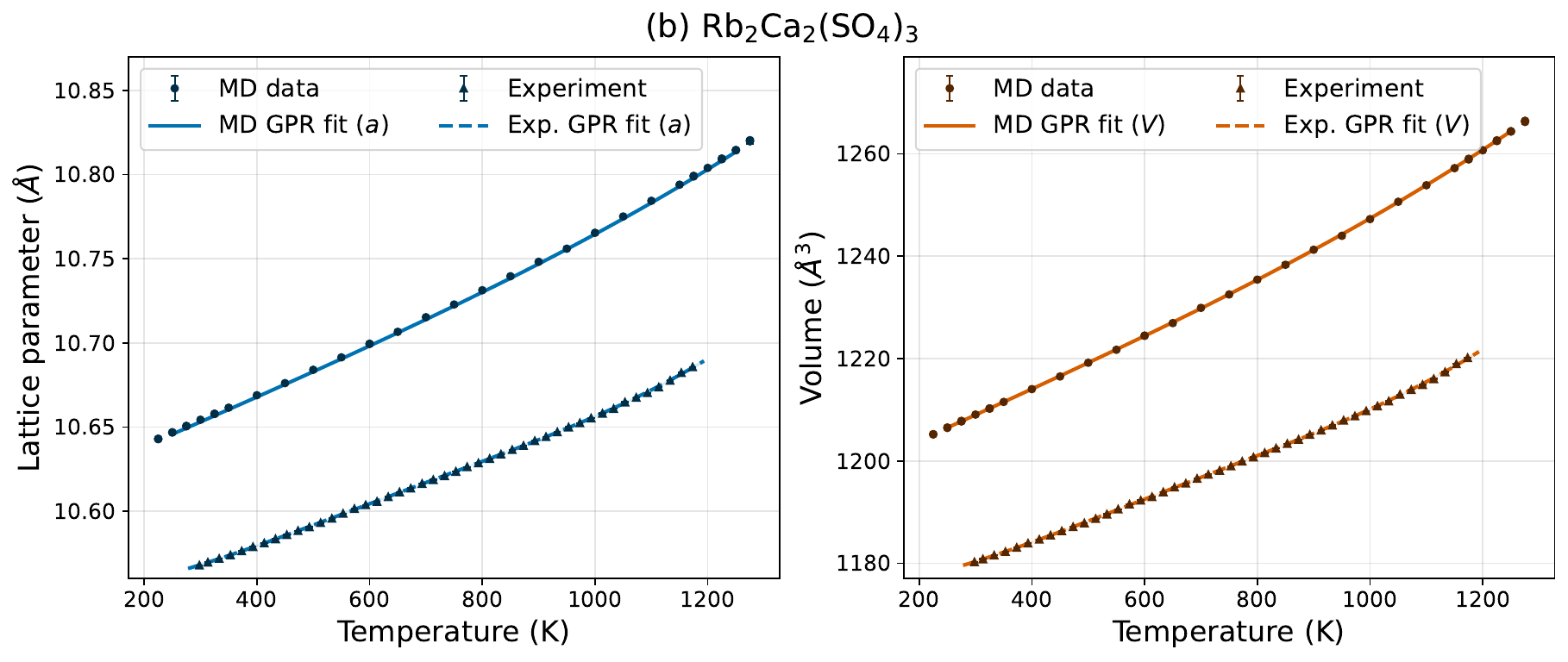} \\[0.5ex]
\includegraphics[width=0.84\textwidth]{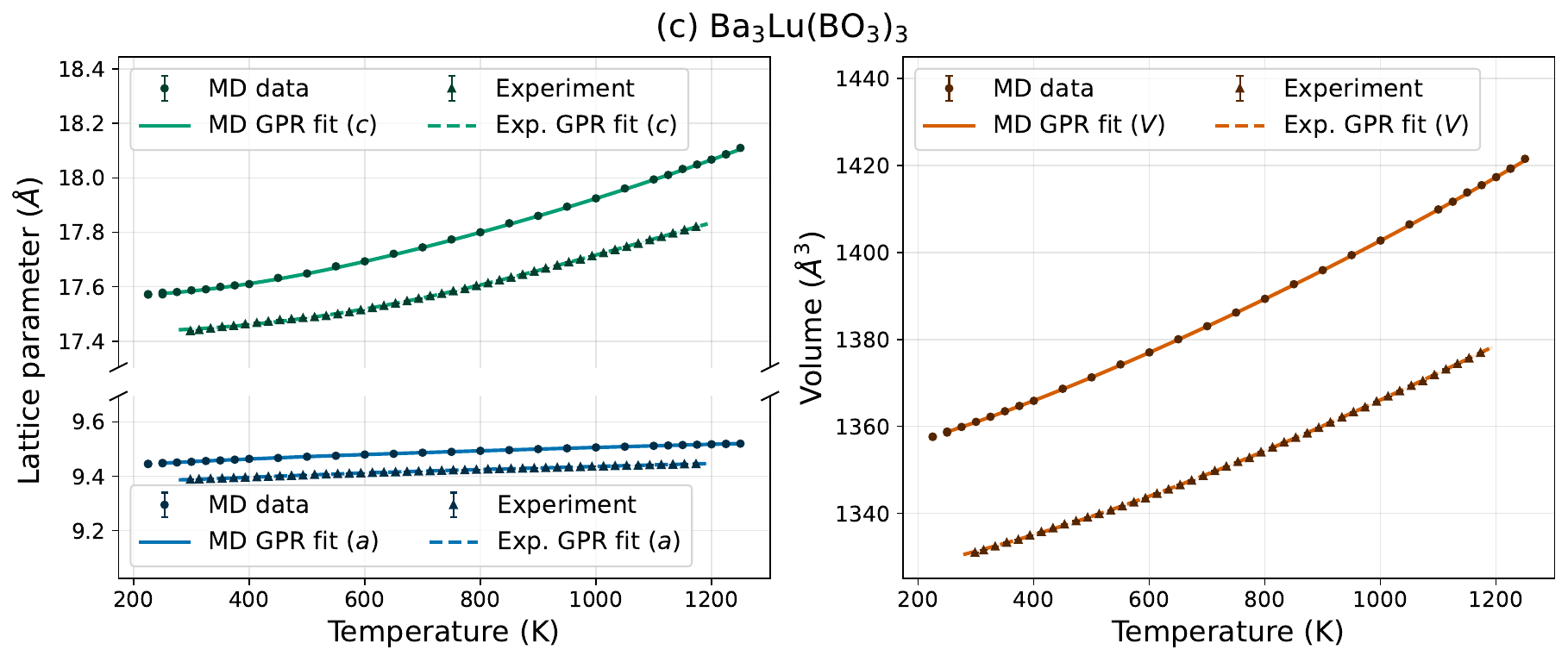}
\end{tabular}
\caption{Temperature dependence of the lattice parameters (left) and unit-cell volumes (right) of (a) Rb$_2$SO$_4$, (b) Rb$_2$Ca$_2$(SO$_4$)$_3$, and (c) Ba$_3$Lu(BO$_3$)$_3$. Symbols show MD results and high-temperature X-ray diffraction data from Refs.~\cite{shablinskii2025structural,biryukov2023thermal}. Solid and dashed curves show GPR fits to the MD and experimental data, respectively. Shaded regions show $2\sigma$ confidence intervals, which are mostly too narrow to be visible for the lattice parameters at this scale. In (a), $b_\beta/\sqrt{3}$ is shown alongside $a_\beta$ and $a_\alpha$, and the volumes $V_\beta/2$ and $V_\alpha$ both correspond to two formula units. Vertical dashed lines mark the chosen boundary between the analyzed temperature ranges (850~K) and the experimental transition temperature (936~K).}
\label{fig:lattice_parameters_all}
\end{figure*}

The MD simulations reproduce the main features of the thermal expansion observed by high-temperature powder x-ray diffraction~\cite{biryukov2023thermal,shablinskii2025structural}: strongly anisotropic expansion of Rb$_2$SO$_4$ and Ba$_3$Lu(BO$_3$)$_3$ and positive expansion of cubic Rb$_2$Ca$_2$(SO$_4$)$_3$ (Fig.~\ref{fig:lattice_parameters_all}). The calculated lattice parameters are systematically larger than the experimental ones. This offset is inherited from the reference method: the 0~K PBE+D3 volumes already exceed the room-temperature experimental volumes by 0.8--1.5\% (Table~S1). Because we focus on high temperatures, we use classical MD without quantum corrections to the equilibrium cell, which may contribute to the discrepancies in the expansion coefficients at the lower end of the studied temperature range. The expansion coefficients were obtained from GPR fits to both the MD and the experimental data, whereas the experimental studies used polynomial fits. For each compound, we first compare the calculated and experimental expansion, then test the mechanisms proposed from diffraction, and finally quantify them with the rod decomposition of Eq.~(\ref{eq:rod_decomposition}).

\subsubsection{Rb$_2$SO$_4$}

\begin{figure}[htbp]
\centering
\includegraphics[width=\columnwidth]{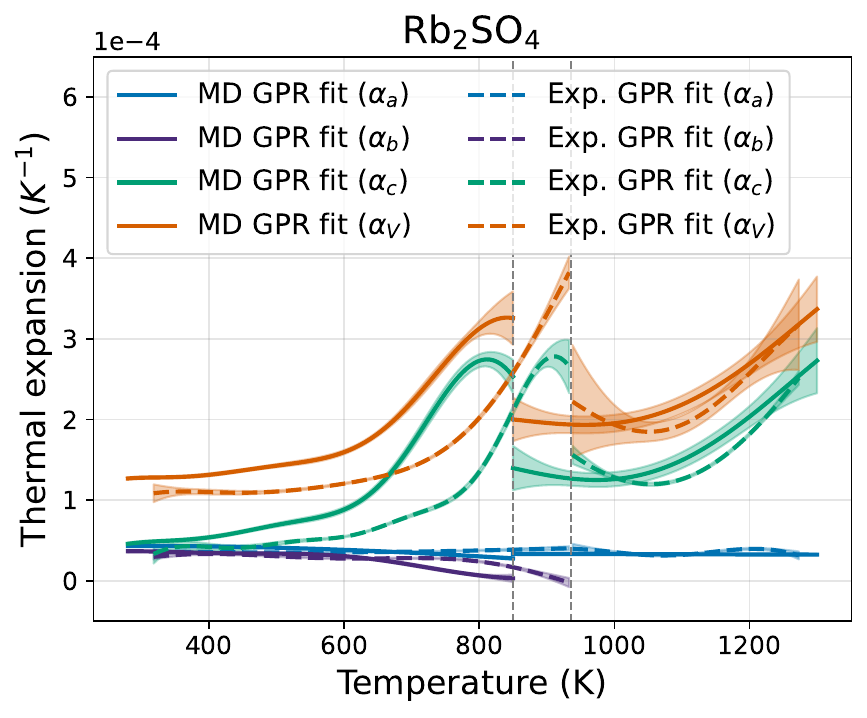}
\caption{Linear and volumetric thermal expansion coefficients of $\beta$- and $\alpha$-Rb$_2$SO$_4$ from GPR fits to MD (solid) and experimental (dashed) data~\cite{shablinskii2025structural}. Vertical lines mark the chosen boundary between the analyzed temperature ranges (850~K) and the experimental transition temperature (936~K). Shaded regions indicate $2\sigma$ confidence intervals from the GPR fits.}
\label{fig:rb2so4_alpha}
\end{figure}

The $\beta$ and $\alpha$ phases were simulated with separate MTPs, and we take 850~K as the boundary between the analyzed temperature ranges; the experimental transition occurs at 936~K~\cite{shablinskii2025structural}. In both phases, Rb$_2$SO$_4$ expands much more strongly along $c$ than along $a$ and $b$ (Fig.~\ref{fig:lattice_parameters_all}(a)), and this anisotropy grows with temperature within each phase (Fig.~\ref{fig:rb2so4_alpha}).

\begin{figure}[h!]
    \centering
    \includegraphics[width=\linewidth]{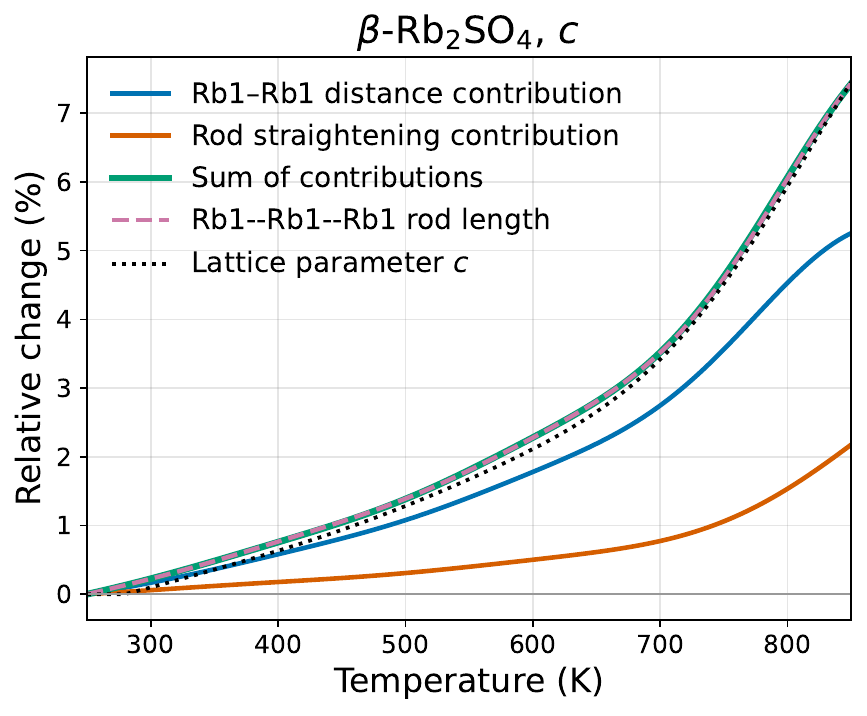}
    \caption{Decomposition of thermal expansion along $c$ in
    $\beta$-Rb$_2$SO$_4$ into contributions from increasing
    Rb1--Rb1 distances and rod straightening. Their sum is compared
    with the relative changes in the rod length and lattice
    parameter $c$, with 250~K as the reference temperature.}
    \label{fig:rb2so4_row_end}
\end{figure}

The experimental study attributes this anisotropy to the straightening of corrugated Rb1--Rb1--Rb1 rods along $c$ (Fig.~\ref{fig:structures}), associated with rotation of the SO$_4$ tetrahedra~\cite{shablinskii2025structural}. In the $\beta$ phase, the mean Rb1--Rb1--Rb1 angle increases from $151.6^\circ$ at 250~K to $163.9^\circ$ at 850~K. The decomposition (Fig.~\ref{fig:rb2so4_row_end}) shows that increasing Rb1--Rb1 distances contribute 5.3\% to the rod-length increase and rod straightening 2.2\%; straightening thus accounts for about 30\% of the expansion along $c$, supporting and refining the experimental picture. The straightening is accompanied by an increasing inclination of the S--O bond closest to $c$, consistent with the proposed SO$_4$ rotation (Fig.~S8). In the $\alpha$ phase, the Rb1--Rb1--Rb1 angle remains close to $170^\circ$, and the expansion along $c$ is almost entirely due to increasing Rb1--Rb1 distances (Fig.~S6). Here the SO$_4$ tetrahedra reorient over time, and the oxygen atoms occupy several regions around each S atom, consistent with the orientational disorder reported experimentally.

\begin{figure}[h!]
    \centering
    \includegraphics[width=\linewidth]{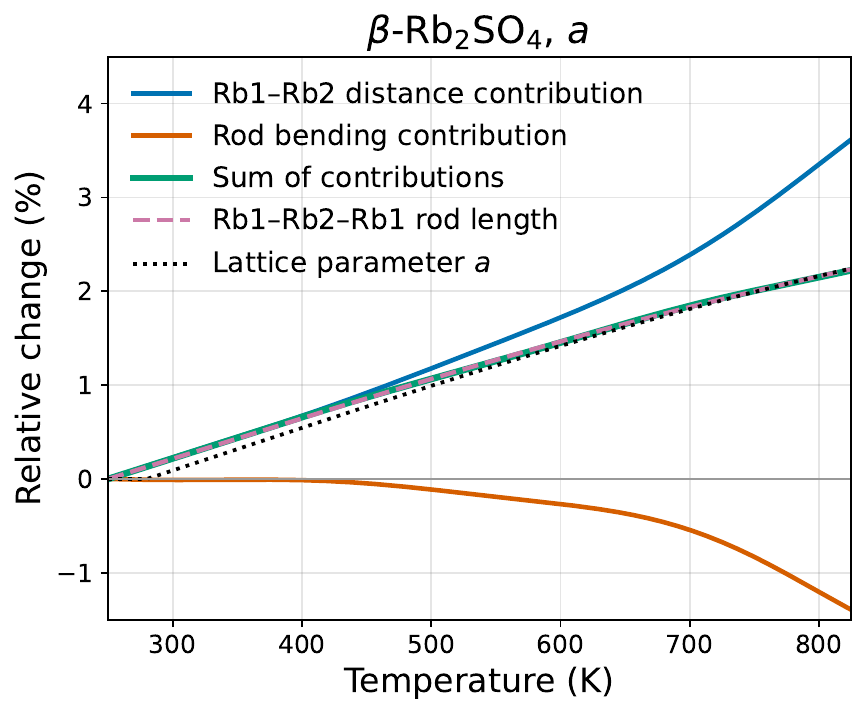}
    \caption{Decomposition of thermal expansion along
    $a$ in $\beta$-Rb$_2$SO$_4$ into contributions
    from increasing Rb1--Rb2 distances and rod bending.
    Their sum is compared with the relative changes in the
    Rb1--Rb2--Rb1 rod length and lattice parameter $a$, with 250~K as the
    reference temperature.}
    \label{fig:rb2so4_ab_beta}
\end{figure}

Along $a$ and $b$, we analyze rods connecting translationally equivalent Rb1 atoms through Rb2, Rb1--Rb2--Rb1 along $a$ and Rb1--Rb2--Rb1--Rb2--Rb1 along $b$ in the $\beta$ phase. Increasing Rb1--Rb2 distances lengthen these rods, whereas rod bending partly compensates this expansion (Fig.~\ref{fig:rb2so4_ab_beta}; Fig.~S7). Along $b$, the compensation grows on heating, and the rod length and lattice parameter approach a plateau. Straightening along $c$ and bending along $a$ and $b$ thus together produce the strong anisotropy of Rb$_2$SO$_4$.

\subsubsection{Rb$_2$Ca$_2$(SO$_4$)$_3$}

\begin{figure}[htbp]
\centering
\includegraphics[width=\columnwidth]{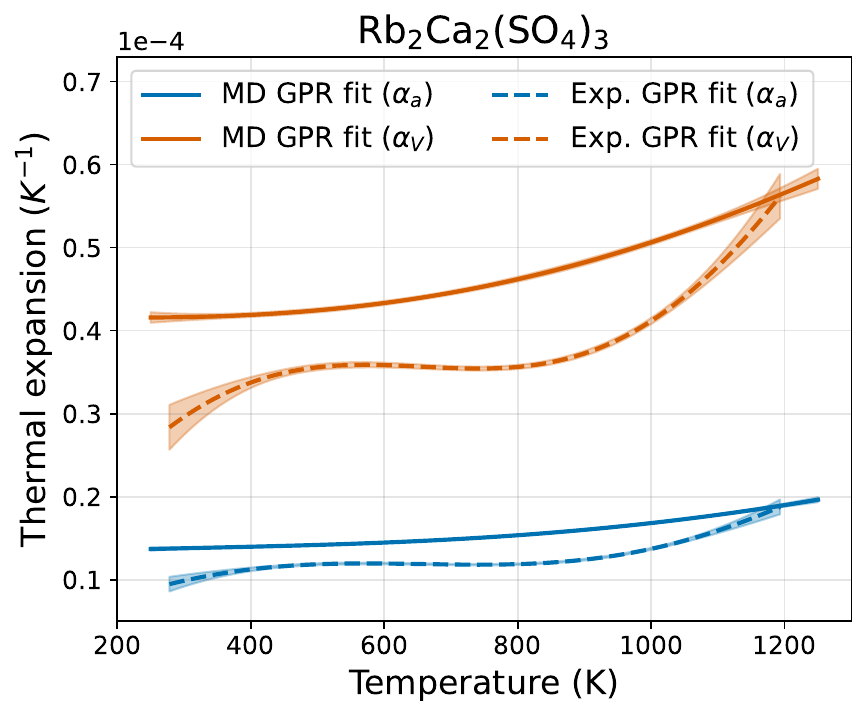}
\caption{Linear and volumetric thermal expansion coefficients of Rb$_2$Ca$_2$(SO$_4$)$_3$ from GPR fits to MD (solid) and experimental (dashed) data~\cite{shablinskii2025structural}. Shaded regions indicate $2\sigma$ confidence intervals from the GPR fits.}
\label{fig:rb2ca2_alpha}
\end{figure}

The lattice parameter and volume of cubic Rb$_2$Ca$_2$(SO$_4$)$_3$ increase throughout the studied temperature range (Fig.~\ref{fig:lattice_parameters_all}(b)), and both the calculated and experimental expansion coefficients remain positive (Fig.~\ref{fig:rb2ca2_alpha}). The calculated $\alpha_a$ exceeds the experimental one at low temperatures, but the two approach each other at high temperatures.

The experimental study describes the structure in terms of $M$(SO$_4$)$_6$ microblocks, each centered on a Ca or Rb atom and surrounded by six SO$_4$ tetrahedra~\cite{shablinskii2025structural} (Fig.~\ref{fig:structures}). Neighboring microblocks share three tetrahedra and form $L_3$ rods along the body diagonals of the cubic cell, with the repeat Rb1--Ca2--Ca1--Rb2--Rb1. Two competing contributions to the rod expansion were proposed: elongation of the $M$--O bonds, and angular motion of the connected polyhedra, i.e., hinge deformation at the shared oxygen atoms (changes in the $M$--O--S angles) and rocking of the SO$_4$ tetrahedra.

The MD trajectories show all three effects. Between 250 and 1250~K, the Ca--O bonds elongate while the mean Ca--O--S angles decrease by about $5^\circ$ at both Ca sites. This bending reduces the Ca--S separation and partly offsets the bond elongation, so the Ca-centered microblocks expand less than the CaO$_6$ polyhedra (Figs.~S9 and S11). The RMS rocking of the SO$_4$ tetrahedra grows from about $12.5^\circ$ to $23^\circ$, whereas the mean S--O distance increases by less than 1\%.

\begin{figure}[h!]
\centering
\includegraphics[width=\linewidth]{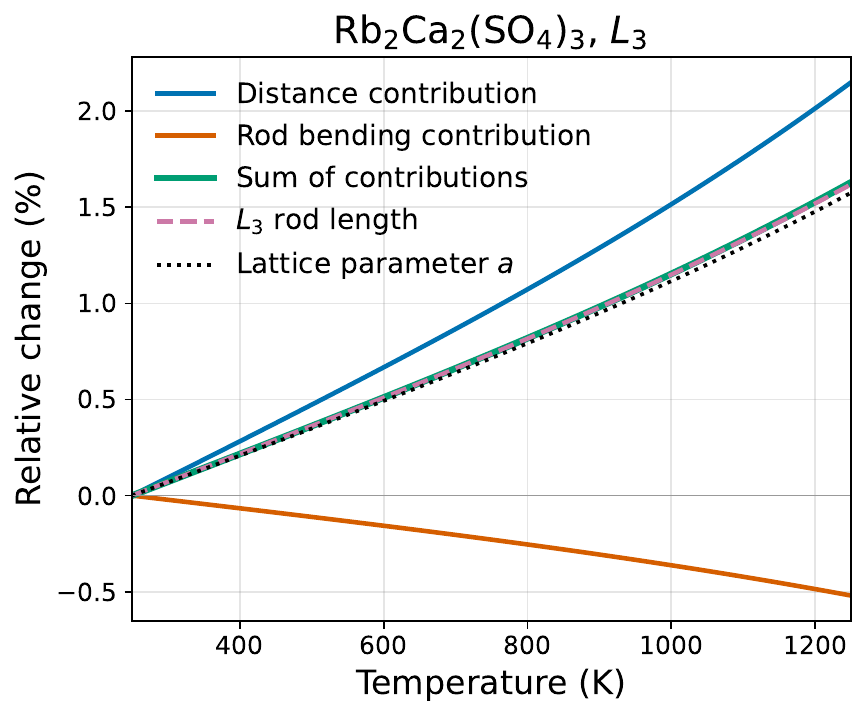}
\caption{Relative change in the $L_3$ rod length, compared with the lattice parameter $a$ and the sum of the contributions from changing distances between neighboring centers and rod bending, relative to 250~K. Curves show GPR estimates in the thermodynamic limit using all available system sizes.}
\label{fig:rb2ca2_l3}
\end{figure}

The rod decomposition quantifies this balance (Fig.~\ref{fig:rb2ca2_l3}). At 1250~K relative to 250~K, increasing distances between neighboring microblock centers contribute $+2.15$ percentage points and changes in the link directions $-0.52$ points. Their sum reproduces the 1.62\% increase in the $L_3$ rod length, which is close to the relative change in $a$. The MD thus confirms the local mechanisms proposed experimentally and shows that the net expansion is driven by increasing distances between microblock centers and partly limited by angular changes.

\subsubsection{Ba$_3$Lu(BO$_3$)$_3$}

\begin{figure}[htbp]
\centering
\includegraphics[width=\columnwidth]{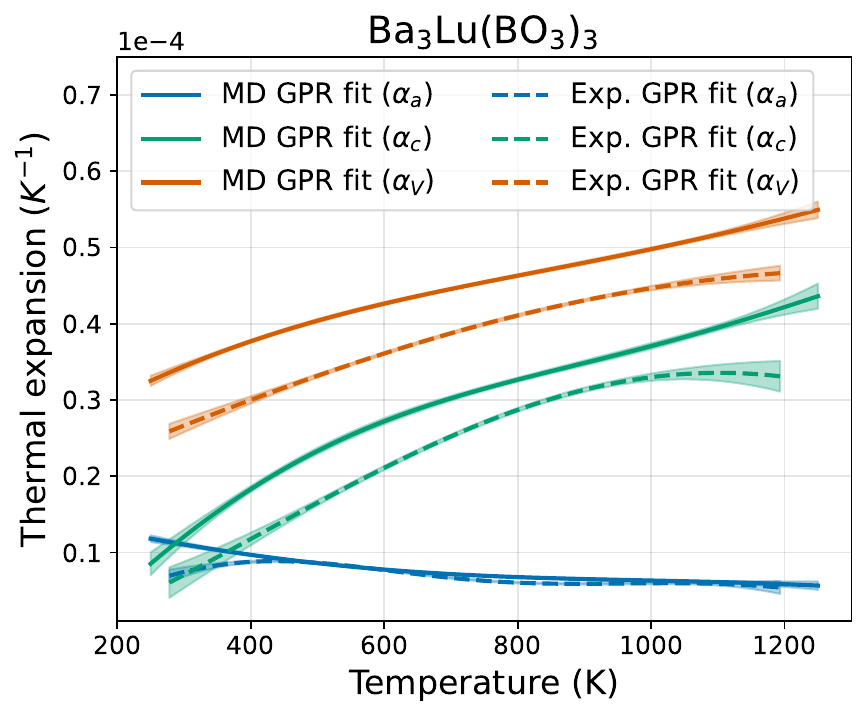}
\caption{Linear and volumetric thermal expansion coefficients of Ba$_3$Lu(BO$_3$)$_3$ from GPR fits to MD (solid) and experimental (dashed) data~\cite{biryukov2023thermal}. Shaded regions indicate $2\sigma$ confidence intervals from the GPR fits.}
\label{fig:thermal_expansion}
\end{figure}

Ba$_3$Lu(BO$_3$)$_3$ expands more strongly along $c$ than within the $ab$ plane, and the anisotropy grows on heating~\cite{biryukov2023thermal} (Fig.~\ref{fig:lattice_parameters_all}(c)): in the MD, $\alpha_a$ decreases and $\alpha_c$ increases with temperature (Fig.~\ref{fig:thermal_expansion}). The calculated $\alpha_a$ follows the experimental trend closely, whereas $\alpha_c$ is higher than experiment at high temperatures.

The experimental study related this behavior to the layered structure of LuO$_6$ and BaO$_9$ polyhedra connected by BO$_3$ triangles (Fig.~\ref{fig:structures})~\cite{biryukov2023thermal}. It proposed preferential atomic motion perpendicular to the BO$_3$ planes and a decreasing inclination of the B(1)O$_3$ and B(3)O$_3$ triangles, which, together with elongation of the Ba--O bonds along $c$, would expand BaO$_9$ along $c$ and contract it within $ab$.

The MD trajectories support this picture. The BO$_3$ triangles behave as nearly rigid units: between 250 and 1250~K, their RMS rocking grows from about 5--$6^\circ$ to $13^\circ$, whereas the mean B--O distance increases by only about 1\% (Fig.~S12). Atomic displacements perpendicular to the triangle planes are about four times larger than the in-plane displacements per direction. The mean inclinations of B(1)O$_3$ and B(3)O$_3$, which share oxygen vertices with the LuO$_6$ octahedra, decrease to a minimum at about 600--700~K and then increase slightly; this minimum lies close to the slight bend in the experimental $c(T)$ near 573~K~\cite{biryukov2023thermal}. In contrast, the B(2)O$_3$ triangles, located between the Ba--O polyhedral layers, become slightly more inclined, and at high temperatures the three groups approach similar inclinations and rocking amplitudes.

The BaO$_9$ polyhedra expand much more strongly along $c$ than within $ab$: between 250 and 1250~K, their directional RMS sizes increase by about 6--9\% along $c$ and only 0.3--1.5\% within $ab$, depending on the Ba site (Fig.~S13). The LuO$_6$ octahedra show the opposite anisotropy, with increases of about 0.6\% along $c$ and 2.8\% within $ab$ (Fig.~S14), possibly because they are coupled to the reorienting B(1)O$_3$ and B(3)O$_3$ triangles through shared oxygen atoms (see the Supplementary Material).

\begin{figure}[htbp]
\centering
\includegraphics[width=\linewidth]{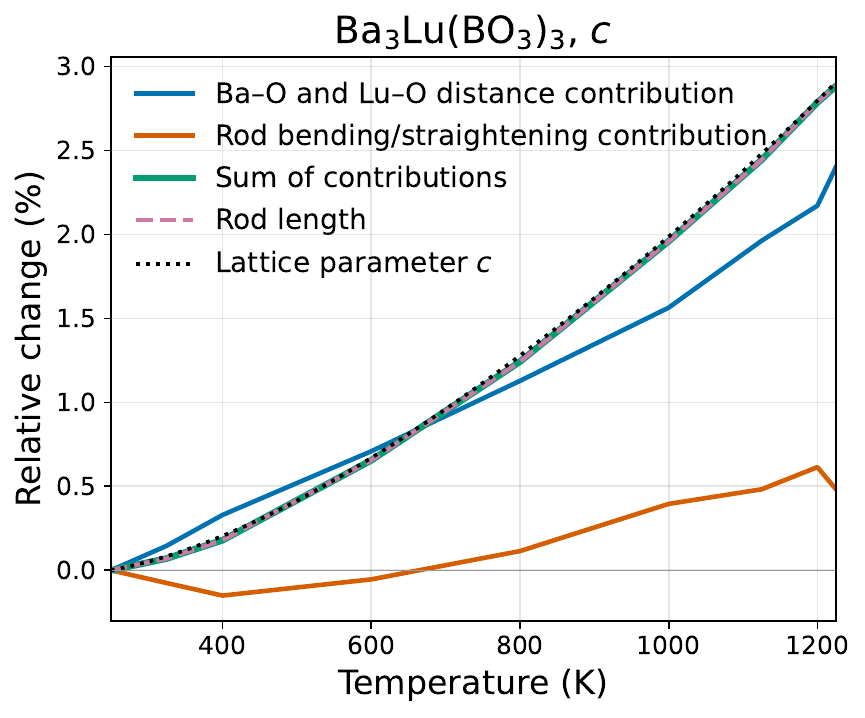}\par\vspace{1ex}
\includegraphics[width=\linewidth]{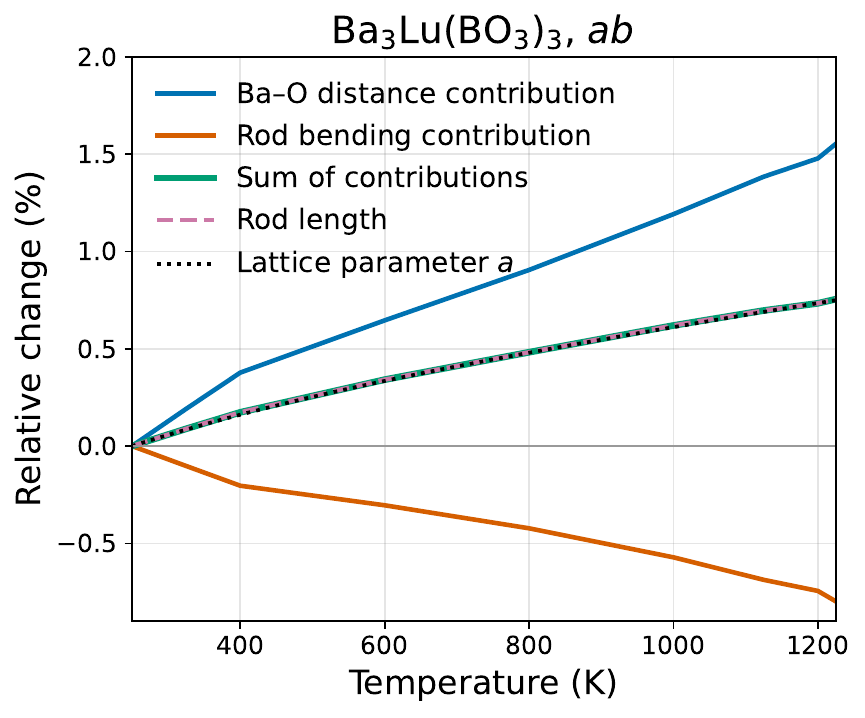}
\caption{Geometric contributions to thermal expansion in Ba$_3$Lu(BO$_3$)$_3$: upper panel, along $c$ for a periodic rod through Ba--O and Lu--O contacts; lower panel, within $ab$ for Ba(1)--O--Ba(1)--O--Ba(1) rods, averaged over $a$ and $b$. Curves show GPR estimates in the thermodynamic limit using all three supercell sizes. The sums of the contributions are compared with the relative changes in rod lengths and corresponding lattice parameters, all relative to 250~K.}
\label{fig:ba3lu_path_contributions}
\end{figure}

To connect these local changes with the lattice expansion, we decompose periodic rods connecting translationally equivalent Ba(1) atoms through Ba--O contacts within $ab$ and through Ba--O and Lu--O contacts along $c$ (Fig.~\ref{fig:ba3lu_path_contributions}). Along $c$, increasing segment lengths dominate the expansion, and directional changes enhance it further at high temperatures, consistent with the strong growth of BaO$_9$ along this axis. Within $ab$, directional changes partly offset the increase in Ba--O segment lengths, consistent with the small in-plane response of BaO$_9$. Together with the BO$_3$ reorientation and out-of-plane motion, these results support the experimental picture: expansion is favored along $c$, while competing geometric contributions limit it within $ab$.

\subsection{Lattice thermal conductivity}\label{sec:ltc}

\begin{figure}[htbp]
\centering
\begin{minipage}[t]{0.49\linewidth}
\centering
\includegraphics[width=\linewidth]{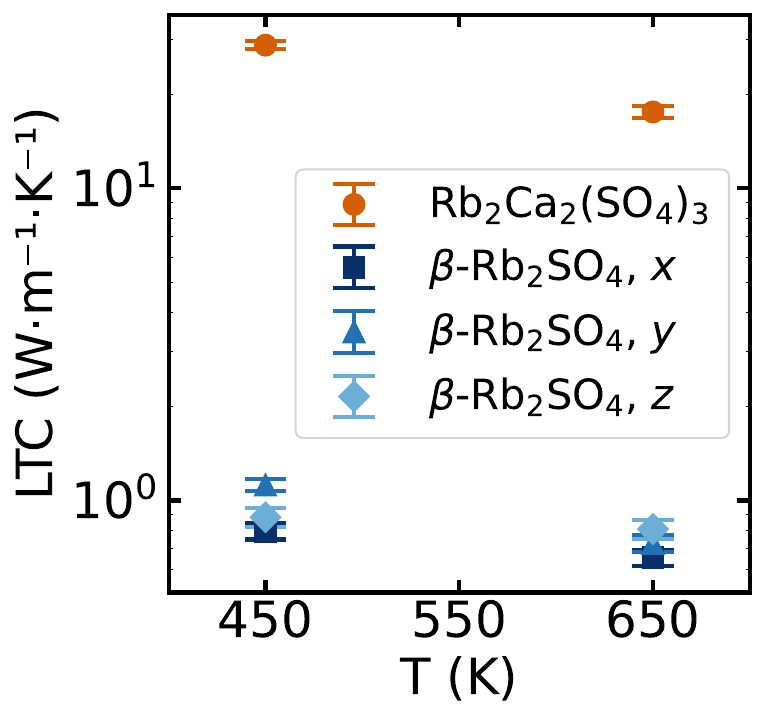}\\ (a)
\end{minipage}\hfill
\begin{minipage}[t]{0.49\linewidth}
\centering
\includegraphics[width=\linewidth]{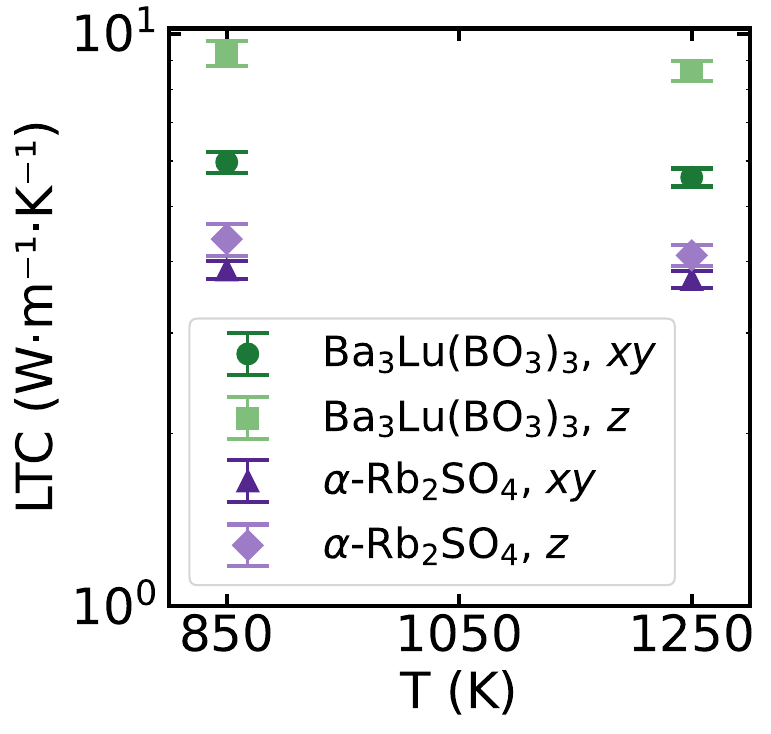}\\ (b)
\end{minipage}
\caption{Green--Kubo lattice thermal conductivity in (a) the low-temperature range, showing the three diagonal components for $\beta$-Rb$_2$SO$_4$ and their average for cubic Rb$_2$Ca$_2$(SO$_4$)$_3$, and (b) the high-temperature range, showing the in-plane average $(\kappa_{xx}+\kappa_{yy})/2$ ($xy$) and $\kappa_{zz}$ ($z$) for $\alpha$-Rb$_2$SO$_4$ and Ba$_3$Lu(BO$_3$)$_3$. Error bars show standard errors over independent trajectories.}
\label{fig:ltc}
\end{figure}

The Green--Kubo LTCs are shown in Fig.~\ref{fig:ltc}. In the low-temperature range, $\beta$-Rb$_2$SO$_4$ has the lowest conductivity, 0.6--1.1~W\,m$^{-1}$\,K$^{-1}$ at 450--650~K with weak anisotropy, whereas Rb$_2$Ca$_2$(SO$_4$)$_3$ reaches about 29 and 18~W\,m$^{-1}$\,K$^{-1}$ at 450 and 650~K, respectively. In the high-temperature range, the conductivity of $\alpha$-Rb$_2$SO$_4$ is about 4~W\,m$^{-1}$\,K$^{-1}$ in all directions, whereas Ba$_3$Lu(BO$_3$)$_3$ has an in-plane conductivity of about 6~W\,m$^{-1}$\,K$^{-1}$ and $\kappa_{zz}\approx9$~W\,m$^{-1}$\,K$^{-1}$ at 850--1250~K.

Among the high-temperature phases, the LTC follows the anharmonicity measure: $\alpha$-Rb$_2$SO$_4$, with $\sigma^A\approx0.75$--$0.78$, conducts heat less well than Ba$_3$Lu(BO$_3$)$_3$, with $\sigma^A\approx0.40$--$0.51$. However, $\sigma^A$ alone does not determine the LTC. $\beta$-Rb$_2$SO$_4$ and Rb$_2$Ca$_2$(SO$_4$)$_3$ have similar $\sigma^A$ (0.30--0.39) and similar acoustic bandwidths in their sTDEP dispersions (Figs.~\ref{fig:phonon_renormalization_all} and S5), yet their LTCs differ by a factor of about 25. Similarly, $\alpha$-Rb$_2$SO$_4$ conducts heat several times better than $\beta$-Rb$_2$SO$_4$ despite its twice larger $\sigma^A$. Because $\sigma^A$ averages the anharmonic forces over all modes, it does not resolve the scattering of the low-frequency, heat-carrying phonons; a mode-resolved analysis is needed to identify the origin of these differences.

\subsection{Band-gap renormalization}\label{sec:bgr}

\begin{figure}[h!]
	\centering
	\begin{minipage}[h]{\linewidth}
		\center{\includegraphics[trim={0cm 0cm 0cm 0cm},clip, width=1\linewidth]{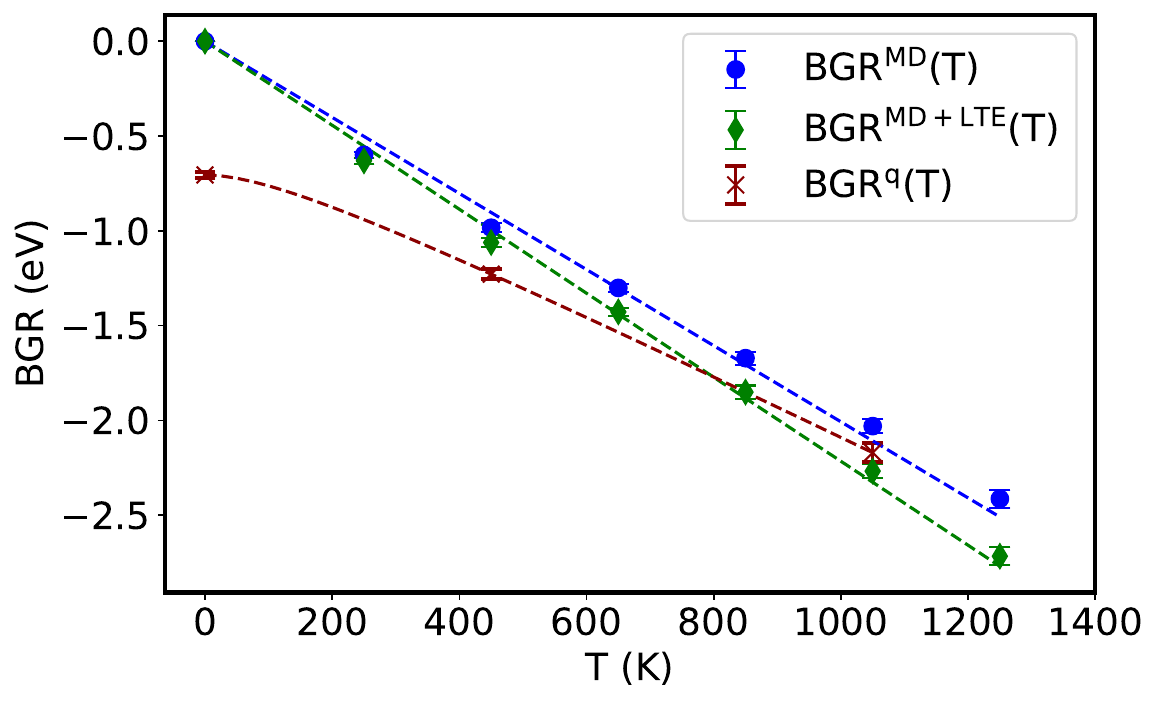}} \\ (a)
	\end{minipage}
    \begin{minipage}[h]{\linewidth}
		\center{\includegraphics[trim={0cm 0cm 0cm 0cm},clip, width=1\linewidth]{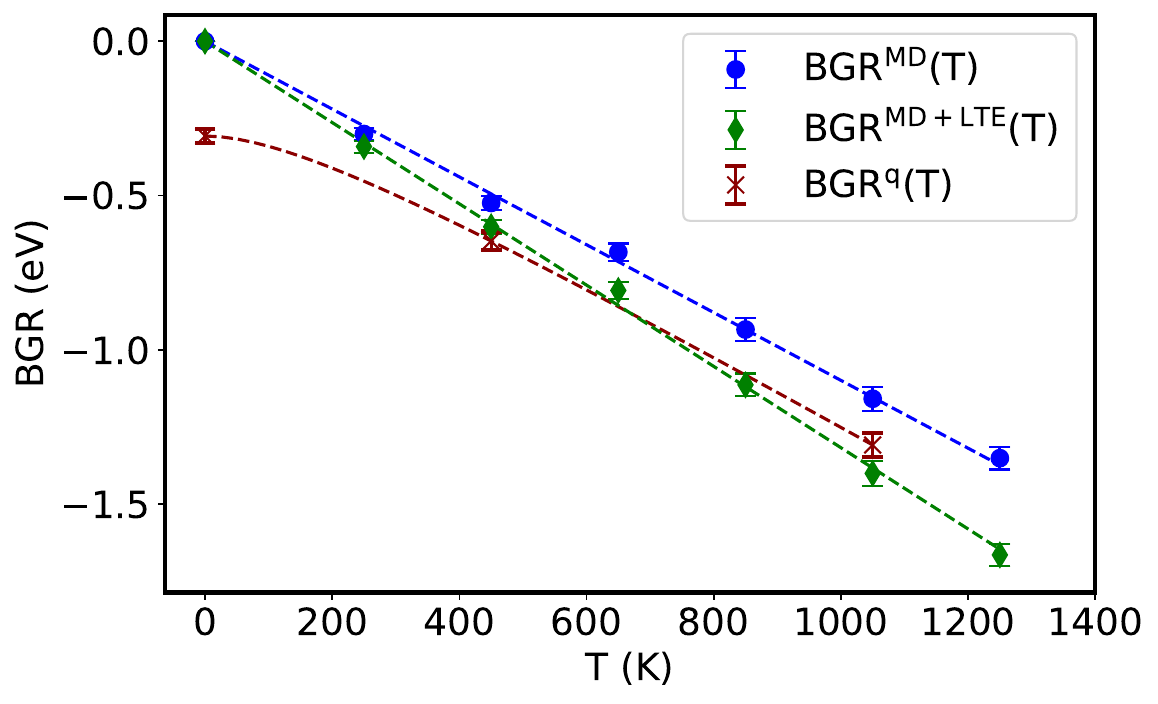}} \\ (b)
	\end{minipage}
    \begin{minipage}[h]{\linewidth}
		\center{\includegraphics[trim={0cm 0cm 0cm 0cm},clip, width=1\linewidth]{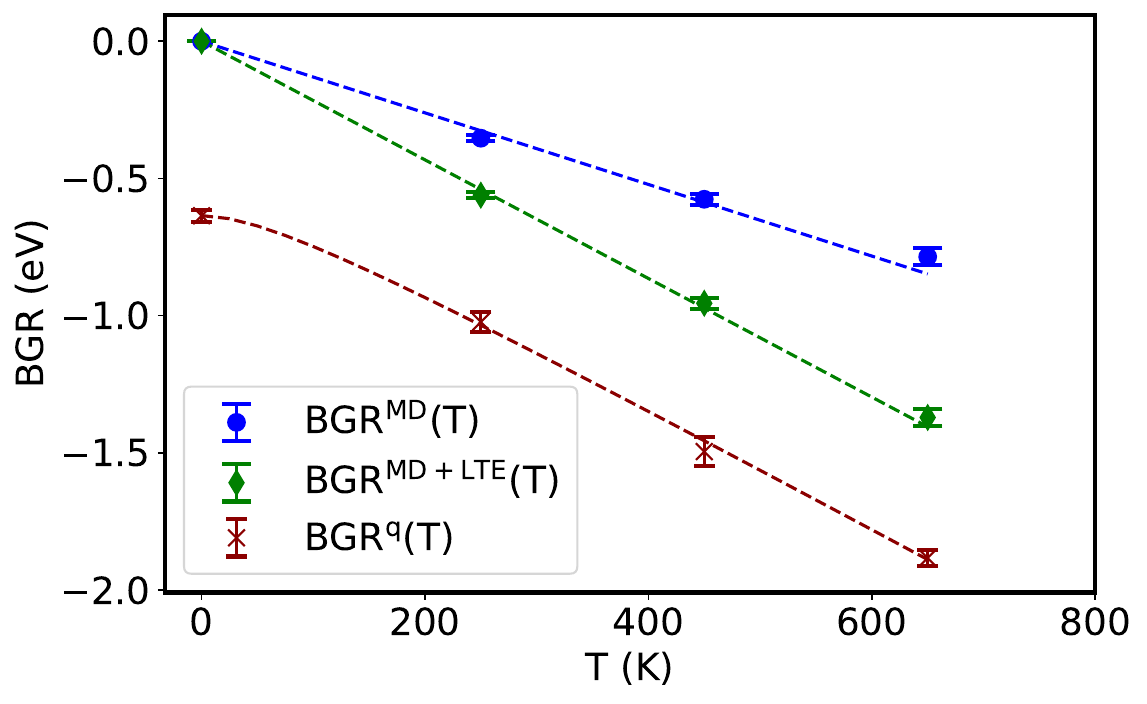}} \\ (c)
	\end{minipage}
    \caption{Band-gap renormalization in (a) Rb$_2$Ca$_2$(SO$_4$)$_3$, (b) Ba$_3$Lu(BO$_3$)$_3$, and (c) $\beta$-Rb$_2$SO$_4$: ${\rm BGR^{MD}}$ (classical nuclear motion), ${\rm BGR^{MD+LTE}}$ (classical nuclear motion and lattice thermal expansion), and ${\rm BGR^{q}}$ (quantum nuclear motion and lattice thermal expansion); see Sec.~\ref{sec:bgr_methods}. Dashed lines are linear fits to ${\rm BGR^{MD}}$ and ${\rm BGR^{MD+LTE}}$ and Varshni-type interpolations ${\rm BGR^{q}}(T)={\rm ZPR}-aT^2/(T+b)$ of ${\rm BGR^{q}}$ (see the Supplementary Material).}
    \label{fig:bgr}
\end{figure}

The PBE band gaps of the static structures are 5.57, 4.08, and 5.24~eV for Rb$_2$Ca$_2$(SO$_4$)$_3$, Ba$_3$Lu(BO$_3$)$_3$, and $\beta$-Rb$_2$SO$_4$, respectively (band structures in Fig.~S31). Nuclear motion reduces these gaps strongly (Fig.~\ref{fig:bgr}). The classical ${\rm BGR^{MD}}$ is nearly linear in temperature and already amounts to 0.3--0.6~eV at 250~K. With lattice thermal expansion included, ${\rm BGR^{MD+LTE}}$ reaches about 2.7 and 1.7~eV at 1250~K in Rb$_2$Ca$_2$(SO$_4$)$_3$ and Ba$_3$Lu(BO$_3$)$_3$, respectively, and 1.4~eV at 650~K in $\beta$-Rb$_2$SO$_4$. These values are large fractions of the PBE gaps; their absolute magnitudes carry the uncertainties of PBE and of finite supercells sampled at the $\Gamma$ point.

Thermal expansion contributes most in $\beta$-Rb$_2$SO$_4$, where it adds 58--75\% to ${\rm BGR^{MD}}$, which may reflect the strong and anisotropic expansion of this phase (Sec.~\ref{sec:lte}). In Rb$_2$Ca$_2$(SO$_4$)$_3$ and Ba$_3$Lu(BO$_3$)$_3$, the LTE contribution is smaller but still significant, 5--23\% of ${\rm BGR^{MD}}$.

The ZPR values are $0.64\pm0.02$, $0.70\pm0.02$, and $0.31\pm0.02$~eV for $\beta$-Rb$_2$SO$_4$, Rb$_2$Ca$_2$(SO$_4$)$_3$, and Ba$_3$Lu(BO$_3$)$_3$, respectively, comparable to the values obtained with non-adiabatic perturbation theory for BN (0.41~eV), diamond (0.33~eV), SiO$_2$ (0.59~eV), and BeO (0.70~eV)~\cite{gonze_bgr_database}. At finite temperature, quantum effects remain important in $\beta$-Rb$_2$SO$_4$, where ${\rm BGR^{q}}$ exceeds ${\rm BGR^{MD+LTE}}$ in magnitude by about 0.5~eV between 250 and 650~K. In Rb$_2$Ca$_2$(SO$_4$)$_3$ and Ba$_3$Lu(BO$_3$)$_3$, the difference decreases to 0.16 and 0.04~eV, respectively, at 450~K, and at 1050~K ${\rm BGR^{q}}$ becomes about 0.1~eV smaller in magnitude than ${\rm BGR^{MD+LTE}}$. Because, for the same force constants, the quantum amplitudes in Eq.~(\ref{eq:amplitudes}) always exceed the classical ones, this crossing suggests that the Gaussian sTDEP sampling misses large-amplitude, non-Gaussian configurations present in the MD at high temperatures, such as strong SO$_4$ and BO$_3$ rocking (Sec.~\ref{sec:lte}); the additive treatment of thermal expansion in ${\rm BGR^{MD+LTE}}$ may also contribute.

\section{Conclusions}\label{sec:concl}

We combined actively learned MTPs with MD, sTDEP, and DFT to study anharmonic lattice dynamics and its thermal and electronic consequences in $\beta$- and $\alpha$-Rb$_2$SO$_4$, Rb$_2$Ca$_2$(SO$_4$)$_3$, and Ba$_3$Lu(BO$_3$)$_3$.
All four phases are strongly anharmonic. The anharmonicity measure is $\sigma^A\approx0.3$--$0.5$ for $\beta$-Rb$_2$SO$_4$, Rb$_2$Ca$_2$(SO$_4$)$_3$, and Ba$_3$Lu(BO$_3$)$_3$, and $\sigma^A\approx0.75$--$0.78$ for orientationally disordered $\alpha$-Rb$_2$SO$_4$, which is above SrTiO$_3$. Anharmonic renormalization at finite temperature largely removes the harmonic instabilities of Rb$_2$Ca$_2$(SO$_4$)$_3$ and $\alpha$-Rb$_2$SO$_4$.

The MD reproduces the measured thermal expansion, with lattice parameters systematically larger than experiment because of the PBE+D3 reference, and supports the mechanisms proposed from diffraction. Straightening of the Rb1 rods contributes about 30\% of the expansion along $c$ in $\beta$-Rb$_2$SO$_4$. Bending of polyhedral linkages partly compensates bond elongation along $a$ and $b$ in Rb$_2$SO$_4$, within the $ab$ plane of Ba$_3$Lu(BO$_3$)$_3$, and along the $L_3$ rods of Rb$_2$Ca$_2$(SO$_4$)$_3$. In Ba$_3$Lu(BO$_3$)$_3$, nearly rigid BO$_3$ groups rock and reorient, and the minimum in the inclination of B(1)O$_3$ and B(3)O$_3$ near 600--700~K is close to the bend in the experimental $c(T)$.

The Green--Kubo LTC ranges from about 1~W\,m$^{-1}$\,K$^{-1}$ in $\beta$-Rb$_2$SO$_4$ to about 29~W\,m$^{-1}$\,K$^{-1}$ in Rb$_2$Ca$_2$(SO$_4$)$_3$ at 450~K. At high temperatures, the more anharmonic $\alpha$-Rb$_2$SO$_4$ conducts heat less well than Ba$_3$Lu(BO$_3$)$_3$. However, phases with similar $\sigma^A$ can differ in LTC by more than an order of magnitude, so $\sigma^A$ alone does not predict heat transport, and mode-resolved analyses are needed.

Nuclear motion reduces the PBE band gaps by 0.3--0.6~eV already at 250~K. With thermal expansion, the BGR reaches 2.7 and 1.7~eV at 1250~K in Rb$_2$Ca$_2$(SO$_4$)$_3$ and Ba$_3$Lu(BO$_3$)$_3$, respectively, and 1.4~eV at 650~K in $\beta$-Rb$_2$SO$_4$. Thermal expansion increases the classical BGR by 58--75\% in $\beta$-Rb$_2$SO$_4$ and by 5--23\% in the other two compounds. The ZPRs of 0.64, 0.70, and 0.31~eV are comparable to those of light-element wide-gap insulators, and quantum effects remain significant at finite temperature, most strongly in $\beta$-Rb$_2$SO$_4$.

For optical applications, these results suggest that the absorption edges of these crystals shift substantially with temperature and that heat dissipation in $\beta$-Rb$_2$SO$_4$ is poor. The workflow used here, with active learning, MD extrapolated to the thermodynamic limit, and sampling-based BGR, can be transferred to other structurally complex optical crystals.

\section*{Conflicts of interest}
The authors declare no conflicts of interest.

\section*{Data availability}
The MTPs and training sets that support the findings of this article will be made openly available in a Git repository upon publication. The Supplementary Material contains additional computational details, convergence tests, relaxed lattice parameters, structural analyses, phonon dispersions, and electronic band structures.

\begin{acknowledgments}
E.S., R.A., N.R., A.S. acknowledge funding from the Russian Science Foundation (RSF) project No.~23-13-00332-P.
\end{acknowledgments}

\clearpage
\bibliographystyle{apsrev4-2}

\bibliography{main}

\begin{thebibliography}{10}

\bibitem{shablinskii2025structural}
AP~Shablinskii, SV~Demina, YP~Biryukov, RS~Bubnova, MG~Krzhizhanovskaya, and
  SK~Filatov.
\newblock A structural origin of both positive and negative thermal expansion
  in langbeinite-, arcanite-and metathenardite-type and related $\rm
  {R}b_2{SO}_4$ and $\rm {R}b_2{C}a_2 ({SO}_4)_3$ compounds.
\newblock {\em Ceramics International}, 51(26-C):51342--51350, 2025.

\bibitem{biryukov2023thermal}
Ya~P Biryukov and RS~Bubnova.
\newblock Thermal expansion of {B}a$_3${L}u({BO}$_3$)$_3$ borate.
\newblock {\em Glass Physics and Chemistry}, 49(4):374--378, 2023.

\bibitem{mtp}
Alexander~V Shapeev.
\newblock Moment tensor potentials: A class of systematically improvable
  interatomic potentials.
\newblock {\em Multiscale Modeling \& Simulation}, 14(3):1153--1173, 2016.

\bibitem{mlip}
I.~S. Novikov, K.~Gubaev, E.~V. Podryabinkin, and A.~V. Shapeev.
\newblock The {MLIP} package: moment tensor potentials with {MPI} and active
  learning.
\newblock {\em Machine Learning: Science and Technology}, 2(2):025002, 2020.

\bibitem{podryabinkin2017active}
Evgeny~V. Podryabinkin and Alexander~V. Shapeev.
\newblock Active learning of linearly parametrized interatomic potentials.
\newblock {\em Computational Materials Science}, 140:171--180, 2017.

\bibitem{gubaev2019accelerating}
Konstantin Gubaev, Evgeny~V. Podryabinkin, Gus L.~W. Hart, and Alexander~V.
  Shapeev.
\newblock Accelerating high-throughput searches for new alloys with active
  learning of interatomic potentials.
\newblock {\em Computational Materials Science}, 156:148--156, 2019.

\bibitem{goreinov2010submatrix}
S.~A. Goreinov, I.~V. Oseledets, D.~V. Savostyanov, E.~E. Tyrtyshnikov, and
  N.~L. Zamarashkin.
\newblock How to find a good submatrix.
\newblock In {\em Matrix Methods: Theory, Algorithms and Applications}, pages
  247--256. World Scientific, 2010.

\bibitem{wisesa2016efficient}
Pandu Wisesa, Kyle~A. McGill, and Tim Mueller.
\newblock Efficient generation of generalized {M}onkhorst-{P}ack grids through
  the use of informatics.
\newblock {\em Physical Review B}, 93:155109, 2016.

\bibitem{wang2021rapid}
Yunzhe Wang, Pandu Wisesa, Adarsh Balasubramanian, Shyam Dwaraknath, and Tim
  Mueller.
\newblock Rapid generation of optimal generalized {M}onkhorst-{P}ack grids.
\newblock {\em Computational Materials Science}, 187:110100, 2021.

\bibitem{ilyukhin1993crystal}
AB~Ilyukhin and BF~Dzhurinskij.
\newblock Crystal structure of double borates $\rm {L}n{B}a_3({BO}_3)_3$ ({L}n:
  {D}y, {H}o, {T}m, {Y}b, {L}u).
\newblock {\em ChemInform}, 25(13), 1994.

\bibitem{boujelben2007langbeinite}
Mohamed Boujelben, Mohamed Toumi, and Tahar Mhiri.
\newblock Langbeinite-type $\rm {R}b_2{C}a_2({SO}_4)_3$.
\newblock {\em Structure Reports}, 63(7):i157--i157, 2007.

\bibitem{weber1989determination}
H~J Weber, M~Schulz, S~Schmitz, J~Granzin, and H~Siegert.
\newblock Determination and structural application of anisotropic bond
  polarisabilities in complex crystals.
\newblock {\em Journal of Physics: Condensed Matter}, 1(44):8543--8547, 1989.

\bibitem{LAMMPS}
A.~P. Thompson, H.~M. Aktulga, R.~Berger, D.~S. Bolintineanu, W.~M. Brown,
  P.~S. Crozier, P.~J. in~'t Veld, A.~Kohlmeyer, S.~G. Moore, T.~D. Nguyen,
  R.~Shan, M.~J. Stevens, J.~Tranchida, C.~Trott, and S.~J. Plimpton.
\newblock {LAMMPS} - a flexible simulation tool for particle-based materials
  modeling at the atomic, meso, and continuum scales.
\newblock {\em Comp. Phys. Comm.}, 271:108171, 2022.

\bibitem{linear_regres}
Geoffrey~S Watson.
\newblock Linear least squares regression.
\newblock {\em The Annals of Mathematical Statistics}, pages 1679--1699, 1967.

\bibitem{gp_rasmussen}
C.~E. Rasmussen and C.~K.~I. Williams.
\newblock {\em Gaussian Processes for Machine Learning}.
\newblock MIT Press, 2006.

\bibitem{spirande2025automated}
Ekaterina Spirande, Timofei Miryashkin, Andrei Kolmakov, and Alexander Shapeev.
\newblock Automated prediction of thermodynamic properties via {B}ayesian
  free-energy reconstruction from molecular dynamics.
\newblock {\em Computational Condensed Matter}, 45:e01163, 2025.

\bibitem{miryashkin2023bayesian}
Timofei Miryashkin, Olga Klimanova, Vladimir Ladygin, and Alexander Shapeev.
\newblock Bayesian inference of composition-dependent phase diagrams.
\newblock {\em Physical Review B}, 108:174103, 2023.

\bibitem{lammps-mlip-gk}
Siu~Ting Tai, Chen Wang, Ruihuan Cheng, and Yue Chen.
\newblock Revisiting many-body interaction heat current and thermal
  conductivity calculations using the moment tensor potential/{LAMMPS}
  interface.
\newblock {\em Journal of Chemical Theory and Computation}, 21(7):3649, 2025.

\bibitem{fhiaims}
Volker Blum, Ralf Gehrke, Felix Hanke, Paula Havu, Ville Havu, Xinguo Ren,
  Karsten Reuter, and Matthias Scheffler.
\newblock Ab initio molecular simulations with numeric atom-centered orbitals.
\newblock {\em Computer Physics Communications}, 180(11):2175--2196, 2009.

\bibitem{pbe}
John~P Perdew, Kieron Burke, and Matthias Ernzerhof.
\newblock Generalized gradient approximation made simple.
\newblock {\em Physical review letters}, 77(18):3865--3868, 1996.

\bibitem{varshni}
Yatendra~Pal Varshni.
\newblock Temperature dependence of the energy gap in semiconductors.
\newblock {\em Physica}, 34(1):149--154, 1967.

\end{thebibliography}


\begin{thebibliography}{52}%
\makeatletter
\providecommand \@ifxundefined [1]{%
 \@ifx{#1\undefined}
}%
\providecommand \@ifnum [1]{%
 \ifnum #1\expandafter \@firstoftwo
 \else \expandafter \@secondoftwo
 \fi
}%
\providecommand \@ifx [1]{%
 \ifx #1\expandafter \@firstoftwo
 \else \expandafter \@secondoftwo
 \fi
}%
\providecommand \natexlab [1]{#1}%
\providecommand \enquote  [1]{``#1''}%
\providecommand \bibnamefont  [1]{#1}%
\providecommand \bibfnamefont [1]{#1}%
\providecommand \citenamefont [1]{#1}%
\providecommand \href@noop [0]{\@secondoftwo}%
\providecommand \href [0]{\begingroup \@sanitize@url \@href}%
\providecommand \@href[1]{\@@startlink{#1}\@@href}%
\providecommand \@@href[1]{\endgroup#1\@@endlink}%
\providecommand \@sanitize@url [0]{\catcode `\\12\catcode `\$12\catcode
  `\&12\catcode `\#12\catcode `\^12\catcode `\_12\catcode `\%12\relax}%
\providecommand \@@startlink[1]{}%
\providecommand \@@endlink[0]{}%
\providecommand \url  [0]{\begingroup\@sanitize@url \@url }%
\providecommand \@url [1]{\endgroup\@href {#1}{\urlprefix }}%
\providecommand \urlprefix  [0]{URL }%
\providecommand \Eprint [0]{\href }%
\providecommand \doibase [0]{https://doi.org/}%
\providecommand \selectlanguage [0]{\@gobble}%
\providecommand \bibinfo  [0]{\@secondoftwo}%
\providecommand \bibfield  [0]{\@secondoftwo}%
\providecommand \translation [1]{[#1]}%
\providecommand \BibitemOpen [0]{}%
\providecommand \bibitemStop [0]{}%
\providecommand \bibitemNoStop [0]{.\EOS\space}%
\providecommand \EOS [0]{\spacefactor3000\relax}%
\providecommand \BibitemShut  [1]{\csname bibitem#1\endcsname}%
\let\auto@bib@innerbib\@empty
\bibitem [{\citenamefont {Halasyamani}\ and\ \citenamefont
  {Rondinelli}(2018)}]{halasyamani2018must}%
  \BibitemOpen
  \bibfield  {author} {\bibinfo {author} {\bibfnamefont {P.~S.}\ \bibnamefont
  {Halasyamani}}\ and\ \bibinfo {author} {\bibfnamefont {J.~M.}\ \bibnamefont
  {Rondinelli}},\ }\href {https://doi.org/10.1038/s41467-018-05411-1}
  {\bibfield  {journal} {\bibinfo  {journal} {Nature Communications}\ }\textbf
  {\bibinfo {volume} {9}},\ \bibinfo {pages} {2972} (\bibinfo {year}
  {2018})}\BibitemShut {NoStop}%
\bibitem [{\citenamefont {Stefanovich}\ and\ \citenamefont
  {Khamaganova}(2002)}]{ba3lu_nlo}%
  \BibitemOpen
  \bibfield  {author} {\bibinfo {author} {\bibfnamefont {S.~Y.}\ \bibnamefont
  {Stefanovich}}\ and\ \bibinfo {author} {\bibfnamefont {T.}~\bibnamefont
  {Khamaganova}},\ }\href@noop {} {\bibfield  {journal} {\bibinfo  {journal}
  {Inorganic materials}\ }\textbf {\bibinfo {volume} {38}},\ \bibinfo {pages}
  {66} (\bibinfo {year} {2002})}\BibitemShut {NoStop}%
\bibitem [{\citenamefont {Xiao-Yu}\ \emph {et~al.}(2021)\citenamefont
  {Xiao-Yu}, \citenamefont {En-Qiang}, \citenamefont {Sheng-Dong},
  \citenamefont {Hao-Hang}, \citenamefont {Xiao-Xin},\ and\ \citenamefont
  {Yao-Guo}}]{rb2ca2_nlo}%
  \BibitemOpen
  \bibfield  {author} {\bibinfo {author} {\bibfnamefont {Z.}~\bibnamefont
  {Xiao-Yu}}, \bibinfo {author} {\bibfnamefont {W.}~\bibnamefont {En-Qiang}},
  \bibinfo {author} {\bibfnamefont {Y.}~\bibnamefont {Sheng-Dong}}, \bibinfo
  {author} {\bibfnamefont {W.}~\bibnamefont {Hao-Hang}}, \bibinfo {author}
  {\bibfnamefont {L.}~\bibnamefont {Xiao-Xin}},\ and\ \bibinfo {author}
  {\bibfnamefont {S.}~\bibnamefont {Yao-Guo}},\ }\href@noop {} {\bibfield
  {journal} {\bibinfo  {journal} {CHINESE JOURNAL OF STRUCTURAL CHEMISTRY}\
  }\textbf {\bibinfo {volume} {40}},\ \bibinfo {pages} {949} (\bibinfo {year}
  {2021})}\BibitemShut {NoStop}%
\bibitem [{\citenamefont {Shablinskii}\ \emph {et~al.}(2025)\citenamefont
  {Shablinskii}, \citenamefont {Demina}, \citenamefont {Biryukov},
  \citenamefont {Bubnova}, \citenamefont {Krzhizhanovskaya},\ and\
  \citenamefont {Filatov}}]{shablinskii2025structural}%
  \BibitemOpen
  \bibfield  {author} {\bibinfo {author} {\bibfnamefont {A.}~\bibnamefont
  {Shablinskii}}, \bibinfo {author} {\bibfnamefont {S.}~\bibnamefont {Demina}},
  \bibinfo {author} {\bibfnamefont {Y.}~\bibnamefont {Biryukov}}, \bibinfo
  {author} {\bibfnamefont {R.}~\bibnamefont {Bubnova}}, \bibinfo {author}
  {\bibfnamefont {M.}~\bibnamefont {Krzhizhanovskaya}},\ and\ \bibinfo {author}
  {\bibfnamefont {S.}~\bibnamefont {Filatov}},\ }\href
  {https://doi.org/10.1016/j.ceramint.2025.08.360} {\bibfield  {journal}
  {\bibinfo  {journal} {Ceramics International}\ }\textbf {\bibinfo {volume}
  {51}},\ \bibinfo {pages} {51342} (\bibinfo {year} {2025})}\BibitemShut
  {NoStop}%
\bibitem [{\citenamefont {Biryukov}\ and\ \citenamefont
  {Bubnova}(2023)}]{biryukov2023thermal}%
  \BibitemOpen
  \bibfield  {author} {\bibinfo {author} {\bibfnamefont {Y.~P.}\ \bibnamefont
  {Biryukov}}\ and\ \bibinfo {author} {\bibfnamefont {R.}~\bibnamefont
  {Bubnova}},\ }\href@noop {} {\bibfield  {journal} {\bibinfo  {journal} {Glass
  Physics and Chemistry}\ }\textbf {\bibinfo {volume} {49}},\ \bibinfo {pages}
  {374} (\bibinfo {year} {2023})}\BibitemShut {NoStop}%
\bibitem [{\citenamefont {Bubnova}\ and\ \citenamefont
  {Filatov}(2013)}]{bubnova2013high}%
  \BibitemOpen
  \bibfield  {author} {\bibinfo {author} {\bibfnamefont {R.~S.}\ \bibnamefont
  {Bubnova}}\ and\ \bibinfo {author} {\bibfnamefont {S.~K.}\ \bibnamefont
  {Filatov}},\ }\href@noop {} {\bibfield  {journal} {\bibinfo  {journal}
  {Zeitschrift f{\"u}r Kristallographie -- Crystalline Materials}\ }\textbf
  {\bibinfo {volume} {228}},\ \bibinfo {pages} {395} (\bibinfo {year}
  {2013})}\BibitemShut {NoStop}%
\bibitem [{\citenamefont {Baroni}\ \emph {et~al.}(2001)\citenamefont {Baroni},
  \citenamefont {de~Gironcoli}, \citenamefont {Dal~Corso},\ and\ \citenamefont
  {Giannozzi}}]{baroni2001phonons}%
  \BibitemOpen
  \bibfield  {author} {\bibinfo {author} {\bibfnamefont {S.}~\bibnamefont
  {Baroni}}, \bibinfo {author} {\bibfnamefont {S.}~\bibnamefont
  {de~Gironcoli}}, \bibinfo {author} {\bibfnamefont {A.}~\bibnamefont
  {Dal~Corso}},\ and\ \bibinfo {author} {\bibfnamefont {P.}~\bibnamefont
  {Giannozzi}},\ }\href {https://doi.org/10.1103/RevModPhys.73.515} {\bibfield
  {journal} {\bibinfo  {journal} {Reviews of Modern Physics}\ }\textbf
  {\bibinfo {volume} {73}},\ \bibinfo {pages} {515} (\bibinfo {year}
  {2001})}\BibitemShut {NoStop}%
\bibitem [{\citenamefont {Togo}(2023)}]{ld_theor}%
  \BibitemOpen
  \bibfield  {author} {\bibinfo {author} {\bibfnamefont {A.}~\bibnamefont
  {Togo}},\ }\href@noop {} {\bibfield  {journal} {\bibinfo  {journal} {Journal
  of the Physical Society of Japan}\ }\textbf {\bibinfo {volume} {92}},\
  \bibinfo {pages} {012001} (\bibinfo {year} {2023})}\BibitemShut {NoStop}%
\bibitem [{\citenamefont {Hellman}\ \emph {et~al.}(2011)\citenamefont
  {Hellman}, \citenamefont {Abrikosov},\ and\ \citenamefont
  {Simak}}]{hellman2011lattice}%
  \BibitemOpen
  \bibfield  {author} {\bibinfo {author} {\bibfnamefont {O.}~\bibnamefont
  {Hellman}}, \bibinfo {author} {\bibfnamefont {I.~A.}\ \bibnamefont
  {Abrikosov}},\ and\ \bibinfo {author} {\bibfnamefont {S.~I.}\ \bibnamefont
  {Simak}},\ }\href {https://doi.org/10.1103/PhysRevB.84.180301} {\bibfield
  {journal} {\bibinfo  {journal} {Physical Review B}\ }\textbf {\bibinfo
  {volume} {84}},\ \bibinfo {pages} {180301} (\bibinfo {year}
  {2011})}\BibitemShut {NoStop}%
\bibitem [{\citenamefont {Zacharias}\ \emph {et~al.}(2023)\citenamefont
  {Zacharias}, \citenamefont {Volonakis}, \citenamefont {Giustino},\ and\
  \citenamefont {Even}}]{anh_ld_theor}%
  \BibitemOpen
  \bibfield  {author} {\bibinfo {author} {\bibfnamefont {M.}~\bibnamefont
  {Zacharias}}, \bibinfo {author} {\bibfnamefont {G.}~\bibnamefont
  {Volonakis}}, \bibinfo {author} {\bibfnamefont {F.}~\bibnamefont
  {Giustino}},\ and\ \bibinfo {author} {\bibfnamefont {J.}~\bibnamefont
  {Even}},\ }\href@noop {} {\bibfield  {journal} {\bibinfo  {journal} {Physical
  Review B}\ }\textbf {\bibinfo {volume} {108}},\ \bibinfo {pages} {035155}
  (\bibinfo {year} {2023})}\BibitemShut {NoStop}%
\bibitem [{\citenamefont {Knoop}\ \emph {et~al.}(2020)\citenamefont {Knoop},
  \citenamefont {Purcell}, \citenamefont {Scheffler},\ and\ \citenamefont
  {Carbogno}}]{knoop2020anharmonicity}%
  \BibitemOpen
  \bibfield  {author} {\bibinfo {author} {\bibfnamefont {F.}~\bibnamefont
  {Knoop}}, \bibinfo {author} {\bibfnamefont {T.~A.}\ \bibnamefont {Purcell}},
  \bibinfo {author} {\bibfnamefont {M.}~\bibnamefont {Scheffler}},\ and\
  \bibinfo {author} {\bibfnamefont {C.}~\bibnamefont {Carbogno}},\ }\href@noop
  {} {\bibfield  {journal} {\bibinfo  {journal} {Physical Review Materials}\
  }\textbf {\bibinfo {volume} {4}},\ \bibinfo {pages} {083809} (\bibinfo {year}
  {2020})}\BibitemShut {NoStop}%
\bibitem [{\citenamefont {Knoop}\ \emph {et~al.}(2023)\citenamefont {Knoop},
  \citenamefont {Purcell}, \citenamefont {Scheffler},\ and\ \citenamefont
  {Carbogno}}]{knoop2023anharmonicity}%
  \BibitemOpen
  \bibfield  {author} {\bibinfo {author} {\bibfnamefont {F.}~\bibnamefont
  {Knoop}}, \bibinfo {author} {\bibfnamefont {T.~A.~R.}\ \bibnamefont
  {Purcell}}, \bibinfo {author} {\bibfnamefont {M.}~\bibnamefont {Scheffler}},\
  and\ \bibinfo {author} {\bibfnamefont {C.}~\bibnamefont {Carbogno}},\ }\href
  {https://doi.org/10.1103/PhysRevLett.130.236301} {\bibfield  {journal}
  {\bibinfo  {journal} {Physical Review Letters}\ }\textbf {\bibinfo {volume}
  {130}},\ \bibinfo {pages} {236301} (\bibinfo {year} {2023})}\BibitemShut
  {NoStop}%
\bibitem [{\citenamefont {Hellman}\ \emph {et~al.}(2013)\citenamefont
  {Hellman}, \citenamefont {Steneteg}, \citenamefont {Abrikosov},\ and\
  \citenamefont {Simak}}]{hellman2013temperature}%
  \BibitemOpen
  \bibfield  {author} {\bibinfo {author} {\bibfnamefont {O.}~\bibnamefont
  {Hellman}}, \bibinfo {author} {\bibfnamefont {P.}~\bibnamefont {Steneteg}},
  \bibinfo {author} {\bibfnamefont {I.~A.}\ \bibnamefont {Abrikosov}},\ and\
  \bibinfo {author} {\bibfnamefont {S.~I.}\ \bibnamefont {Simak}},\ }\href
  {https://doi.org/10.1103/PhysRevB.87.104111} {\bibfield  {journal} {\bibinfo
  {journal} {Physical Review B}\ }\textbf {\bibinfo {volume} {87}},\ \bibinfo
  {pages} {104111} (\bibinfo {year} {2013})}\BibitemShut {NoStop}%
\bibitem [{\citenamefont {Shulumba}\ \emph {et~al.}(2017)\citenamefont
  {Shulumba}, \citenamefont {Hellman},\ and\ \citenamefont
  {Minnich}}]{stdep_paper}%
  \BibitemOpen
  \bibfield  {author} {\bibinfo {author} {\bibfnamefont {N.}~\bibnamefont
  {Shulumba}}, \bibinfo {author} {\bibfnamefont {O.}~\bibnamefont {Hellman}},\
  and\ \bibinfo {author} {\bibfnamefont {A.~J.}\ \bibnamefont {Minnich}},\
  }\href@noop {} {\bibfield  {journal} {\bibinfo  {journal} {Physical Review
  B}\ }\textbf {\bibinfo {volume} {95}},\ \bibinfo {pages} {014302} (\bibinfo
  {year} {2017})}\BibitemShut {NoStop}%
\bibitem [{\citenamefont {Green}(1954)}]{green1954markoff}%
  \BibitemOpen
  \bibfield  {author} {\bibinfo {author} {\bibfnamefont {M.~S.}\ \bibnamefont
  {Green}},\ }\href {https://doi.org/10.1063/1.1740082} {\bibfield  {journal}
  {\bibinfo  {journal} {The Journal of Chemical Physics}\ }\textbf {\bibinfo
  {volume} {22}},\ \bibinfo {pages} {398} (\bibinfo {year} {1954})}\BibitemShut
  {NoStop}%
\bibitem [{\citenamefont {Kubo}(1957)}]{kubo1957statistical}%
  \BibitemOpen
  \bibfield  {author} {\bibinfo {author} {\bibfnamefont {R.}~\bibnamefont
  {Kubo}},\ }\href {https://doi.org/10.1143/JPSJ.12.570} {\bibfield  {journal}
  {\bibinfo  {journal} {Journal of the Physical Society of Japan}\ }\textbf
  {\bibinfo {volume} {12}},\ \bibinfo {pages} {570} (\bibinfo {year}
  {1957})}\BibitemShut {NoStop}%
\bibitem [{\citenamefont {Carbogno}\ \emph {et~al.}(2017)\citenamefont
  {Carbogno}, \citenamefont {Ramprasad},\ and\ \citenamefont
  {Scheffler}}]{carbogno2017ab}%
  \BibitemOpen
  \bibfield  {author} {\bibinfo {author} {\bibfnamefont {C.}~\bibnamefont
  {Carbogno}}, \bibinfo {author} {\bibfnamefont {R.}~\bibnamefont
  {Ramprasad}},\ and\ \bibinfo {author} {\bibfnamefont {M.}~\bibnamefont
  {Scheffler}},\ }\href {https://doi.org/10.1103/PhysRevLett.118.175901}
  {\bibfield  {journal} {\bibinfo  {journal} {Physical Review Letters}\
  }\textbf {\bibinfo {volume} {118}},\ \bibinfo {pages} {175901} (\bibinfo
  {year} {2017})}\BibitemShut {NoStop}%
\bibitem [{\citenamefont {Giustino}(2017)}]{giustino2017electron}%
  \BibitemOpen
  \bibfield  {author} {\bibinfo {author} {\bibfnamefont {F.}~\bibnamefont
  {Giustino}},\ }\href {https://doi.org/10.1103/RevModPhys.89.015003}
  {\bibfield  {journal} {\bibinfo  {journal} {Reviews of Modern Physics}\
  }\textbf {\bibinfo {volume} {89}},\ \bibinfo {pages} {015003} (\bibinfo
  {year} {2017})}\BibitemShut {NoStop}%
\bibitem [{\citenamefont {Zacharias}\ and\ \citenamefont
  {Giustino}(2016)}]{zacharias2016one}%
  \BibitemOpen
  \bibfield  {author} {\bibinfo {author} {\bibfnamefont {M.}~\bibnamefont
  {Zacharias}}\ and\ \bibinfo {author} {\bibfnamefont {F.}~\bibnamefont
  {Giustino}},\ }\href {https://doi.org/10.1103/PhysRevB.94.075125} {\bibfield
  {journal} {\bibinfo  {journal} {Physical Review B}\ }\textbf {\bibinfo
  {volume} {94}},\ \bibinfo {pages} {075125} (\bibinfo {year}
  {2016})}\BibitemShut {NoStop}%
\bibitem [{\citenamefont {Monserrat}(2018)}]{monserrat2018electron}%
  \BibitemOpen
  \bibfield  {author} {\bibinfo {author} {\bibfnamefont {B.}~\bibnamefont
  {Monserrat}},\ }\href@noop {} {\bibfield  {journal} {\bibinfo  {journal}
  {Journal of Physics: Condensed Matter}\ }\textbf {\bibinfo {volume} {30}},\
  \bibinfo {pages} {083001} (\bibinfo {year} {2018})}\BibitemShut {NoStop}%
\bibitem [{\citenamefont {Karsai}\ \emph {et~al.}(2018)\citenamefont {Karsai},
  \citenamefont {Engel}, \citenamefont {Flage-Larsen},\ and\ \citenamefont
  {Kresse}}]{karsai2018electron}%
  \BibitemOpen
  \bibfield  {author} {\bibinfo {author} {\bibfnamefont {F.}~\bibnamefont
  {Karsai}}, \bibinfo {author} {\bibfnamefont {M.}~\bibnamefont {Engel}},
  \bibinfo {author} {\bibfnamefont {E.}~\bibnamefont {Flage-Larsen}},\ and\
  \bibinfo {author} {\bibfnamefont {G.}~\bibnamefont {Kresse}},\ }\href@noop {}
  {\bibfield  {journal} {\bibinfo  {journal} {New Journal of Physics}\ }\textbf
  {\bibinfo {volume} {20}},\ \bibinfo {pages} {123008} (\bibinfo {year}
  {2018})}\BibitemShut {NoStop}%
\bibitem [{\citenamefont {Kang}\ \emph {et~al.}(2025)\citenamefont {Kang},
  \citenamefont {Purcell}, \citenamefont {Carbogno},\ and\ \citenamefont
  {Scheffler}}]{PhysRevMaterials.9.063801}%
  \BibitemOpen
  \bibfield  {author} {\bibinfo {author} {\bibfnamefont {K.}~\bibnamefont
  {Kang}}, \bibinfo {author} {\bibfnamefont {T.~A.~R.}\ \bibnamefont
  {Purcell}}, \bibinfo {author} {\bibfnamefont {C.}~\bibnamefont {Carbogno}},\
  and\ \bibinfo {author} {\bibfnamefont {M.}~\bibnamefont {Scheffler}},\ }\href
  {https://doi.org/10.1103/PhysRevMaterials.9.063801} {\bibfield  {journal}
  {\bibinfo  {journal} {Phys. Rev. Mater.}\ }\textbf {\bibinfo {volume} {9}},\
  \bibinfo {pages} {063801} (\bibinfo {year} {2025})}\BibitemShut {NoStop}%
\bibitem [{\citenamefont {Shapeev}(2016)}]{mtp}%
  \BibitemOpen
  \bibfield  {author} {\bibinfo {author} {\bibfnamefont {A.~V.}\ \bibnamefont
  {Shapeev}},\ }\href@noop {} {\bibfield  {journal} {\bibinfo  {journal}
  {Multiscale Modeling \& Simulation}\ }\textbf {\bibinfo {volume} {14}},\
  \bibinfo {pages} {1153} (\bibinfo {year} {2016})}\BibitemShut {NoStop}%
\bibitem [{\citenamefont {Novikov}\ \emph {et~al.}(2020)\citenamefont
  {Novikov}, \citenamefont {Gubaev}, \citenamefont {Podryabinkin},\ and\
  \citenamefont {Shapeev}}]{mlip}%
  \BibitemOpen
  \bibfield  {author} {\bibinfo {author} {\bibfnamefont {I.~S.}\ \bibnamefont
  {Novikov}}, \bibinfo {author} {\bibfnamefont {K.}~\bibnamefont {Gubaev}},
  \bibinfo {author} {\bibfnamefont {E.~V.}\ \bibnamefont {Podryabinkin}},\ and\
  \bibinfo {author} {\bibfnamefont {A.~V.}\ \bibnamefont {Shapeev}},\
  }\href@noop {} {\bibfield  {journal} {\bibinfo  {journal} {Machine Learning:
  Science and Technology}\ }\textbf {\bibinfo {volume} {2}},\ \bibinfo {pages}
  {025002} (\bibinfo {year} {2020})}\BibitemShut {NoStop}%
\bibitem [{\citenamefont {Podryabinkin}\ and\ \citenamefont
  {Shapeev}(2017)}]{podryabinkin2017active}%
  \BibitemOpen
  \bibfield  {author} {\bibinfo {author} {\bibfnamefont {E.~V.}\ \bibnamefont
  {Podryabinkin}}\ and\ \bibinfo {author} {\bibfnamefont {A.~V.}\ \bibnamefont
  {Shapeev}},\ }\href {https://doi.org/10.1016/j.commatsci.2017.08.031}
  {\bibfield  {journal} {\bibinfo  {journal} {Computational Materials Science}\
  }\textbf {\bibinfo {volume} {140}},\ \bibinfo {pages} {171} (\bibinfo {year}
  {2017})}\BibitemShut {NoStop}%
\bibitem [{\citenamefont {Gubaev}\ \emph {et~al.}(2019)\citenamefont {Gubaev},
  \citenamefont {Podryabinkin}, \citenamefont {Hart},\ and\ \citenamefont
  {Shapeev}}]{gubaev2019accelerating}%
  \BibitemOpen
  \bibfield  {author} {\bibinfo {author} {\bibfnamefont {K.}~\bibnamefont
  {Gubaev}}, \bibinfo {author} {\bibfnamefont {E.~V.}\ \bibnamefont
  {Podryabinkin}}, \bibinfo {author} {\bibfnamefont {G.~L.~W.}\ \bibnamefont
  {Hart}},\ and\ \bibinfo {author} {\bibfnamefont {A.~V.}\ \bibnamefont
  {Shapeev}},\ }\href {https://doi.org/10.1016/j.commatsci.2018.09.031}
  {\bibfield  {journal} {\bibinfo  {journal} {Computational Materials Science}\
  }\textbf {\bibinfo {volume} {156}},\ \bibinfo {pages} {148} (\bibinfo {year}
  {2019})}\BibitemShut {NoStop}%
\bibitem [{\citenamefont {Ladygin}\ \emph {et~al.}(2020)\citenamefont
  {Ladygin}, \citenamefont {Korotaev}, \citenamefont {Yanilkin},\ and\
  \citenamefont {Shapeev}}]{ladygin2020lattice}%
  \BibitemOpen
  \bibfield  {author} {\bibinfo {author} {\bibfnamefont {V.~V.}\ \bibnamefont
  {Ladygin}}, \bibinfo {author} {\bibfnamefont {P.~Y.}\ \bibnamefont
  {Korotaev}}, \bibinfo {author} {\bibfnamefont {A.~V.}\ \bibnamefont
  {Yanilkin}},\ and\ \bibinfo {author} {\bibfnamefont {A.~V.}\ \bibnamefont
  {Shapeev}},\ }\href@noop {} {\bibfield  {journal} {\bibinfo  {journal}
  {Computational Materials Science}\ }\textbf {\bibinfo {volume} {172}},\
  \bibinfo {pages} {109333} (\bibinfo {year} {2020})}\BibitemShut {NoStop}%
\bibitem [{\citenamefont {Mortazavi}\ \emph {et~al.}(2020)\citenamefont
  {Mortazavi}, \citenamefont {Podryabinkin}, \citenamefont {Roche},
  \citenamefont {Rabczuk}, \citenamefont {Zhuang},\ and\ \citenamefont
  {Shapeev}}]{mortazavi2020machine}%
  \BibitemOpen
  \bibfield  {author} {\bibinfo {author} {\bibfnamefont {B.}~\bibnamefont
  {Mortazavi}}, \bibinfo {author} {\bibfnamefont {E.~V.}\ \bibnamefont
  {Podryabinkin}}, \bibinfo {author} {\bibfnamefont {S.}~\bibnamefont {Roche}},
  \bibinfo {author} {\bibfnamefont {T.}~\bibnamefont {Rabczuk}}, \bibinfo
  {author} {\bibfnamefont {X.}~\bibnamefont {Zhuang}},\ and\ \bibinfo {author}
  {\bibfnamefont {A.~V.}\ \bibnamefont {Shapeev}},\ }\href
  {https://doi.org/10.1039/D0MH00787K} {\bibfield  {journal} {\bibinfo
  {journal} {Materials Horizons}\ }\textbf {\bibinfo {volume} {7}},\ \bibinfo
  {pages} {2359} (\bibinfo {year} {2020})}\BibitemShut {NoStop}%
\bibitem [{\citenamefont {Rybin}\ and\ \citenamefont
  {Shapeev}(2024)}]{rybin2024moment}%
  \BibitemOpen
  \bibfield  {author} {\bibinfo {author} {\bibfnamefont {N.}~\bibnamefont
  {Rybin}}\ and\ \bibinfo {author} {\bibfnamefont {A.}~\bibnamefont
  {Shapeev}},\ }\href {https://doi.org/10.1063/5.0206215} {\bibfield  {journal}
  {\bibinfo  {journal} {Journal of Applied Physics}\ }\textbf {\bibinfo
  {volume} {135}},\ \bibinfo {pages} {205108} (\bibinfo {year}
  {2024})}\BibitemShut {NoStop}%
\bibitem [{\citenamefont {Arabov}\ \emph
  {et~al.}(2026{\natexlab{a}})\citenamefont {Arabov}, \citenamefont {Li},
  \citenamefont {Chen}, \citenamefont {Rybin},\ and\ \citenamefont
  {Shapeev}}]{arabov2026thermal}%
  \BibitemOpen
  \bibfield  {author} {\bibinfo {author} {\bibfnamefont {R.}~\bibnamefont
  {Arabov}}, \bibinfo {author} {\bibfnamefont {J.}~\bibnamefont {Li}}, \bibinfo
  {author} {\bibfnamefont {X.}~\bibnamefont {Chen}}, \bibinfo {author}
  {\bibfnamefont {N.}~\bibnamefont {Rybin}},\ and\ \bibinfo {author}
  {\bibfnamefont {A.}~\bibnamefont {Shapeev}},\ }\href
  {https://doi.org/10.1063/5.0336982} {\bibfield  {journal} {\bibinfo
  {journal} {Applied Physics Letters}\ }\textbf {\bibinfo {volume} {129}},\
  \bibinfo {pages} {152202} (\bibinfo {year} {2026}{\natexlab{a}})}\BibitemShut
  {NoStop}%
\bibitem [{\citenamefont {Arabov}\ \emph
  {et~al.}(2026{\natexlab{b}})\citenamefont {Arabov}, \citenamefont {Rybin},
  \citenamefont {Demin}, \citenamefont {Polovinkin}, \citenamefont {Kvashnin},
  \citenamefont {Chernozatonskii},\ and\ \citenamefont
  {Shapeev}}]{arabov2026tuning}%
  \BibitemOpen
  \bibfield  {author} {\bibinfo {author} {\bibfnamefont {R.}~\bibnamefont
  {Arabov}}, \bibinfo {author} {\bibfnamefont {N.}~\bibnamefont {Rybin}},
  \bibinfo {author} {\bibfnamefont {V.}~\bibnamefont {Demin}}, \bibinfo
  {author} {\bibfnamefont {M.}~\bibnamefont {Polovinkin}}, \bibinfo {author}
  {\bibfnamefont {A.}~\bibnamefont {Kvashnin}}, \bibinfo {author}
  {\bibfnamefont {L.}~\bibnamefont {Chernozatonskii}},\ and\ \bibinfo {author}
  {\bibfnamefont {A.}~\bibnamefont {Shapeev}},\ }\href
  {https://doi.org/10.1016/j.apsusc.2025.164801} {\bibfield  {journal}
  {\bibinfo  {journal} {Applied Surface Science}\ }\textbf {\bibinfo {volume}
  {717}},\ \bibinfo {pages} {164801} (\bibinfo {year}
  {2026}{\natexlab{b}})}\BibitemShut {NoStop}%
\bibitem [{\citenamefont {Solovykh}\ \emph {et~al.}(2025)\citenamefont
  {Solovykh}, \citenamefont {Rybin}, \citenamefont {Novikov},\ and\
  \citenamefont {Shapeev}}]{solovykh2025path}%
  \BibitemOpen
  \bibfield  {author} {\bibinfo {author} {\bibfnamefont {A.~A.}\ \bibnamefont
  {Solovykh}}, \bibinfo {author} {\bibfnamefont {N.~E.}\ \bibnamefont {Rybin}},
  \bibinfo {author} {\bibfnamefont {I.~S.}\ \bibnamefont {Novikov}},\ and\
  \bibinfo {author} {\bibfnamefont {A.~V.}\ \bibnamefont {Shapeev}},\
  }\href@noop {} {\bibfield  {journal} {\bibinfo  {journal} {Computer Physics
  Communications}\ ,\ \bibinfo {pages} {109902}} (\bibinfo {year}
  {2025})}\BibitemShut {NoStop}%
\bibitem [{\citenamefont {Kresse}\ and\ \citenamefont
  {Furthm{\"u}ller}(1996)}]{VASP19961}%
  \BibitemOpen
  \bibfield  {author} {\bibinfo {author} {\bibfnamefont {G.}~\bibnamefont
  {Kresse}}\ and\ \bibinfo {author} {\bibfnamefont {J.}~\bibnamefont
  {Furthm{\"u}ller}},\ }\href@noop {} {\bibfield  {journal} {\bibinfo
  {journal} {Physical Review B}\ }\textbf {\bibinfo {volume} {54}},\ \bibinfo
  {pages} {11169} (\bibinfo {year} {1996})}\BibitemShut {NoStop}%
\bibitem [{\citenamefont {Bl{\"o}chl}(1994)}]{PAW}%
  \BibitemOpen
  \bibfield  {author} {\bibinfo {author} {\bibfnamefont {P.~E.}\ \bibnamefont
  {Bl{\"o}chl}},\ }\href@noop {} {\bibfield  {journal} {\bibinfo  {journal}
  {Physical review B}\ }\textbf {\bibinfo {volume} {50}},\ \bibinfo {pages}
  {17953} (\bibinfo {year} {1994})}\BibitemShut {NoStop}%
\bibitem [{\citenamefont {Kresse}\ and\ \citenamefont {Joubert}(1999)}]{gga}%
  \BibitemOpen
  \bibfield  {author} {\bibinfo {author} {\bibfnamefont {G.}~\bibnamefont
  {Kresse}}\ and\ \bibinfo {author} {\bibfnamefont {D.}~\bibnamefont
  {Joubert}},\ }\href@noop {} {\bibfield  {journal} {\bibinfo  {journal}
  {Physical review b}\ }\textbf {\bibinfo {volume} {59}},\ \bibinfo {pages}
  {1758} (\bibinfo {year} {1999})}\BibitemShut {NoStop}%
\bibitem [{\citenamefont {Perdew}\ \emph {et~al.}(1996)\citenamefont {Perdew},
  \citenamefont {Burke},\ and\ \citenamefont {Ernzerhof}}]{pbe}%
  \BibitemOpen
  \bibfield  {author} {\bibinfo {author} {\bibfnamefont {J.~P.}\ \bibnamefont
  {Perdew}}, \bibinfo {author} {\bibfnamefont {K.}~\bibnamefont {Burke}},\ and\
  \bibinfo {author} {\bibfnamefont {M.}~\bibnamefont {Ernzerhof}},\ }\href@noop
  {} {\bibfield  {journal} {\bibinfo  {journal} {Physical review letters}\
  }\textbf {\bibinfo {volume} {77}},\ \bibinfo {pages} {3865} (\bibinfo {year}
  {1996})}\BibitemShut {NoStop}%
\bibitem [{\citenamefont {Grimme}\ \emph {et~al.}(2010)\citenamefont {Grimme},
  \citenamefont {Antony}, \citenamefont {Ehrlich},\ and\ \citenamefont
  {Krieg}}]{grimme2010consistent}%
  \BibitemOpen
  \bibfield  {author} {\bibinfo {author} {\bibfnamefont {S.}~\bibnamefont
  {Grimme}}, \bibinfo {author} {\bibfnamefont {J.}~\bibnamefont {Antony}},
  \bibinfo {author} {\bibfnamefont {S.}~\bibnamefont {Ehrlich}},\ and\ \bibinfo
  {author} {\bibfnamefont {H.}~\bibnamefont {Krieg}},\ }\href@noop {}
  {\bibfield  {journal} {\bibinfo  {journal} {The Journal of Chemical Physics}\
  }\textbf {\bibinfo {volume} {132}},\ \bibinfo {pages} {154104} (\bibinfo
  {year} {2010})}\BibitemShut {NoStop}%
\bibitem [{\citenamefont {Wisesa}\ \emph {et~al.}(2016)\citenamefont {Wisesa},
  \citenamefont {McGill},\ and\ \citenamefont {Mueller}}]{wisesa2016efficient}%
  \BibitemOpen
  \bibfield  {author} {\bibinfo {author} {\bibfnamefont {P.}~\bibnamefont
  {Wisesa}}, \bibinfo {author} {\bibfnamefont {K.~A.}\ \bibnamefont {McGill}},\
  and\ \bibinfo {author} {\bibfnamefont {T.}~\bibnamefont {Mueller}},\ }\href
  {https://doi.org/10.1103/PhysRevB.93.155109} {\bibfield  {journal} {\bibinfo
  {journal} {Physical Review B}\ }\textbf {\bibinfo {volume} {93}},\ \bibinfo
  {pages} {155109} (\bibinfo {year} {2016})}\BibitemShut {NoStop}%
\bibitem [{\citenamefont {Wang}\ \emph {et~al.}(2021)\citenamefont {Wang},
  \citenamefont {Wisesa}, \citenamefont {Balasubramanian}, \citenamefont
  {Dwaraknath},\ and\ \citenamefont {Mueller}}]{wang2021rapid}%
  \BibitemOpen
  \bibfield  {author} {\bibinfo {author} {\bibfnamefont {Y.}~\bibnamefont
  {Wang}}, \bibinfo {author} {\bibfnamefont {P.}~\bibnamefont {Wisesa}},
  \bibinfo {author} {\bibfnamefont {A.}~\bibnamefont {Balasubramanian}},
  \bibinfo {author} {\bibfnamefont {S.}~\bibnamefont {Dwaraknath}},\ and\
  \bibinfo {author} {\bibfnamefont {T.}~\bibnamefont {Mueller}},\ }\href
  {https://doi.org/10.1016/j.commatsci.2020.110100} {\bibfield  {journal}
  {\bibinfo  {journal} {Computational Materials Science}\ }\textbf {\bibinfo
  {volume} {187}},\ \bibinfo {pages} {110100} (\bibinfo {year}
  {2021})}\BibitemShut {NoStop}%
\bibitem [{\citenamefont {Weber}\ \emph {et~al.}(1989)\citenamefont {Weber},
  \citenamefont {Schulz}, \citenamefont {Schmitz}, \citenamefont {Granzin},\
  and\ \citenamefont {Siegert}}]{weber1989determination}%
  \BibitemOpen
  \bibfield  {author} {\bibinfo {author} {\bibfnamefont {H.~J.}\ \bibnamefont
  {Weber}}, \bibinfo {author} {\bibfnamefont {M.}~\bibnamefont {Schulz}},
  \bibinfo {author} {\bibfnamefont {S.}~\bibnamefont {Schmitz}}, \bibinfo
  {author} {\bibfnamefont {J.}~\bibnamefont {Granzin}},\ and\ \bibinfo {author}
  {\bibfnamefont {H.}~\bibnamefont {Siegert}},\ }\href@noop {} {\bibfield
  {journal} {\bibinfo  {journal} {Journal of Physics: Condensed Matter}\
  }\textbf {\bibinfo {volume} {1}},\ \bibinfo {pages} {8543} (\bibinfo {year}
  {1989})}\BibitemShut {NoStop}%
\bibitem [{\citenamefont {Boujelben}\ \emph {et~al.}(2007)\citenamefont
  {Boujelben}, \citenamefont {Toumi},\ and\ \citenamefont
  {Mhiri}}]{boujelben2007langbeinite}%
  \BibitemOpen
  \bibfield  {author} {\bibinfo {author} {\bibfnamefont {M.}~\bibnamefont
  {Boujelben}}, \bibinfo {author} {\bibfnamefont {M.}~\bibnamefont {Toumi}},\
  and\ \bibinfo {author} {\bibfnamefont {T.}~\bibnamefont {Mhiri}},\
  }\href@noop {} {\bibfield  {journal} {\bibinfo  {journal} {Structure
  Reports}\ }\textbf {\bibinfo {volume} {63}},\ \bibinfo {pages} {i157}
  (\bibinfo {year} {2007})}\BibitemShut {NoStop}%
\bibitem [{\citenamefont {Ilyukhin}\ and\ \citenamefont
  {Dzhurinskij}(1994)}]{ilyukhin1993crystal}%
  \BibitemOpen
  \bibfield  {author} {\bibinfo {author} {\bibfnamefont {A.}~\bibnamefont
  {Ilyukhin}}\ and\ \bibinfo {author} {\bibfnamefont {B.}~\bibnamefont
  {Dzhurinskij}},\ }\href@noop {} {\bibfield  {journal} {\bibinfo  {journal}
  {ChemInform}\ }\textbf {\bibinfo {volume} {25}} (\bibinfo {year}
  {1994})}\BibitemShut {NoStop}%
\bibitem [{\citenamefont {Goreinov}\ \emph {et~al.}(2010)\citenamefont
  {Goreinov}, \citenamefont {Oseledets}, \citenamefont {Savostyanov},
  \citenamefont {Tyrtyshnikov},\ and\ \citenamefont
  {Zamarashkin}}]{goreinov2010submatrix}%
  \BibitemOpen
  \bibfield  {author} {\bibinfo {author} {\bibfnamefont {S.~A.}\ \bibnamefont
  {Goreinov}}, \bibinfo {author} {\bibfnamefont {I.~V.}\ \bibnamefont
  {Oseledets}}, \bibinfo {author} {\bibfnamefont {D.~V.}\ \bibnamefont
  {Savostyanov}}, \bibinfo {author} {\bibfnamefont {E.~E.}\ \bibnamefont
  {Tyrtyshnikov}},\ and\ \bibinfo {author} {\bibfnamefont {N.~L.}\ \bibnamefont
  {Zamarashkin}},\ }in\ \href {https://doi.org/10.1142/9789812836021_0015}
  {\emph {\bibinfo {booktitle} {Matrix Methods: Theory, Algorithms and
  Applications}}}\ (\bibinfo  {publisher} {World Scientific},\ \bibinfo {year}
  {2010})\ pp.\ \bibinfo {pages} {247--256}\BibitemShut {NoStop}%
\bibitem [{\citenamefont {Parlinski}\ \emph {et~al.}(1997)\citenamefont
  {Parlinski}, \citenamefont {Li},\ and\ \citenamefont
  {Kawazoe}}]{supercell-method}%
  \BibitemOpen
  \bibfield  {author} {\bibinfo {author} {\bibfnamefont {K.}~\bibnamefont
  {Parlinski}}, \bibinfo {author} {\bibfnamefont {Z.~Q.}\ \bibnamefont {Li}},\
  and\ \bibinfo {author} {\bibfnamefont {Y.}~\bibnamefont {Kawazoe}},\
  }\href@noop {} {\bibfield  {journal} {\bibinfo  {journal} {Physical Review
  Letters}\ }\textbf {\bibinfo {volume} {78}},\ \bibinfo {pages} {4063}
  (\bibinfo {year} {1997})}\BibitemShut {NoStop}%
\bibitem [{\citenamefont {Togo}\ \emph {et~al.}(2023)\citenamefont {Togo},
  \citenamefont {Chaput}, \citenamefont {Tadano},\ and\ \citenamefont
  {Tanaka}}]{phonopy}%
  \BibitemOpen
  \bibfield  {author} {\bibinfo {author} {\bibfnamefont {A.}~\bibnamefont
  {Togo}}, \bibinfo {author} {\bibfnamefont {L.}~\bibnamefont {Chaput}},
  \bibinfo {author} {\bibfnamefont {T.}~\bibnamefont {Tadano}},\ and\ \bibinfo
  {author} {\bibfnamefont {I.}~\bibnamefont {Tanaka}},\ }\href
  {https://doi.org/10.1088/1361-648X/acd831} {\bibfield  {journal} {\bibinfo
  {journal} {Journal of Physics: Condensed Matter}\ }\textbf {\bibinfo {volume}
  {35}},\ \bibinfo {pages} {353001} (\bibinfo {year} {2023})}\BibitemShut
  {NoStop}%
\bibitem [{\citenamefont {Thompson}\ \emph {et~al.}(2022)\citenamefont
  {Thompson}, \citenamefont {Aktulga}, \citenamefont {Berger}, \citenamefont
  {Bolintineanu}, \citenamefont {Brown}, \citenamefont {Crozier}, \citenamefont
  {in~'t Veld}, \citenamefont {Kohlmeyer}, \citenamefont {Moore}, \citenamefont
  {Nguyen}, \citenamefont {Shan}, \citenamefont {Stevens}, \citenamefont
  {Tranchida}, \citenamefont {Trott},\ and\ \citenamefont {Plimpton}}]{LAMMPS}%
  \BibitemOpen
  \bibfield  {author} {\bibinfo {author} {\bibfnamefont {A.~P.}\ \bibnamefont
  {Thompson}}, \bibinfo {author} {\bibfnamefont {H.~M.}\ \bibnamefont
  {Aktulga}}, \bibinfo {author} {\bibfnamefont {R.}~\bibnamefont {Berger}},
  \bibinfo {author} {\bibfnamefont {D.~S.}\ \bibnamefont {Bolintineanu}},
  \bibinfo {author} {\bibfnamefont {W.~M.}\ \bibnamefont {Brown}}, \bibinfo
  {author} {\bibfnamefont {P.~S.}\ \bibnamefont {Crozier}}, \bibinfo {author}
  {\bibfnamefont {P.~J.}\ \bibnamefont {in~'t Veld}}, \bibinfo {author}
  {\bibfnamefont {A.}~\bibnamefont {Kohlmeyer}}, \bibinfo {author}
  {\bibfnamefont {S.~G.}\ \bibnamefont {Moore}}, \bibinfo {author}
  {\bibfnamefont {T.~D.}\ \bibnamefont {Nguyen}}, \bibinfo {author}
  {\bibfnamefont {R.}~\bibnamefont {Shan}}, \bibinfo {author} {\bibfnamefont
  {M.~J.}\ \bibnamefont {Stevens}}, \bibinfo {author} {\bibfnamefont
  {J.}~\bibnamefont {Tranchida}}, \bibinfo {author} {\bibfnamefont
  {C.}~\bibnamefont {Trott}},\ and\ \bibinfo {author} {\bibfnamefont {S.~J.}\
  \bibnamefont {Plimpton}},\ }\href {https://doi.org/10.1016/j.cpc.2021.108171}
  {\bibfield  {journal} {\bibinfo  {journal} {Comp. Phys. Comm.}\ }\textbf
  {\bibinfo {volume} {271}},\ \bibinfo {pages} {108171} (\bibinfo {year}
  {2022})}\BibitemShut {NoStop}%
\bibitem [{\citenamefont {Rasmussen}\ and\ \citenamefont
  {Williams}(2006)}]{gp_rasmussen}%
  \BibitemOpen
  \bibfield  {author} {\bibinfo {author} {\bibfnamefont {C.~E.}\ \bibnamefont
  {Rasmussen}}\ and\ \bibinfo {author} {\bibfnamefont {C.~K.~I.}\ \bibnamefont
  {Williams}},\ }\href@noop {} {\emph {\bibinfo {title} {Gaussian Processes for
  Machine Learning}}}\ (\bibinfo  {publisher} {MIT Press},\ \bibinfo {year}
  {2006})\BibitemShut {NoStop}%
\bibitem [{\citenamefont {Spirande}\ \emph {et~al.}(2025)\citenamefont
  {Spirande}, \citenamefont {Miryashkin}, \citenamefont {Kolmakov},\ and\
  \citenamefont {Shapeev}}]{spirande2025automated}%
  \BibitemOpen
  \bibfield  {author} {\bibinfo {author} {\bibfnamefont {E.}~\bibnamefont
  {Spirande}}, \bibinfo {author} {\bibfnamefont {T.}~\bibnamefont
  {Miryashkin}}, \bibinfo {author} {\bibfnamefont {A.}~\bibnamefont
  {Kolmakov}},\ and\ \bibinfo {author} {\bibfnamefont {A.}~\bibnamefont
  {Shapeev}},\ }\href {https://doi.org/10.1016/j.cocom.2025.e01163} {\bibfield
  {journal} {\bibinfo  {journal} {Computational Condensed Matter}\ }\textbf
  {\bibinfo {volume} {45}},\ \bibinfo {pages} {e01163} (\bibinfo {year}
  {2025})}\BibitemShut {NoStop}%
\bibitem [{\citenamefont {Miryashkin}\ \emph {et~al.}(2023)\citenamefont
  {Miryashkin}, \citenamefont {Klimanova}, \citenamefont {Ladygin},\ and\
  \citenamefont {Shapeev}}]{miryashkin2023bayesian}%
  \BibitemOpen
  \bibfield  {author} {\bibinfo {author} {\bibfnamefont {T.}~\bibnamefont
  {Miryashkin}}, \bibinfo {author} {\bibfnamefont {O.}~\bibnamefont
  {Klimanova}}, \bibinfo {author} {\bibfnamefont {V.}~\bibnamefont {Ladygin}},\
  and\ \bibinfo {author} {\bibfnamefont {A.}~\bibnamefont {Shapeev}},\ }\href
  {https://doi.org/10.1103/PhysRevB.108.174103} {\bibfield  {journal} {\bibinfo
   {journal} {Physical Review B}\ }\textbf {\bibinfo {volume} {108}},\ \bibinfo
  {pages} {174103} (\bibinfo {year} {2023})}\BibitemShut {NoStop}%
\bibitem [{\citenamefont {Tai}\ \emph {et~al.}(2025)\citenamefont {Tai},
  \citenamefont {Wang}, \citenamefont {Cheng},\ and\ \citenamefont
  {Chen}}]{lammps-mlip-gk}%
  \BibitemOpen
  \bibfield  {author} {\bibinfo {author} {\bibfnamefont {S.~T.}\ \bibnamefont
  {Tai}}, \bibinfo {author} {\bibfnamefont {C.}~\bibnamefont {Wang}}, \bibinfo
  {author} {\bibfnamefont {R.}~\bibnamefont {Cheng}},\ and\ \bibinfo {author}
  {\bibfnamefont {Y.}~\bibnamefont {Chen}},\ }\href@noop {} {\bibfield
  {journal} {\bibinfo  {journal} {Journal of Chemical Theory and Computation}\
  }\textbf {\bibinfo {volume} {21}},\ \bibinfo {pages} {3649} (\bibinfo {year}
  {2025})}\BibitemShut {NoStop}%
\bibitem [{\citenamefont {Blum}\ \emph {et~al.}(2009)\citenamefont {Blum},
  \citenamefont {Gehrke}, \citenamefont {Hanke}, \citenamefont {Havu},
  \citenamefont {Havu}, \citenamefont {Ren}, \citenamefont {Reuter},\ and\
  \citenamefont {Scheffler}}]{fhiaims}%
  \BibitemOpen
  \bibfield  {author} {\bibinfo {author} {\bibfnamefont {V.}~\bibnamefont
  {Blum}}, \bibinfo {author} {\bibfnamefont {R.}~\bibnamefont {Gehrke}},
  \bibinfo {author} {\bibfnamefont {F.}~\bibnamefont {Hanke}}, \bibinfo
  {author} {\bibfnamefont {P.}~\bibnamefont {Havu}}, \bibinfo {author}
  {\bibfnamefont {V.}~\bibnamefont {Havu}}, \bibinfo {author} {\bibfnamefont
  {X.}~\bibnamefont {Ren}}, \bibinfo {author} {\bibfnamefont {K.}~\bibnamefont
  {Reuter}},\ and\ \bibinfo {author} {\bibfnamefont {M.}~\bibnamefont
  {Scheffler}},\ }\href@noop {} {\bibfield  {journal} {\bibinfo  {journal}
  {Computer Physics Communications}\ }\textbf {\bibinfo {volume} {180}},\
  \bibinfo {pages} {2175} (\bibinfo {year} {2009})}\BibitemShut {NoStop}%
\bibitem [{\citenamefont {Miglio}\ \emph {et~al.}(2020)\citenamefont {Miglio},
  \citenamefont {Brousseau-Couture}, \citenamefont {Godbout}, \citenamefont
  {Antonius}, \citenamefont {Chan}, \citenamefont {Louie}, \citenamefont
  {C{\^o}t{\'e}}, \citenamefont {Giantomassi},\ and\ \citenamefont
  {Gonze}}]{gonze_bgr_database}%
  \BibitemOpen
  \bibfield  {author} {\bibinfo {author} {\bibfnamefont {A.}~\bibnamefont
  {Miglio}}, \bibinfo {author} {\bibfnamefont {V.}~\bibnamefont
  {Brousseau-Couture}}, \bibinfo {author} {\bibfnamefont {E.}~\bibnamefont
  {Godbout}}, \bibinfo {author} {\bibfnamefont {G.}~\bibnamefont {Antonius}},
  \bibinfo {author} {\bibfnamefont {Y.-H.}\ \bibnamefont {Chan}}, \bibinfo
  {author} {\bibfnamefont {S.~G.}\ \bibnamefont {Louie}}, \bibinfo {author}
  {\bibfnamefont {M.}~\bibnamefont {C{\^o}t{\'e}}}, \bibinfo {author}
  {\bibfnamefont {M.}~\bibnamefont {Giantomassi}},\ and\ \bibinfo {author}
  {\bibfnamefont {X.}~\bibnamefont {Gonze}},\ }\href@noop {} {\bibfield
  {journal} {\bibinfo  {journal} {npj Computational Materials}\ }\textbf
  {\bibinfo {volume} {6}},\ \bibinfo {pages} {167} (\bibinfo {year}
  {2020})}\BibitemShut {NoStop}%
\end{thebibliography}%
\end{document}


\maketitle
\tableofcontents
\clearpage

\section{Relaxed lattice parameters}

Table~\ref{tab:relaxed_lattice_parameters} compares the lattice parameters and unit-cell volumes of $\beta$-Rb$_2$SO$_4$, Rb$_2$Ca$_2$(SO$_4$)$_3$, and Ba$_3$Lu(BO$_3$)$_3$ obtained by 0~K relaxations with DFT (PBE+D3) and the MTPs with x-ray diffraction data measured at 298~K~\cite{shablinskii2025structural,biryukov2023thermal}. The MTPs reproduce the DFT lattice parameters and volumes within 0.4\%, and the DFT volumes exceed the experimental ones by 0.8--1.5\%.

\begin{table}[!h]
\centering
\caption{Lattice parameters and unit-cell volumes obtained from DFT (PBE+D3) and MTP relaxations at 0~K, compared with x-ray diffraction data at 298~K~\cite{shablinskii2025structural,biryukov2023thermal}. Lengths are in \AA{} and volumes in \AA$^3$. All cell angles agree with the crystallographic values within $0.003^\circ$: $\alpha=\beta=\gamma=90^\circ$ for the sulfates and $\alpha=\beta=90^\circ$, $\gamma=120^\circ$ for the borate.}
\label{tab:relaxed_lattice_parameters}
\begin{tabular}{llrrr}
\toprule
Compound & Parameter & DFT (PBE+D3) & MTP & Experiment \\
\midrule
\multirow{4}{*}{$\beta$-Rb$_2$SO$_4$}
 & $a$ & 5.9935 & 5.9926 & 5.9777 \\
 & $b$ & 10.4785 & 10.4763 & 10.4388 \\
 & $c$ & 7.8333 & 7.8388 & 7.8205 \\
 & $V$ & 491.95 & 492.12 & 488.00 \\
\midrule
\multirow{2}{*}{Rb$_2$Ca$_2$(SO$_4$)$_3$}
 & $a$ & 10.6194 & 10.6255 & 10.5679 \\
 & $V$ & 1197.56 & 1199.64 & 1180.24 \\
\midrule
\multirow{3}{*}{Ba$_3$Lu(BO$_3$)$_3$}
 & $a$ & 9.4222 & 9.4217 & 9.3882 \\
 & $c$ & 17.4546 & 17.5251 & 17.4375 \\
 & $V$ & 1341.97 & 1347.25 & 1331.01 \\
\bottomrule
\end{tabular}
\end{table}

\section{Moment tensor potentials}

\subsection{Potential form and active learning}

In the moment tensor potential (MTP)~\cite{mtp,mlip}, the energy of a configuration is a sum of contributions of the local atomic environments $\mathcal{N}_i$,
\begin{equation}
E_{\mathrm{MTP}}=\sum_{i=1}^{N}V(\mathcal{N}_i),
\qquad
V(\mathcal{N}_i)=\sum_{\alpha}\xi_{\alpha}B_{\alpha}(\mathcal{N}_i),
\end{equation}
where the basis functions $B_\alpha$ are scalar contractions of the moment tensor descriptors
\begin{equation}
M_{\mu,\nu}(\mathcal{N}_i)=\sum_{j}f_{\mu}(|\mathbf r_{ij}|,z_i,z_j)\,\underbrace{\mathbf r_{ij}\otimes\cdots\otimes\mathbf r_{ij}}_{\nu\ \text{times}}.
\end{equation}
Here $\mathbf r_{ij}$ is the vector from atom $i$ to a neighbor $j$ within the cutoff radius $R_{\mathrm{cut}}$, and $z_i$ and $z_j$ are the atomic types. The radial functions $f_\mu(r,z_i,z_j)=\sum_k c^{(k)}_{\mu,z_i,z_j}Q^{(k)}(r)$ are expansions in Chebyshev polynomials multiplied by $(R_{\mathrm{cut}}-r)^2$, so that they vanish smoothly at $R_{\mathrm{cut}}$. The complexity of the potential is controlled by its level: the level of $M_{\mu,\nu}$ is defined as $2+4\mu+\nu$, the level of a contraction of several moment tensors is the sum of their levels, and an MTP of level $\mathrm{lev}_{\max}$ contains all basis functions with level not exceeding $\mathrm{lev}_{\max}$. The fitting parameters are the linear coefficients $\xi_\alpha$ and the radial coefficients $c^{(k)}_{\mu,z_i,z_j}$, which were fitted to DFT energies, forces, and stresses. All potentials in this work are of level 16; their cutoff radii and numbers of parameters are listed in Table~\ref{tab:mtp_training}.

The training sets were constructed by active learning based on the D-optimality criterion~\cite{podryabinkin2017active,gubaev2019accelerating}. For each configuration visited in MD, the extrapolation grade $\gamma$ was evaluated as the largest absolute coefficient obtained when the configuration's vector of basis-function values is expressed through those of the active set, i.e., the configurations selected from the training set by the MaxVol algorithm~\cite{goreinov2010submatrix}; $\gamma\le1$ corresponds to interpolation. Configurations with $\gamma>\gamma_{\mathrm{select}}=2$ were added to a candidate set, and an MD run was interrupted when $\gamma>\gamma_{\mathrm{break}}=10$. The most informative candidates were selected with MaxVol, their energies, forces, and stresses were computed with DFT, the data were added to the training set, and the MTP was refitted. The cycle was repeated until the MD simulations produced no configurations with $\gamma>\gamma_{\mathrm{select}}$.

\subsection{Reference calculations and training sets}

The reference DFT calculations were performed with VASP using the PAW datasets B, O, S, Rb\_sv, Ca\_sv, Ba\_sv, and Lu\_3, a plane-wave cutoff of 520~eV, and an electronic convergence criterion of $3\times10^{-6}\sqrt{N}$~eV, where $N$ is the number of atoms in the cell. The $k$-point grids were generated with the generalized Monkhorst--Pack grid generator of Refs.~\cite{wisesa2016efficient, wang2021rapid} using a minimum real-space distance of 21.7~\AA.

\begin{figure*}[t!]
    \centering
    \includegraphics[width=\textwidth]{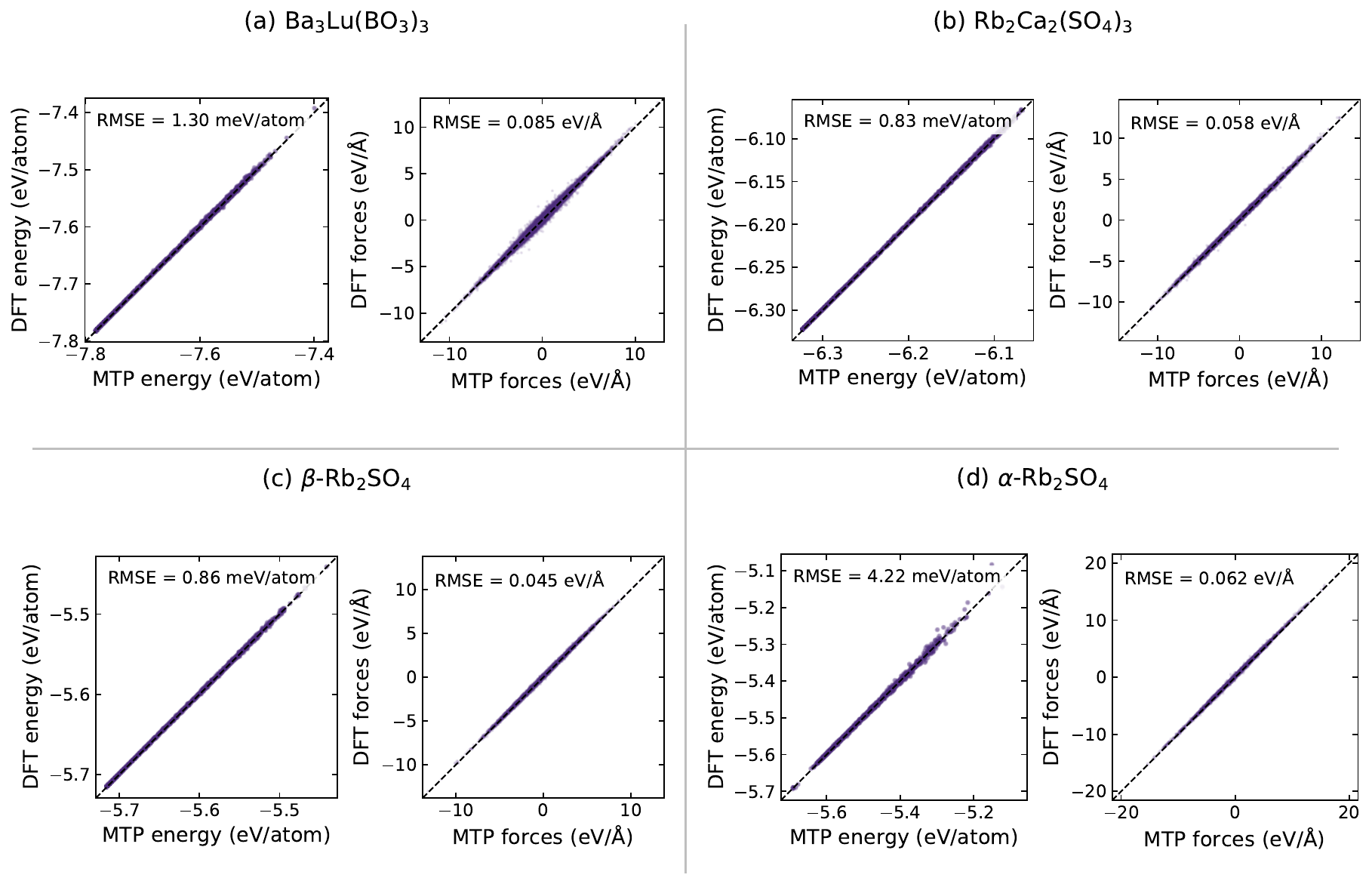}
    \caption{Comparison between DFT- and MTP-calculated energies (left panels) and forces (right panels) for (a) Ba$_3$Lu(BO$_3$)$_3$, (b) Rb$_2$Ca$_2$(SO$_4$)$_3$, (c) $\beta$-Rb$_2$SO$_4$, and (d) $\alpha$-Rb$_2$SO$_4$. The dashed black lines represent perfect agreement.}
    \label{fig:mtp_training_parity}
\end{figure*}

The starting structures were the experimental structures of hexagonal Ba$_3$Lu(BO$_3$)$_3$ ($P6_3cm$, $a = 9.382$~\AA, $c = 17.421$~\AA, $V = 1327.99$~\AA$^3$, $Z = 6$)~\cite{ilyukhin1993crystal}, cubic langbeinite-type Rb$_2$Ca$_2$(SO$_4$)$_3$ ($P2_13$, $a = 10.553(3)$~\AA, $V = 1175.2(6)$~\AA$^3$, $Z = 4$)~\cite{boujelben2007langbeinite}, orthorhombic arcanite-type $\beta$-Rb$_2$SO$_4$ ($Pmcn$ setting, $a = 5.9694(5)$~\AA, $b = 10.4255(7)$~\AA, $c = 7.8128(5)$~\AA, $V = 486.22$~\AA$^3$, $Z = 4$)~\cite{weber1989determination}, and hexagonal $\alpha$-Rb$_2$SO$_4$ refined at 700 and 1000~$^\circ$C ($P6_3/mmc$, $a = 6.13513(2)$~\AA, $c = 8.49002(5)$~\AA, $V = 277.749(4)$~\AA$^3$, $Z = 2$)~\cite{shablinskii2025structural}. For $\alpha$-Rb$_2$SO$_4$, fully occupied structures corresponding to the apex- and edge-oriented models of the SO$_4$ orientational disorder were reconstructed, and supercells in which each SO$_4$ group was randomly assigned one of the allowed orientations were generated. The structures were relaxed with DFT, and the short $NPT$ AIMD simulations listed in Table~\ref{tab:aimd} provided the initial training sets. The MTP levels and the final training-set sizes are given in Table~\ref{tab:mtp_training}, and the DFT--MTP parity plots for the training sets are shown in Fig.~\ref{fig:mtp_training_parity}. The larger energy error of the $\alpha$-Rb$_2$SO$_4$ potential is likely related to the wider range of configurations in its training set, which includes both apex- and edge-oriented structures and different realizations of the orientational disorder.

\begin{table}[htbp]
\centering
\small
\setlength{\tabcolsep}{4pt}
\caption{Levels, cutoff radii, numbers of fitting parameters, and training-set sizes of the MTPs.}
\label{tab:mtp_training}
\begin{tabular}{lrrrr}
\toprule
Compound & Level & $R_{\mathrm{cut}}$ (\AA) & Fitting parameters & Training configurations \\
\midrule
$\beta$-Rb$_2$SO$_4$ & 16 & 5.5 & 383 & 1246 \\
$\alpha$-Rb$_2$SO$_4$ & 16 & 5.5 & 383 & 1415  \\
Rb$_2$Ca$_2$(SO$_4$)$_3$ & 16 & 5.5 & 608 & 1523  \\
Ba$_3$Lu(BO$_3$)$_3$ & 16 & 5.0 & 608 & 1846 \\
\bottomrule
\end{tabular}
\end{table}

\begin{table}[htbp]
\centering
\small
\setlength{\tabcolsep}{4pt}
\caption{$NPT$ AIMD simulations used to generate the initial training sets: temperatures, numbers of MD steps, and simulation cells (numbers of atoms in parentheses).}
\label{tab:aimd}
\begin{tabular}{llll}
\toprule
Compound & $T$ (K) & Steps & Simulation cells \\
\midrule
$\beta$-Rb$_2$SO$_4$ & 800 & 200 & $1\times1\times1$ (28) \\
$\alpha$-Rb$_2$SO$_4$ & 1000 & 50 & $1\times1\times1$ (14), $2\times1\times1$ (28), $1\times2\times1$ (28), $2\times2\times1$ (56) \\
Rb$_2$Ca$_2$(SO$_4$)$_3$ & 100, 800 & 200 & $1\times1\times1$ (76) \\
Ba$_3$Lu(BO$_3$)$_3$ & 300, 900 & 200 & $1\times1\times1$ (96) \\
\bottomrule
\end{tabular}
\end{table}

\clearpage

\section{Stochastic TDEP calculations}\label{sec:stdep_si}

\begin{figure}[h!]
	\centering
	\begin{minipage}[h]{\linewidth}
		\center{\includegraphics[trim={0cm 0cm 0cm 0cm},clip, width=1\linewidth]{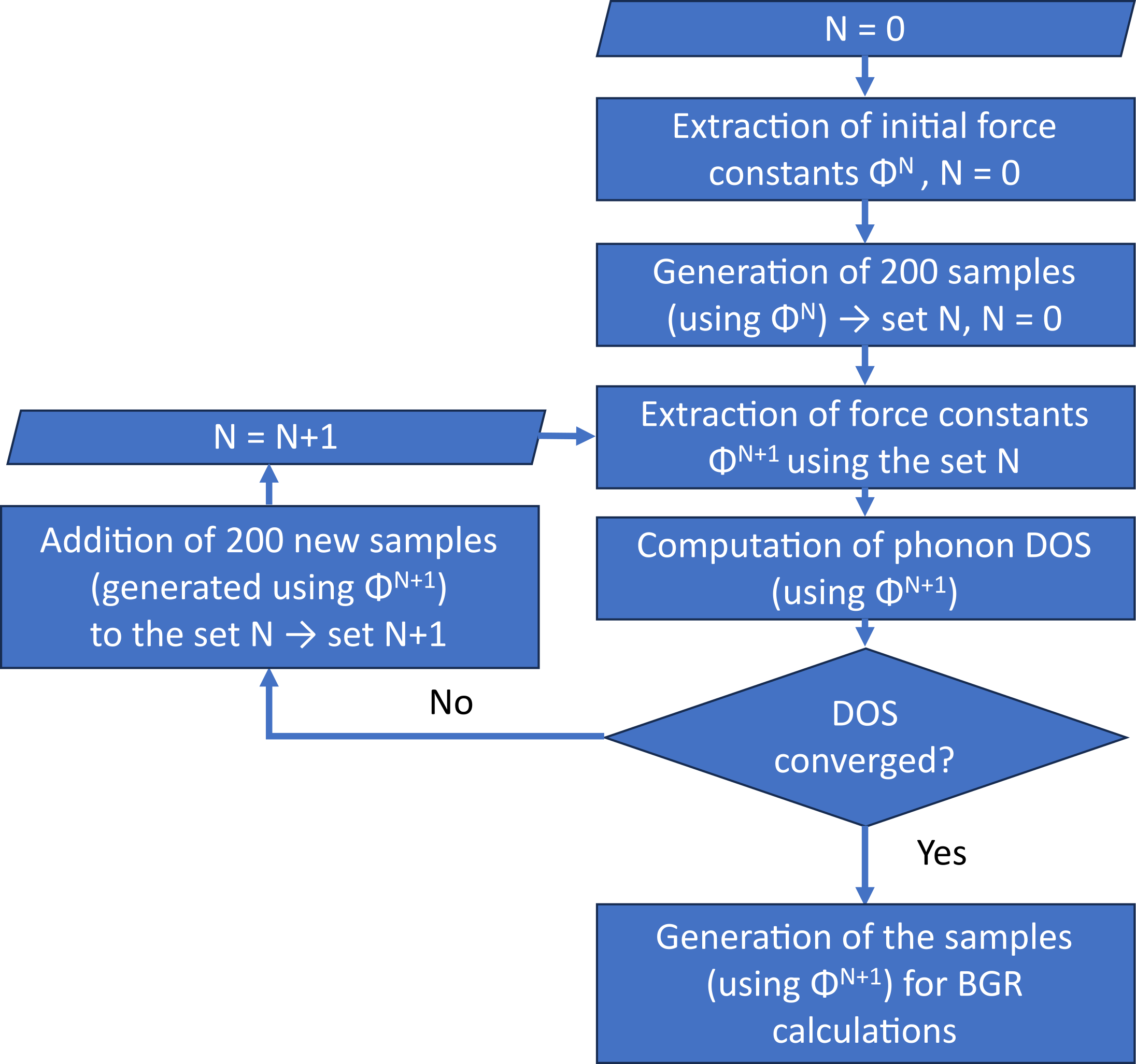}} \\ 
	\end{minipage}

    \caption{Schematic representation of the sTDEP self-consistent loop.}
    \label{fig:scheme_stdep}
\end{figure}

We used sTDEP to obtain the finite-temperature effective force constants, from which the phonon dispersions and the anharmonicity measure $\sigma^A$ were computed. The self-consistent loop is shown schematically in Fig.~\ref{fig:scheme_stdep}. Only second-order force constants were fitted, using for each compound the largest pair cutoff compatible with the supercell (Table~\ref{tab:stdep}).

\begin{table}[htbp]
\centering
\small
\caption{Supercells and cutoff radii of the second-order force constants used in the sTDEP calculations.}
\label{tab:stdep}
\begin{tabular}{lccc}
\toprule
Compound & Supercell & Atoms & Cutoff (\AA) \\
\midrule
$\beta$-Rb$_2$SO$_4$ & $2\times3\times2$ & 336 & 7.84 \\
$\alpha$-Rb$_2$SO$_4$ & $3\times4\times3$ & 504 & 7.85 \\
Rb$_2$Ca$_2$(SO$_4$)$_3$ & $2\times2\times2$ & 608 & 10.63 \\
Ba$_3$Lu(BO$_3$)$_3$ & $2\times2\times1$ & 384 & 8.20 \\
\bottomrule
\end{tabular}
\end{table}

Each sTDEP loop started from a 125~ps $NVT$ MD simulation with the MTP in LAMMPS~\cite{LAMMPS} (25~ps of equilibration and 100~ps of production, timestep 1~fs). Forces and displacements of 1001 frames separated by 100~fs were extracted from the trajectory, and initial effective force constants were obtained by minimizing $\Delta F$ [Eq.~(\ref{eq:tdep_fit_si})],
\begin{equation}
\Delta F=\frac{1}{N_t}\sum_{t=1}^{N_t}\left\|\mathbf F_t+\bar{\Phi}\,\mathbf U_t\right\|_2^2,
\label{eq:tdep_fit_si}
\end{equation}
by linear least-squares regression~\cite{linear_regres}. Phonon frequencies $\omega_s$, eigenvectors $\epsilon^{is}_\alpha$, and the phonon density of states (DOS) were then computed from these force constants, and 200 supercells with displaced atoms were generated by stochastic sampling,
\begin{equation}
u^{i}_{\alpha} = \sum_{s=1}^{3N_a}\epsilon^{is}_\alpha A_{is}\sqrt{-2\ln\xi_{1,s}}\,\sin(2\pi\xi_{2,s}),
\label{eq:tdep_displ_si}
\end{equation}
where $u^{i}_{\alpha}$ is the displacement of atom $i$ along direction $\alpha$, $A_{is}$ is the mode amplitude defined in the main text, and $\xi_{1,s}$ and $\xi_{2,s}$ are independent random numbers uniformly distributed in $(0,1]$. In the classical calculations, $A_{is}=\omega_s^{-1}\sqrt{k_{\mathrm B}T/m_i}$. The MTP forces of these samples were used to refit the force constants. In each subsequent iteration, another 200 samples were generated with the updated force constants and added to the existing set, and the force constants were refitted on the enlarged set.

The iterations were continued until the phonon DOS converged. Convergence was monitored by the cosine distance between the DOS obtained in two subsequent iterations,
\begin{equation}
D[{\rm DOS}(N), {\rm DOS}(N-1)] = 1 - \frac{{\rm DOS}(N)\cdot {\rm DOS}(N-1)}{\|{\rm DOS}(N)\|\,\|{\rm DOS}(N-1)\|},
\end{equation}
where ${\rm DOS}(N)=({\rm DOS}_1(N),\ldots,{\rm DOS}_K(N))$ is the vector of DOS values on a frequency grid obtained in the $N$-th iteration and $\|{\rm DOS}(N)\|=\big[\sum_{k=1}^{K}{\rm DOS}_k(N)^2\big]^{1/2}$. The force constants were considered converged when $D$ dropped below $10^{-2}$.

\begin{figure*}[h!]
	\centering
	\begin{minipage}[h]{0.49\linewidth}
		\center{\includegraphics[trim={0cm 0cm 0cm 0cm},clip, width=1\linewidth]{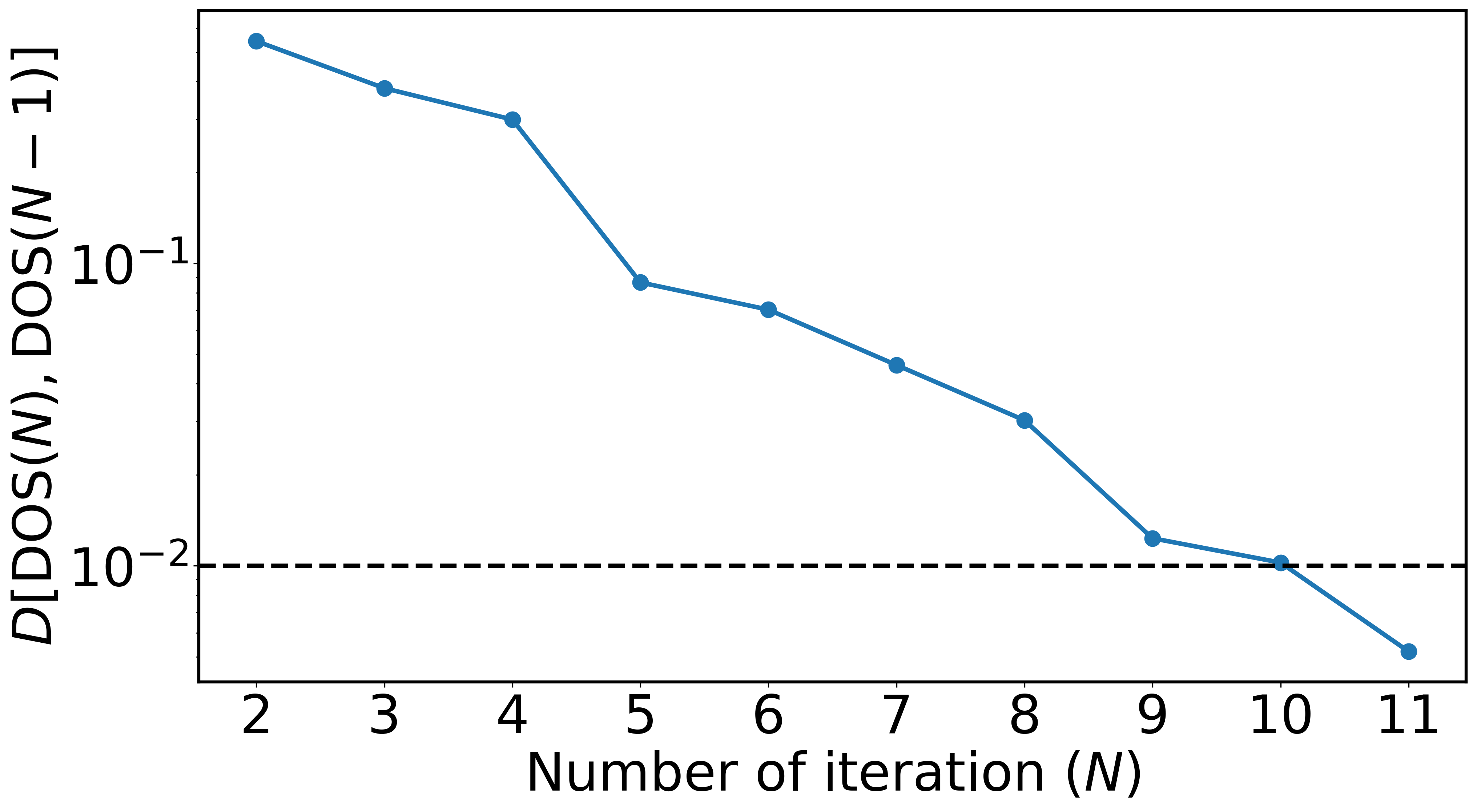}} \\ (a)
	\end{minipage}
    \begin{minipage}[h]{0.49\linewidth}
		\center{\includegraphics[trim={0cm 0cm 0cm 0cm},clip, width=1\linewidth]{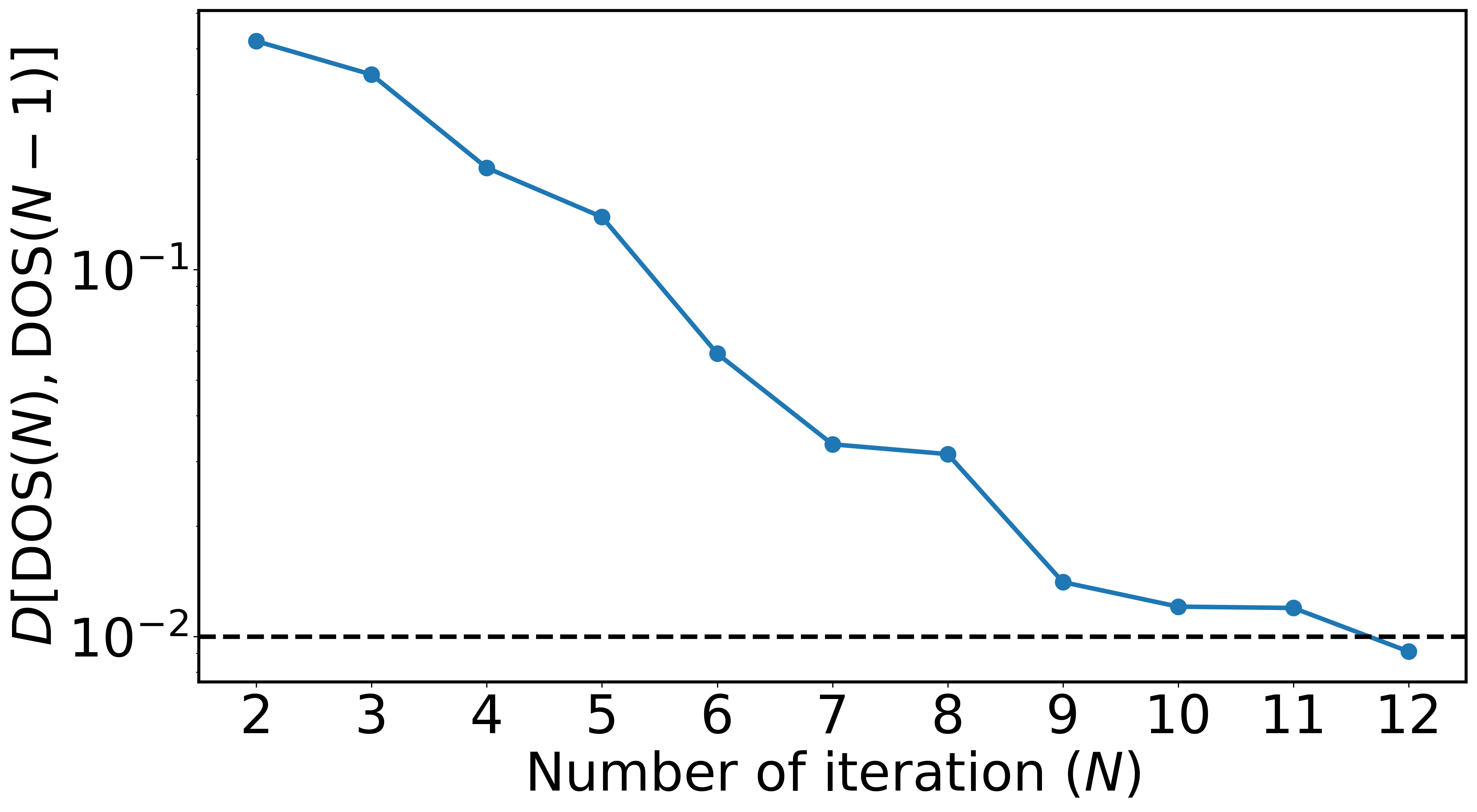}} \\ (b)
	\end{minipage}
    \begin{minipage}[h]{0.49\linewidth}
		\center{\includegraphics[trim={0cm 0cm 0cm 0cm},clip, width=1\linewidth]{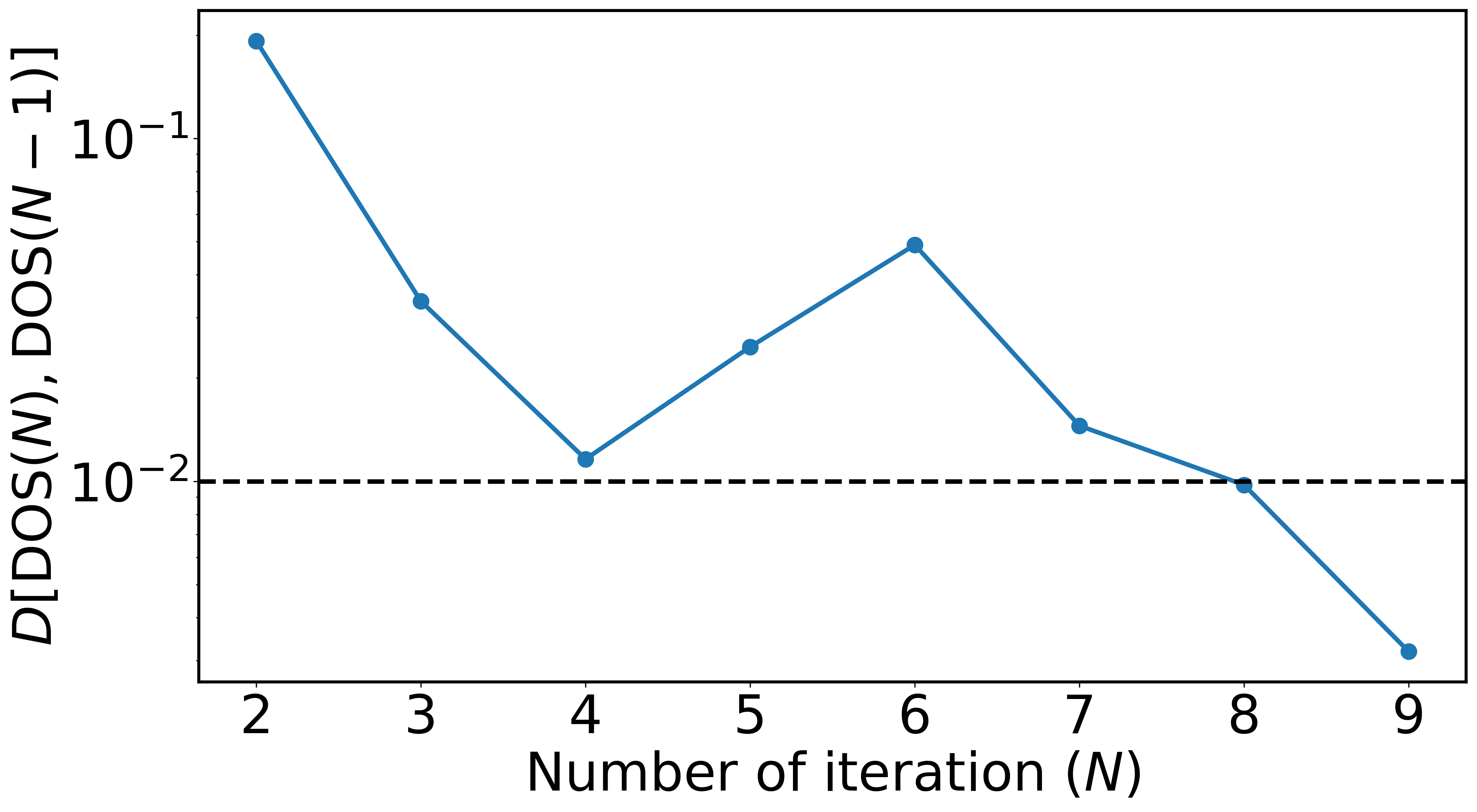}} \\ (c)
	\end{minipage}
    
    \caption{Cosine distance $D[{\rm DOS}(N),{\rm DOS}(N-1)]$ between the phonon DOS of consecutive classical sTDEP iterations for Rb$_2$Ca$_2$(SO$_4$)$_3$ at (a) 250~K, (b) 450~K, and (c) 1050~K. Dashed lines mark the convergence threshold $D=10^{-2}$.}
    \label{fig:conv_stdep_rbca_cl}
\end{figure*}

\begin{figure*}[h!]
	\centering
	\begin{minipage}[h]{0.49\linewidth}
		\center{\includegraphics[trim={0cm 0cm 0cm 0cm},clip, width=1\linewidth]{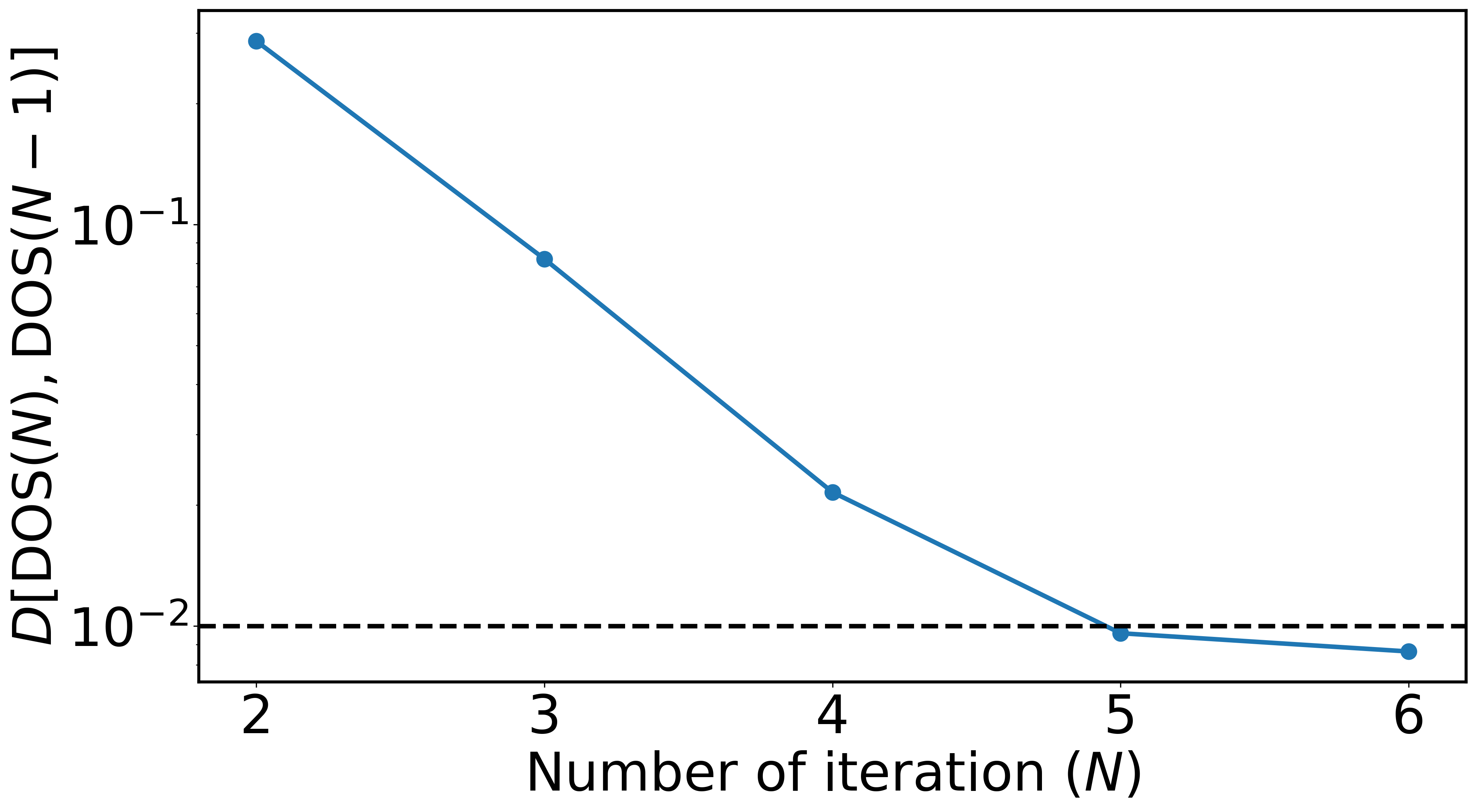}} \\ (a)
	\end{minipage}
    \begin{minipage}[h]{0.49\linewidth}
		\center{\includegraphics[trim={0cm 0cm 0cm 0cm},clip, width=1\linewidth]{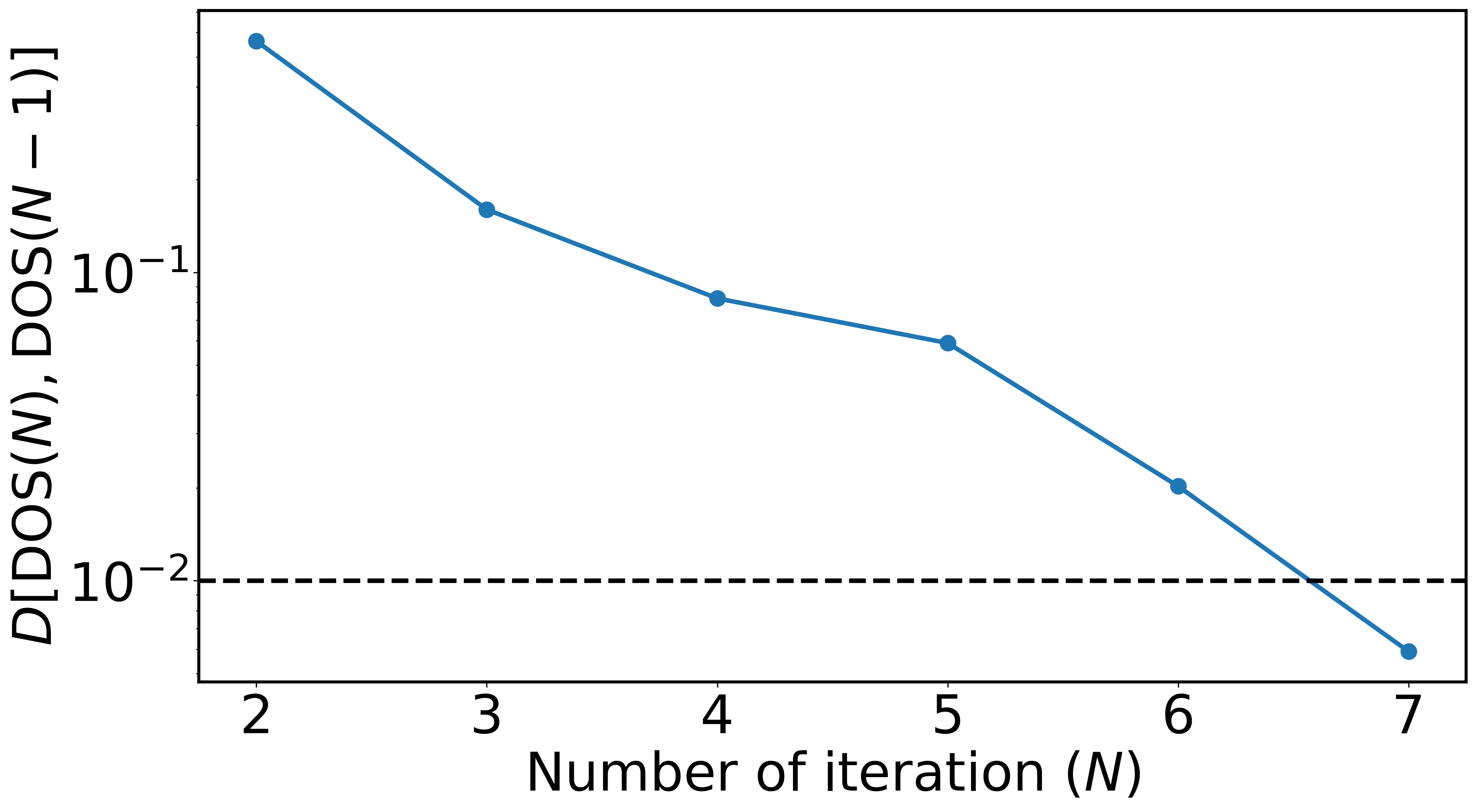}} \\ (b)
	\end{minipage}
    \begin{minipage}[h]{0.49\linewidth}
		\center{\includegraphics[trim={0cm 0cm 0cm 0cm},clip, width=1\linewidth]{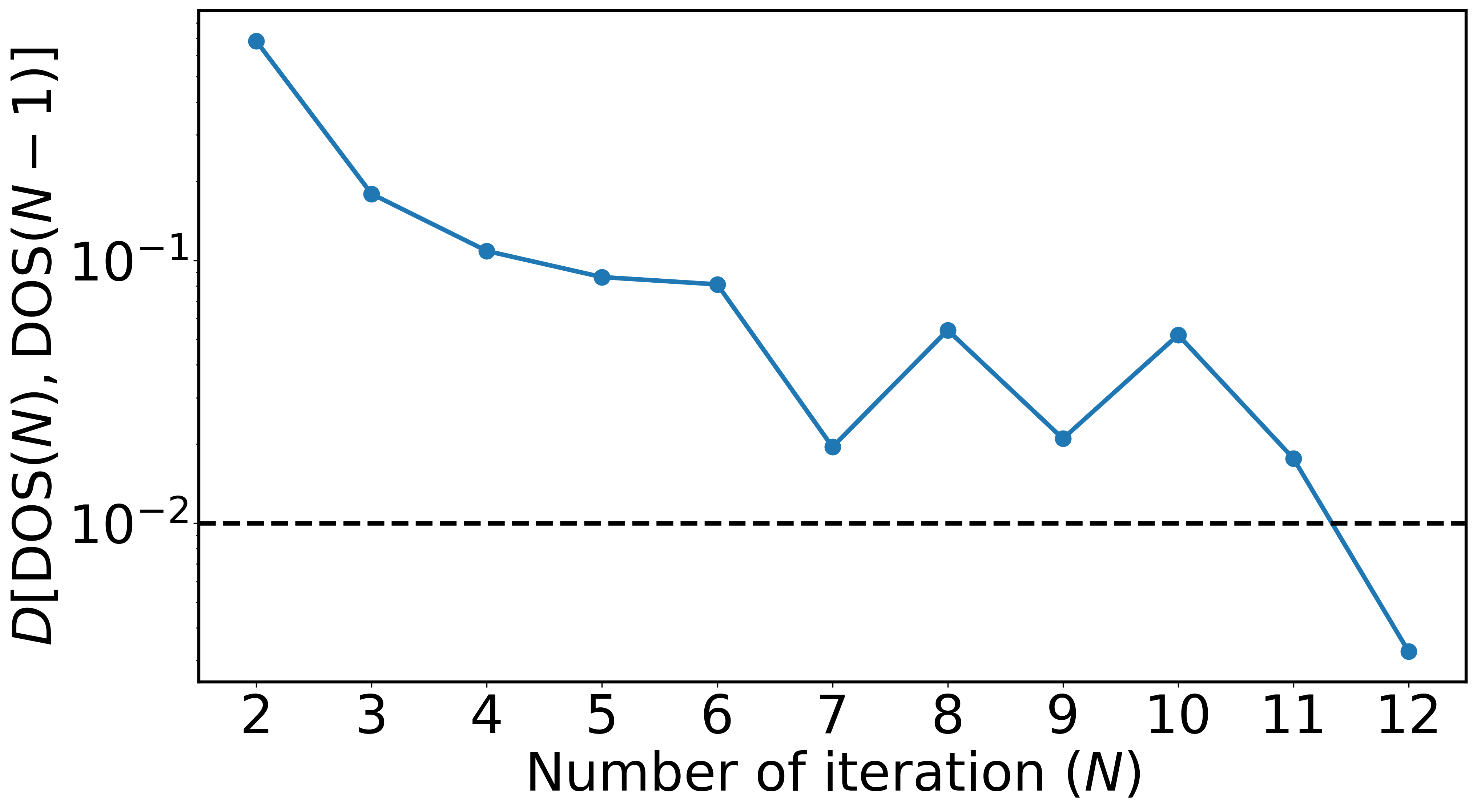}} \\ (c)
	\end{minipage}
   
    \caption{Cosine distance $D[{\rm DOS}(N),{\rm DOS}(N-1)]$ between the phonon DOS of consecutive classical sTDEP iterations for $\beta$-Rb$_2$SO$_4$ at (a) 250~K, (b) 450~K, and (c) 650~K. Dashed lines mark the convergence threshold $D=10^{-2}$.}
    \label{fig:conv_stdep_rbso_cl}
\end{figure*}

\begin{figure*}[t!]
	\centering
	\begin{minipage}[h]{1\linewidth}
		\center{\includegraphics[trim={0cm 0cm 0cm 0cm},clip, width=1\linewidth]{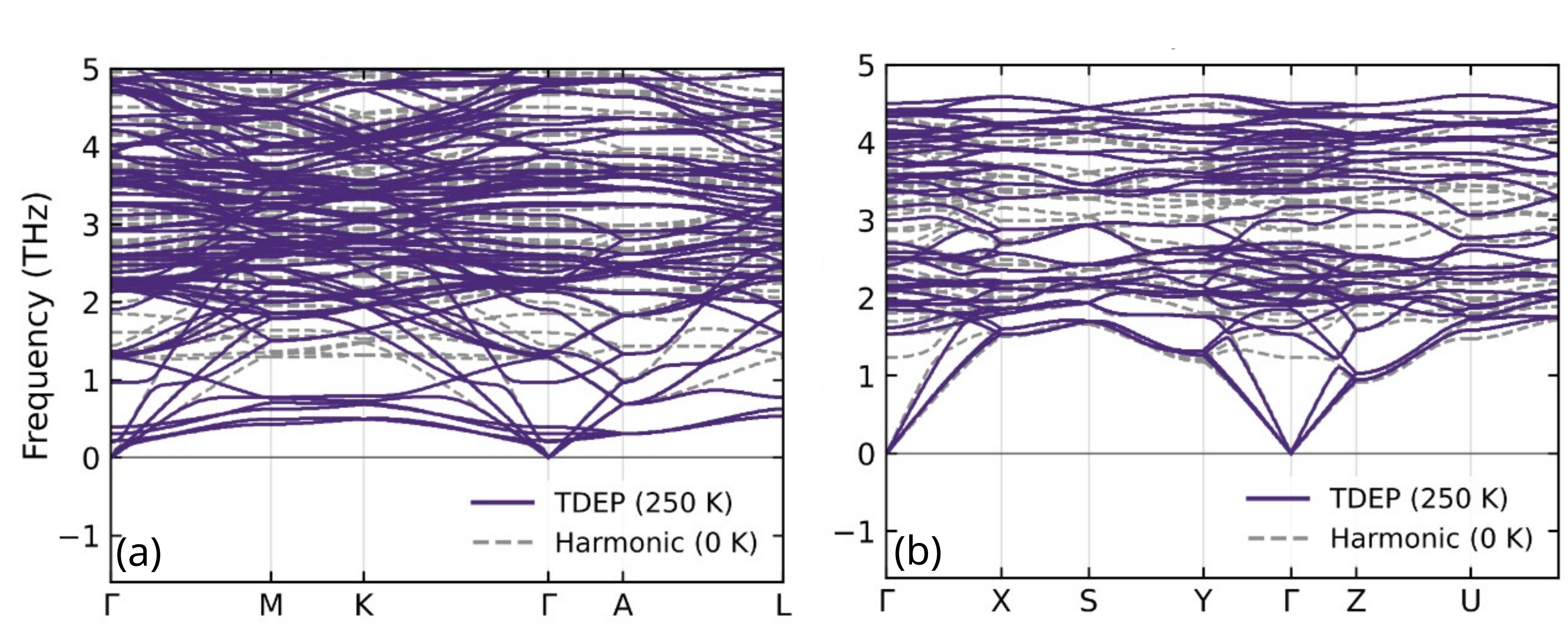}}
	\end{minipage}
    \caption{Low-frequency phonon dispersions of (a) Ba$_3$Lu(BO$_3$)$_3$ and (b) $\beta$-Rb$_2$SO$_4$ in the harmonic approximation (dashed) and from classical sTDEP at 250~K (solid).}
    \label{fig:phonons_si}
\end{figure*}

The convergence of the classical sTDEP loops is shown in Figs.~\ref{fig:conv_stdep_rbca_cl} and \ref{fig:conv_stdep_rbso_cl} for Rb$_2$Ca$_2$(SO$_4$)$_3$ and $\beta$-Rb$_2$SO$_4$. Figure~\ref{fig:phonons_si} compares the harmonic and sTDEP phonon dispersions of Ba$_3$Lu(BO$_3$)$_3$ and $\beta$-Rb$_2$SO$_4$ at 250~K; the corresponding results for Rb$_2$Ca$_2$(SO$_4$)$_3$ and $\alpha$-Rb$_2$SO$_4$ are shown in the main text.

\clearpage

\section{Thermal expansion and structural analysis}

The $NPT$ calculations used $2\times2\times2$, $3\times3\times2$, and $4\times4\times3$ supercells for Ba$_3$Lu(BO$_3$)$_3$; $2\times2\times2$, $3\times3\times3$, and $4\times4\times4$ for Rb$_2$Ca$_2$(SO$_4$)$_3$; $3\times2\times4$, $4\times3\times5$, $5\times4\times6$, and $5\times4\times7$ for $\beta$-Rb$_2$SO$_4$; and $5\times5\times4$, $6\times6\times4$, and $6\times6\times5$ for $\alpha$-Rb$_2$SO$_4$. The $\beta$-phase repeats refer to the simulation-cell axis order; its first and third axes are interchanged for comparison with the experimental crystallographic setting.

All $NPT$ simulations used a timestep of 1~fs. For each compound, temperature, and supercell size, the lattice parameters, volume, and structural quantities are first averaged over 30\,000 production steps (30~ps) following 20\,000 equilibration steps (20~ps) within each MD trajectory. The resulting values are then averaged over eight independent trajectories. For scalar structural quantities, the uncertainties supplied to GPR are the standard errors of the trajectory averages, $s/\sqrt{8}$, where $s$ is their sample standard deviation.

\subsection{Gaussian process regression}\label{sec:gpr}

Each lattice parameter, the unit-cell volume, and each structural quantity were fitted independently as functions of temperature $T$ and inverse number of atoms $N^{-1}$ using GPR~\cite{gp_rasmussen} with the kernel~\cite{spirande2025automated,miryashkin2023bayesian}
\begin{equation}
\begin{split}
k\left((T_1,N_1), (T_2,N_2)\right)
={}&\theta_0^2+\theta_1^2
\exp\!\left[-\frac{(T_1-T_2)^2}{2\theta_2^2}\right]\\
&\times\exp\!\left[-\frac{1}{2\theta_3^2}
\left(\frac{1}{N_1}-\frac{1}{N_2}\right)^2\right],
\end{split}
\label{eq:gp_kernel_si}
\end{equation}
where $\theta_0,\ldots,\theta_3$ are hyperparameters optimized separately for each quantity by maximizing the log marginal likelihood. The explicit dependence on the inverse number of atoms allows the fitted quantities to be extrapolated to the thermodynamic limit, $N\rightarrow\infty$.

The linear and volumetric thermal expansion coefficients were calculated as
\begin{equation}
\alpha_i(T)=\frac{1}{L_i(T)}\frac{dL_i(T)}{dT},
\qquad
\alpha_V(T)=\frac{1}{V(T)}\frac{dV(T)}{dT},
\label{eq:thermal_expansion_si}
\end{equation}
where $L_i(T)$ denotes the lattice parameter along direction $i$, and $V(T)$ is the unit-cell volume. The temperature derivatives were obtained directly from the GPR model, which provides predictions for both the fitted quantities and their derivatives~\cite{gp_rasmussen}. For all compounds and phases, $\alpha_V$ was calculated from the independently fitted volume, and all thermal expansion coefficients were evaluated in the thermodynamic limit.

The experimental lattice parameters and volumes reported in Refs.~\cite{biryukov2023thermal,shablinskii2025structural} were also fitted using GPR, with a kernel depending only on temperature. To estimate the experimental uncertainties, we included an additional contribution describing independent Gaussian errors with zero mean:
\begin{equation}
\begin{split}
k_{\mathrm{exp}}(T_i,T_j)
={}&\theta_0^2+\theta_1^2
\exp\!\left[-\frac{(T_i-T_j)^2}{2\theta_2^2}\right]\\
&+\left[\theta_3^2+\theta_4^2(T_i-\bar T)^2\right]\delta_{ij},
\end{split}
\end{equation}
where $\delta_{ij}$ is the Kronecker delta and $\bar T$ is the mean temperature of the corresponding dataset. The last term describes independent noise with a temperature-dependent variance $\theta_3^2+\theta_4^2(T_i-\bar T)^2$, estimated from the experimental data. All hyperparameters were learned independently for each experimental dataset. The estimated noise variances were then used as fixed observation errors in a second GPR fit containing only the smooth temperature-dependent part of the kernel, and the experimental thermal expansion coefficients were obtained from this fit using Eq.~\eqref{eq:thermal_expansion_si}.

For a lattice parameter or volume $Q$, the uncertainty of $\alpha_Q=Q'/Q$ is estimated from the joint GPR uncertainties of $Q$ and its temperature derivative $Q'=dQ/dT$:
\begin{equation}
\operatorname{Var}(\alpha_Q)\simeq
\frac{\operatorname{Var}(Q')}{Q^2}
+\frac{(Q')^2\operatorname{Var}(Q)}{Q^4}
-\frac{2Q'\operatorname{Cov}(Q,Q')}{Q^3}.
\end{equation}
We use the fitted means of $Q$ and $Q'$ in this expression for both linear and volumetric expansion coefficients.

\subsection{Common geometric decomposition}\label{sec:geometric_decomposition}

We separate rod expansion into contributions from changes in segment lengths and directions. The rod length is the straight-line distance between its first and last atoms; the segments are the individual connections forming the rod. For example, an Rb1--Rb2--Rb1 rod contains two Rb1--Rb2 segments. Fixed atom indices and periodic images are retained throughout the analysis, with the initial configuration used as the reference geometry.

For a rod $r$ with segment vectors $\mathbf v_{r,i}$, let $\mathbf R_r=\sum_i\mathbf v_{r,i}$, $D_r=|\mathbf R_r|$, and $\hat{\mathbf q}_r=\mathbf R_r/D_r$. Here $\ell_{r,i}=|\mathbf v_{r,i}|$ is the segment length and $u_{r,i}=\mathbf v_{r,i}\cdot\hat{\mathbf q}_r/\ell_{r,i}$ is the cosine of its angle to the rod direction, so $D_r=\sum_i\ell_{r,i}u_{r,i}$. With asterisks denoting the initial geometry, the instantaneous length and directional offsets are
\begin{equation}
\begin{split}
X_{\ell,r}&=\sum_i(\ell_{r,i}-\ell_{r,i}^{*})\frac{u_{r,i}+u_{r,i}^{*}}{2},\\
X_{u,r}&=\sum_i(u_{r,i}-u_{r,i}^{*})\frac{\ell_{r,i}+\ell_{r,i}^{*}}{2},\\
X_{\ell,r}+X_{u,r}&=D_r-D_r^{*}.
\end{split}
\label{eq:instantaneous_path_decomposition}
\end{equation}
We apply the decomposition to each instantaneous configuration, then average the offsets and $D_r$ over equivalent rods and MD trajectories. The relative contributions, normalized by the mean rod length at temperature $T_0$, are
\begin{equation}
C_{\ell}(T)=\frac{\langle X_{\ell}\rangle_T-\langle X_{\ell}\rangle_{T_0}}{\langle D\rangle_{T_0}},
\qquad
C_u(T)=\frac{\langle X_u\rangle_T-\langle X_u\rangle_{T_0}}{\langle D\rangle_{T_0}}.
\end{equation}
Their sum equals the directly calculated relative change in mean rod length before fitting. The starred quantities describe the initial atomic geometry, while $T_0$ sets the temperature from which thermal expansion is measured. The same $T_0$, specified in the figure captions, is used for all system sizes within each phase.

Each contribution and the direct rod-length change are fitted independently as functions of temperature and inverse atom number, using the kernel of Eq.~\eqref{eq:gp_kernel_si} and uncertainties estimated from the independent MD trajectories. All fits are evaluated in the thermodynamic limit. The sum of the fitted contributions is compared with the separately fitted rod-length and lattice-parameter changes.

\subsection{Rb$_2$SO$_4$}\label{sec:rb2so4_structural}

The experimental study describes Rb1 as surrounded by six SO$_4$ tetrahedra. In the $\beta$ phase, this description retains the six shorter Rb1--O contacts of the Rb1O$_9$ coordination environment. These units form corrugated rods along $c$~\cite{shablinskii2025structural}. We follow successive Rb1 atoms using fixed crystallographic indices and consistent periodic images. Each rod connects translationally equivalent Rb1 sites, and its rod length per unit-cell period is compared with $c$.

The same construction is applied along $a$ and $b$ using rods through Rb2. The sequence is Rb1--Rb2--Rb1 along $a$ in both phases and along $b$ in the $\alpha$ phase, and Rb1--Rb2--Rb1--Rb2--Rb1 along $b$ in the $\beta$ phase. Each complete rod is analyzed using the common decomposition in Sec.~\ref{sec:geometric_decomposition} and compared with the corresponding lattice parameter. In the figures, the directional contribution is labeled ``Rod straightening'' along $c$ and ``Rod bending'' along $a$ and $b$, reflecting its positive and negative contributions, respectively.

\begin{figure*}[h!]
\centering
\includegraphics[width=\linewidth]{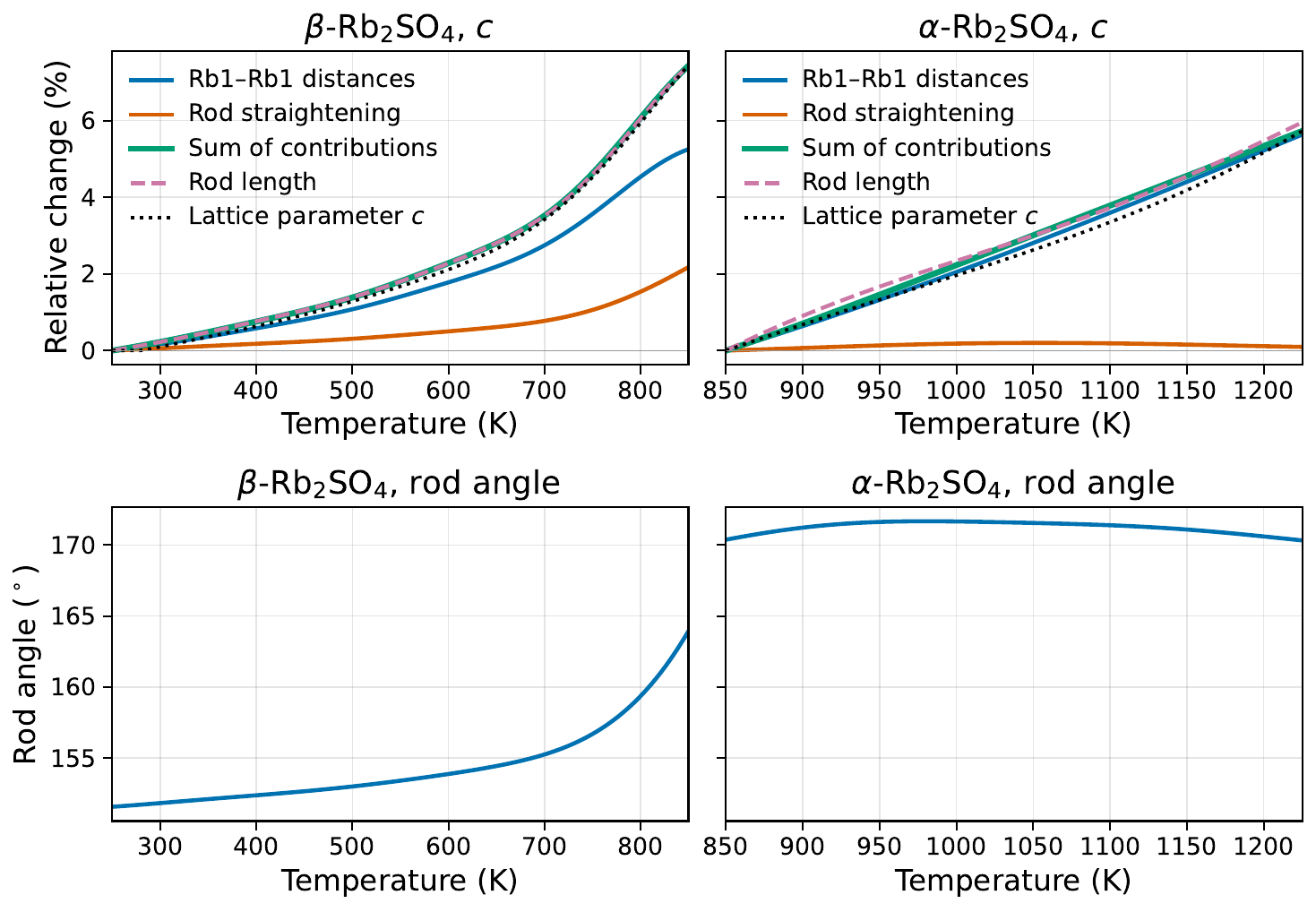}
\caption{Geometric contributions to expansion along $c$ in $\beta$-Rb$_2$SO$_4$ (left) and $\alpha$-Rb$_2$SO$_4$ (right), relative to 250 and 850~K, respectively. Upper panels show the contributions from changing Rb1--Rb1 distances and rod straightening, their sum, the independently fitted change in rod length, and the relative change in the lattice parameter $c$. Lower panels show the mean Rb1--Rb1--Rb1 angles. All curves show GPR estimates at infinite system size.}
\label{fig:rb2so4_c_contributions}
\end{figure*}

Along $c$, the $\beta$-phase mean Rb1--Rb1--Rb1 angle increases from $151.6^\circ$ at 250~K to $163.9^\circ$ at 850~K in the infinite-system-size limit (Fig.~\ref{fig:rb2so4_c_contributions}). The length and directional contributions are $5.26\%$ and $2.18\%$, respectively. Their sum, $7.43\%$, agrees closely with the independently fitted rod-length increase of $7.44\%$.

In the $\alpha$ phase, the endpoint angles are nearly equal: $170.4^\circ$ at 850~K and $170.3^\circ$ at 1225~K. The corresponding length and directional contributions are $5.64\%$ and $0.09\%$, confirming that increasing Rb1--Rb1 distances dominates expansion. Their fitted sum is $5.73\%$, compared with $5.96\%$ from the independent fit of the direct change. At the highest temperatures, some instantaneous nearest-neighbor assignments differ from the fixed rod sequence; the reported angles refer to that fixed sequence.

\begin{figure*}[h!]
\centering
\includegraphics[width=\linewidth]{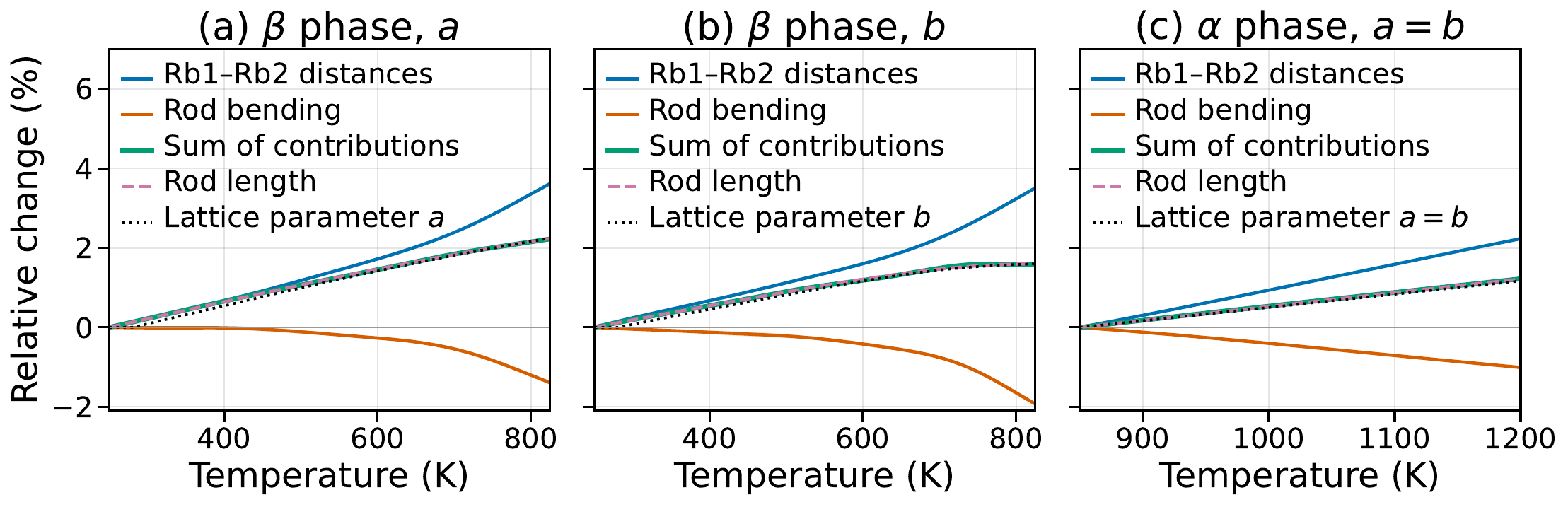}
\caption{Geometric contributions to expansion for rods through Rb2 along (a) $a$ and (b) $b$ in $\beta$-Rb$_2$SO$_4$, and (c) averaged over $a$ and $b$ in $\alpha$-Rb$_2$SO$_4$. Changes are relative to 250~K in the $\beta$ phase and 850~K in the $\alpha$ phase. Curves show the contributions from changing Rb1--Rb2 distances and rod bending, their sum, the independently fitted change in rod length, and the relative change in the corresponding lattice parameter. All curves show GPR estimates at infinite system size. Shaded regions in (a) and (b) indicate $2\sigma$ confidence intervals for the individual contributions.}
\label{fig:rb2so4_ab_contributions}
\end{figure*}

Along $a$ and $b$, increasing Rb1--Rb2 distances promotes expansion, while changes in segment directions partially compensate it (Fig.~\ref{fig:rb2so4_ab_contributions}). Separate results are shown for $a$ and $b$ in the $\beta$ phase, whereas the contributions are averaged over the symmetry-equivalent axes in the $\alpha$ phase. The independently fitted changes in rod length and lattice parameters are included for comparison.

At the highest analyzed temperature for each rod and supercell, the positive length and negative directional contributions persist within twice the standard error. For GPR, we combine only results with a common reference temperature: 250~K for three $\beta$ supercell sizes and 850~K for two $\alpha$ sizes. The common prediction intervals are 250--825~K and 850--1200~K, respectively.

\begin{figure}[h!]
\centering
\includegraphics[width=\linewidth]{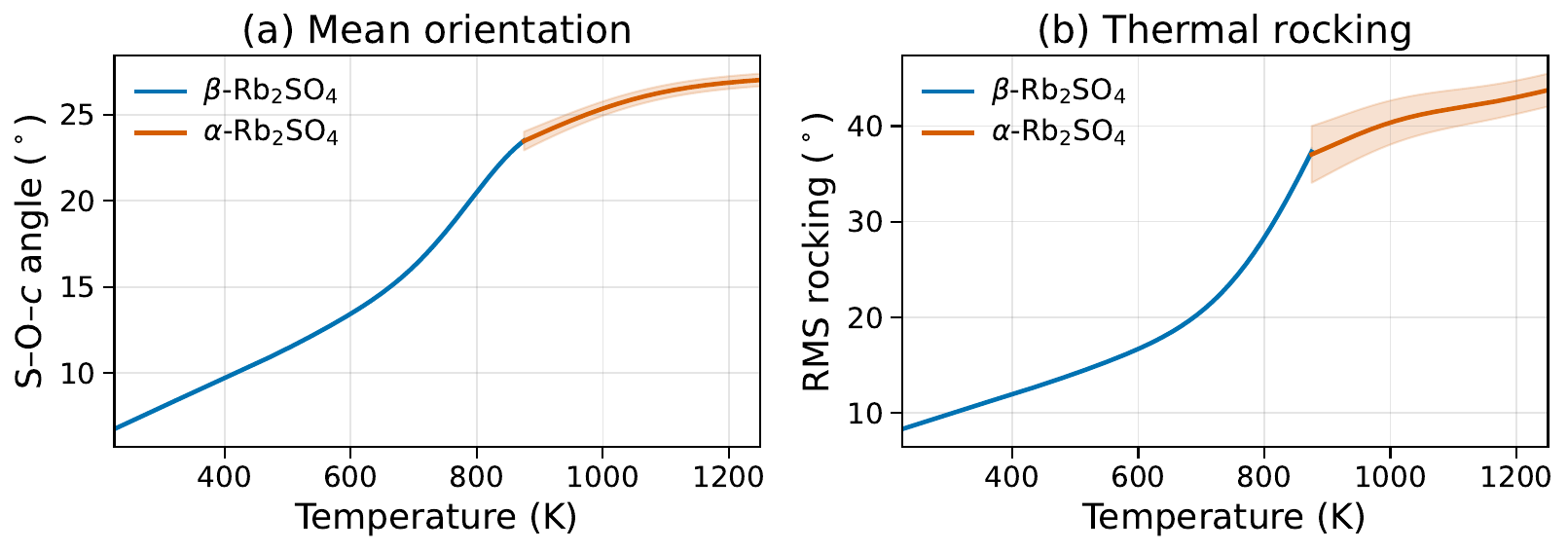}
\caption{SO$_4$ orientation and rocking in $\beta$- and $\alpha$-Rb$_2$SO$_4$. (a) Mean inclination of the S--O bond closest to the $c$ axis. (b) RMS rocking about each tetrahedron's temperature-dependent mean orientation, accounting for symmetry-equivalent oxygen permutations. Curves show GPR estimates at infinite system size, and shaded regions indicate $2\sigma$ confidence intervals.}
\label{fig:rb2so4_so4_orientation}
\end{figure}

To test the experimental interpretation linking rod straightening to SO$_4$ reorientation~\cite{shablinskii2025structural}, we calculate the angle between the $c$ axis and the S--O bond most closely aligned with it. For each tetrahedron, we follow the same four O atoms throughout the MD trajectory, identifying this bond in each atomic configuration. The angles are averaged over tetrahedra and trajectories to obtain the mean inclination shown in Fig.~\ref{fig:rb2so4_so4_orientation}(a). In the $\beta$ phase, the mean inclination increases upon heating together with rod straightening.

We also analyze SO$_4$ rocking, i.e., angular fluctuations about each tetrahedron's mean orientation at a given temperature. Its amplitude is quantified by the root mean square (RMS) angular deviation (Fig.~\ref{fig:rb2so4_so4_orientation}(b)). Orientations are obtained by least-squares alignment of the four unit S--O vectors, accounting for symmetry-equivalent oxygen permutations. RMS values are calculated for each tetrahedron within each trajectory and then averaged. RMS rocking increases strongly upon heating in both phases. In the $\alpha$ phase, individual tetrahedra change orientation over time, and oxygen positions occupy several regions around each S atom, consistent with orientational disorder.

\subsection{Rb$_2$Ca$_2$(SO$_4$)$_3$}

Bond elongation promotes expansion, whereas bending of the connected microblocks can partly compensate it. To examine this balance, we first compare the expansion of the metal--oxygen polyhedra and the $M$(SO$_4$)$_6$ microblocks forming the $L_3$ rods. Their RMS sizes are calculated from the $M$--O and $M$--S vectors, respectively, over 250--1250~K. For each environment, the size is defined as
\begin{equation}
r_{\mathrm{RMS}} =
\sqrt{\frac{1}{3n}\sum_{j=1}^{n}|\mathbf{r}_{Mj}|^2},
\end{equation}
where $M$ denotes Ca or Rb and $n$ is the number of neighboring atoms. The factor of three normalizes the size per spatial direction. The sizes are calculated in each frame and averaged over equivalent sites and MD trajectories. The oxygen environments comprise the nearest six O atoms for Ca1, Ca2, and Rb2 and the nearest nine O atoms for Rb1 in each frame. In the room-temperature experimental structure, both Rb sites have nine oxygen neighbors. The Rb2 environment exhibits a \(6+3\) arrangement, with six shorter Rb--O contacts at 3.017--3.118~\AA{} and three longer contacts at 3.338~\AA{}. Accordingly, the six-neighbor Rb2 environment analyzed here represents the shorter-contact subset of the room-temperature coordination shell. The six sulfate centers defining each microblock retain their reference membership. Both Ca sites retain a distinct six-neighbor oxygen shell: the mean gap between the sixth and seventh neighbors remains approximately $0.6$--$0.7$~\AA\ near the upper end of the range.

\begin{figure}[h!]
\centering
\includegraphics[width=\linewidth]{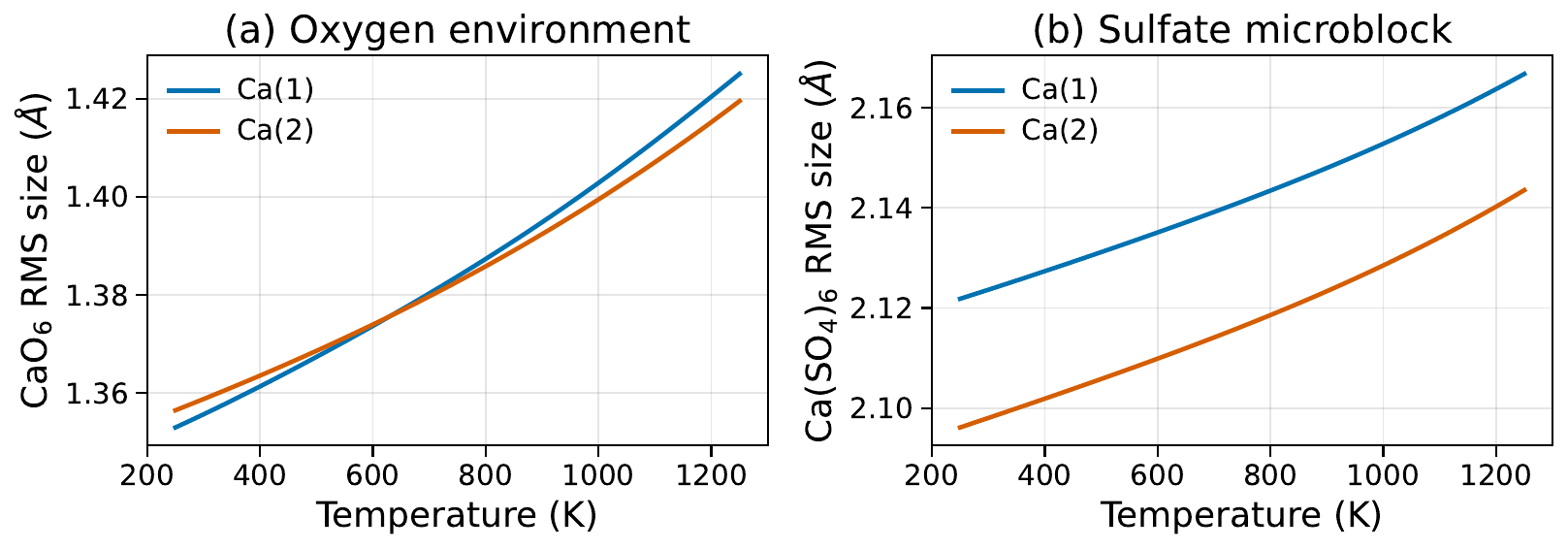}
\caption{Temperature dependence of the RMS sizes of (a) the Ca1O$_6$ and Ca2O$_6$ coordination polyhedra and (b) the corresponding Ca(SO$_4$)$_6$ microblocks, calculated from vectors to the six surrounding sulfate centers. All sizes are normalized per spatial direction. Curves show GPR estimates at infinite system size; shaded regions indicate $2\sigma$ confidence intervals.}
\label{fig:rb2ca2_ca_si}
\end{figure}

\begin{figure}[h!]
\centering
\includegraphics[width=\linewidth]{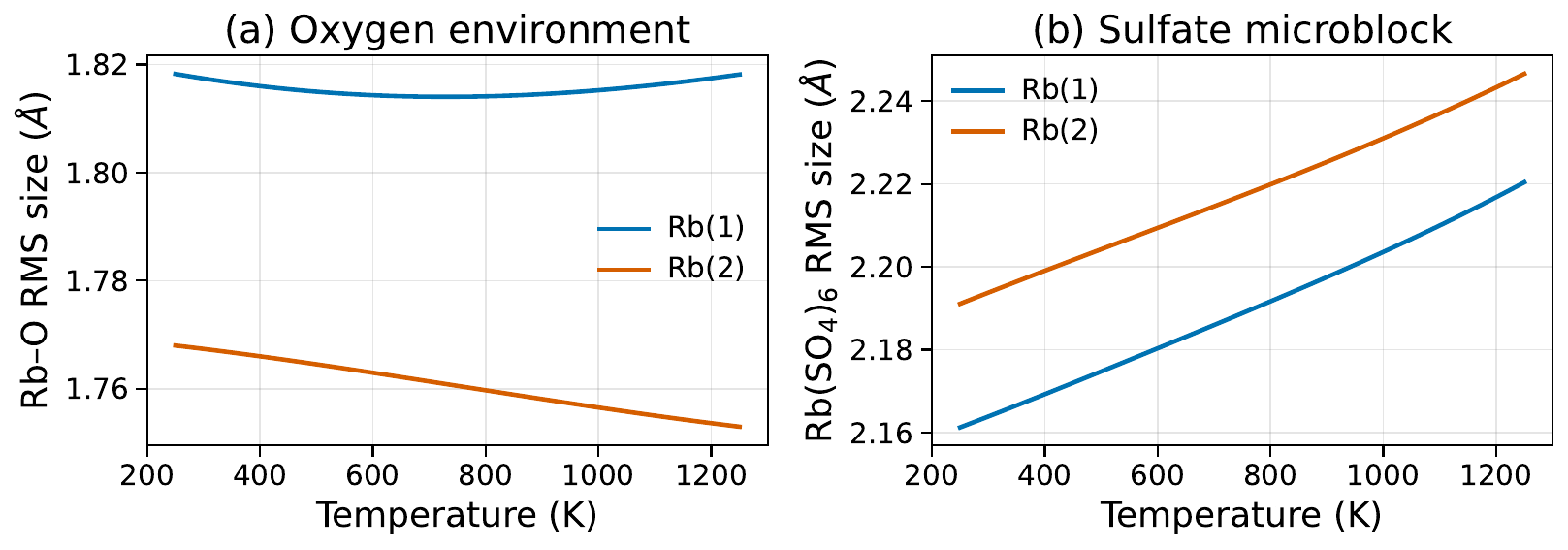}
\caption{Temperature dependence of the RMS sizes of (a) the Rb1O$_9$ and Rb2O$_6$ oxygen environments and (b) the corresponding Rb(SO$_4$)$_6$ microblocks, calculated from vectors to the six surrounding sulfate centers. All sizes are normalized per spatial direction. Curves show GPR estimates at infinite system size; shaded regions indicate $2\sigma$ confidence intervals.}
\label{fig:rb2ca2_rb_si}
\end{figure}

Both the CaO$_6$ polyhedra and the Ca-centered microblocks expand upon heating, but the relative increase in RMS size is smaller for the microblocks (Fig.~\ref{fig:rb2ca2_ca_si}). The Rb-centered microblocks also expand (Fig.~\ref{fig:rb2ca2_rb_si}), although the RMS size of the Rb1O$_9$ oxygen environment changes little and that of Rb2O$_6$ decreases slightly. These quantities describe RMS sizes rather than mean bond distances. The broadening Rb--O distributions and decreasing retention of the initial nearest-neighbor contacts show that these changes accompany rearrangement of the oxygen environments.

\begin{figure*}[h!]
\centering
\includegraphics[width=0.49\linewidth]{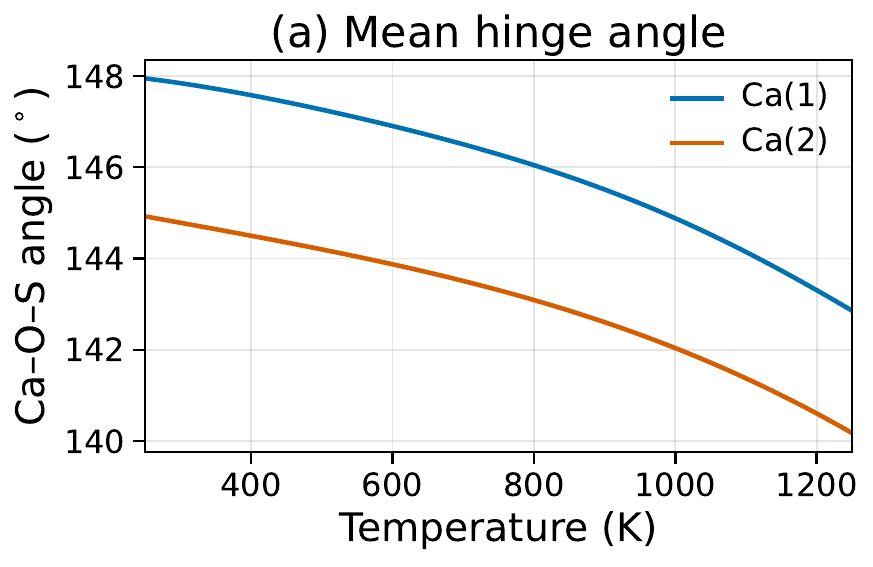}
\hfill
\includegraphics[width=0.49\linewidth]{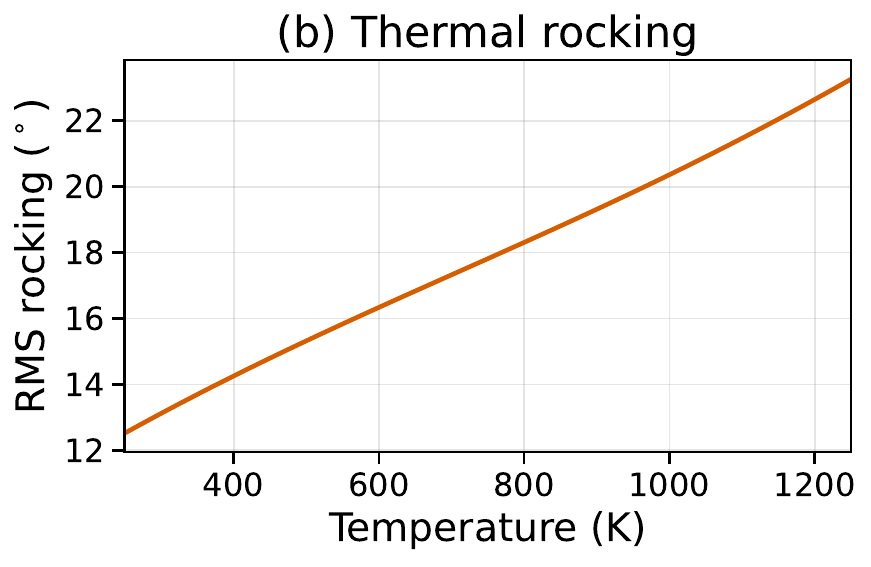}
\caption{Temperature dependence of (a) the mean Ca--O--S angles for Ca1 and Ca2, calculated using the six nearest oxygen neighbors in each frame, and (b) the RMS SO$_4$ rocking about the temperature-dependent mean orientation. Curves show GPR estimates at infinite system size; shaded regions indicate $2\sigma$ confidence intervals.}
\label{fig:rb2ca2_angular}
\end{figure*}

To understand why the Ca-centered microblocks expand less strongly than the CaO$_6$ polyhedra, we examine the Ca--O--S hinge angles. Their mean values decrease upon heating (Fig.~\ref{fig:rb2ca2_angular}(a)), while their distributions broaden. At fixed bond lengths, this bending reduces the Ca--S separation, partly compensating Ca--O elongation. The decrease persists when the same reference Ca--O pairs are followed, confirming that bending also occurs within the original connections. This hinge bending is accompanied by increasing SO$_4$ rocking (Fig.~\ref{fig:rb2ca2_angular}(b)), quantified using the RMS angular deviation defined for Rb$_2$SO$_4$ (Sec.~\ref{sec:rb2so4_structural}). Larger orientational deviations correlate with smaller Ca--O--S angles at both Ca sites, with a stronger negative correlation on heating, consistent with coupled rocking and hinge deformation.

Finally, we apply the decomposition in Sec.~\ref{sec:geometric_decomposition} to complete Rb1--Ca2--Ca1--Rb2--Rb1 repeats of the $L_3$ rod, selecting the least corrugated sequence through face-sharing microblocks from each reference Rb1 site. Increasing interatomic distances promotes expansion, while rod bending partly compensates it. GPR fits use all three supercell sizes; the sum of the contributions is compared with the relative changes in rod length and lattice parameter $a$, measured from 250~K.

\subsection{Ba$_3$Lu(BO$_3$)$_3$}

To test the proposed reorientation of BO$_3$ units, we distinguish changes in mean orientation from thermal rocking, introduced in Sec.~\ref{sec:rb2so4_structural}. The mean inclination relative to the $ab$ plane is calculated from the trajectory-averaged unit vector perpendicular to each triangle. RMS rocking measures the angular deviations of this vector from its mean orientation.

\begin{figure}[h!]
\centering
\includegraphics[width=\linewidth]{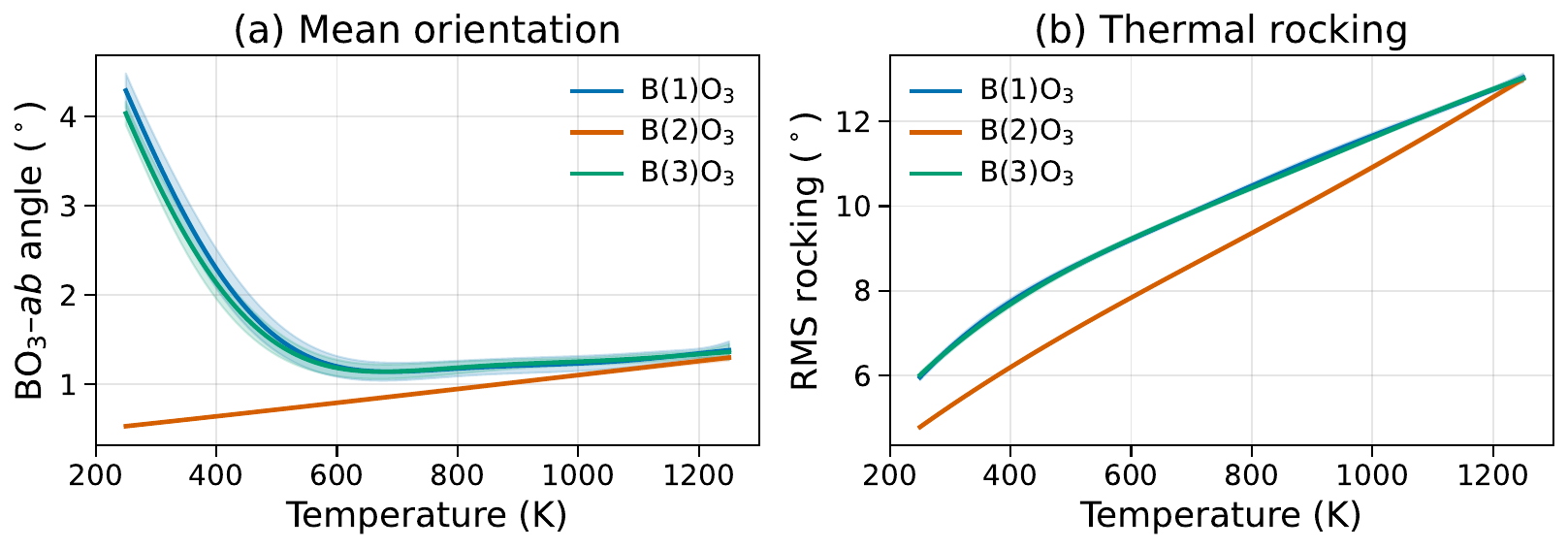}
\caption{Temperature dependence of the crystallographically distinct B(1)O$_3$, B(2)O$_3$, and B(3)O$_3$ triangles. (a) Mean inclination relative to the $ab$ plane. (b) RMS rocking about the temperature-dependent mean orientation. Curves show GPR estimates at infinite system size; shaded regions indicate $2\sigma$ confidence intervals.}
\label{fig:bo3_orientation}
\end{figure}

Between 250 and 1250~K, B(1)O$_3$ and B(3)O$_3$ show an overall decrease in mean inclination, while the nearly parallel B(2)O$_3$ triangles become slightly more inclined (Fig.~\ref{fig:bo3_orientation}). RMS rocking increases strongly for all three types.

We analyze atomic displacements to test whether motion is preferentially perpendicular to the BO$_3$ planes. We first subtract the center of each BO$_3$ group to exclude translation, then measure deviations from the trajectory-mean coordinates. The displacements are resolved into components normal to and within the mean triangle plane, with the in-plane mean-square displacement normalized per direction. The normal component is approximately four times larger than the in-plane component. Removing the overall rotation of each triangle substantially reduces the displacements, consistent with rocking of nearly rigid BO$_3$ groups. Between 250 and 1250~K, the mean B--O distance and triangle area increase by only 0.97\% and 1.18\%, respectively.

To compare polyhedral expansion along $c$ and within $ab$, we calculate the directional RMS sizes of LuO$_6$ and BaO$_9$ from the metal--oxygen vectors $\mathbf{r}_j$:
\begin{equation}
r_c =
\left[
\frac{1}{n}\sum_{j=1}^{n}
(\mathbf{r}_j\cdot\hat{\mathbf{c}})^2
\right]^{1/2},
\qquad
r_{ab} =
\left[
\frac{1}{2n}\sum_{j=1}^{n}
\left(
|\mathbf{r}_j|^2-
(\mathbf{r}_j\cdot\hat{\mathbf{c}})^2
\right)
\right]^{1/2},
\end{equation}
where $\hat{\mathbf{c}}$ is the unit vector along $c$, and $n=6$ for LuO$_6$ and $n=9$ for BaO$_9$. The factor of two normalizes the in-plane size per direction. These sizes are calculated in each MD frame and then averaged over the production trajectories. They include changes in bond lengths, shape, orientation, and thermal fluctuations of the oxygen environment.

The same six oxygen neighbors are followed around each Lu atom throughout a trajectory. For Ba, the coordination shell is broader, so the nine nearest oxygen atoms are identified separately in each frame. The BaO$_9$ sizes therefore describe the instantaneous coordination environment rather than a fixed set of Ba--O pairs.

\begin{figure}[h!]
\centering
\includegraphics[width=\linewidth]{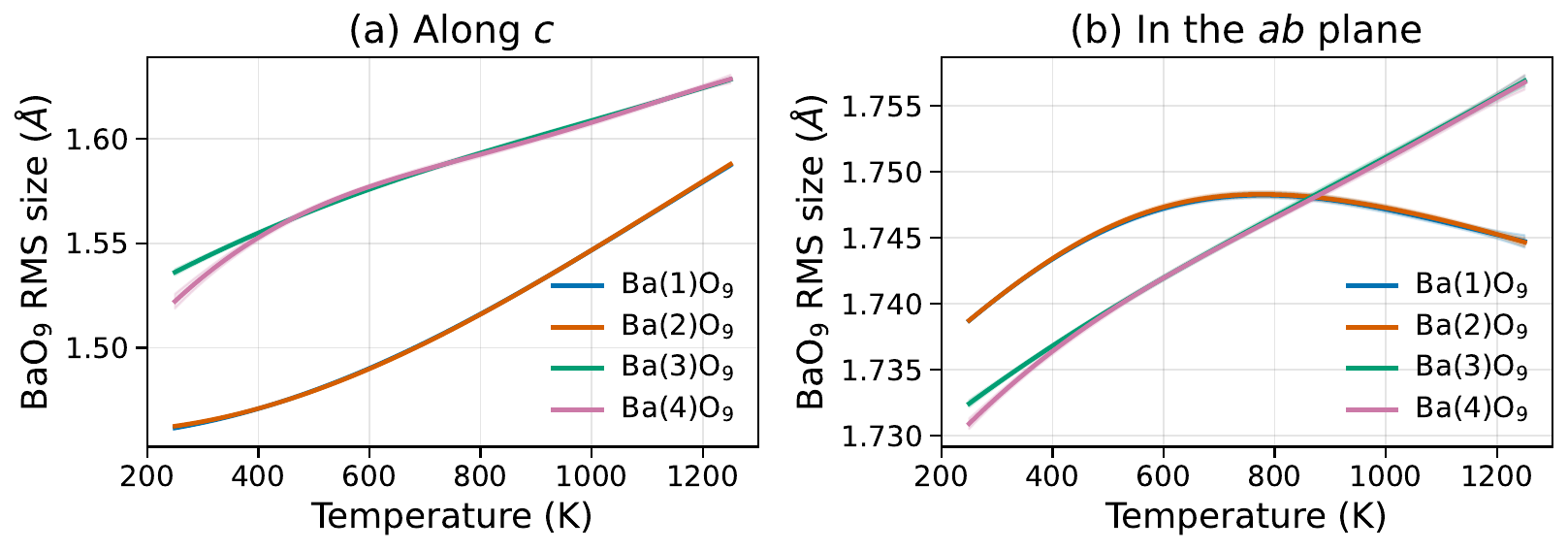}
\caption{Temperature dependence of the directional RMS sizes of the crystallographically distinct Ba(1)O$_9$--Ba(4)O$_9$ polyhedra: (a) along $c$ and (b) within the $ab$ plane. The sizes are calculated from the nine nearest Ba--O vectors along each MD trajectory; the in-plane quantity is normalized per direction. Curves show GPR estimates at infinite system size; shaded regions indicate $2\sigma$ confidence intervals.}
\label{fig:bao9_size}
\end{figure}

The directional RMS sizes of BaO$_9$ increase strongly along $c$ for all four Ba sites (Fig.~\ref{fig:bao9_size}). The changes within $ab$ are much smaller and depend on the Ba site. The RMS sizes of Ba(1)O$_9$ and Ba(2)O$_9$ initially increase and then weakly decrease at high temperatures, while those of Ba(3)O$_9$ and Ba(4)O$_9$ increase slightly.

\begin{figure}[h!]
\centering
\includegraphics[width=\linewidth]{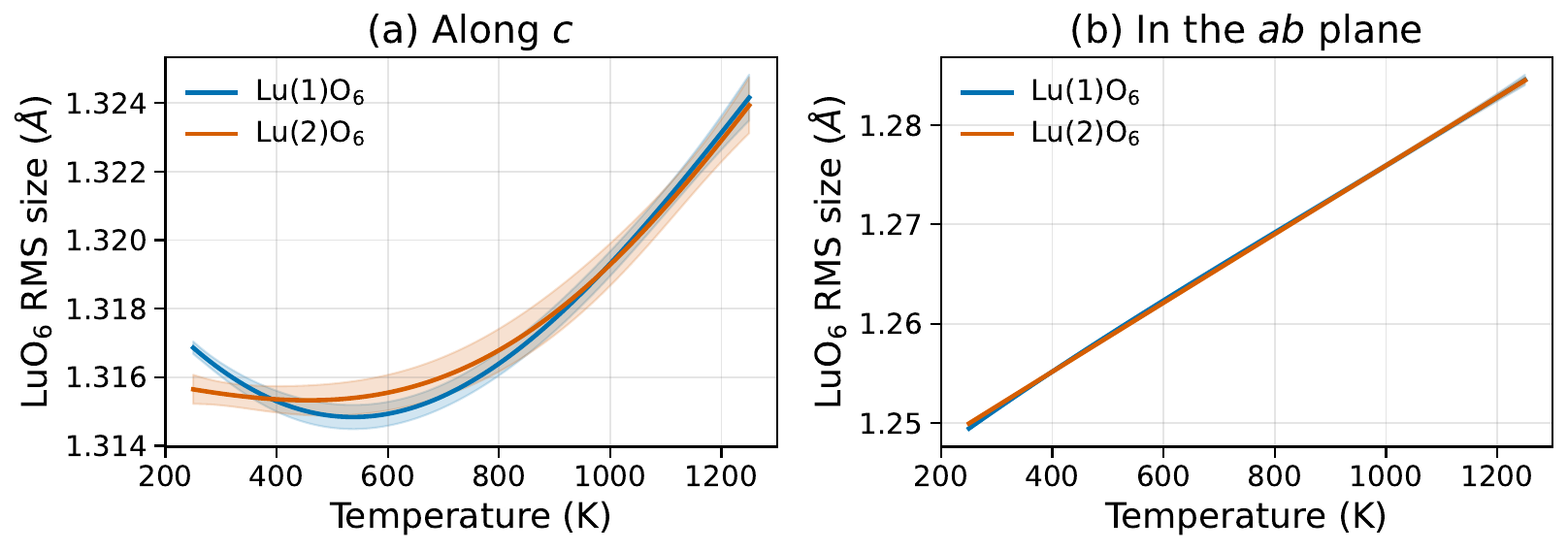}
\caption{Temperature dependence of the directional RMS sizes of the Lu(1)O$_6$ and Lu(2)O$_6$ octahedra: (a) along $c$ and (b) within the $ab$ plane, normalized per direction. Curves show GPR estimates at infinite system size; shaded regions indicate $2\sigma$ confidence intervals.}
\label{fig:luo6_size_si}
\end{figure}

Both Lu sites show similar temperature dependence (Fig.~\ref{fig:luo6_size_si}). The RMS size along $c$ initially decreases slightly and then increases. Between 250 and 1250~K, its net increase is approximately 0.6\%, whereas $r_{ab}$ increases by about 2.8\%. The comparatively small change along $c$ contrasts with the strong response of BaO$_9$ along this direction (Fig.~\ref{fig:bao9_size}). The anisotropic response of the LuO$_6$ octahedra may be coupled to the reorientation of the B(1)O$_3$ and B(3)O$_3$ triangles through their shared oxygen vertices, because displacements of these oxygen atoms change both the orientations and the lengths of the Lu--O bonds. The weaker in-plane response of Ba(1)O$_9$ and Ba(2)O$_9$ may be related to their face-sharing connections with the LuO$_6$ octahedra, with the shared faces parallel to the $ab$ plane; compared with the edge-sharing connections of Ba(3)O$_9$ and Ba(4)O$_9$, this connectivity may impose stronger constraints on angular rearrangements.

To connect these polyhedral changes with lattice expansion, we analyze periodic rods joining a Ba(1) atom to its periodic image along each lattice axis. In the reference coordination graph, we select the rod with the smallest sum of segment lengths. These rods pass through Ba--O contacts within $ab$ and through Ba--O and Lu--O contacts along $c$. One reference Ba(1) site and all its equivalent translations are used at each supercell size, with the same atom indices and periodic images followed in every production frame.

We apply the decomposition and GPR procedure in Sec.~\ref{sec:geometric_decomposition} to all three supercell sizes, using 250~K as the reference temperature. The $a$ and $b$ contributions are averaged. All fits are evaluated at infinite system size and compared with independent GPR estimates of the lattice-parameter changes. Along $c$, the increase in Ba--O and Lu--O distances dominates rod expansion. Within $ab$, rod bending partly compensates Ba--O elongation, consistent with the smaller in-plane expansion.

\section{Lattice thermal conductivity}
\label{sec:ltc_convergence}

\subsection{Molecular dynamics and Green--Kubo analysis}

All MD simulations used a timestep of 1~fs. For each temperature and supercell, an initial 100~ps of $NPT$ equilibration was followed by a further 200~ps of $NPT$ sampling with Langevin temperature control. The barostat was isotropic for cubic Rb$_2$Ca$_2$(SO$_4$)$_3$ and anisotropic for Rb$_2$SO$_4$ and Ba$_3$Lu(BO$_3$)$_3$. The mean cell matrix and mean volume from the sampling interval were used to construct the fixed cell for the $NVE$ calculations.

Independent $NVE$ trajectories started from the same atomic configuration with independently sampled Maxwell--Boltzmann velocities. The mass-weighted center-of-mass velocity was removed, and the velocities were rescaled so that the total energy per atom matched the mean $NPT$ potential energy plus $3k_{\mathrm B}T/2$. The $NVE$ production runs lasted 200~ps for $\beta$-Rb$_2$SO$_4$ and Rb$_2$Ca$_2$(SO$_4$)$_3$ at 450 and 650~K and 250~ps for $\alpha$-Rb$_2$SO$_4$ and Ba$_3$Lu(BO$_3$)$_3$ at 1250~K, 600~ps for $\alpha$-Rb$_2$SO$_4$ at 850~K, and 800~ps for Ba$_3$Lu(BO$_3$)$_3$ at 850~K.

The heat flux was computed as~\cite{lammps-mlip-gk}
\begin{equation}
\mathbf J=\frac{1}{V}\left[\sum_i e_i\mathbf v_i+\sum_i\sum_{j\neq i}\mathbf r_{ij}\left(\frac{\partial U_j}{\partial\mathbf r_{ji}}\cdot\mathbf v_i\right)\right],
\label{eq:heat_flux_si}
\end{equation}
where $e_i$, $\mathbf v_i$, and $U_i$ are the total energy, velocity, and MTP site energy of atom $i$, $\mathbf r_{ij}=\mathbf r_j-\mathbf r_i$, and $V$ is the cell volume. The first term is the convective contribution, and the second is the many-body potential contribution.

The $xx$, $yy$, and $zz$ heat-flux autocorrelation functions (HCACFs) were evaluated with a lag spacing of 1~fs. Their running integrals were calculated using the trapezoidal rule, with the mean kinetic temperature and total cell volume of each $NVE$ trajectory used in the Green--Kubo prefactor. Table~\ref{tab:ltc_parameters} lists the supercells, the numbers of trajectories retained for averaging, and the maximum correlation times present in the saved analysis. The correlation time is the maximum HCACF lag and is distinct from the duration of an $NVE$ trajectory.

For each trajectory and Cartesian direction, the conductivity was obtained by averaging its running integral over the interval $[t_1,t_2)$ in Table~\ref{tab:ltc_parameters}. For Ba$_3$Lu(BO$_3$)$_3$, the averaging interval was selected within the correlation-time range up to 20~ps, giving $[14,20)$~ps at both temperatures. The uncertainty was calculated as $s/\sqrt{n}$, where $s$ is the sample standard deviation and $n$ is the number of retained trajectories. Directional averages were formed within each trajectory before calculating their ensemble mean and standard error: $(\kappa_{xx}+\kappa_{yy}+\kappa_{zz})/3$ for cubic Rb$_2$Ca$_2$(SO$_4$)$_3$, and $(\kappa_{xx}+\kappa_{yy})/2$ for the in-plane conductivity of $\alpha$-Rb$_2$SO$_4$ and Ba$_3$Lu(BO$_3$)$_3$.

\begin{table}[htbp]
\centering
\small
\setlength{\tabcolsep}{4pt}
\caption{Green--Kubo calculation parameters. $n$ is the number of trajectories included in the final average; $t_{\max}$ is the maximum correlation time in the saved analysis; $[t_1,t_2)$ is the interval used to average each running integral.}
\label{tab:ltc_parameters}
\begin{tabular}{lrrrrrr}
\toprule
Compound & $T$ (K) & Supercell & Atoms & $n$ & $t_{\max}$ (ps) & $[t_1,t_2)$ (ps) \\
\midrule
$\beta$-Rb$_2$SO$_4$ & 450 & $4\times5\times3$ & 1680 & 25 & 30 & $[6,30)$ \\
$\beta$-Rb$_2$SO$_4$ & 650 & $4\times5\times3$ & 1680 & 25 & 30 & $[6,30)$ \\
$\alpha$-Rb$_2$SO$_4$ & 850 & $5\times6\times4$ & 1680 & 25 & 30 & $[8,30)$ \\
$\alpha$-Rb$_2$SO$_4$ & 1250 & $4\times5\times3$ & 840 & 25 & 25 & $[8,25)$ \\
Rb$_2$Ca$_2$(SO$_4$)$_3$ & 450 & $3\times3\times3$ & 2052 & 35 & 20 & $[12,18)$ \\
Rb$_2$Ca$_2$(SO$_4$)$_3$ & 650 & $3\times3\times3$ & 2052 & 35 & 20 & $[12,18)$ \\
Ba$_3$Lu(BO$_3$)$_3$ & 850 & $3\times3\times2$ & 1728 & 45 & 40 & $[14,20)$ \\
Ba$_3$Lu(BO$_3$)$_3$ & 1250 & $3\times3\times2$ & 1728 & 40 & 25 & $[14,20)$ \\
\bottomrule
\end{tabular}
\end{table}

\subsection{Convergence}

\begin{figure}[!h]
\centering
\includegraphics[width=\linewidth]{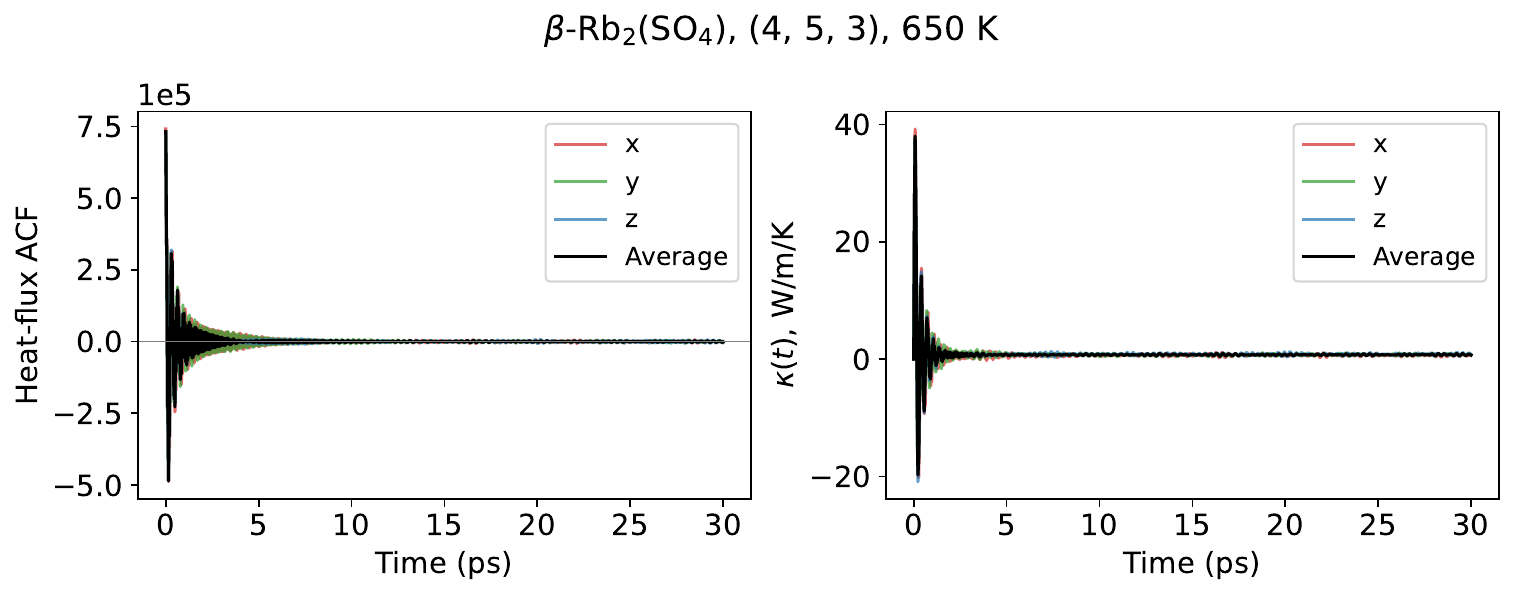}
\caption{Heat-flux autocorrelation functions (left) and running Green--Kubo lattice thermal conductivity (right) of $\beta$-Rb$_2$SO$_4$ at 650~K for the $4\times5\times3$ supercell (1680 atoms). Colored curves show the $x$, $y$, and $z$ components; black curves show their average.}
\label{fig:gk_beta_rb2so4}
\end{figure}

\begin{figure}[!h]
\centering
\includegraphics[width=\linewidth]{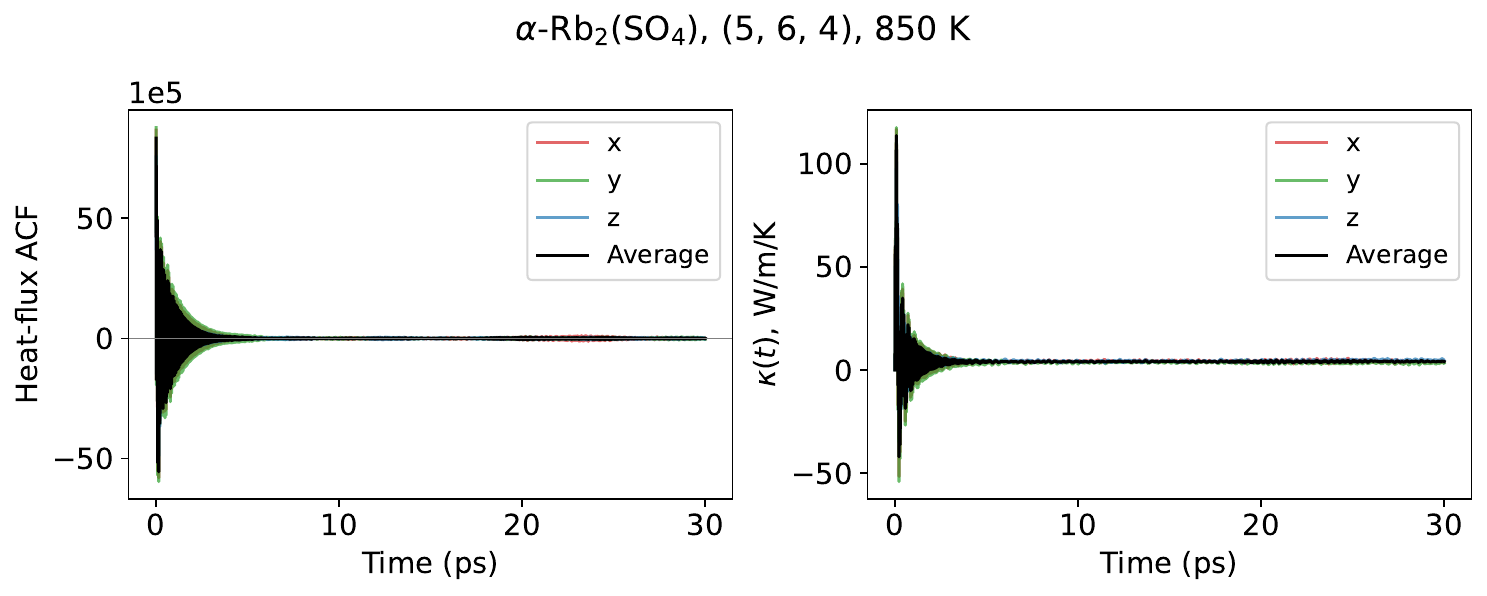}
\caption{Heat-flux autocorrelation functions (left) and running Green--Kubo lattice thermal conductivity (right) of $\alpha$-Rb$_2$SO$_4$ at 850~K for the $5\times6\times4$ supercell (1680 atoms). Colored curves show the $x$, $y$, and $z$ components; black curves show their average.}
\label{fig:gk_alpha_rb2so4}
\end{figure}

\begin{figure}[!h]
\centering
\includegraphics[width=\linewidth]{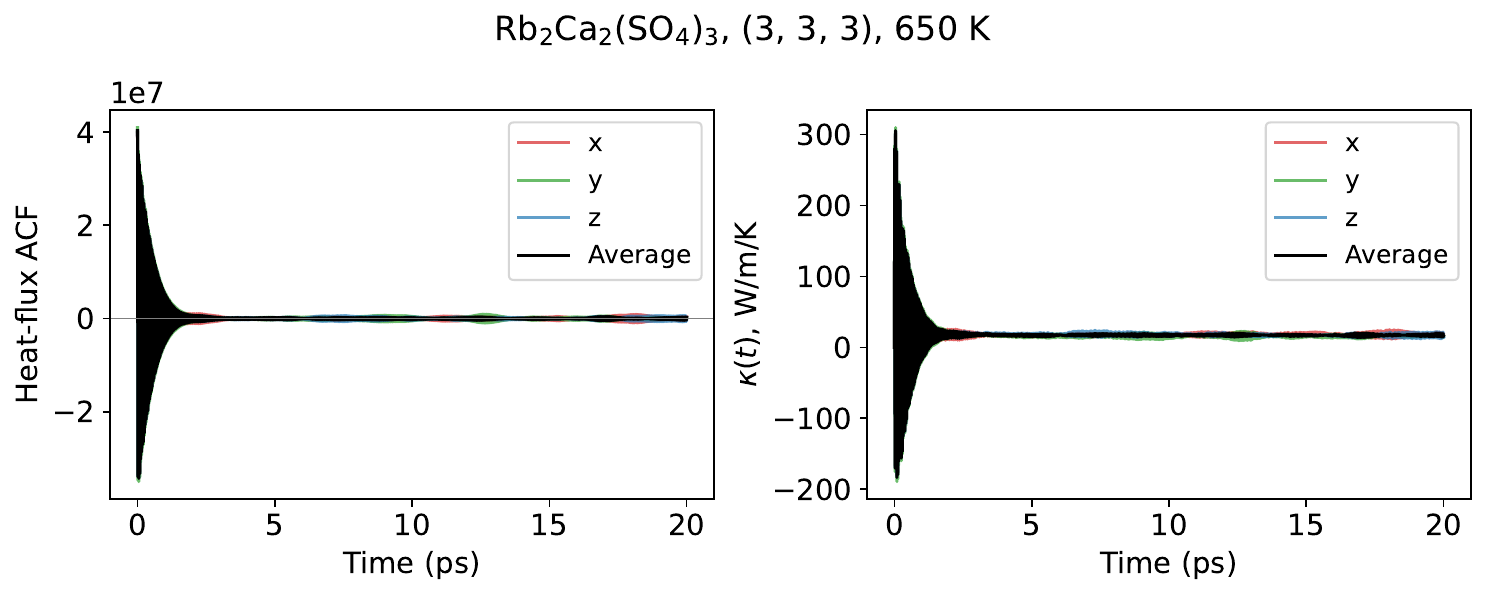}
\caption{Heat-flux autocorrelation functions (left) and running Green--Kubo lattice thermal conductivity (right) of Rb$_2$Ca$_2$(SO$_4$)$_3$ at 650~K for the $3\times3\times3$ supercell (2052 atoms). Colored curves show the $x$, $y$, and $z$ components; black curves show their average.}
\label{fig:gk_rb2ca2}
\end{figure}

\begin{figure}[!h]
\centering
\includegraphics[width=\linewidth]{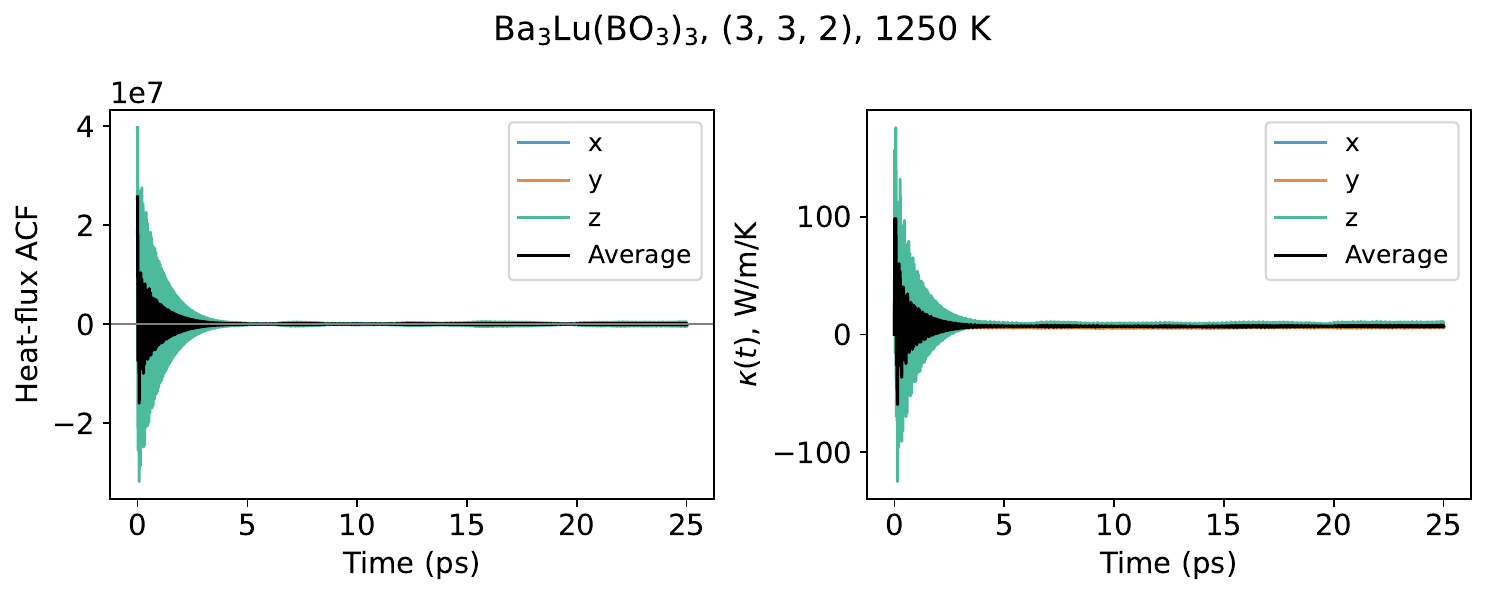}
\caption{Heat-flux autocorrelation functions (left) and running Green--Kubo lattice thermal conductivity (right) of Ba$_3$Lu(BO$_3$)$_3$ at 1250~K for the $3\times3\times2$ supercell (1728 atoms). Colored curves show the $x$, $y$, and $z$ components; black curves show their average.}
\label{fig:gk_ba3lu}
\end{figure}

We first examine the heat-flux autocorrelation functions (left) and running Green--Kubo conductivities (right) at representative temperatures for the studied compounds (Figs.~\ref{fig:gk_beta_rb2so4}--\ref{fig:gk_ba3lu}). Curves show the three Cartesian components and their average. The final conductivity values are obtained by averaging the running integrals over the intervals listed in Table~\ref{tab:ltc_parameters}.

\begin{figure}[htbp]
\centering
\begin{subfigure}{0.48\linewidth}
\includegraphics[width=\linewidth]{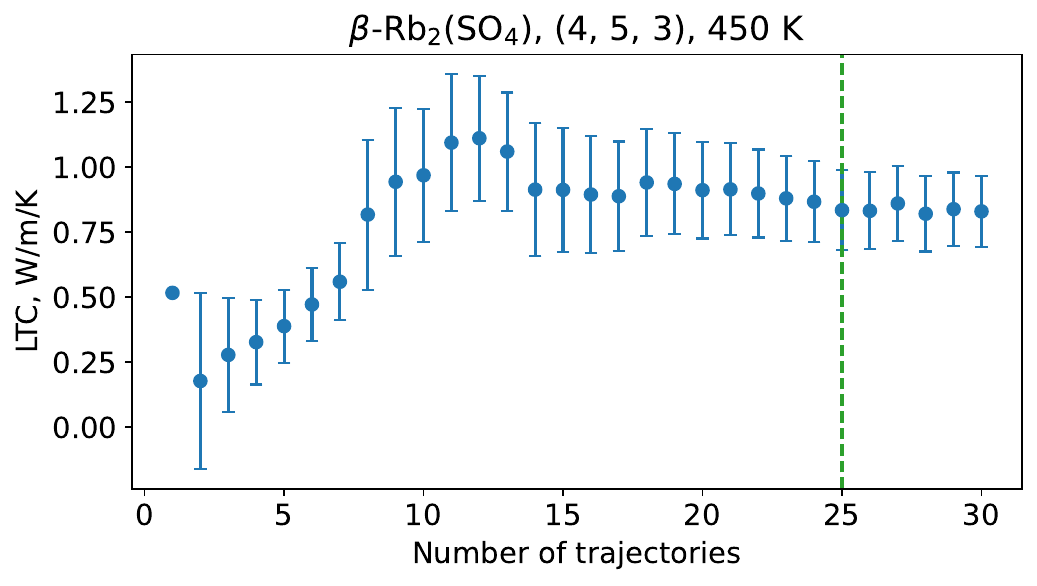}
\caption{$\beta$ phase, 450~K.}
\end{subfigure}\hfill
\begin{subfigure}{0.48\linewidth}
\includegraphics[width=\linewidth]{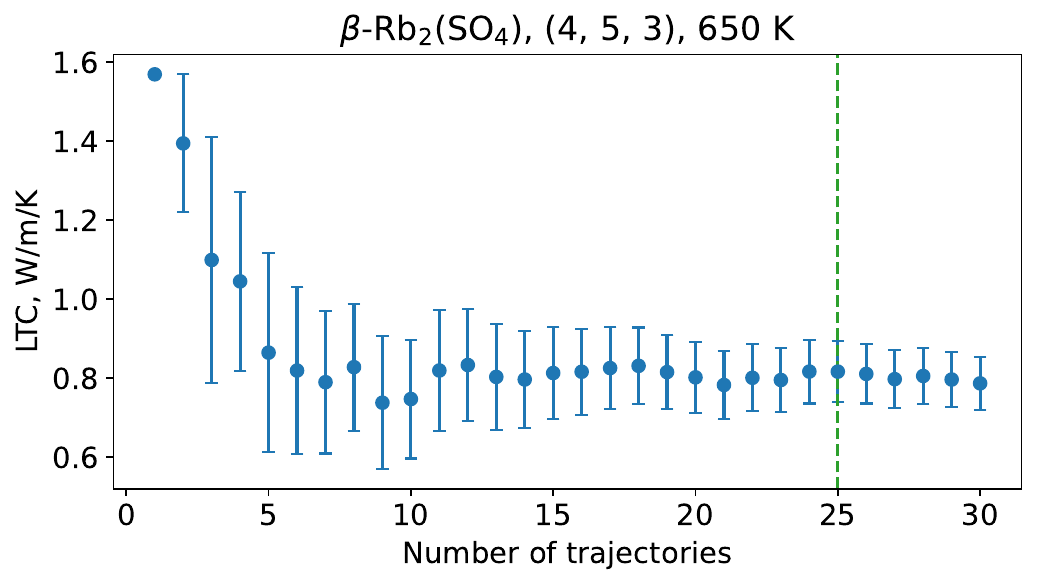}
\caption{$\beta$ phase, 650~K.}
\end{subfigure}
\par\medskip
\begin{subfigure}{0.48\linewidth}
\includegraphics[width=\linewidth]{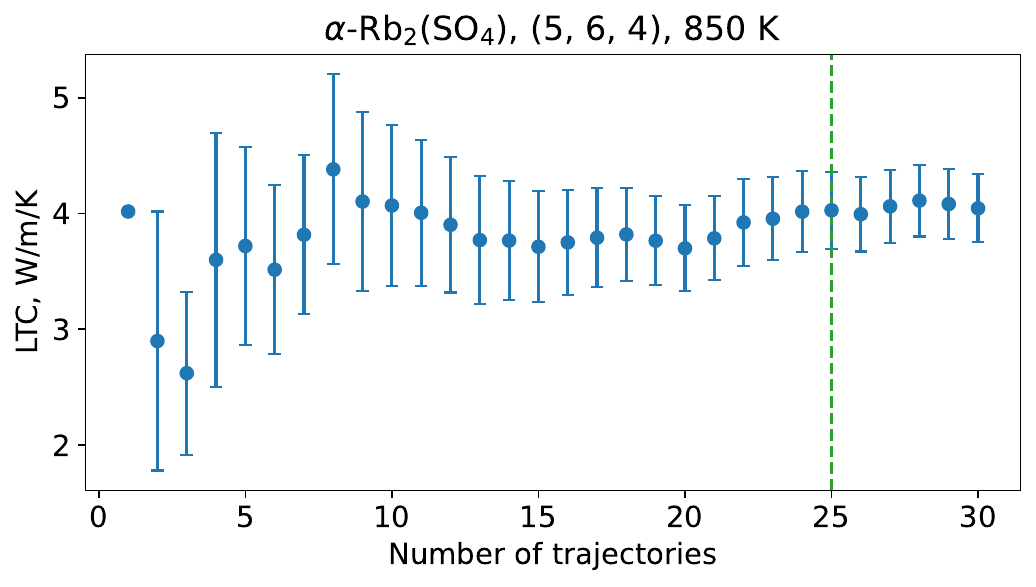}
\caption{$\alpha$ phase, 850~K.}
\end{subfigure}\hfill
\begin{subfigure}{0.48\linewidth}
\includegraphics[width=\linewidth]{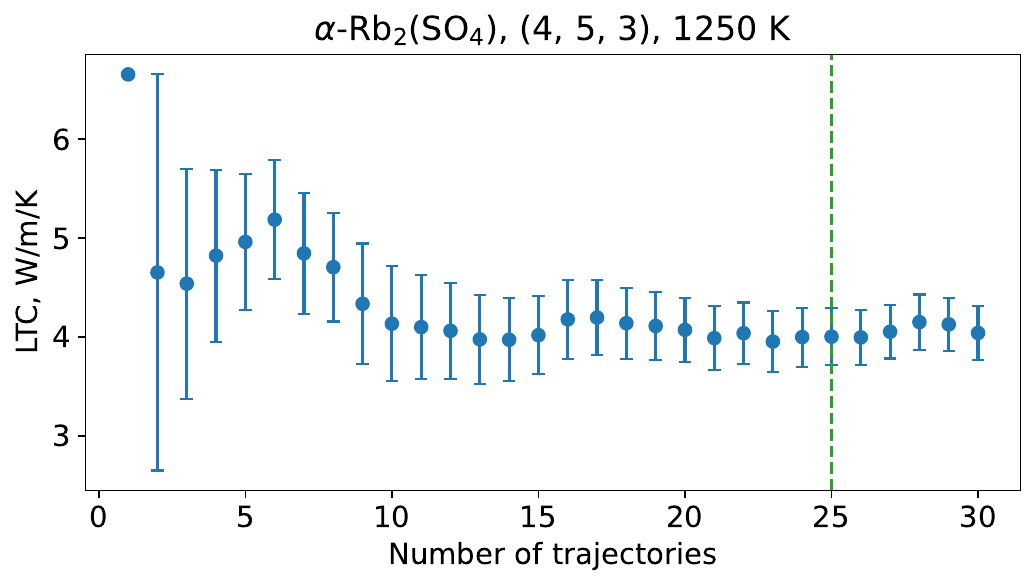}
\caption{$\alpha$ phase, 1250~K.}
\end{subfigure}
\caption{Convergence of the lattice thermal conductivity of Rb$_2$SO$_4$ with the number of trajectories. The supercells contain 1680 atoms ($4\times5\times3$) for the $\beta$ phase, 1680 atoms ($5\times6\times4$) for the $\alpha$ phase at 850~K, and 840 atoms ($4\times5\times3$) for the $\alpha$ phase at 1250~K. Green dashed lines mark 25 trajectories.}
\label{fig:ltc_convergence_rb2so4}
\end{figure}

\begin{figure}[htbp]
\centering
\begin{subfigure}{0.48\linewidth}
\includegraphics[width=\linewidth]{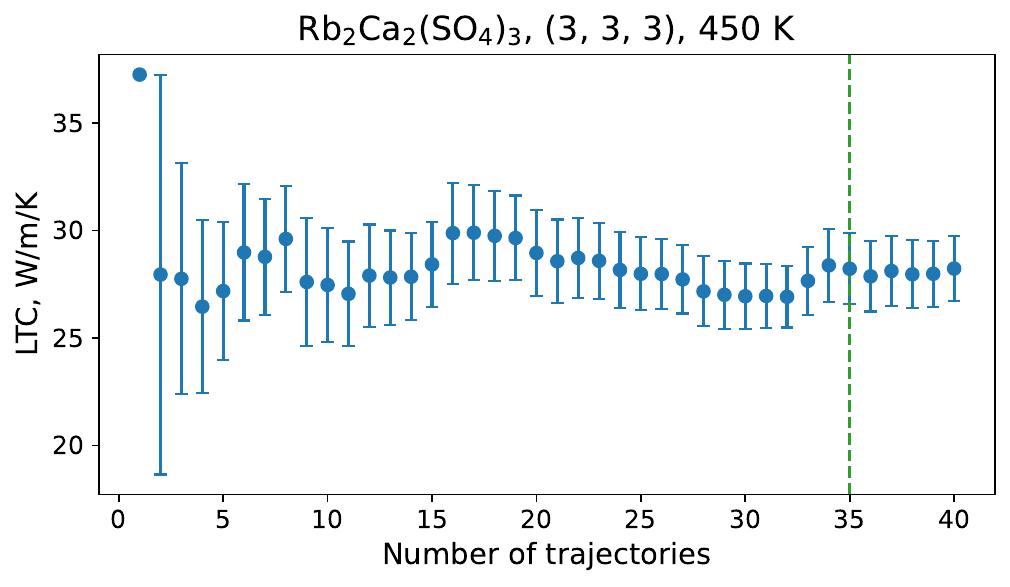}
\caption{450~K.}
\end{subfigure}\hfill
\begin{subfigure}{0.48\linewidth}
\includegraphics[width=\linewidth]{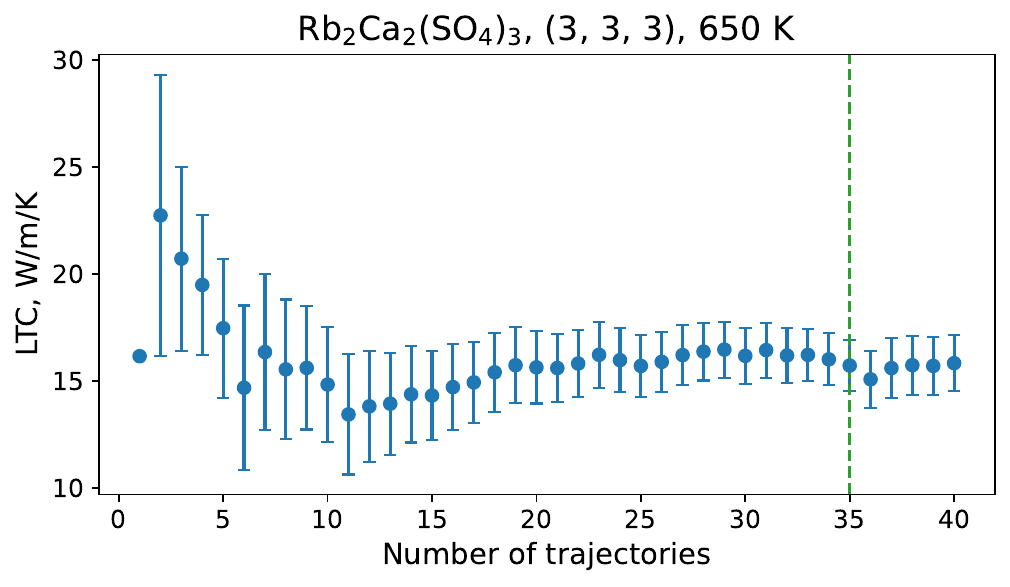}
\caption{650~K.}
\end{subfigure}
\caption{Convergence of the lattice thermal conductivity of Rb$_2$Ca$_2$(SO$_4$)$_3$ with the number of trajectories for the $3\times3\times3$ supercell (2052 atoms). Green dashed lines mark 35 trajectories.}
\label{fig:ltc_convergence_rb2ca2}
\end{figure}

\begin{figure}[htbp]
\centering
\begin{subfigure}{0.48\linewidth}
\includegraphics[width=\linewidth]{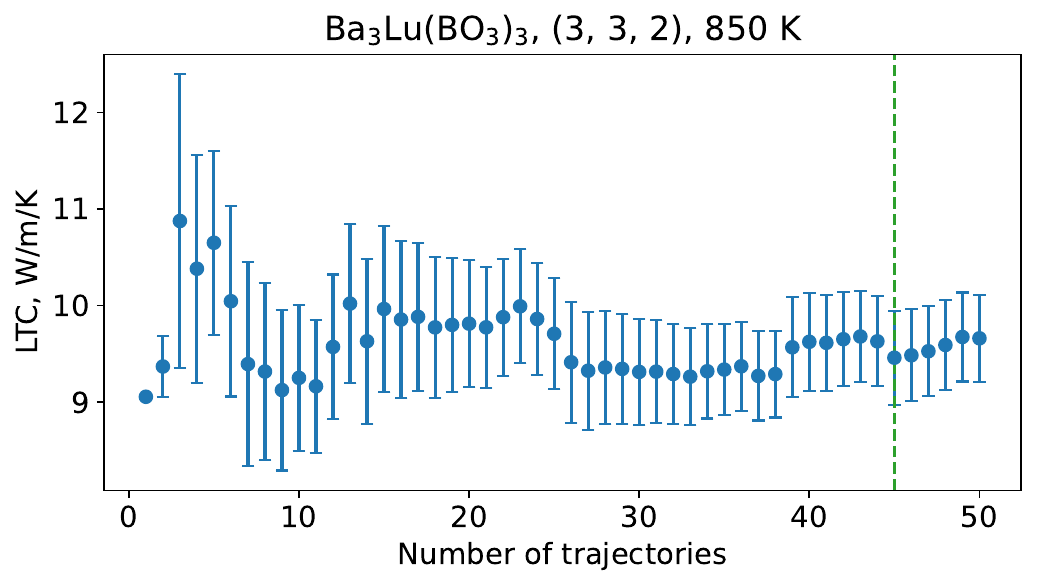}
\caption{850~K.}
\end{subfigure}\hfill
\begin{subfigure}{0.48\linewidth}
\includegraphics[width=\linewidth]{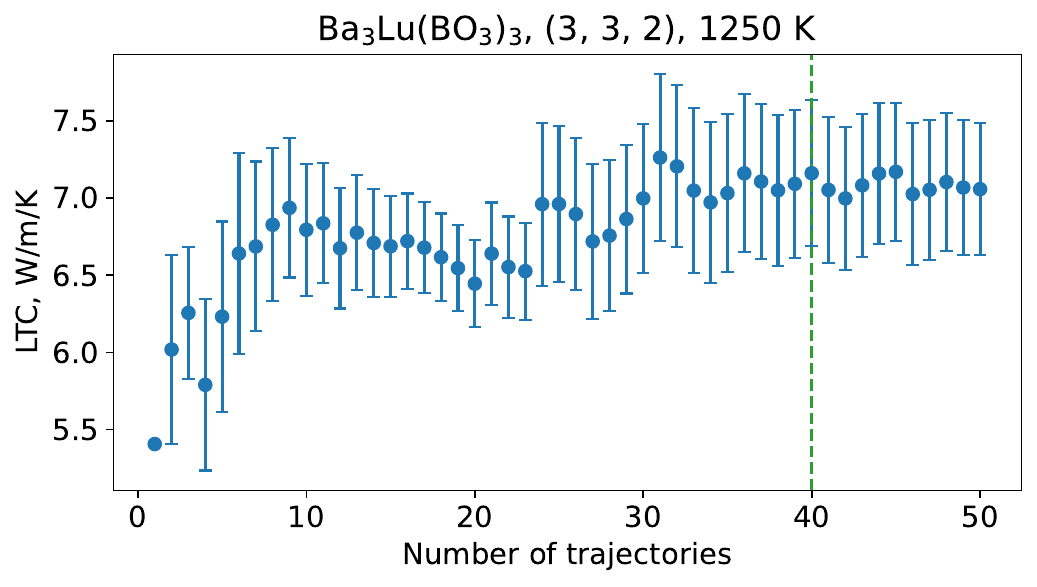}
\caption{1250~K.}
\end{subfigure}
\caption{Convergence of the lattice thermal conductivity of Ba$_3$Lu(BO$_3$)$_3$ with the number of trajectories for the $3\times3\times2$ supercell (1728 atoms). Green dashed lines mark 45 and 40 trajectories at 850 and 1250~K, respectively.}
\label{fig:ltc_convergence_ba3lu}
\end{figure}

Figures~\ref{fig:ltc_convergence_rb2so4}--\ref{fig:ltc_convergence_ba3lu} show the convergence of the Green--Kubo lattice thermal conductivity with the number of independent trajectories included in the average. Each point corresponds to the average over the first $n$ trajectories of the scalar $(\kappa_{xx}+\kappa_{yy}+\kappa_{zz})/3$, evaluated at a correlation time of 20~ps (the last available lag, 19.999~ps, for Rb$_2$Ca$_2$(SO$_4$)$_3$). Error bars show standard errors across trajectories. These figures test trajectory-count convergence at a fixed lag; the final conductivity estimates additionally use the averaging intervals in Table~\ref{tab:ltc_parameters}. The vertical green dashed lines mark the trajectory counts selected in the convergence analysis: 25 for Rb$_2$SO$_4$, 35 for Rb$_2$Ca$_2$(SO$_4$)$_3$, and 45 and 40 for Ba$_3$Lu(BO$_3$)$_3$ at 850 and 1250~K, respectively. The tests include up to 30, 40, and 50 trajectories for these three compounds, respectively.

\clearpage
\section{Band-gap renormalization}

\subsection{Classical sampling}\label{sec:md}

\begin{figure*}[h!]
	\centering
	\begin{minipage}[h]{0.49\linewidth}
		\center{\includegraphics[trim={0cm 0cm 0cm 0cm},clip, width=1\linewidth]{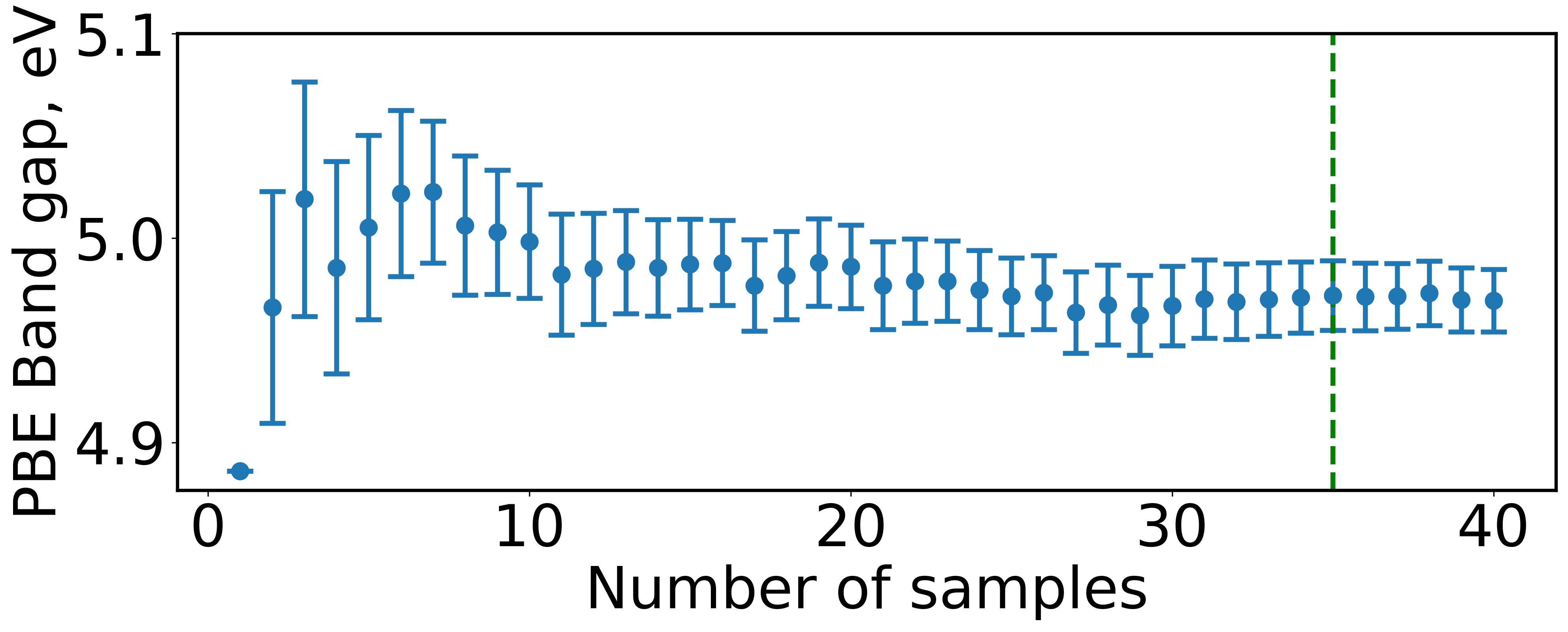}} \\ (a)
	\end{minipage}
    \begin{minipage}[h]{0.49\linewidth}
		\center{\includegraphics[trim={0cm 0cm 0cm 0cm},clip, width=1\linewidth]{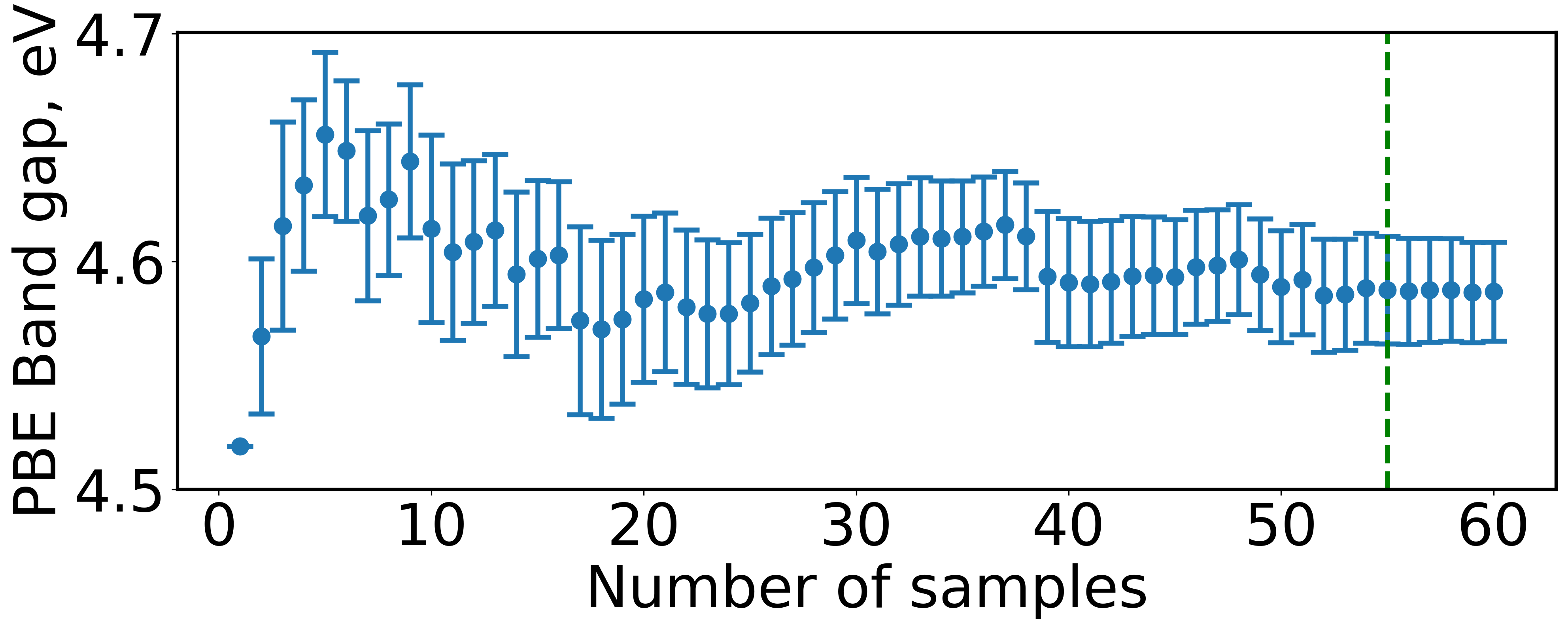}} \\ (b)
	\end{minipage}
    \begin{minipage}[h]{0.49\linewidth}
		\center{\includegraphics[trim={0cm 0cm 0cm 0cm},clip, width=1\linewidth]{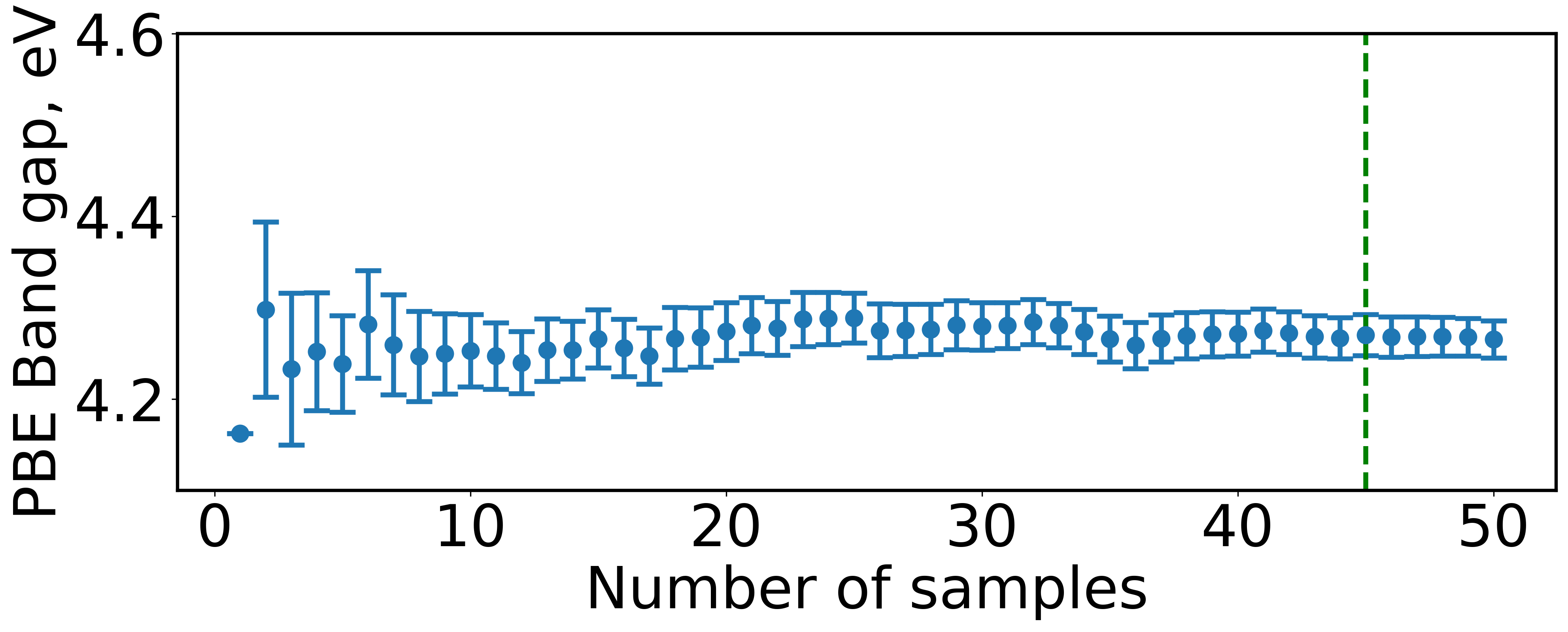}} \\ (c)
	\end{minipage}
    \begin{minipage}[h]{0.49\linewidth}
		\center{\includegraphics[trim={0cm 0cm 0cm 0cm},clip, width=1\linewidth]{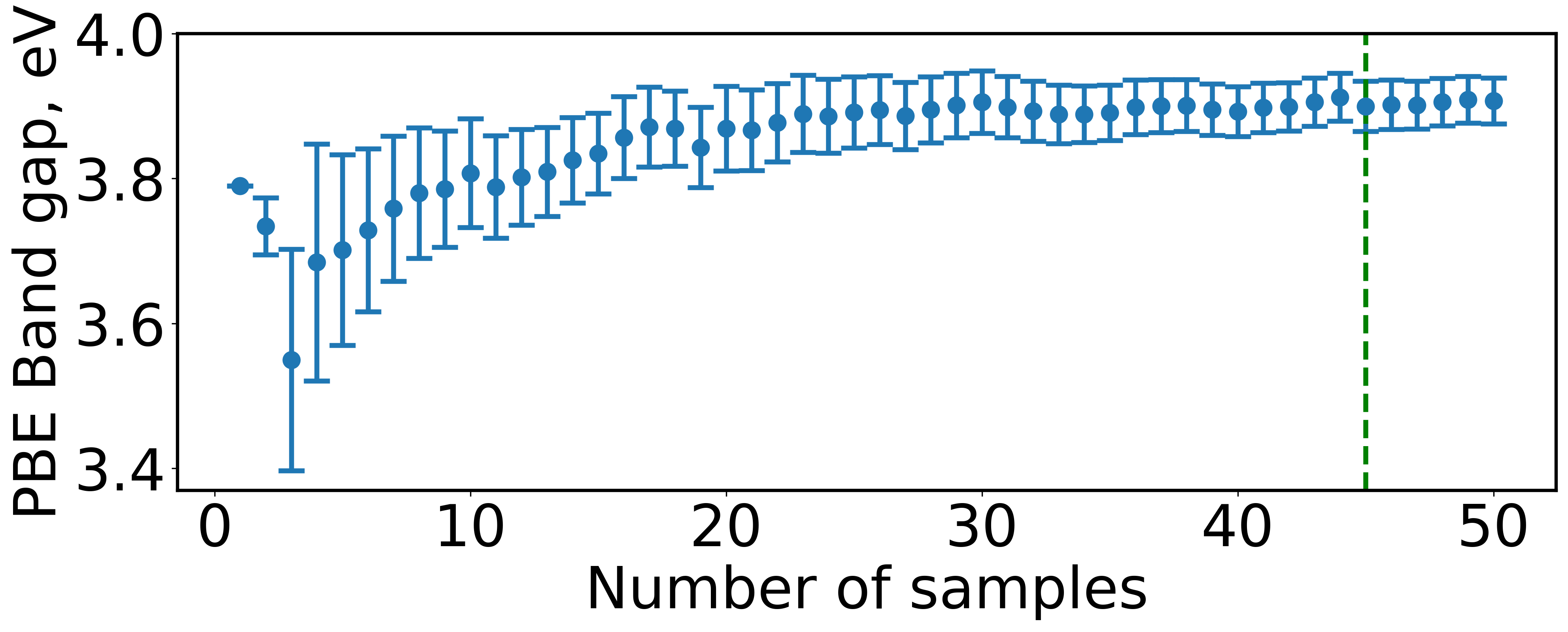}} \\ (d)
	\end{minipage}
    \begin{minipage}[h]{0.49\linewidth}
		\center{\includegraphics[trim={0cm 0cm 0cm 0cm},clip, width=1\linewidth]{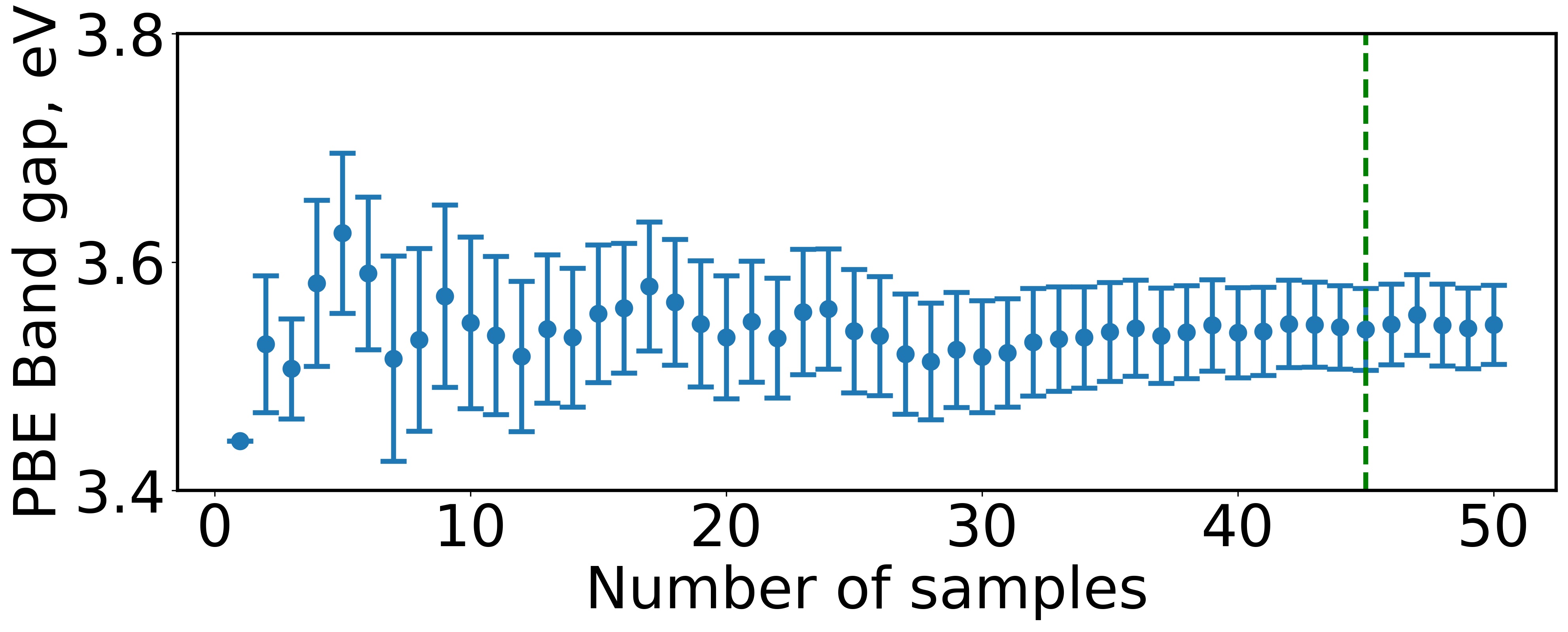}} \\ (e)
	\end{minipage}
    \begin{minipage}[h]{0.49\linewidth}
		\center{\includegraphics[trim={0cm 0cm 0cm 0cm},clip, width=1\linewidth]{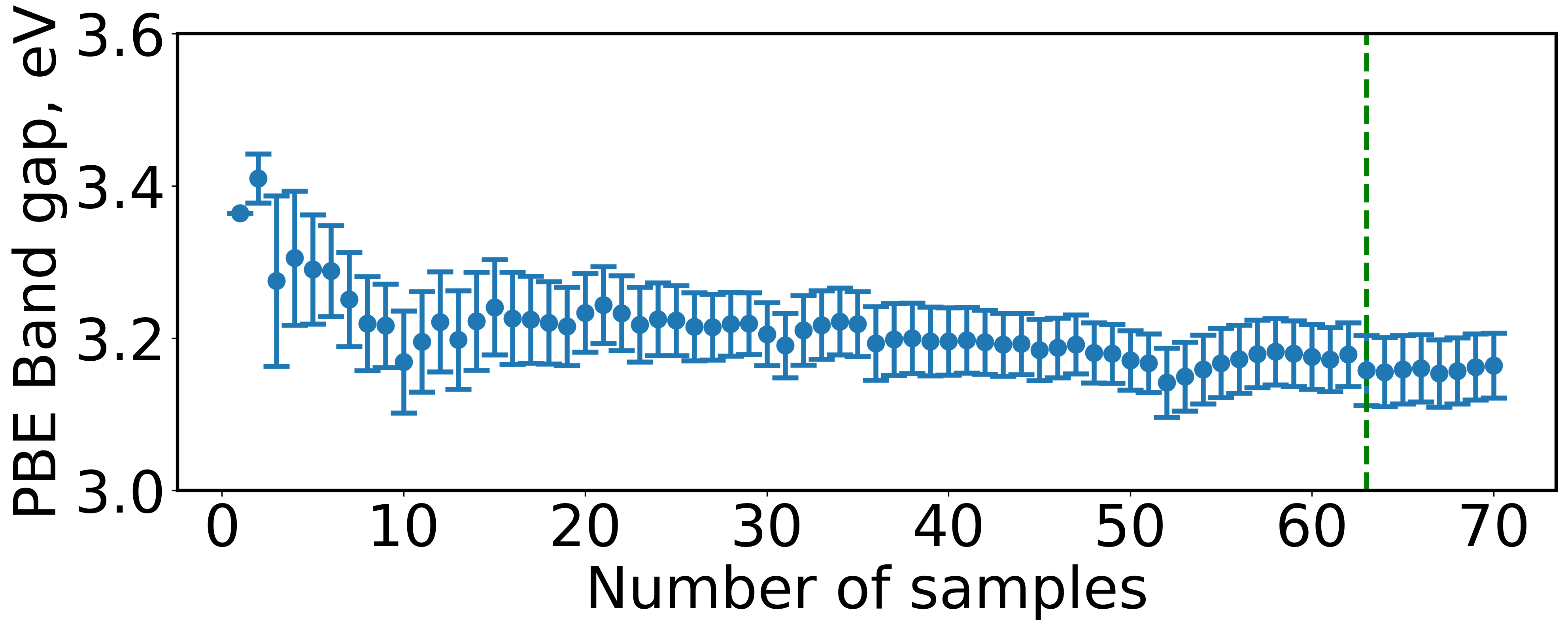}} \\ (f)
	\end{minipage}
    \caption{Convergence of the average PBE band gap of Rb$_2$Ca$_2$(SO$_4$)$_3$ with the number of MD snapshots at (a) 250, (b) 450, (c) 650, (d) 850, (e) 1050, and (f) 1250~K. Dashed green lines mark the number of snapshots at which convergence is reached.}
    \label{fig:conv_cl_rbca}
\end{figure*}

\begin{figure*}[h!]
	\centering
	\begin{minipage}[h]{0.49\linewidth}
		\center{\includegraphics[trim={0cm 0cm 0cm 0cm},clip, width=1\linewidth]{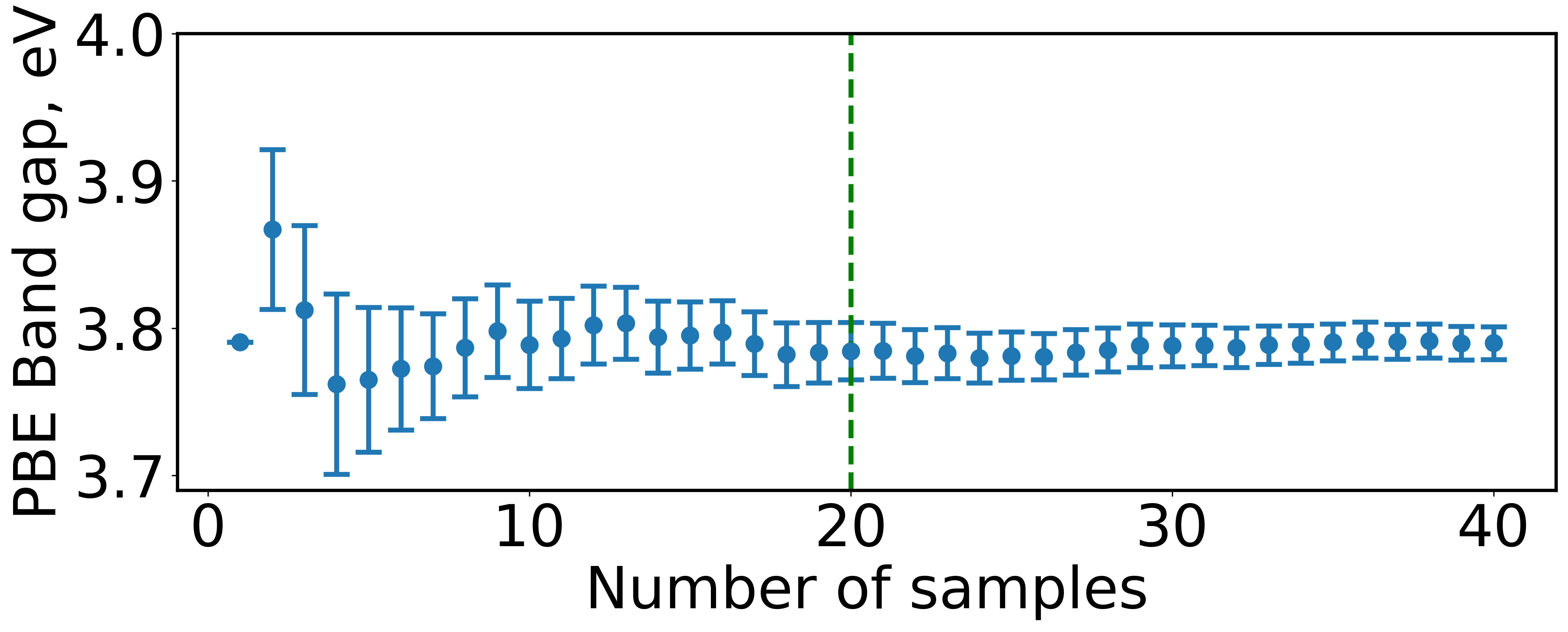}} \\ (a)
	\end{minipage}
    \begin{minipage}[h]{0.49\linewidth}
		\center{\includegraphics[trim={0cm 0cm 0cm 0cm},clip, width=1\linewidth]{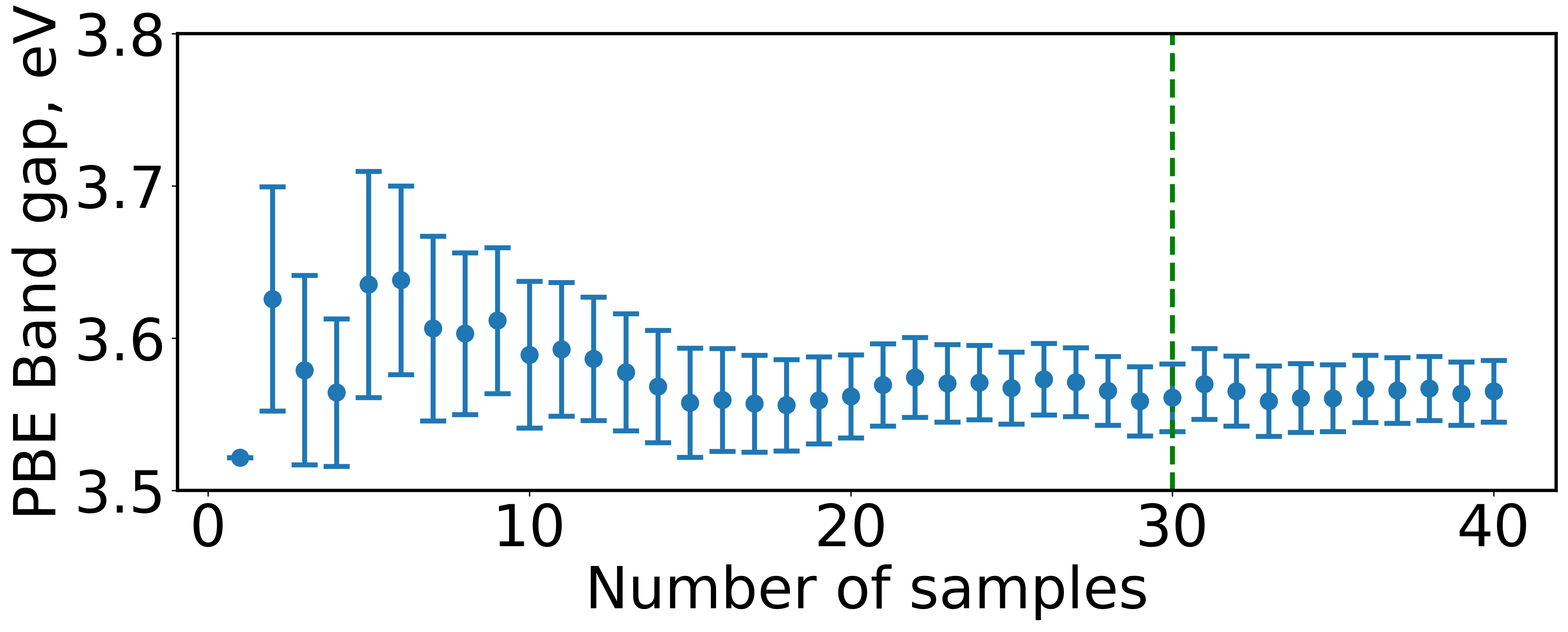}} \\ (b)
	\end{minipage}
    \begin{minipage}[h]{0.49\linewidth}
		\center{\includegraphics[trim={0cm 0cm 0cm 0cm},clip, width=1\linewidth]{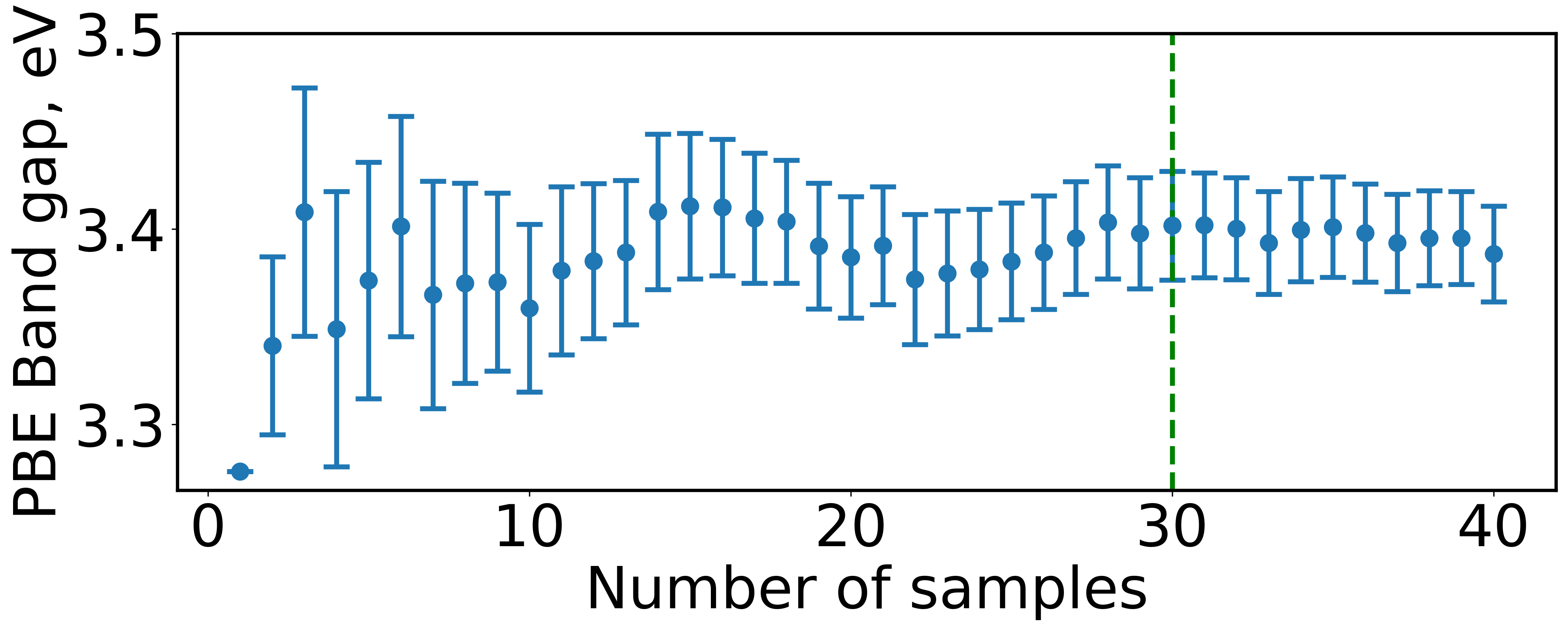}} \\ (c)
	\end{minipage}
    \begin{minipage}[h]{0.49\linewidth}
		\center{\includegraphics[trim={0cm 0cm 0cm 0cm},clip, width=1\linewidth]{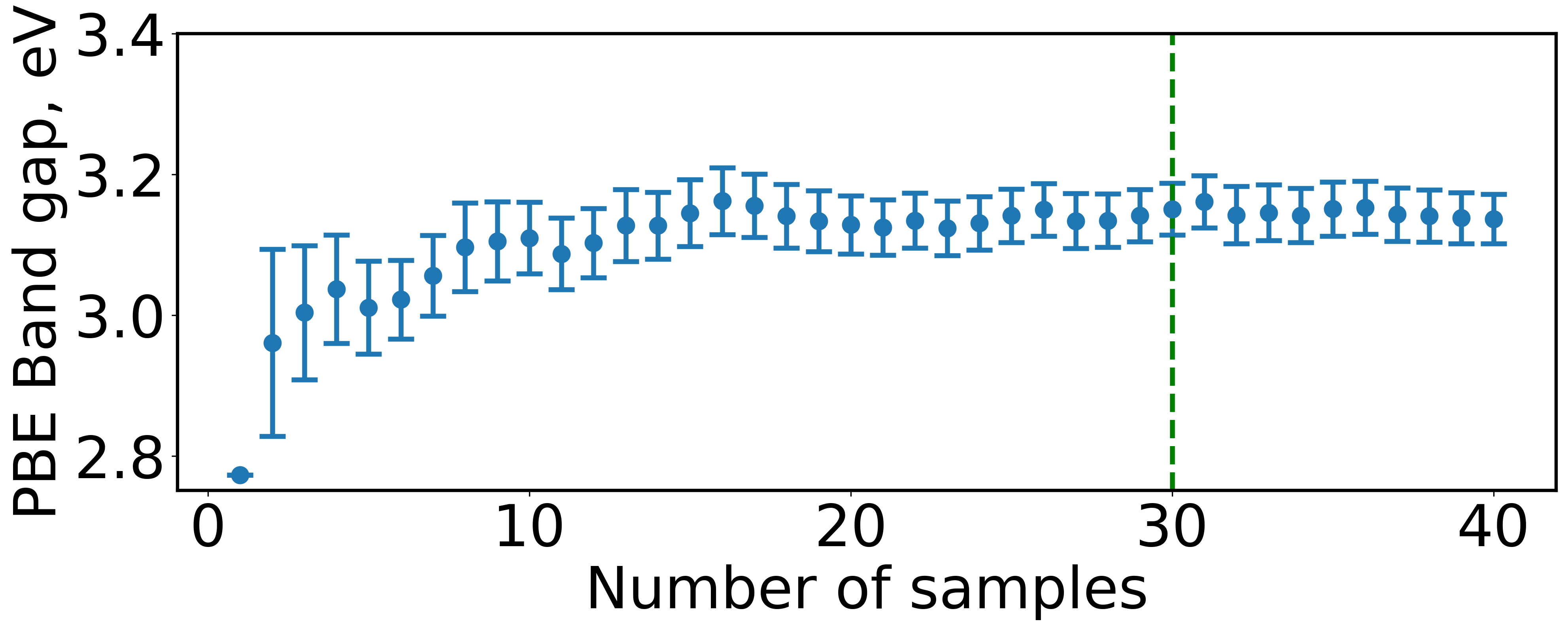}} \\ (d)
	\end{minipage}
    \begin{minipage}[h]{0.49\linewidth}
		\center{\includegraphics[trim={0cm 0cm 0cm 0cm},clip, width=1\linewidth]{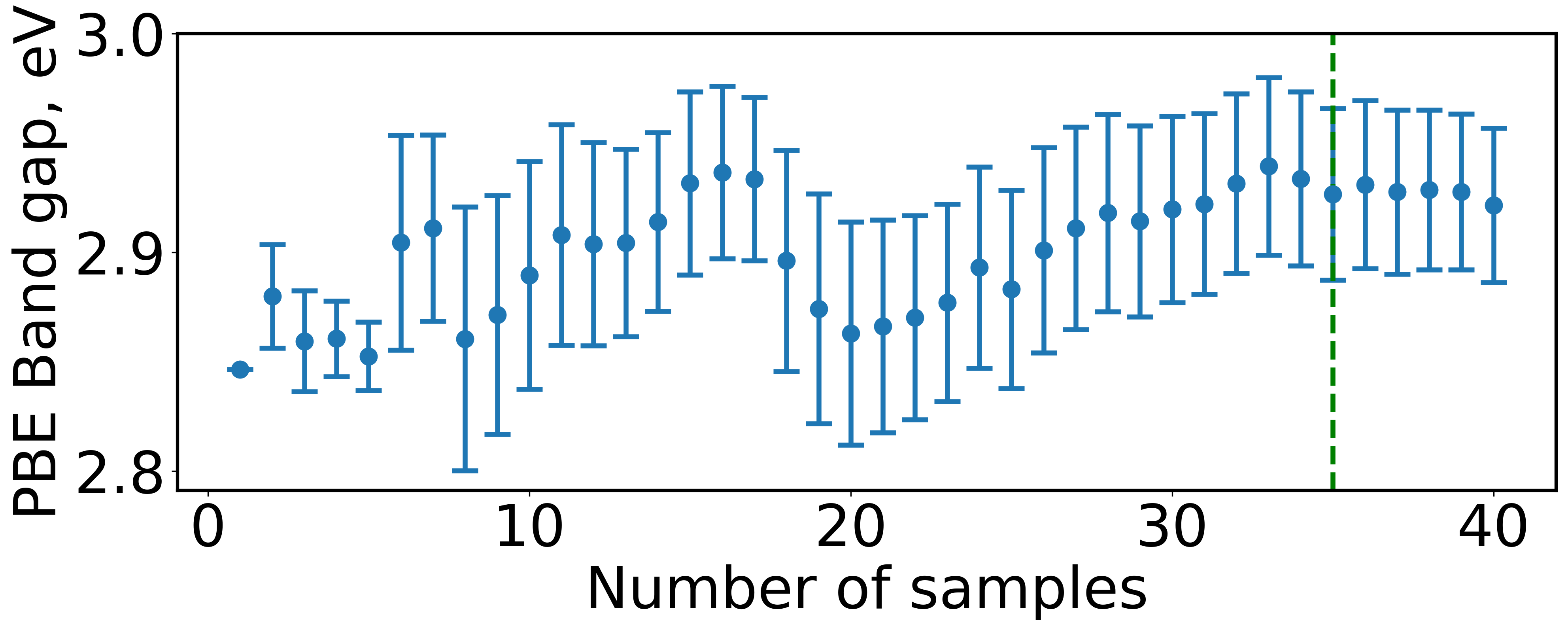}} \\ (e)
	\end{minipage}
    \begin{minipage}[h]{0.49\linewidth}
		\center{\includegraphics[trim={0cm 0cm 0cm 0cm},clip, width=1\linewidth]{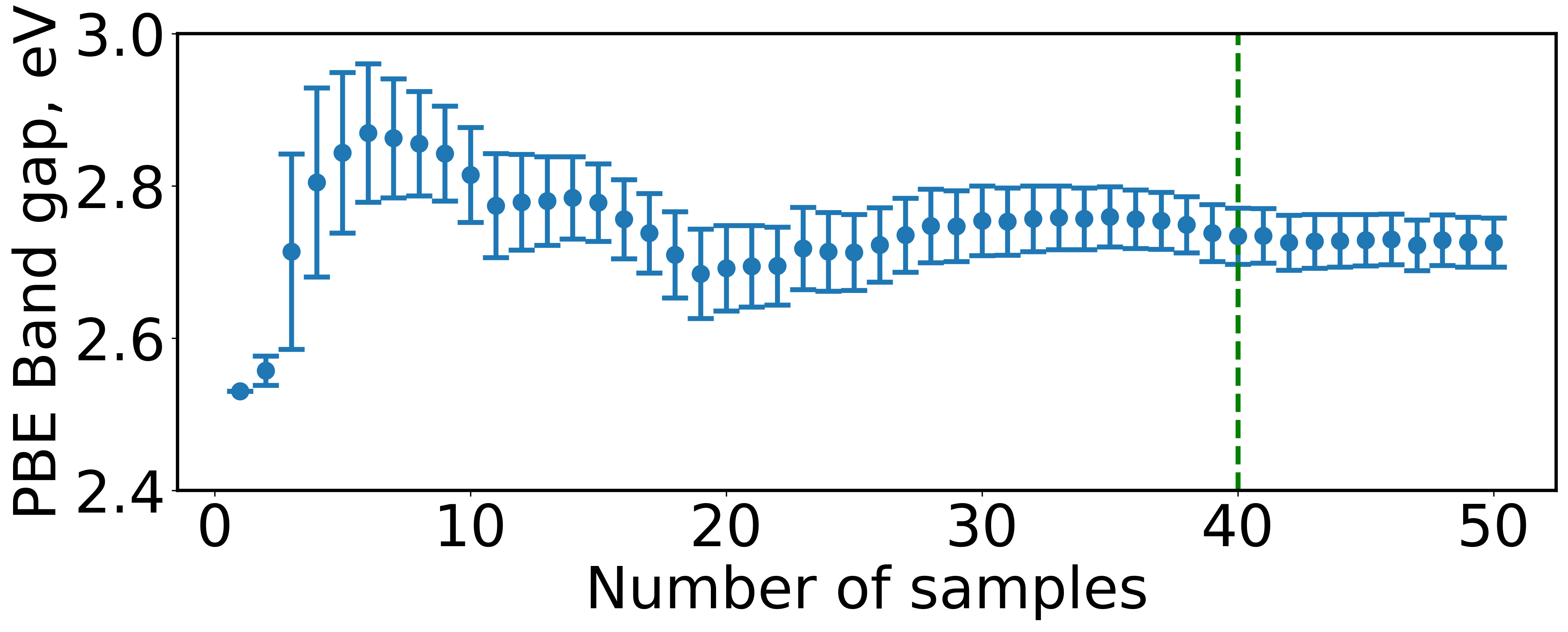}} \\ (f)
	\end{minipage}
    \caption{Convergence of the average PBE band gap of Ba$_3$Lu(BO$_3$)$_3$ with the number of MD snapshots at (a) 250, (b) 450, (c) 650, (d) 850, (e) 1050, and (f) 1250~K. Dashed green lines mark the number of snapshots at which convergence is reached.}
    \label{fig:conv_cl_balu}
\end{figure*}

\begin{figure*}[h!]
	\centering
	\begin{minipage}[h]{0.49\linewidth}
		\center{\includegraphics[trim={0cm 0cm 0cm 0cm},clip, width=1\linewidth]{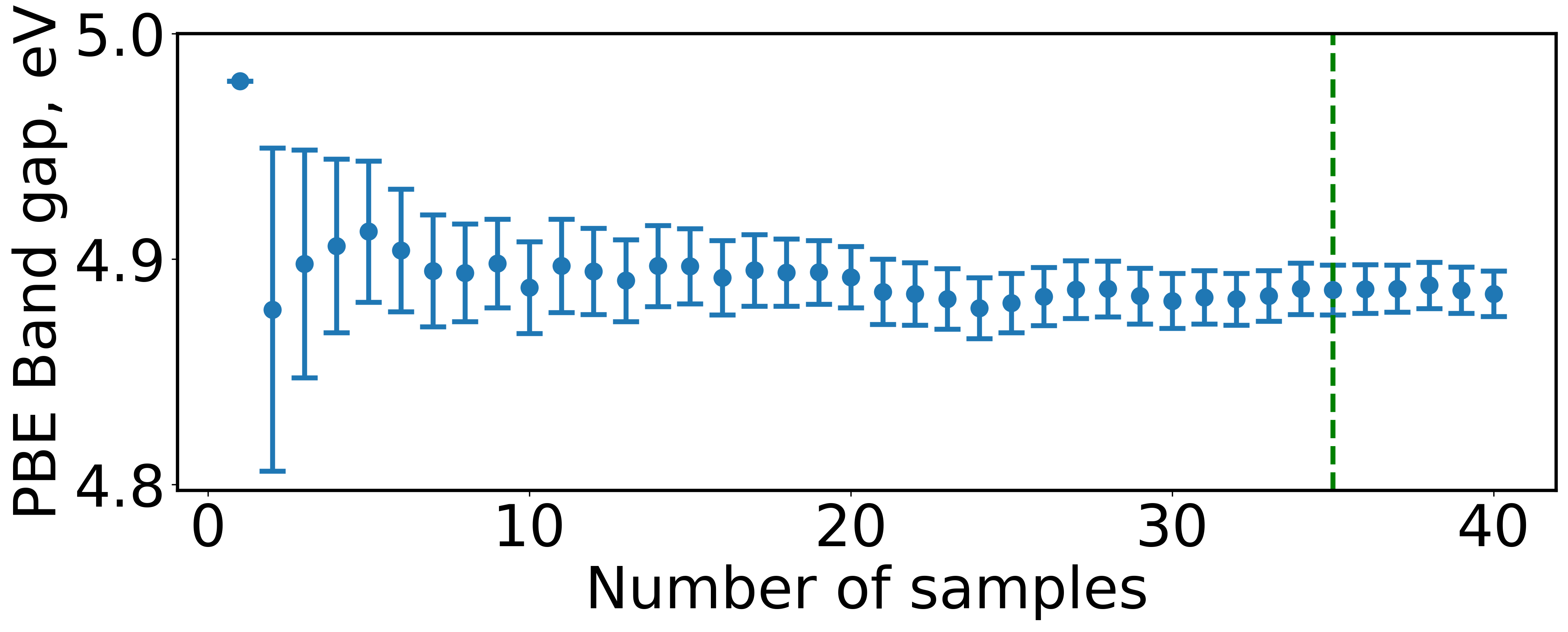}} \\ (a)
	\end{minipage}
    \begin{minipage}[h]{0.49\linewidth}
		\center{\includegraphics[trim={0cm 0cm 0cm 0cm},clip, width=1\linewidth]{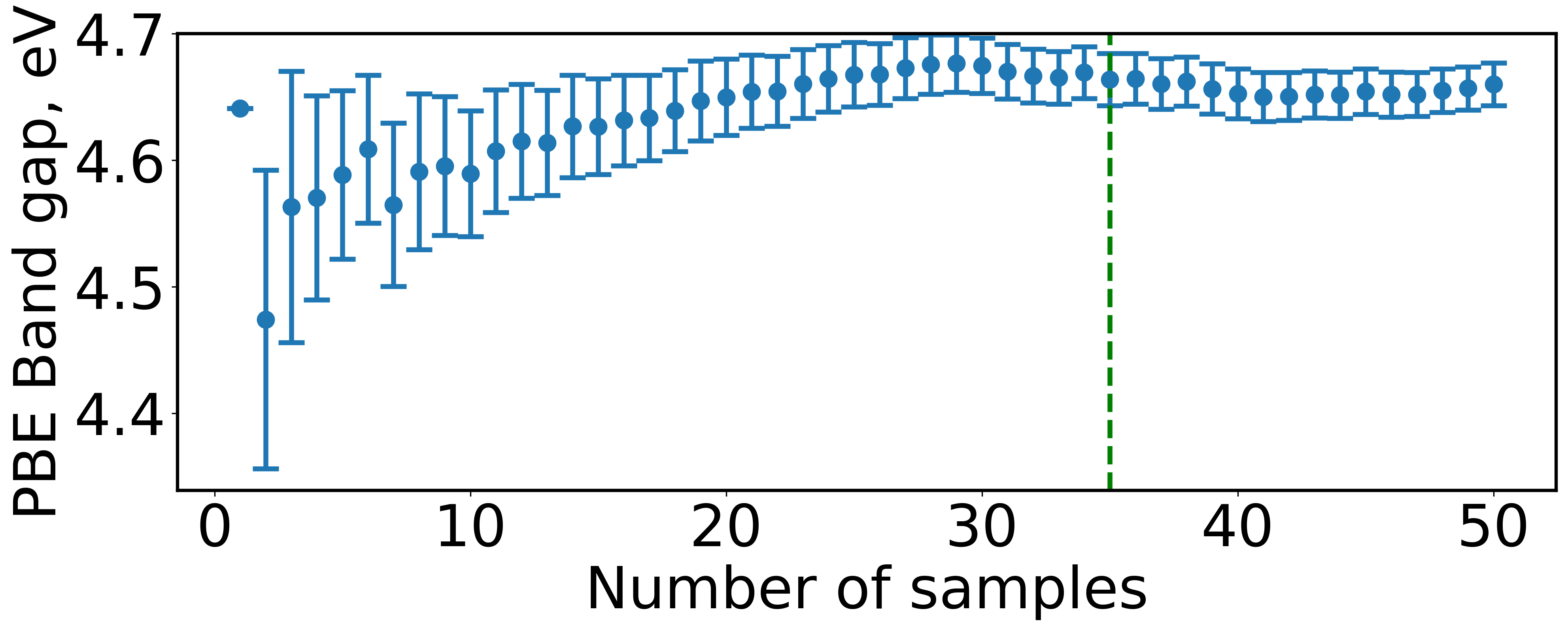}} \\ (b)
	\end{minipage}
    \begin{minipage}[h]{0.49\linewidth}
		\center{\includegraphics[trim={0cm 0cm 0cm 0cm},clip, width=1\linewidth]{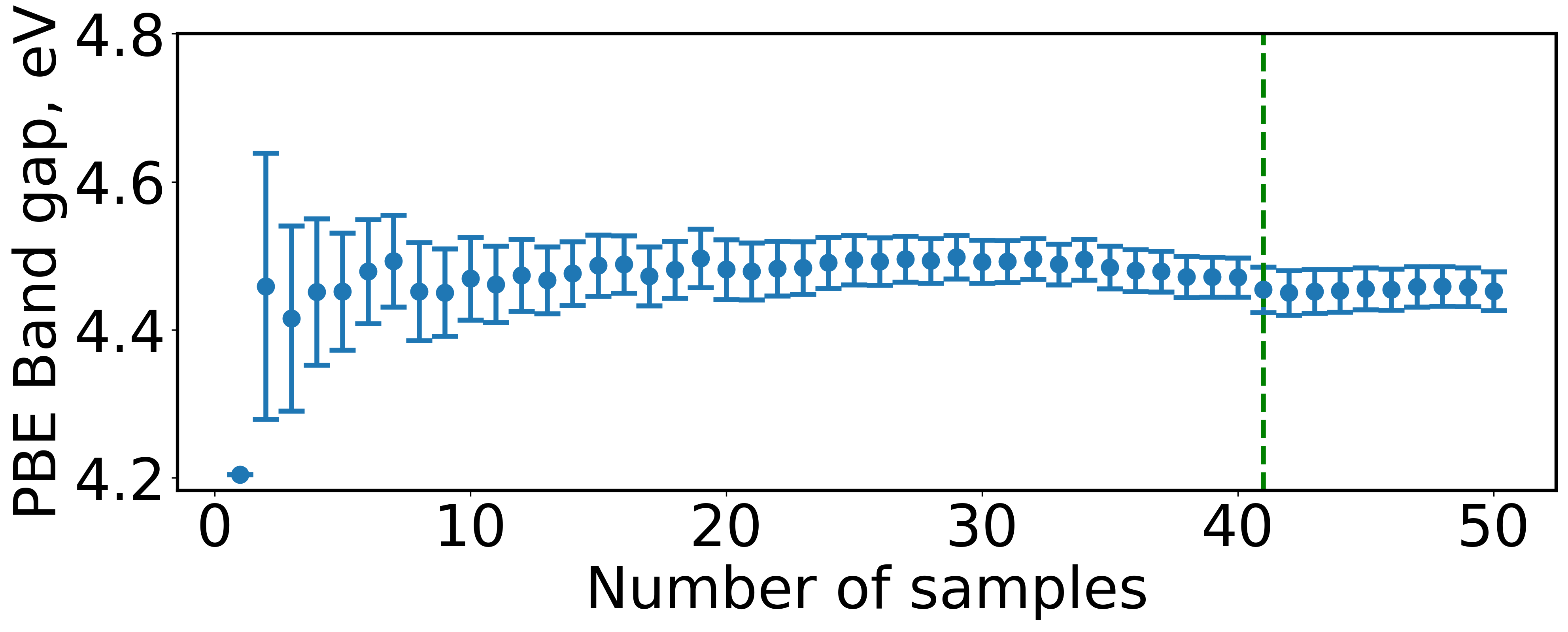}} \\ (c)
	\end{minipage}
   
    \caption{Convergence of the average PBE band gap of $\beta$-Rb$_2$SO$_4$ with the number of MD snapshots at (a) 250, (b) 450, and (c) 650~K. Dashed green lines mark the number of snapshots at which convergence is reached.}
    \label{fig:conv_cl_brso}
\end{figure*}

${\rm BGR^{MD}}$ was obtained from classical $NVT$ MD with the MTPs in $2\times2\times2$ (608 atoms), $2\times2\times1$ (384 atoms), and $3\times4\times2$ (672 atoms) supercells of Rb$_2$Ca$_2$(SO$_4$)$_3$, Ba$_3$Lu(BO$_3$)$_3$, and $\beta$-Rb$_2$SO$_4$, respectively, constructed from the MTP-relaxed unit cells. Each trajectory lasted 125~ps (25~ps of equilibration and 100~ps of production) with a timestep of 1~fs, and snapshots for the DFT calculations were taken every 100~fs. The convergence of the average PBE band gap with the number of snapshots is shown in Figs.~\ref{fig:conv_cl_rbca}--\ref{fig:conv_cl_brso}.

\clearpage
\subsection{DFT calculations}

Band gaps were computed with FHI-aims~\cite{fhiaims} and the PBE functional~\cite{pbe} using the default \textit{light} settings for the basis sets and integration grids. The unit cells were sampled with $2\times2\times2$ (Rb$_2$Ca$_2$(SO$_4$)$_3$) and $2\times2\times1$ ($\beta$-Rb$_2$SO$_4$ and Ba$_3$Lu(BO$_3$)$_3$) $k$-point meshes, and all supercells with displaced atoms were calculated at the $\Gamma$ point only.

\subsection{Quantum sampling}

To include quantum nuclear effects, configurations of $\beta$-Rb$_2$SO$_4$, Rb$_2$Ca$_2$(SO$_4$)$_3$, and Ba$_3$Lu(BO$_3$)$_3$ were generated by sTDEP sampling with quantum amplitudes. The supercells and force-constant cutoffs are the same as in the classical sTDEP calculations (Table~\ref{tab:stdep}), and the procedure is the same as in Sec.~\ref{sec:stdep_si}, except that the displacement amplitudes include the Bose--Einstein occupations,

\begin{equation}
    A_{is} = \sqrt{\frac{\hbar(n_s+1/2)}{m_i\omega_s}},
\end{equation}
where $m_i$ is the mass of atom $i$ and $n_s$ is the Bose--Einstein occupation of mode $s$.

\begin{figure*}[h!]
	\centering
	\begin{minipage}[h]{0.49\linewidth}
		\center{\includegraphics[trim={0cm 0cm 0cm 0cm},clip, width=1\linewidth]{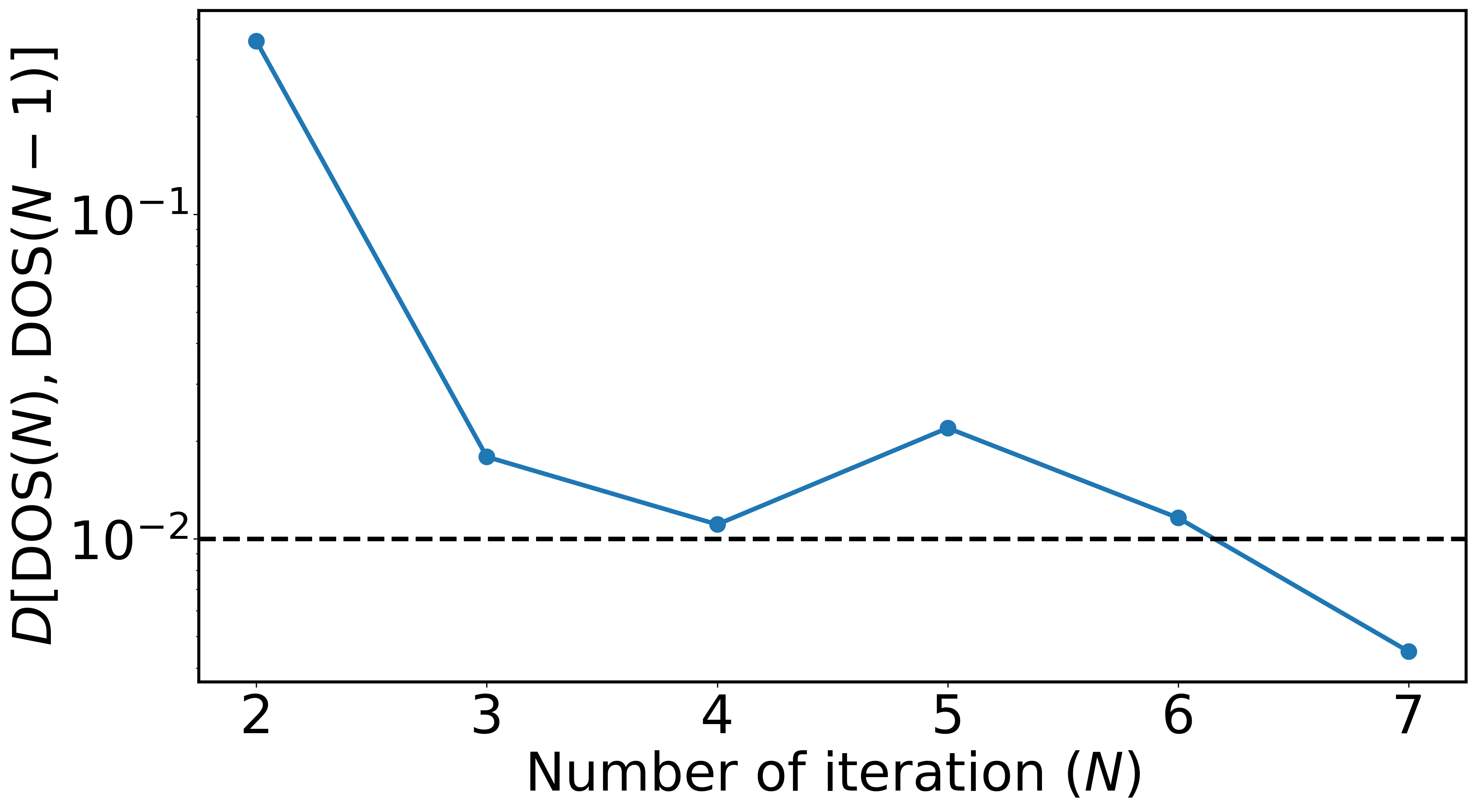}} \\ (a)
	\end{minipage}
    \begin{minipage}[h]{0.49\linewidth}
		\center{\includegraphics[trim={0cm 0cm 0cm 0cm},clip, width=1\linewidth]{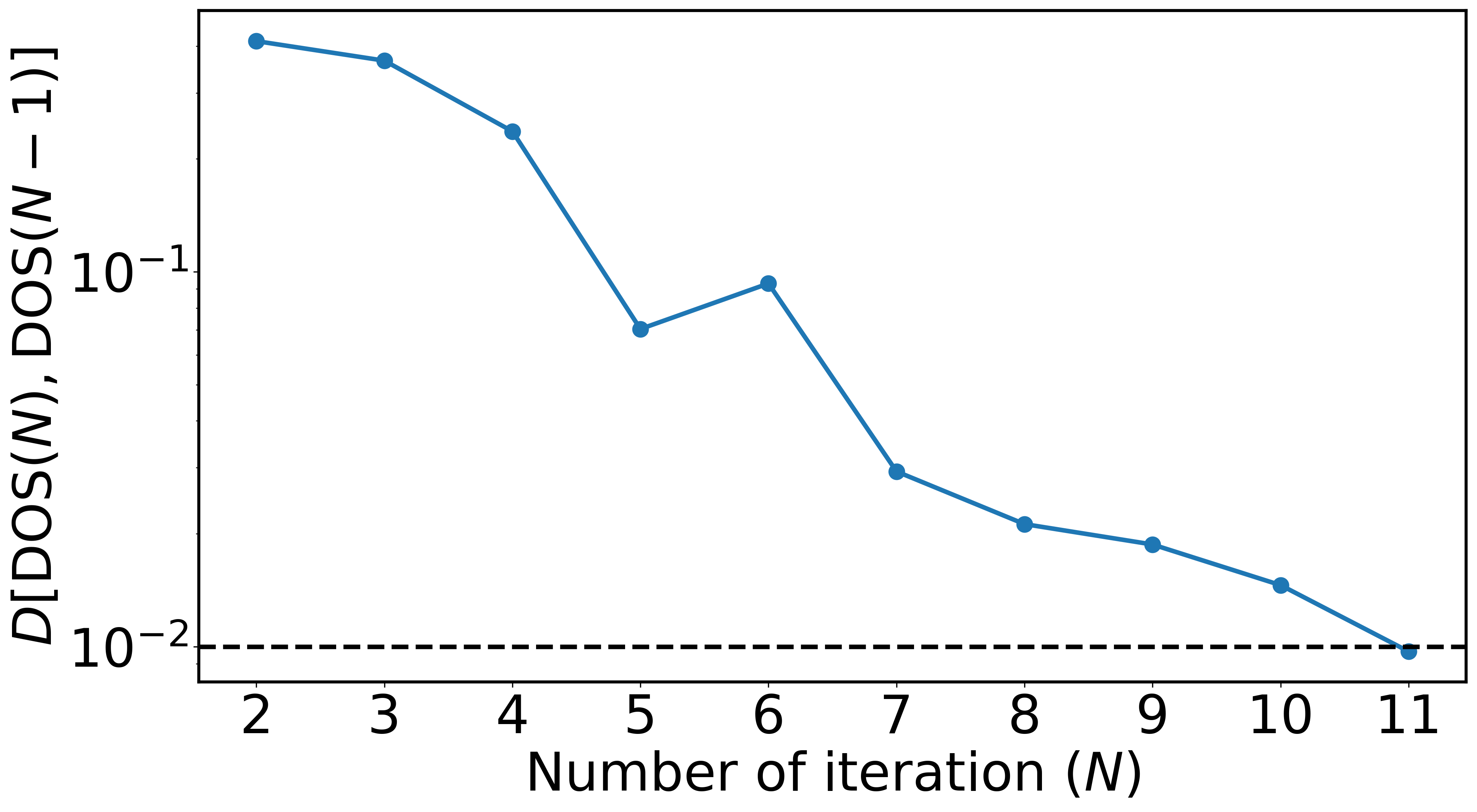}} \\ (b)
	\end{minipage}
    \begin{minipage}[h]{0.49\linewidth}
		\center{\includegraphics[trim={0cm 0cm 0cm 0cm},clip, width=1\linewidth]{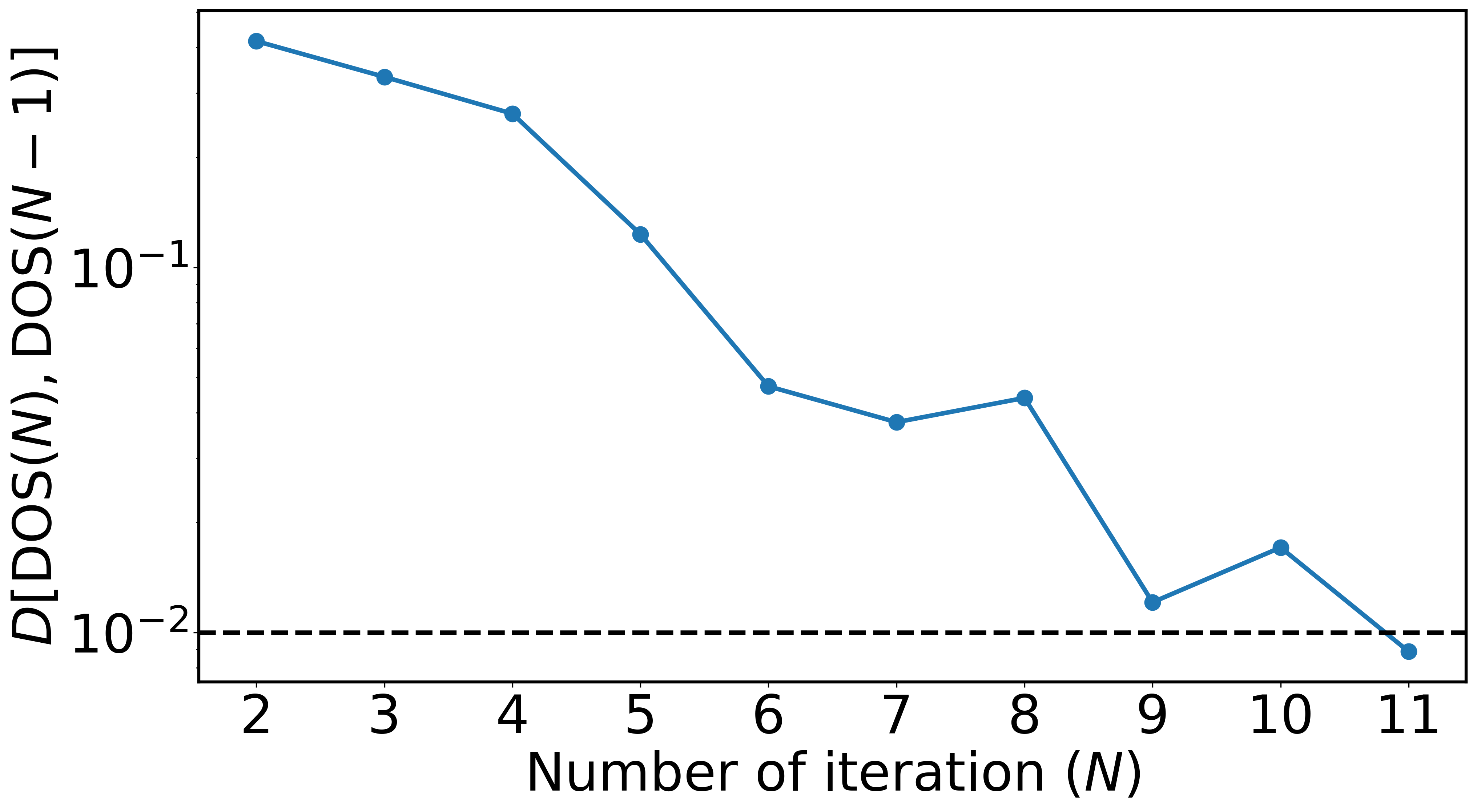}} \\ (c)
	\end{minipage}
        \begin{minipage}[h]{0.49\linewidth}
		\center{\includegraphics[trim={0cm 0cm 0cm 0cm},clip, width=1\linewidth]{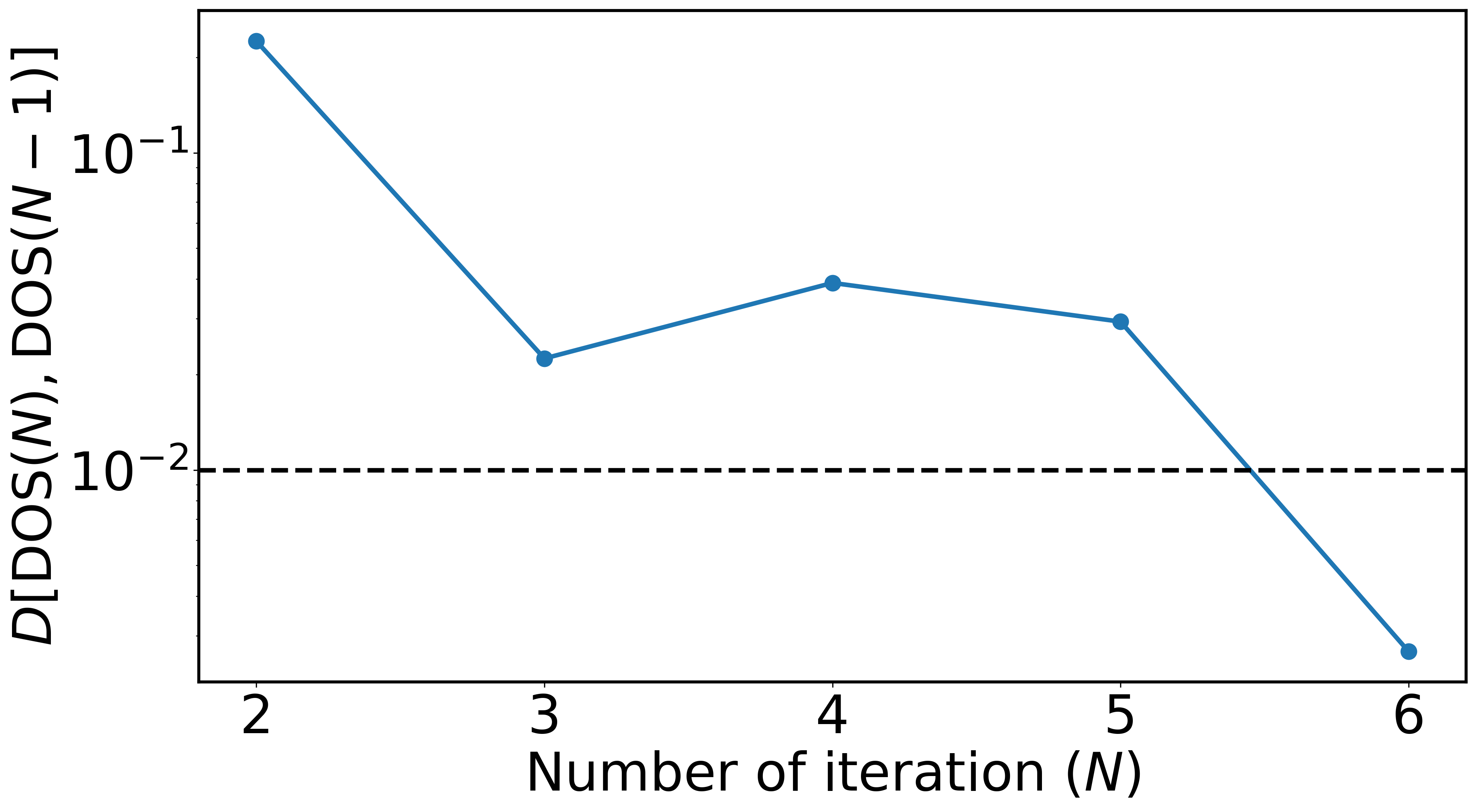}} \\ (d)
	\end{minipage}
    
    \caption{Cosine distance $D[{\rm DOS}(N),{\rm DOS}(N-1)]$ between the phonon DOS of consecutive quantum sTDEP iterations for Rb$_2$Ca$_2$(SO$_4$)$_3$ at (a) 0~K, (b) 250~K, (c) 450~K, and (d) 1050~K. Dashed lines mark the convergence threshold $D=10^{-2}$.}
    \label{fig:conv_stdep_rbca}
\end{figure*}

\begin{figure*}[h!]
	\centering
	\begin{minipage}[h]{0.49\linewidth}
		\center{\includegraphics[trim={0cm 0cm 0cm 0cm},clip, width=1\linewidth]{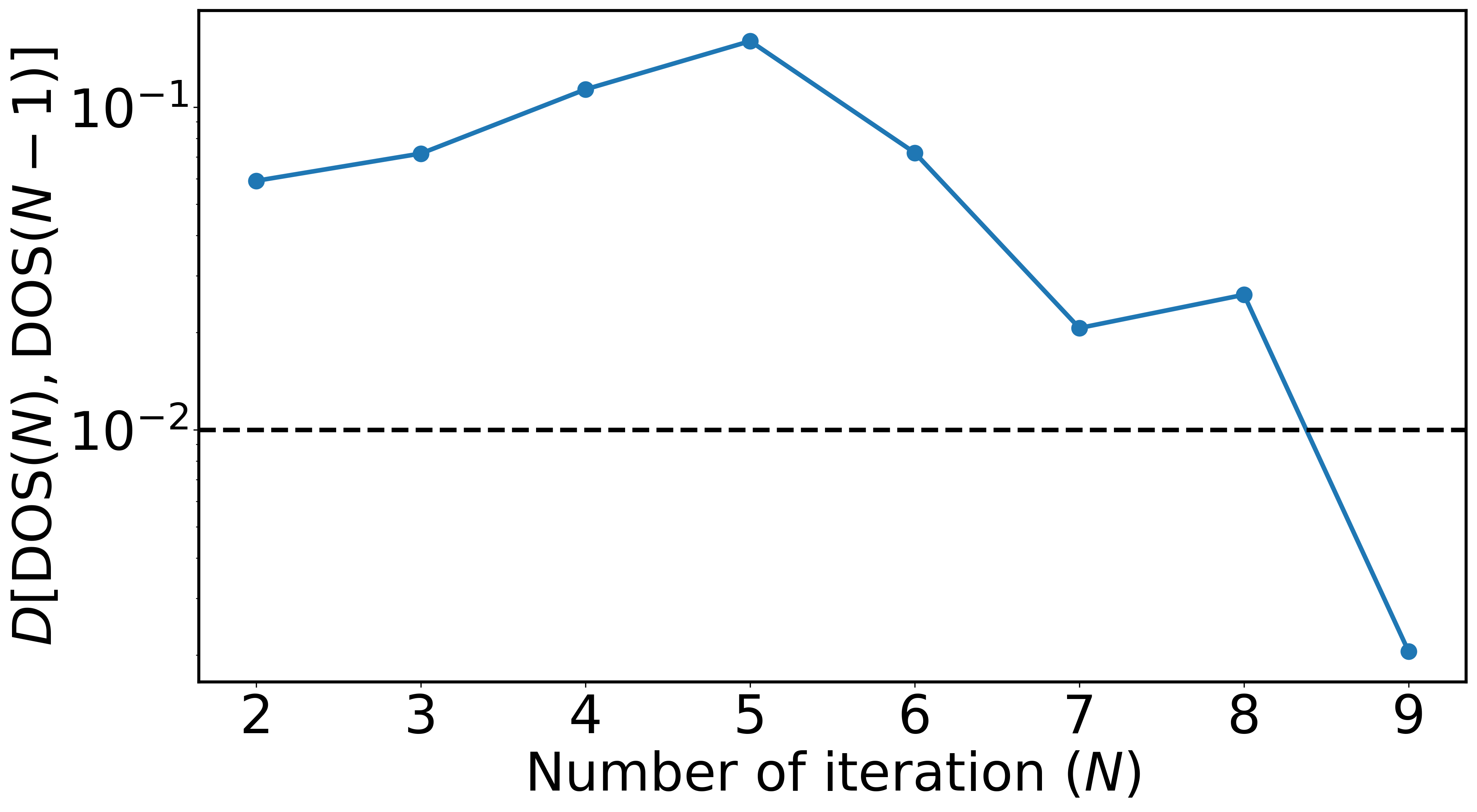}} \\ (a)
	\end{minipage}
    \begin{minipage}[h]{0.49\linewidth}
		\center{\includegraphics[trim={0cm 0cm 0cm 0cm},clip, width=1\linewidth]{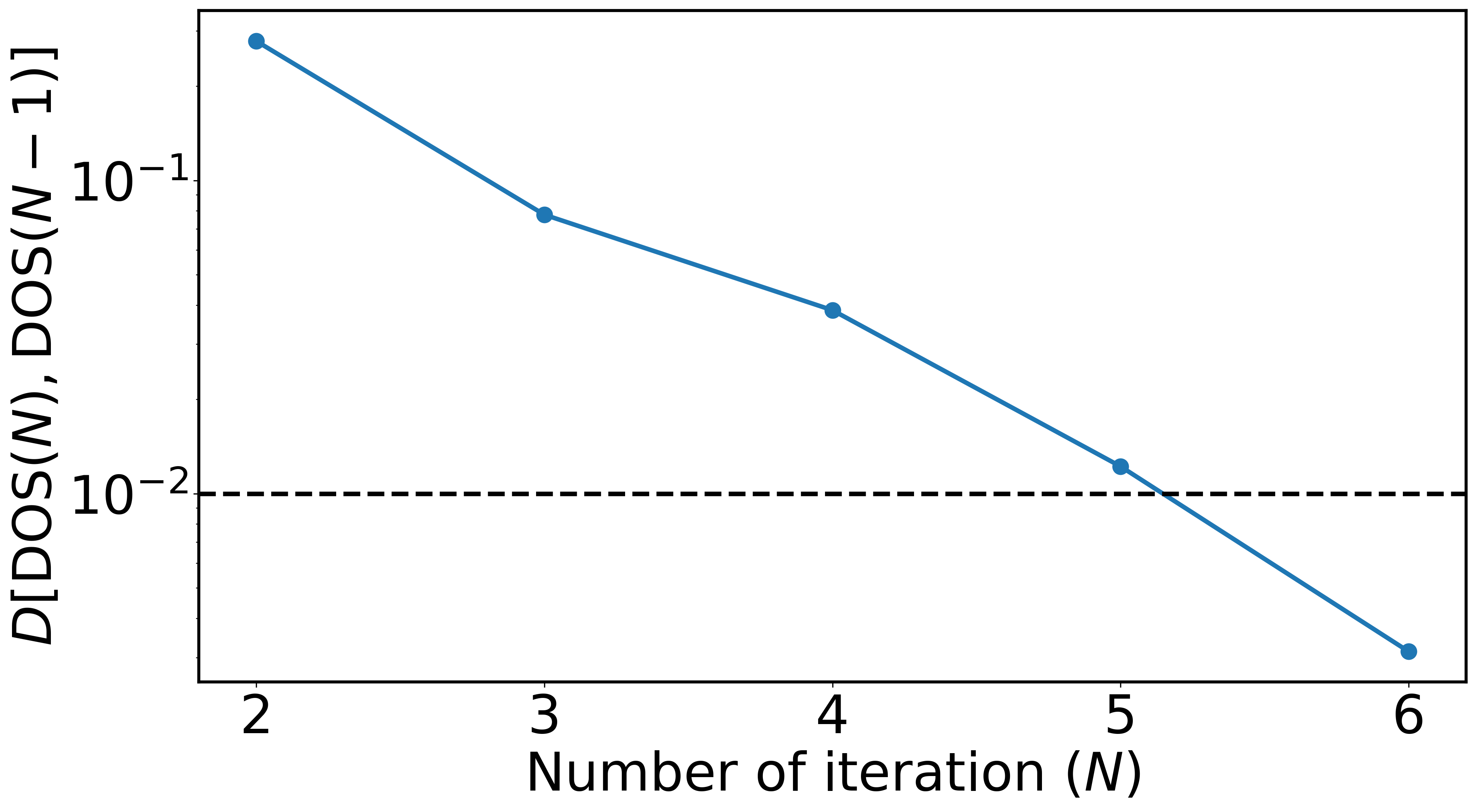}} \\ (b)
	\end{minipage}
    \begin{minipage}[h]{0.49\linewidth}
		\center{\includegraphics[trim={0cm 0cm 0cm 0cm},clip, width=1\linewidth]{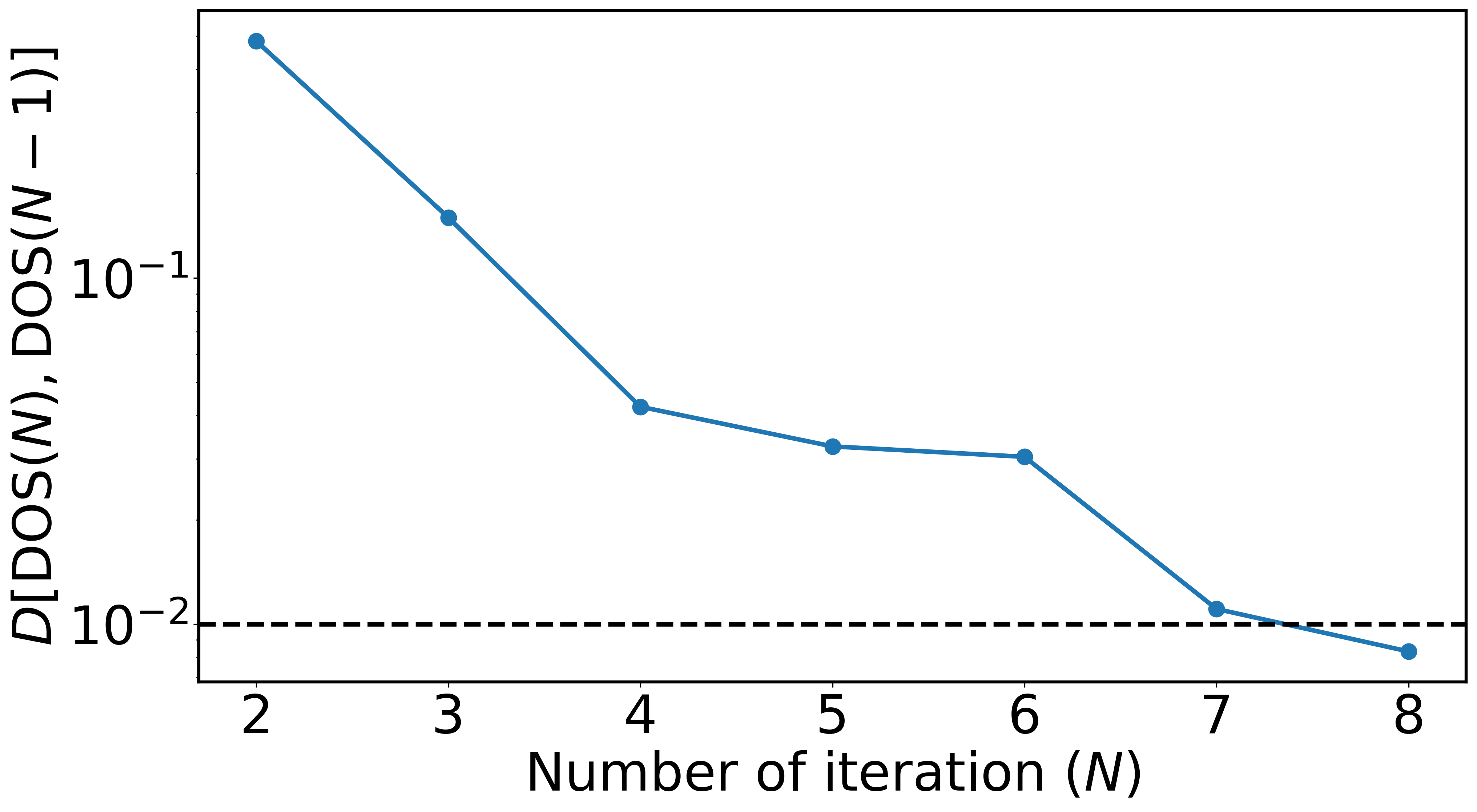}} \\ (c)
	\end{minipage}
        \begin{minipage}[h]{0.49\linewidth}
		\center{\includegraphics[trim={0cm 0cm 0cm 0cm},clip, width=1\linewidth]{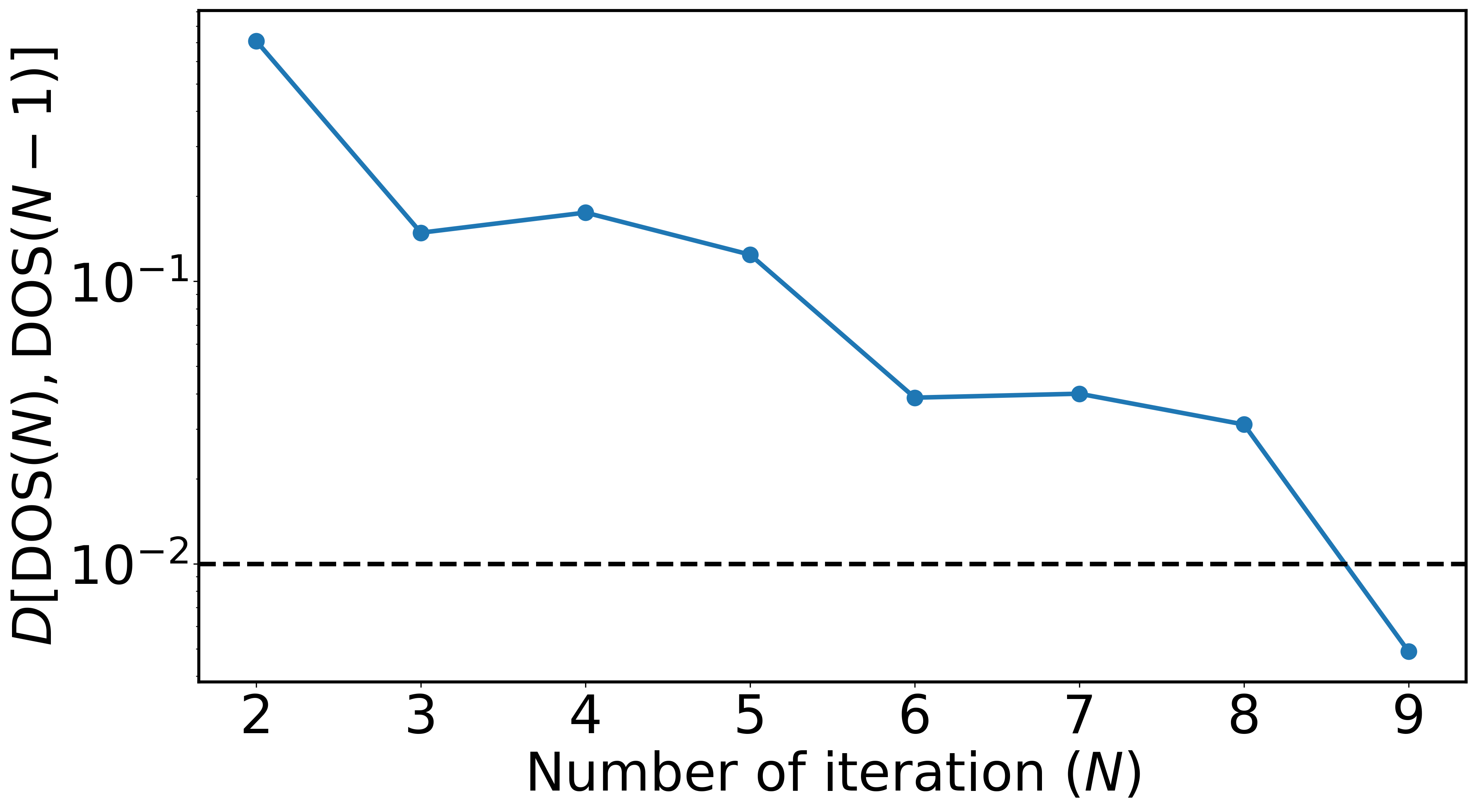}} \\ (d)
	\end{minipage}
   
    \caption{Cosine distance $D[{\rm DOS}(N),{\rm DOS}(N-1)]$ between the phonon DOS of consecutive quantum sTDEP iterations for $\beta$-Rb$_2$SO$_4$ at (a) 0~K, (b) 250~K, (c) 450~K, and (d) 650~K. Dashed lines mark the convergence threshold $D=10^{-2}$.}
    \label{fig:conv_stdep_rbso}
\end{figure*}

The convergence of the quantum sTDEP loops is shown in Figs.~\ref{fig:conv_stdep_rbca} and \ref{fig:conv_stdep_rbso} for Rb$_2$Ca$_2$(SO$_4$)$_3$ and $\beta$-Rb$_2$SO$_4$.

\begin{figure*}[h!]
	\centering
	\begin{minipage}[h]{0.49\linewidth}
		\center{\includegraphics[trim={0cm 0cm 0cm 0cm},clip, width=1\linewidth]{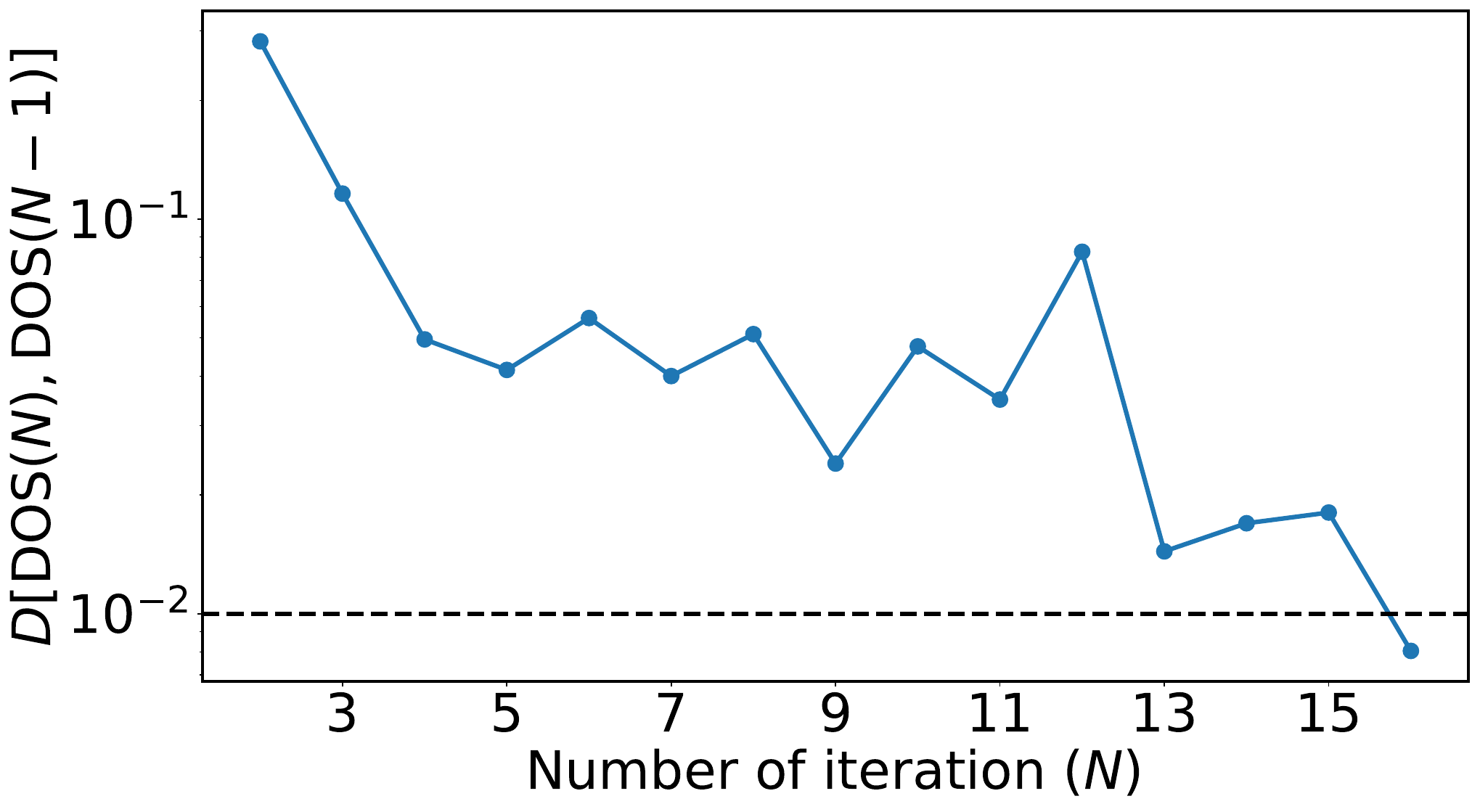}} \\ (a)
	\end{minipage}
    \begin{minipage}[h]{0.49\linewidth}
		\center{\includegraphics[trim={0cm 0cm 0cm 0cm},clip, width=1\linewidth]{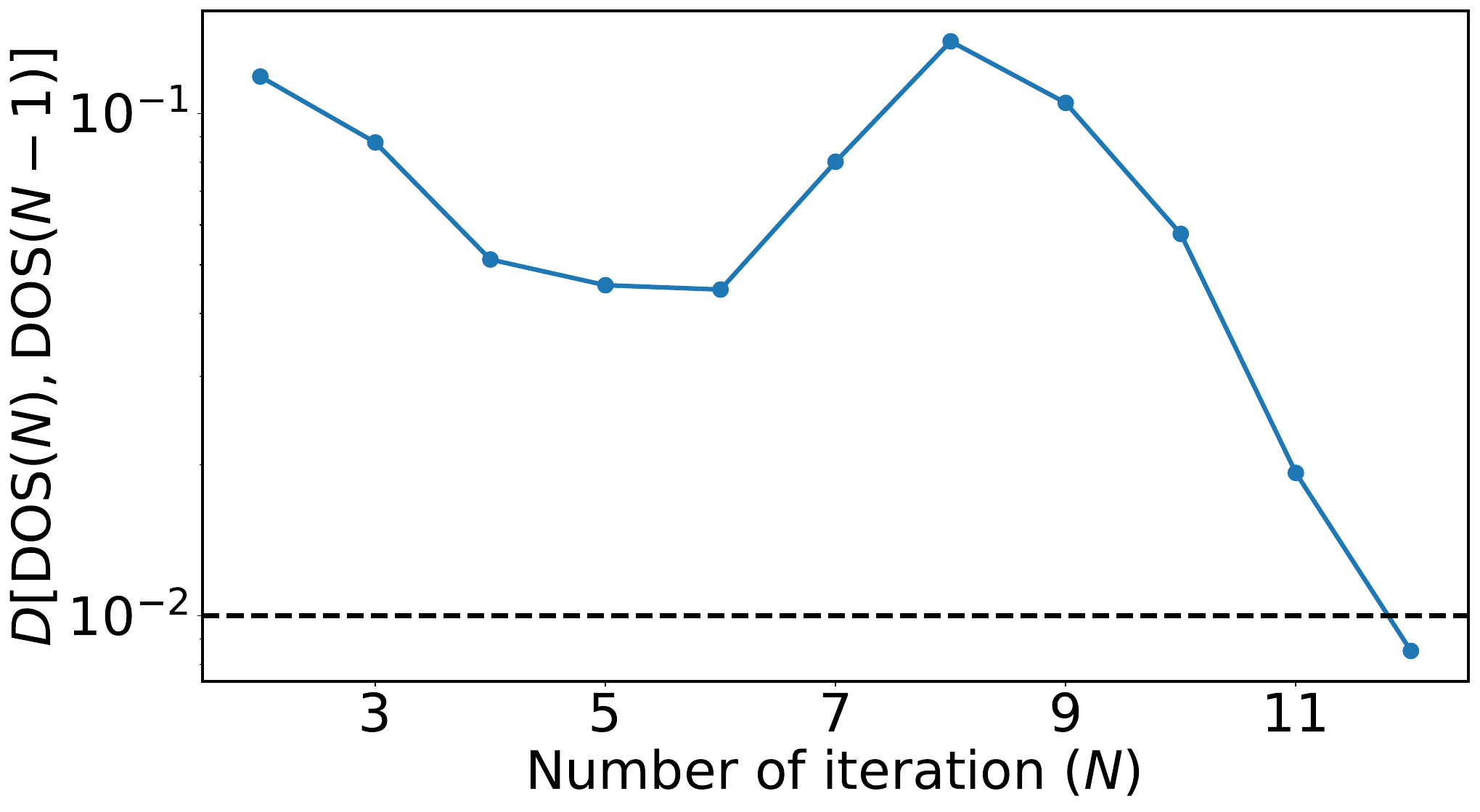}} \\ (b)
	\end{minipage}

    \caption{Cosine distance $D[{\rm DOS}(N),{\rm DOS}(N-1)]$ between the phonon DOS of consecutive quantum sTDEP iterations for Ba$_3$Lu(BO$_3$)$_3$ at (a) 0~K, (b) 450~K. Dashed lines mark the convergence threshold $D=10^{-2}$.}
    \label{fig:conv_stdep_balu}
\end{figure*}

The band gap at a given temperature was obtained by averaging the PBE gaps of configurations generated by quantum sampling with the converged force constants.

\begin{figure}[t]

	\centering
    \begin{minipage}[h]{0.49\linewidth}
		\center{\includegraphics[trim={0cm 0cm 0cm 0cm},clip, width=1\linewidth]{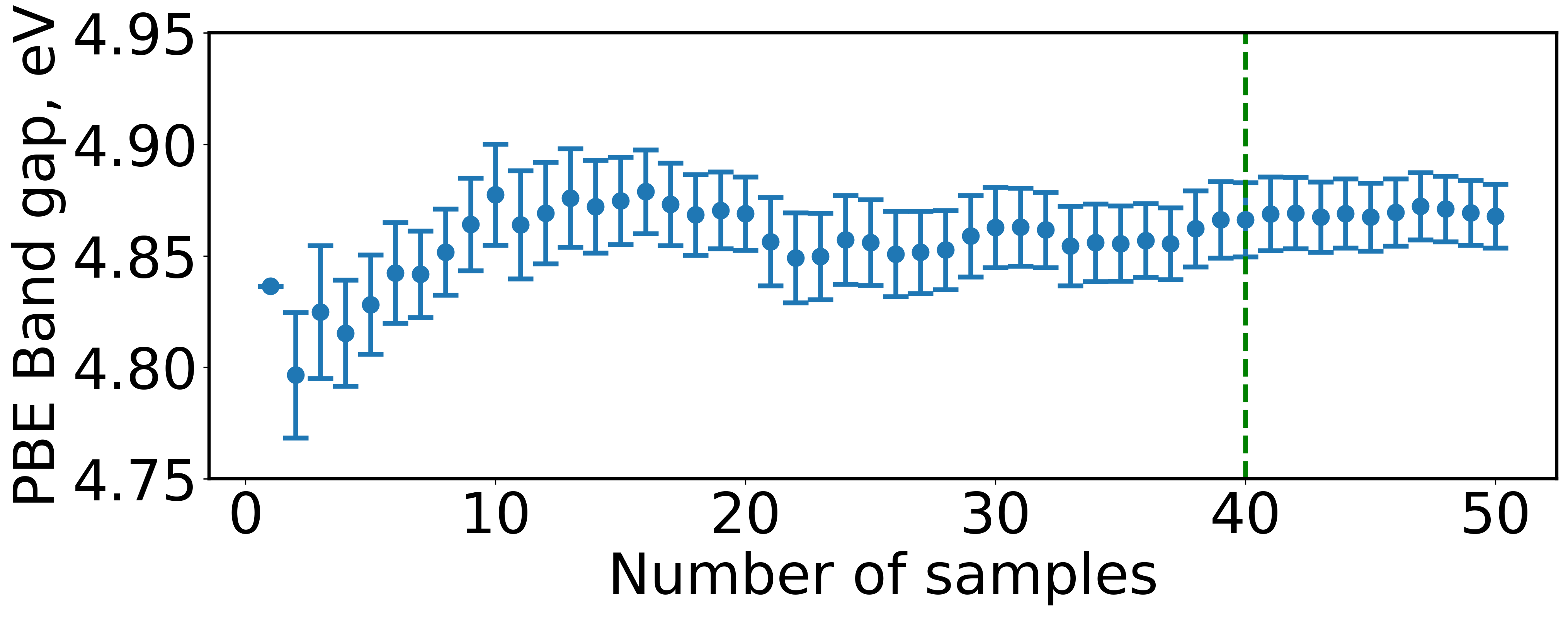}} \\ (a)
	\end{minipage}
	\begin{minipage}[h]{0.49\linewidth}
		\center{\includegraphics[trim={0cm 0cm 0cm 0cm},clip, width=1\linewidth]{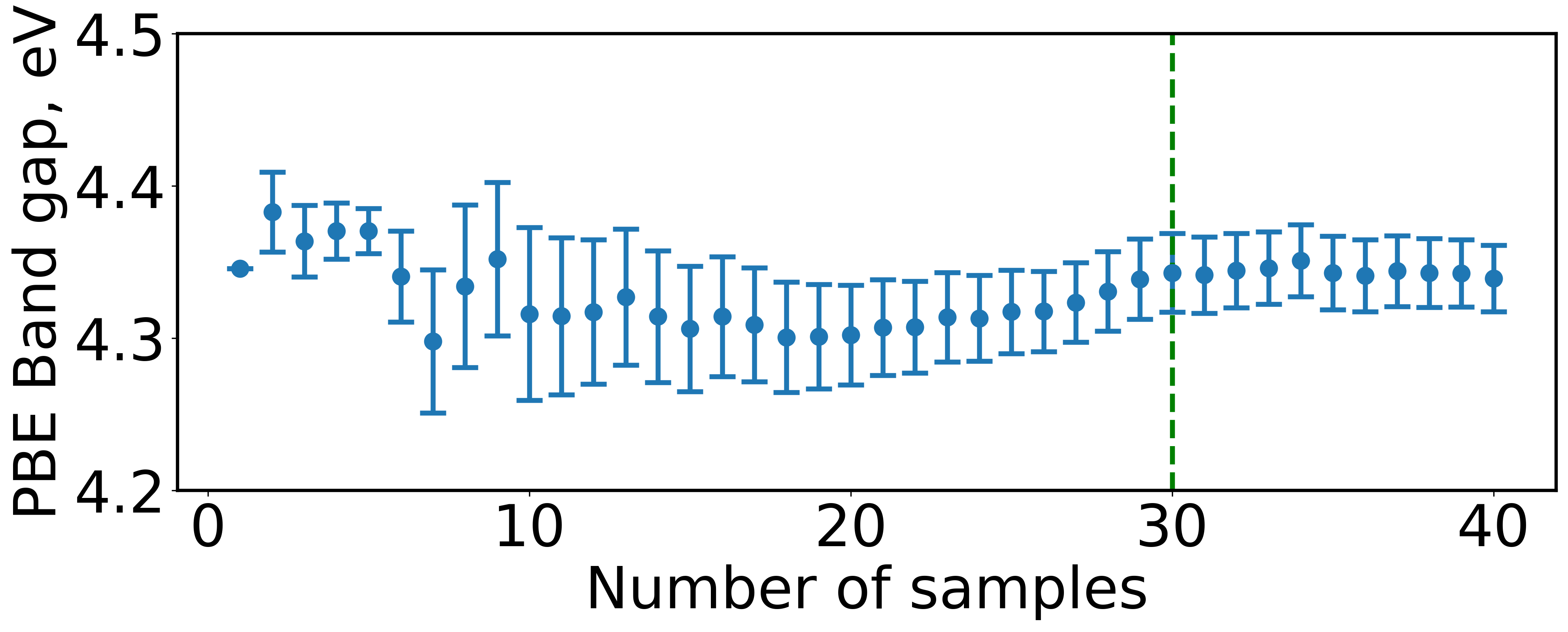}} \\ (b)
	\end{minipage}
    \begin{minipage}[h]{0.49\linewidth}
		\center{\includegraphics[trim={0cm 0cm 0cm 0cm},clip, width=1\linewidth]{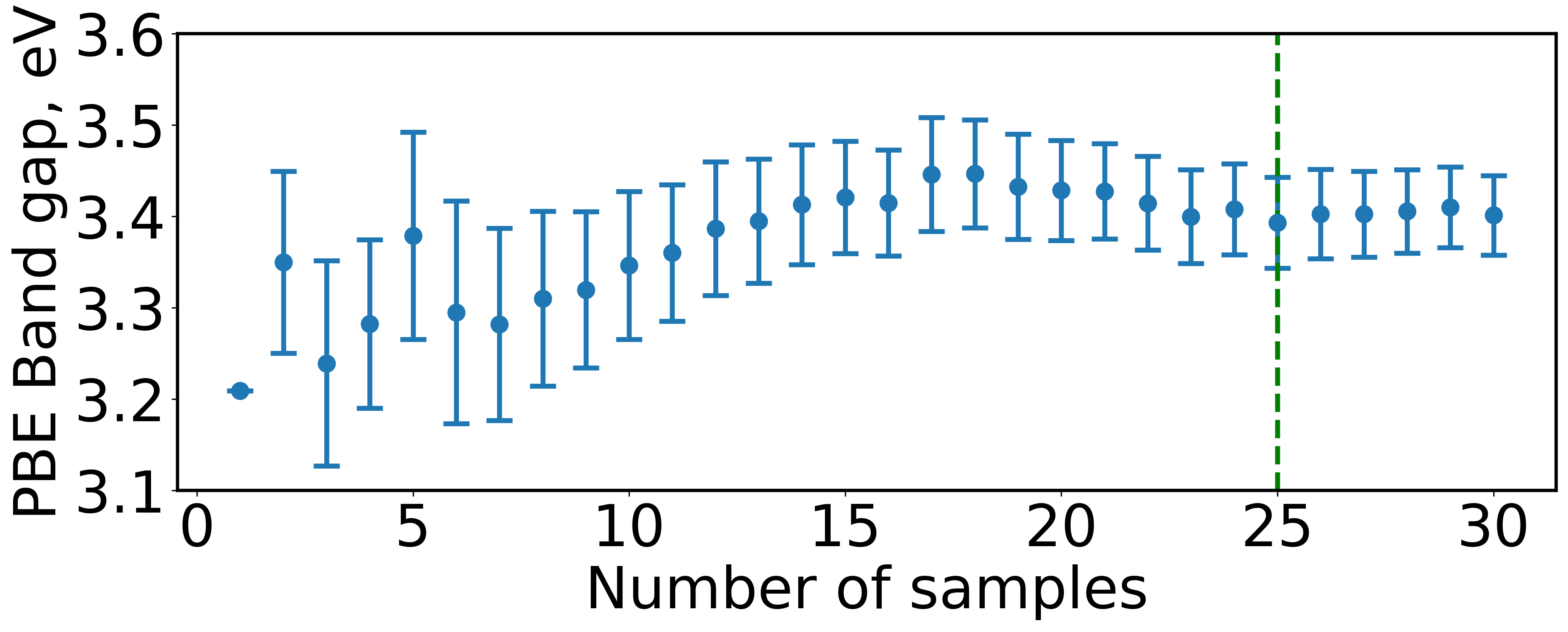}} \\ (c)
	\end{minipage}

\caption{Convergence of the average PBE band gap of Rb$_2$Ca$_2$(SO$_4$)$_3$ with the number of configurations generated by quantum sTDEP sampling at (a) 0, (b) 450, and (c) 1050~K. Dashed green lines mark the number of configurations at which convergence is reached.}
\label{fig:conv_q_rbca}
\end{figure}

\begin{figure}[t]
\centering
	\begin{minipage}[h]{0.49\linewidth}
		\center{\includegraphics[trim={0cm 0cm 0cm 0cm},clip, width=1\linewidth]{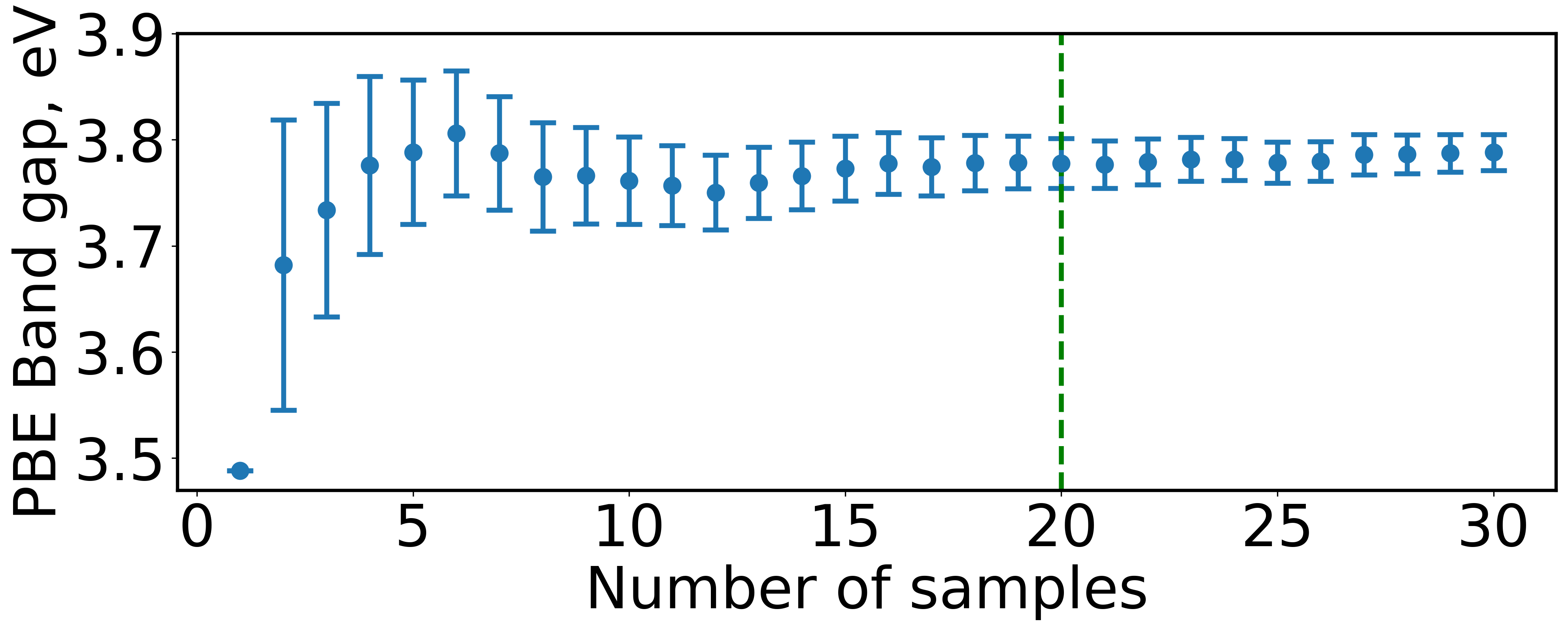}} \\ (a)
	\end{minipage}
    \begin{minipage}[h]{0.49\linewidth}
		\center{\includegraphics[trim={0cm 0cm 0cm 0cm},clip, width=1\linewidth]{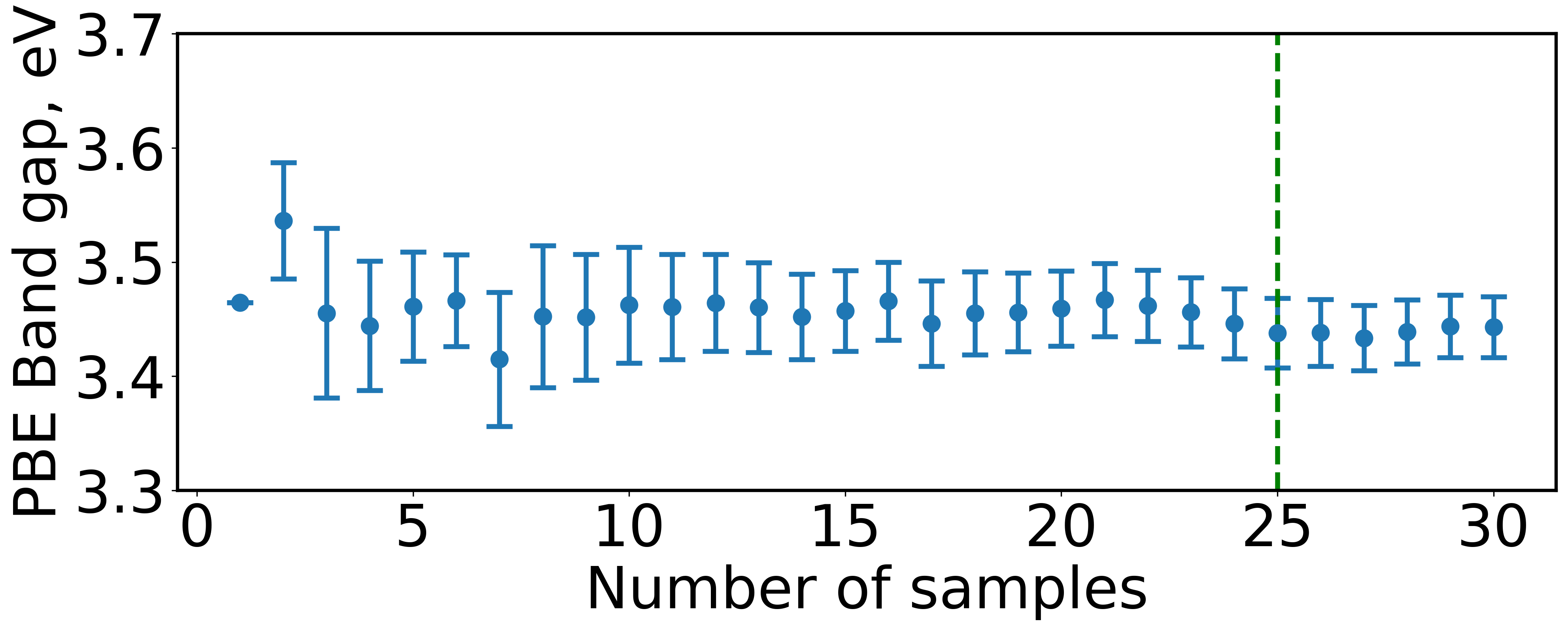}} \\ (b)
	\end{minipage}
    \begin{minipage}[h]{0.49\linewidth}
		\center{\includegraphics[trim={0cm 0cm 0cm 0cm},clip, width=1\linewidth]{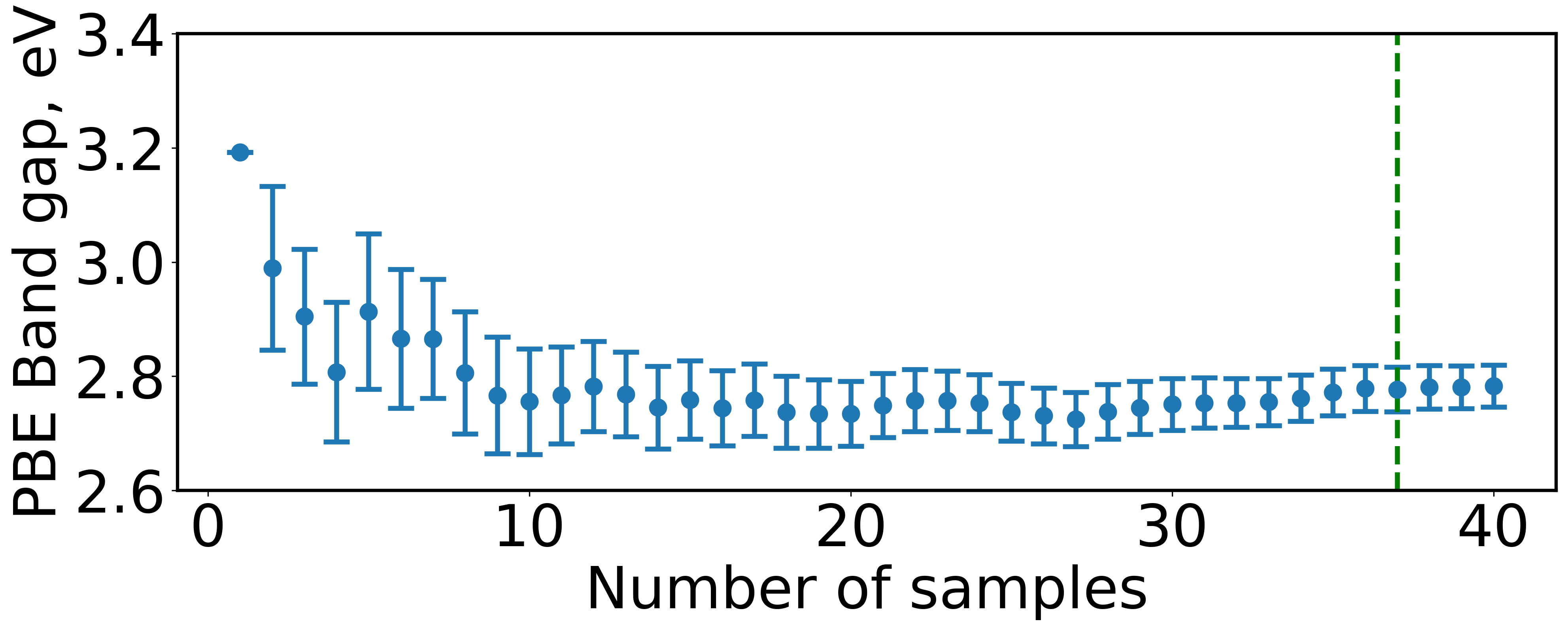}} \\ (c)
	\end{minipage}
\caption{Convergence of the average PBE band gap of Ba$_3$Lu(BO$_3$)$_3$ with the number of configurations generated by quantum sTDEP sampling at (a) 0, (b) 450, and (c) 1050~K. Dashed green lines mark the number of configurations at which convergence is reached.}
\label{fig:conv_q_balu}
\end{figure}

\begin{figure*}[h!]
	\centering
	\begin{minipage}[h]{0.49\linewidth}
		\center{\includegraphics[trim={0cm 0cm 0cm 0cm},clip, width=1\linewidth]{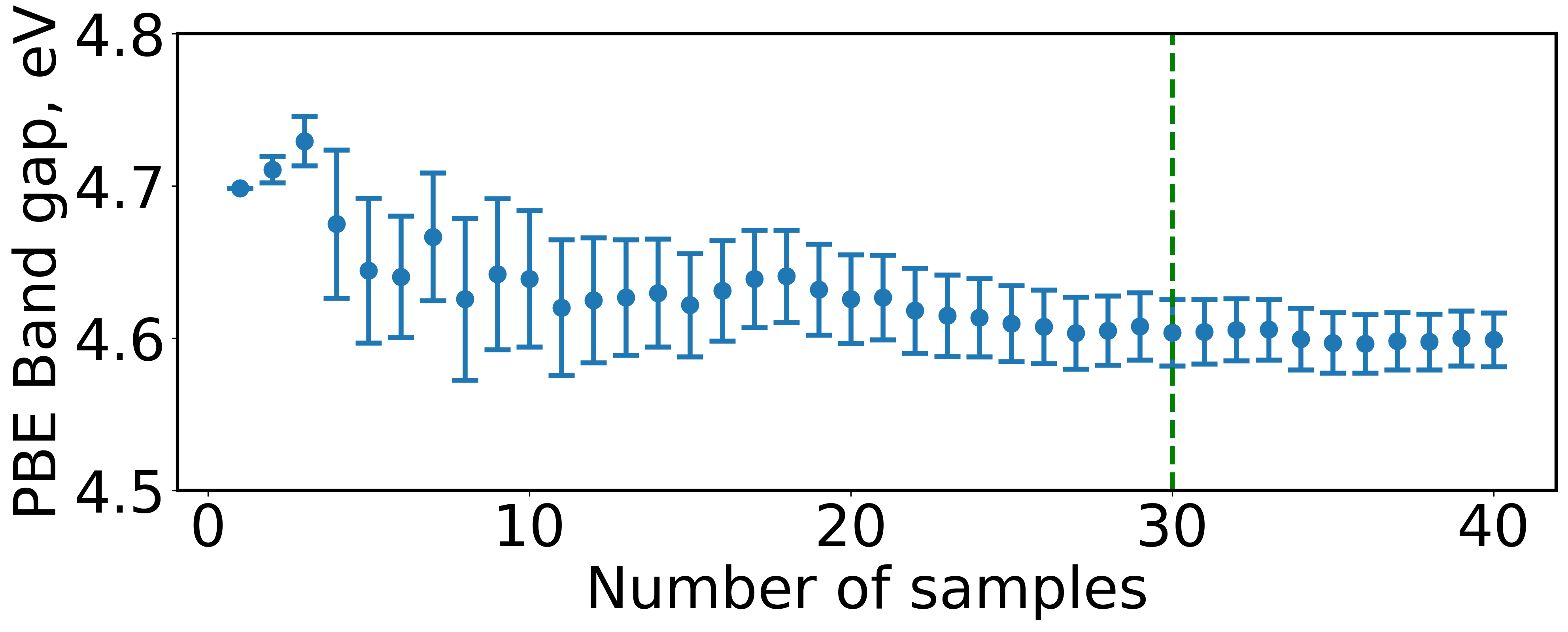}} \\ (a)
	\end{minipage}
    \begin{minipage}[h]{0.49\linewidth}
		\center{\includegraphics[trim={0cm 0cm 0cm 0cm},clip, width=1\linewidth]{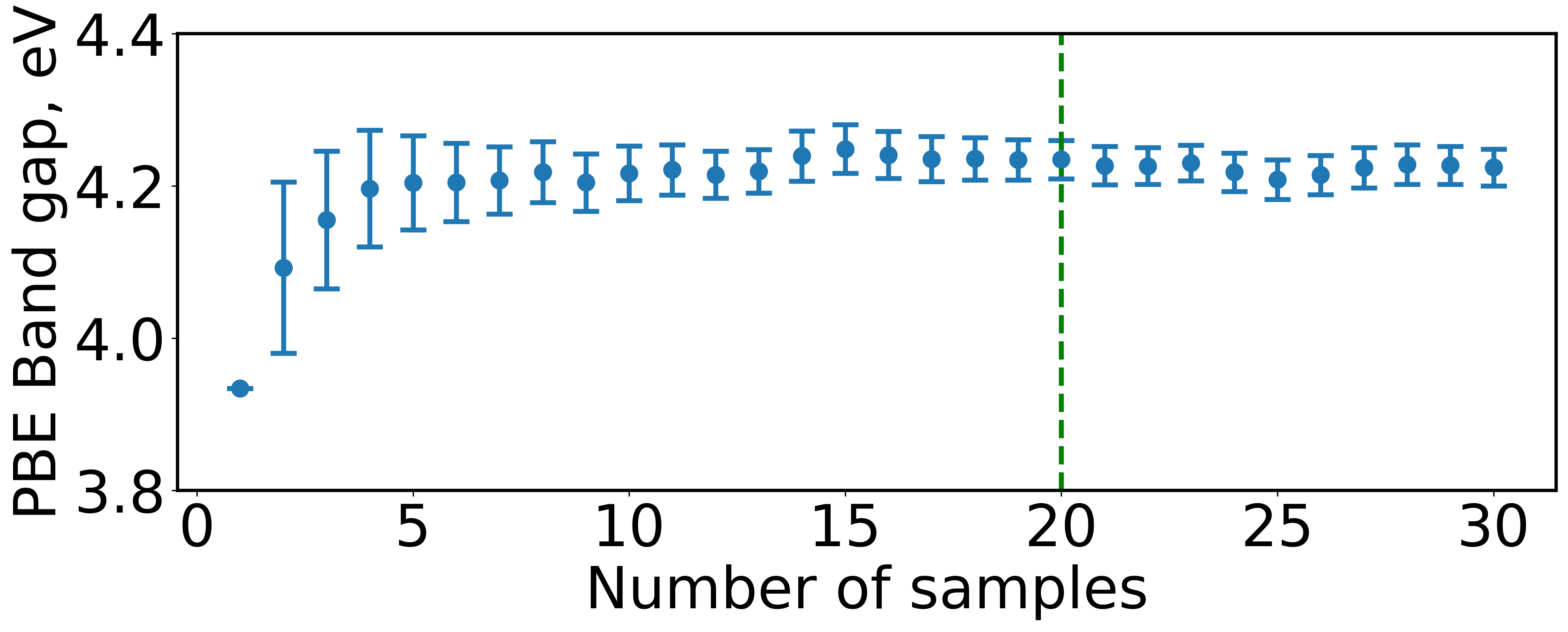}} \\ (b)
	\end{minipage}
    \begin{minipage}[h]{0.49\linewidth}
		\center{\includegraphics[trim={0cm 0cm 0cm 0cm},clip, width=1\linewidth]{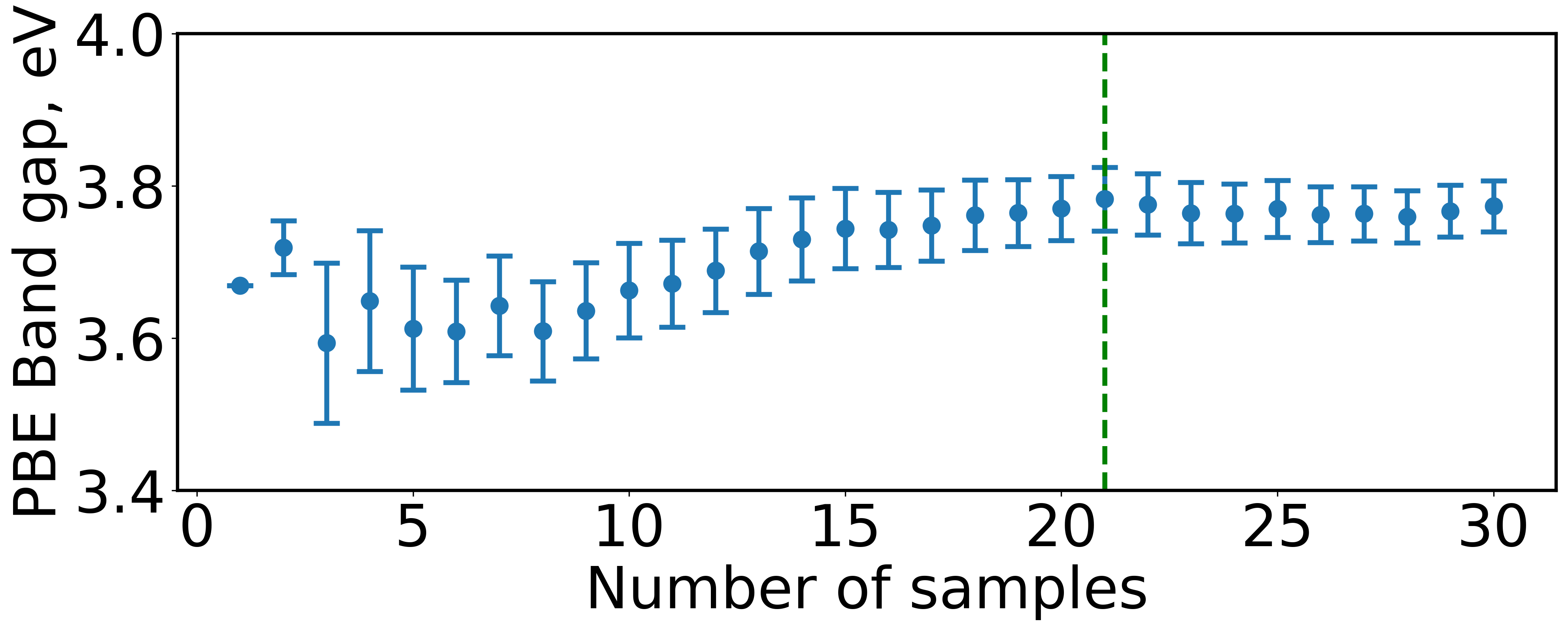}} \\ (c)
	\end{minipage}
        \begin{minipage}[h]{0.49\linewidth}
		\center{\includegraphics[trim={0cm 0cm 0cm 0cm},clip, width=1\linewidth]{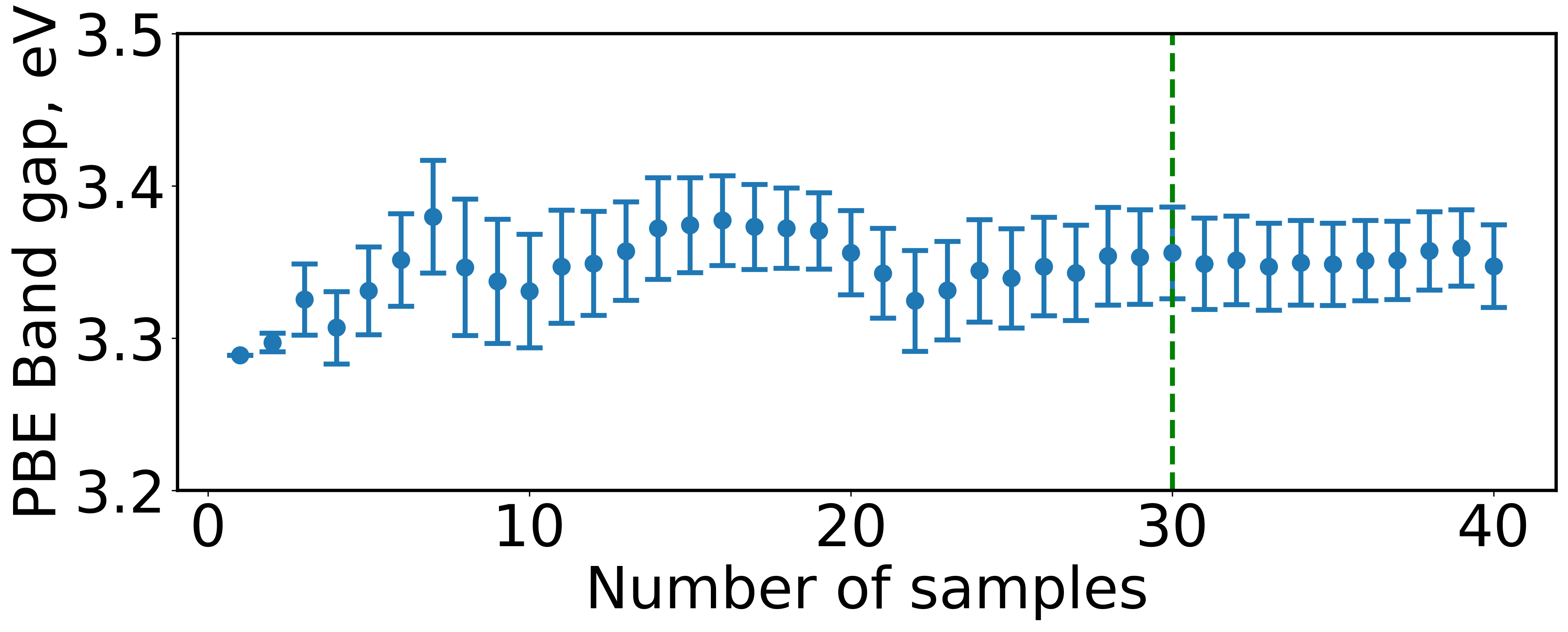}} \\ (d)
	\end{minipage}
   
    \caption{Convergence of the average PBE band gap of $\beta$-Rb$_2$SO$_4$ with the number of configurations generated by quantum sTDEP sampling at (a) 0, (b) 250, (c) 450, and (d) 650~K. Dashed green lines mark the number of configurations at which convergence is reached.}
    \label{fig:conv_q_brso}
\end{figure*}

Quantum sampling was performed at 0, 250, 450, and 650~K for $\beta$-Rb$_2$SO$_4$ and at 0, 450, and 1050~K for Rb$_2$Ca$_2$(SO$_4$)$_3$ and Ba$_3$Lu(BO$_3$)$_3$. For the 0~K calculations, the initial MD of the sTDEP loop was run at 250~K. The convergence of the average band gap with the number of configurations is shown in Figs.~\ref{fig:conv_q_rbca}--\ref{fig:conv_q_brso}.

\clearpage

\subsection{Varshni-type expression}

Varshni's expression~\cite{varshni} for the temperature dependence of a band gap reads
\begin{equation}\label{eq:varsh}
	E^q_g(T) = E^q_g(0) - \frac{aT^2}{T+b},
\end{equation}
where $E^q_g(0)$ is the band gap at $T=0$~K including zero-point motion. In this work, the BGR and the zero-point renormalization (ZPR) are defined as
\begin{equation}
	{\rm BGR^{q}}(T)= E^q_g(T) - E^{cl}_g(0),
	\qquad
	{\rm ZPR} = E^q_g(0) - E^{cl}_g(0),
\end{equation}
where $E^{cl}_g(0)$ is the band gap of the static structure. Subtracting $E^{cl}_g(0)$ from both sides of Eq.~\eqref{eq:varsh} gives
\begin{equation}
	{\rm BGR^{q}}(T) = {\rm ZPR} - \frac{aT^2}{T+b}.
\end{equation}
This expression, with $a$ and $b$ as adjustable parameters, was used to draw the curves through the ${\rm BGR^{q}}(T)$ points in the main text. For Rb$_2$Ca$_2$(SO$_4$)$_3$ and Ba$_3$Lu(BO$_3$)$_3$, which have three points each, it serves as an interpolation rather than a fit.

\clearpage
\section{Electronic band structures}

\begin{figure*}[h!]
	\centering
	\begin{minipage}[h]{0.49\linewidth}
		\center{\includegraphics[trim={0cm 0cm 0cm 0cm},clip, width=1\linewidth]{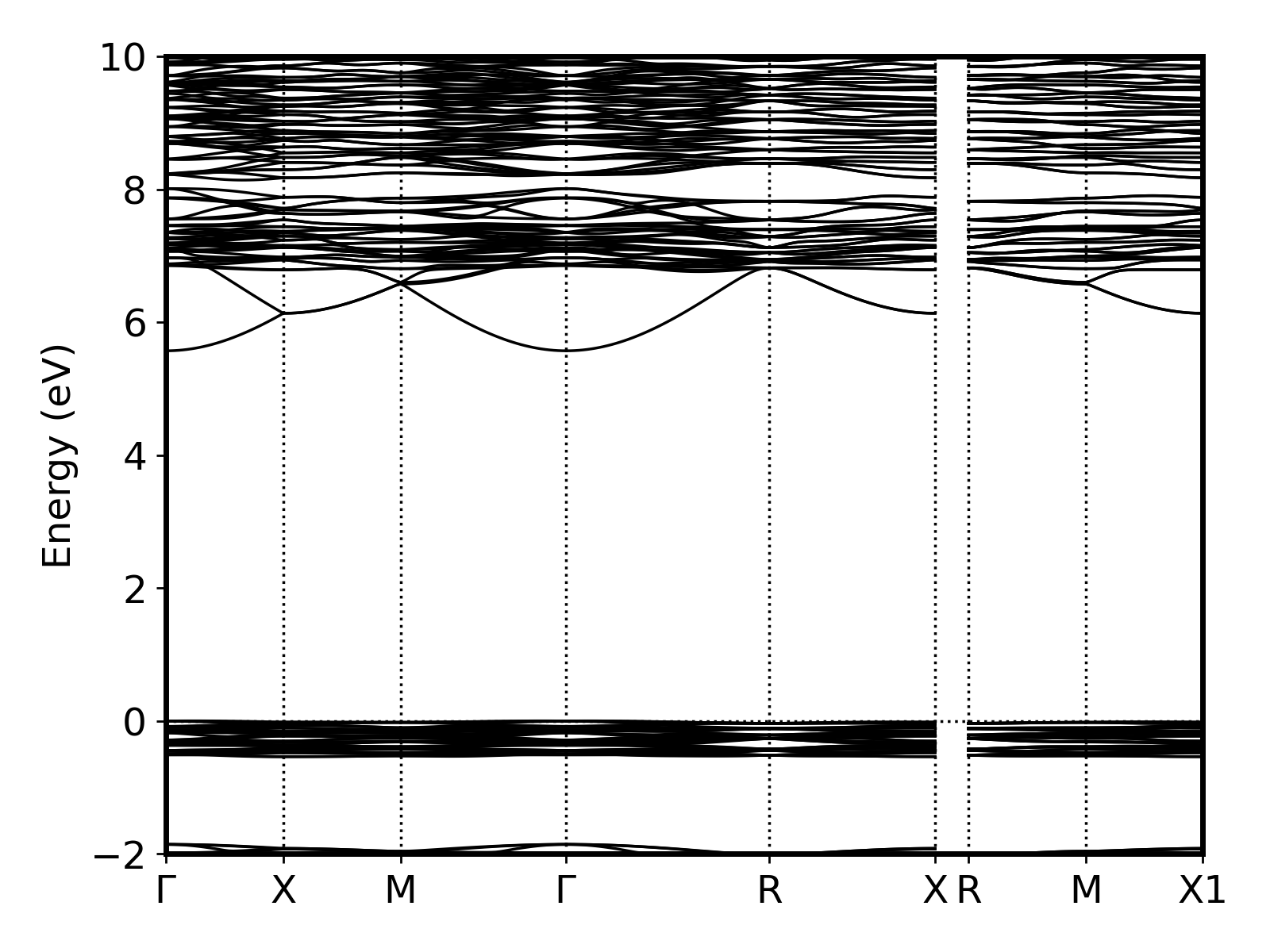}} \\ (a)
	\end{minipage}
    \begin{minipage}[h]{0.49\linewidth}
		\center{\includegraphics[trim={0cm 0cm 0cm 0cm},clip, width=1\linewidth]{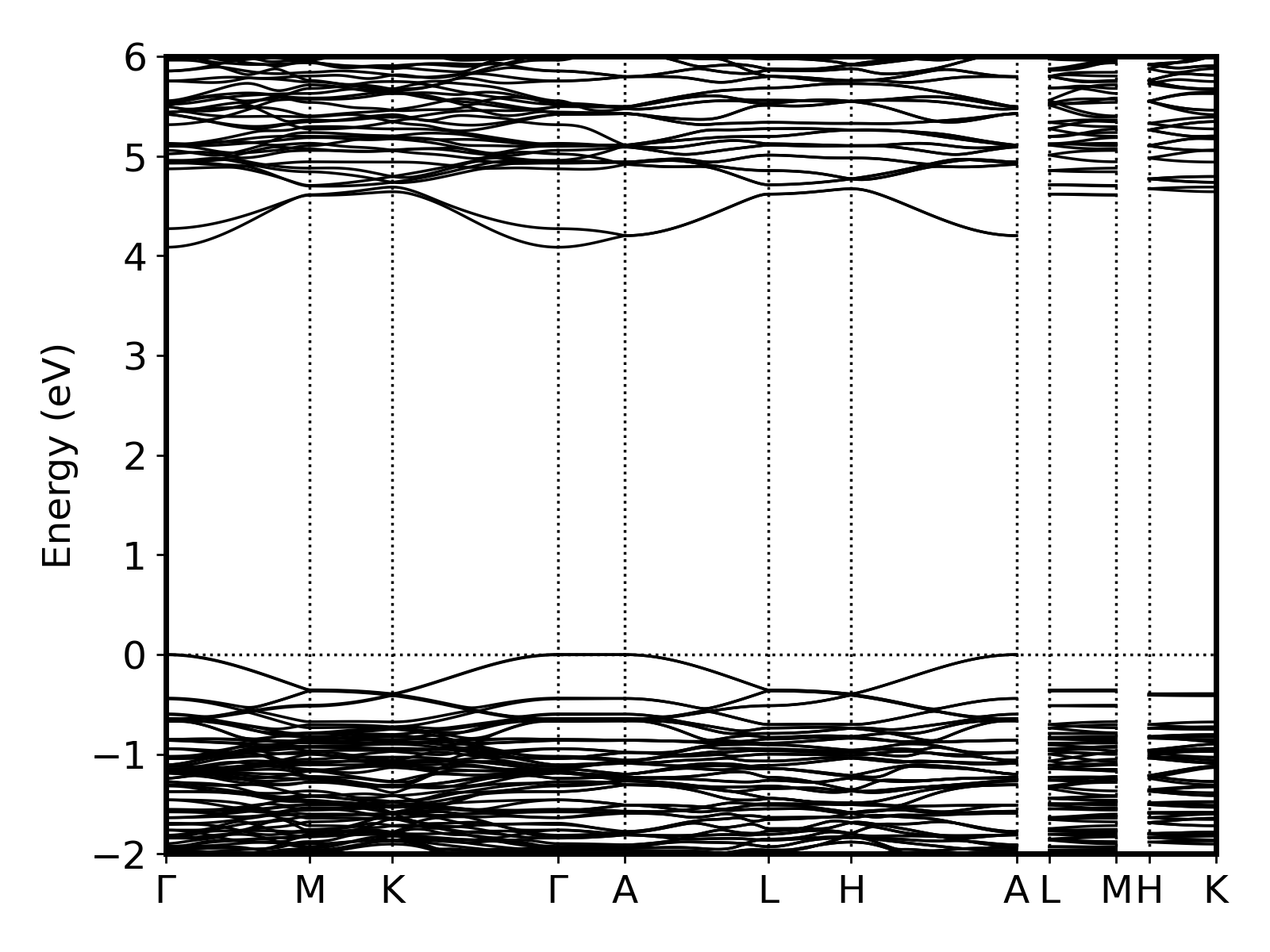}} \\ (b)
	\end{minipage}

    \begin{minipage}[h]{0.49\linewidth}
		\center{\includegraphics[trim={0cm 0cm 0cm 0cm},clip, width=1\linewidth]{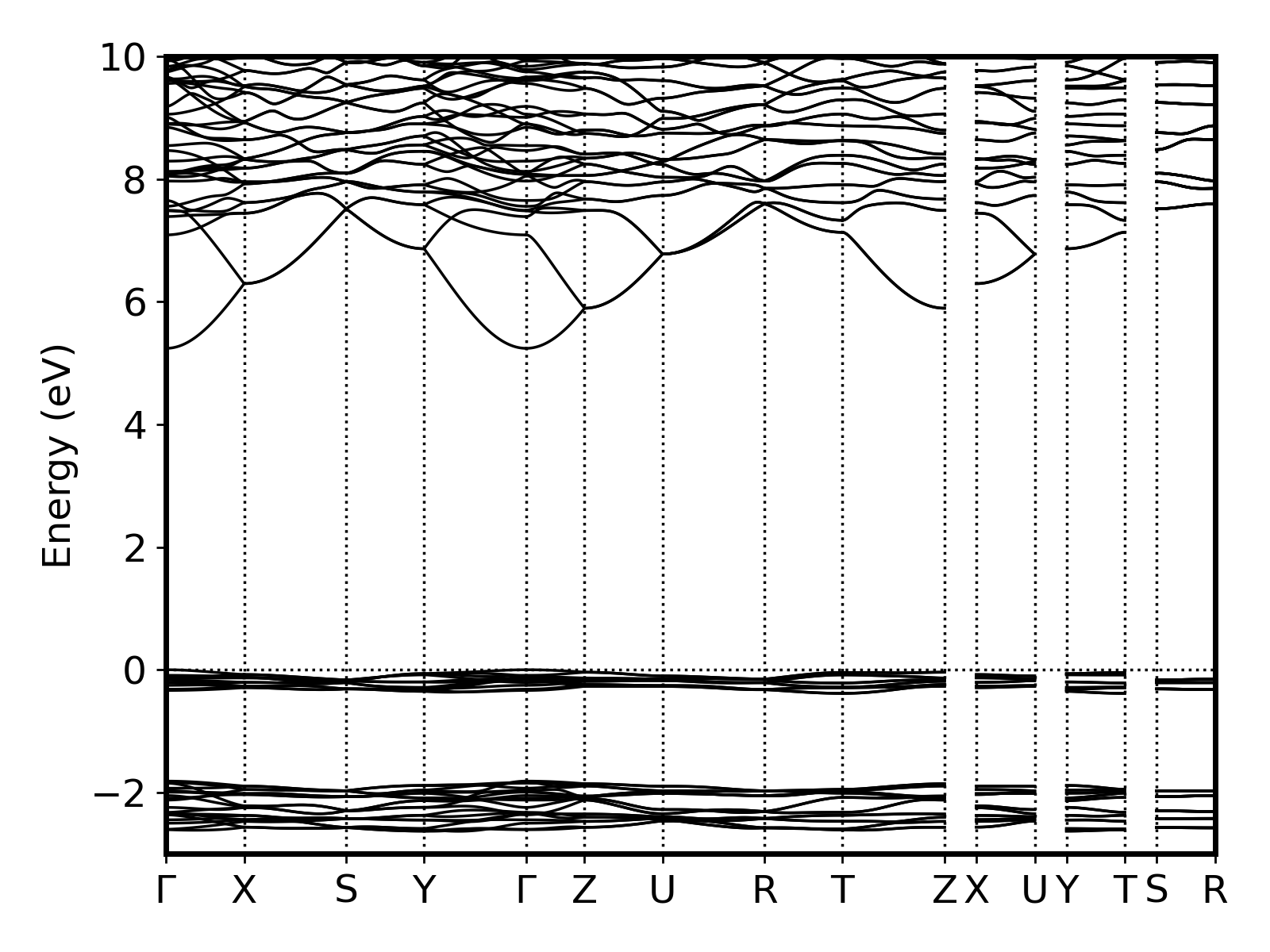}} \\ (c)
	\end{minipage}
    \caption{PBE electronic band structures of (a) Rb$_2$Ca$_2$(SO$_4$)$_3$, (b) Ba$_3$Lu(BO$_3$)$_3$, and (c) $\beta$-Rb$_2$SO$_4$.}
    \label{fig:bandstructures}
\end{figure*}

Figure~\ref{fig:bandstructures} shows the PBE band structures of the static structures. The PBE band gaps are 5.57, 4.08, and 5.24~eV for Rb$_2$Ca$_2$(SO$_4$)$_3$, Ba$_3$Lu(BO$_3$)$_3$, and $\beta$-Rb$_2$SO$_4$, respectively.

\clearpage

\bibliographystyle{unsrt}
\bibliography{main}